%% file: draft.tex
\documentclass[a4paper,11pt]{article}
\pdfoutput=1

\usepackage{jheppub}

\usepackage{subfigure}

\usepackage[T1]{fontenc}
\usepackage[utf8]{inputenc}
\usepackage{lmodern}
\usepackage{comment}
\usepackage{multirow}
\usepackage{multicol}
\usepackage{hhline}
\usepackage[mathscr]{eucal}
\usepackage[normalem]{ulem}

\usepackage{bm,amsmath,amssymb,tensor,mathtools}
\numberwithin{equation}{section}

\usepackage{tikz}
\usetikzlibrary{cd}
\usetikzlibrary{decorations.pathmorphing}
\usetikzlibrary{arrows, arrows.meta}
\usetikzlibrary{shapes.geometric}
\usetikzlibrary{decorations.markings}
\tikzset{>={Stealth[width=3mm,length=3mm]}}

\input{custom.tex}

\def\Res{\qopname\relax m{Res}}

\DeclareMathAlphabet{\mathpzc}{OT1}{pzc}{m}{it}

\title{Residue sums for superconformal indices}

\author[a,b,c]{Sam van Leuven,}
\author[a]{Kayleigh Mathieson,}
\author[a]{Pratik Roy}

\affiliation[a]{Mandelstam Institute for Theoretical Physics, School of Physics, University of the Witwatersrand, Johannesburg 2050, South Africa
}
\affiliation[b]{DSI-NRF Centre of Excellence in Mathematical and Statistical Sciences (CoE-MaSS),  South Africa}
\affiliation[c]{National Institute for Theoretical and Computational Sciences (NITheCS), Gauteng, South Africa}

\emailAdd{sam.vanleuven@wits.ac.za}
\emailAdd{kayleigh.mathieson.wits@gmail.com}
\emailAdd{roy.pratik92@gmail.com}

\abstract{
We study superconformal indices of four-dimensional $SU(N)$ gauge theories with $\mathcal{N}=1,2,4$ supersymmetry.
The usual representation of a gauge theory index involves multiple contour integrals and reflects the BPS spectrum at zero Yang--Mills coupling.
To find an alternative, closed form expression, it is natural to attempt an evaluation of the integrals through residues.
However, the presence of non--isolated essential singularities prevents a straightforward evaluation.
We show how this difficulty can be resolved by fixing the residual Weyl symmetry of the integral.
This allows us to evaluate the residue sums for superconformal indices of $SU(2)$ gauge theories in terms of basic and elliptic hypergeometric series.
For the Macdonald index of the $\mathcal{N}=4$ $SU(2)$ super Yang--Mills theory, we show how known transformation formulas for basic hypergeometric series can be used to simplify the residue sum.
We observe that the simplified form encodes features of the BPS spectrum at non--zero coupling and suggests the absence of fortuitous or non-graviton operators in the Macdonald sector.
Furthermore, we evaluate the residue sums for the Macdonald and full superconformal indices of a general class of $SU(2)$ gauge theories. 
In the process, we find various applications to the theory of basic and elliptic hypergeometric integrals, including a convergent residue sum for Spiridonov's elliptic beta integral.
Finally, we discuss the generalization of our method to higher rank gauge groups and evaluate the $\mathcal{N}=4$ $SU(3)$ Macdonald index in closed form.
}

\begin{document} 
\maketitle
\flushbottom

\section{Introduction}\label{sec:intro}

The superconformal index represents one of a few exactly calculable quantities in superconformal field theories (SCFTs).
It was first defined for four-dimensional $\mathcal{N}=1$ SCFTs and takes the form of a trace over the spectrum of the SCFT on $S^3$ \cite{Romelsberger:2005eg,Kinney:2005ej}.
Similar to the Witten index, this trace is designed such that it receives contributions only from short representations of the superconformal algebra \cite{Dolan:2002zh}.
As a result, the index is protected: it does not change under continuous deformations of the theory, as long as the deformation preserves the supercharge with respect to which the multiplets are short.
This includes exactly marginal deformations, like the gauge coupling of the $\mathcal{N}=4$ super--Yang--Mills (SYM) theory.
The superconformal index is also invariant under renormalization group flows in the sense that one can define an index for a non-conformal UV theory, which equals the superconformal index of the SCFT that emerges as its IR fixed point \cite{Romelsberger:2007ec,Dolan:2008qi,Festuccia:2011ws}.

These features of the index have led to remarkable checks of and predictions for strong-weak coupling dualities of supersymmetric gauge theories \cite{Romelsberger:2005eg,Romelsberger:2007ec,Dolan:2008qi,Spiridonov:2008zr,Spiridonov:2009za,Spiridonov:2010hh,Spiridonov:2011hf}.
The most famous example is the Seiberg, or IR, duality between an $\mathcal{N}=1$ $SU(2)$ gauge theory with $N_f=3$ and a theory of free mesons and (dual) quarks \cite{Seiberg:1994pq}.
The agreement between the superconformal indices follows from a highly non-trivial mathematical identity known as Spiridonov's elliptic beta integral \cite{Spiridonov_2001}.
This identity is given by
\begin{equation}\label{eq:intro-spir-ell-int}
    \frac{(p;p)_\I(q;q)_\I}{2}\oint\frac{du}{2\pi i u}\frac{\prod_{i=1}^{6}\Gamma(y_i u^{\pm};p,q)}{\Gamma(u^{\pm 2};p,q)}=\prod_{1\leq i<j\leq 6}\Gamma(y_iy_j;p,q)\,,
\end{equation}
where the left hand side is identified with the gauge theory index and the right hand side with the superconformal index of the (free) IR SCFT.
The ingredients of this formula will be reviewed in Section \ref{sec:n=2-sci}.

The AdS/CFT correspondence provides another example of a strong-weak coupling duality, although in this case the gravitational side of the duality is not as well understood as the CFT side.
It is in fact in this context that \cite{Kinney:2005ej} originally defined the superconformal index.
In particular, they derived a finite-dimensional integral expression for the exact $\mathcal{N}=4$ $SU(N)$ SYM index and evaluated this integral through a saddle point approximation at large $N$.
The latter agrees precisely with a supergravity index, which counts the Kaluza-Klein towers of (short) multiplets on $AdS_5\times S^5$.
Apart from demonstrating that the superconformal index indeed captures exact information about the strongly coupled spectrum of the $\mathcal{N}=4$ theory, this result also raised a puzzle: on the AdS side, there exist large 1/16 BPS Kerr-Newman black holes with $\mathcal{O}(N^2)$ entropy \cite{Gutowski:2004ez,Gutowski:2004yv,Cvetic:2004ny,Chong:2005da,Chong:2005hr,Kunduri:2006ek}, and one would expect the corresponding CFT states to provide the dominant contribution to the index at large $N$.
So why does it seem that the index does not capture such states?
The authors of \cite{Kinney:2005ej} suggested essentially two possible resolutions.
Firstly, the index only captures differences between bosonic and fermionic degeneracies, so that large cancellations may occur. 
And secondly, when viewed as a path integral, the index imposes periodic boundary conditions on fermions along the temporal circle, which is incompatible with a (smooth) Euclidean black hole geometry.
As such, black holes may simply not contribute to an index at all.
Later on, though, building on \cite{Grant:2008sk}, this puzzle was sharpened when attempts were made to construct the actual BPS operators, or rather their ``$\mathcal{Q}$-cohomology classes'',  corresponding to supersymmetric black holes \cite{Chang:2013fba}.
Indeed, no examples of such classes were found, other than those that could be matched with the supergravity multiplets.

More recently, there has been significant progress in the resolution of this puzzle.
Inspired by observations of \cite{Hosseini:2017mds}, this started with a demonstration of $\mathcal{O}(e^{N^2})$ growth of the superconformal index, after all, and an exact reproduction of the Bekenstein-Hawking entropy of the supersymmetric $AdS_5$ black holes \cite{Cabo-Bizet:2018ehj,Choi:2018hmj,Benini:2018ywd}.
The key insight underlying these works is that the index, as a function of the chemical potentials, has branch cuts and should really be thought of as defined on a multi-sheeted cover of the space of chemical potentials.
While the $\mathcal{O}(N^2)$ entropy is not captured on one sheet, consistent with the analysis of \cite{Kinney:2005ej}, it is captured on a ``second sheet'' \cite{Cassani:2021fyv}.

This development motivated a renewed search for the corresponding BPS operators.
By explicitly evaluating the finite $N$ index to large orders and comparing with an expansion of the ``graviton index'', one can obtain clues on where to look for ``non-graviton'' operators.
Discrepancies between the full and graviton index were first observed in \cite{Murthy:2020scj,Agarwal:2020zwm}.
By now, this has indeed led to the construction of the first examples of ($\mathcal{Q}$-cohomology classes of) non-graviton operators in the $\mathcal{N}=4$ theory at finite $N$ \cite{Chang:2022mjp,Choi:2022caq,Choi:2023znd,Choi:2023vdm,Chang:2024zqi}.
As explained in \cite{Chang:2024zqi}, there appears to be a fortuitous nature to non-graviton operators.
That is, in the $SU(N)$ theory these operators turn out to be BPS only due to trace relations.
As such, they fail to remain BPS for larger values of $N$.
This is in contrast to ``graviton'' operators,\footnote{Graviton operators at finite $N$ refer to operators in the superconformal multiplets of chiral primary operators of the $\mathcal{N}=4$ theory, see, \textit{e.g.}, \cite{Witten:1998qj,Kinney:2005ej}.} which remain BPS for arbitrary $N$ and for this reason are referred to as monotonous.

At present, most of the progress in constructing the non-graviton operators is on a case-by-case basis.
Indeed, by looking at the difference between the full and graviton index up to a finite order in the expansion, it is difficult to determine more structural features of the non-graviton spectrum.
Ideally, one would like to write a closed form formula for the difference between the full and graviton index.
As far as we are aware, this has only been achieved for the so-called BMN subsector of the $\mathcal{N}=4$ $SU(2)$ theory \cite{Choi:2023znd,Choi:2023vdm,Gadde:2025yoa}.

With this motivation in mind, in this paper we focus on developing closed form formulas for the full superconformal index.
We are particularly inspired by Spiridonov's elliptic beta integral \eqref{eq:intro-spir-ell-int}.
This formula allows one to express a highly redundant formula, reflecting the BPS spectrum of the UV-free SQCD description, in a more economical manner, manifesting only the strongly coupled BPS spectrum in the IR (in this case, corresponding to the index of a Seiberg--dual free theory). 
One can ask if similar formulas could exist for the $\mathcal{N}=4$ theory.
A priori, there is an obvious distinction: the coupling of the $\mathcal{N}=4$ theory does not flow and the theory is non-free for all but the ($S$--dual) images of $g=0$.
However, as one tunes the Yang-Mills coupling, say from $g=0$ to $g\neq 0$, operators which were BPS saturated in the free theory can develop anomalous dimensions and therefore lift from the BPS spectrum \cite{Kinney:2005ej,Grant:2008sk}.
One could still ask whether there exists a formula for the index which (more) faithfully represents the BPS spectrum at non-zero coupling.
In more detail, for the $\mathcal{N}=4$ $SU(N)$ theory, the analogue of the left hand side of \eqref{eq:intro-spir-ell-int} is of the form:
\begin{equation}\label{eq:intro-weakly-coupled-int}
    \mathcal{I}_N=\int d\mu_{SU(N)} \,\text{tr}_{\mathcal{H}^{\text{ext}}_{g=0}} (\cdots)\,,
\end{equation}
where $\mathcal{H}^{\text{ext}}_{g=0}$ is the extended space of 1/16 BPS states in the free $\mathcal{N}=4$ theory, containing all matrix degrees of freedom, and the integral over $SU(N)$, with the invariant Haar measure, projects onto gauge singlets.
We are instead after another, more minimal expression for the index which manifests (aspects of) the BPS spectrum at non-zero coupling.
That is, we are after an expression of the schematic form\footnote{See also \cite{Eager:2018czl}, which addresses a similar question in the context of $\mathcal{N}=1$ SQCD.}
\begin{equation}\label{eq:strong-coupling-index}
    \mathcal{I}_N=\text{tr}_{\mathcal{H}^{}_{g\neq 0}} (\cdots)\,.
\end{equation}
Perhaps the most direct approach to develop such an expression is to convert the integral over $SU(N)$ as an integral over eigenvalues, which can be viewed as a (partial) gauge fixing of the $SU(N)$ symmetry of the integrand.
This reduces the integral to $N-1$ contour integrals, each contour being a unit circle, with a non-trivial measure given by Vandermonde determinants.
Given that the integrand is a meromorphic function of the (complexified) eigenvalues, one can then in principle attempt to evaluate these integrals through residues.
It is of course not guaranteed that the resulting residue sum immediately takes on the form suggested above. 
But, as we will demonstrate, it can serve as a useful intermediate step.

However, there turns out to be a key technical difficulty with a naive implementation.
Namely, whenever the integrand in \eqref{eq:intro-weakly-coupled-int} receives contributions from conformal descendants of bosonic matrices, as will be the case for the full 1/16 BPS index, the integrand has a non-isolated essential singularity at the origin in the space of complexified $SU(N)$ eigenvalues.\footnote{The BMN subsector referred to above does not contain conformal descendants. The evaluation of its index is therefore not obstructed by this difficulty.}
For this type of singularity, Cauchy's theorem does not apply and, effectively, one cannot make sense of the residue sum.
This difficulty was noted before in the context of the Schur index of $\mathcal{N}=2$ SCFTs in \cite{Razamat:2012uv}, in the context of $\mathcal{N}=1$ gauge theories in \cite{Peelaers:2014ima} and more recently in the mathematical context of elliptic hypergeometric integrals in \cite{spiri2024}.

\paragraph{Our approach}

In this work, we show how this technical difficulty can be resolved.
Our method applies to the superconformal indices of general $\mathcal{N}=1$ gauge theories, although our main focus is in the context of the $\mathcal{N}=4$ SYM and $\mathcal{N}=2$ superconformal SQCD.
The advantage of considering theories with at least $\mathcal{N}=2$ supersymmetry is that one can define simpler specializations of the index, which capture states with a higher degree of supersymmetry.
These specializations are known as the Schur and Macdonald indices \cite{Gadde:2011uv}. 
They share the issues discussed above with the full index, but are technically easier to handle.
Our method allows us to evaluate closed form expressions for all these indices.

Let us now briefly describe the basic idea on which the method relies.
As mentioned above, the integral over $SU(N)$ can be converted into an integral over eigenvalues.
In particular, the expression \eqref{eq:intro-weakly-coupled-int} becomes
\begin{equation}\label{eq:gen-int-weyl-sym}
    \mathcal{I}=\frac{1}{N!}\prod^{N-1}_{i=1}\oint_{|u_i|=1}\frac{du_i}{2\pi iu_i}\left|\Delta(u)\right|^2\,\text{tr}_{\mathcal{H}^{\text{ext}}_{g=0}} (\cdots)\,,
\end{equation}
with $u_i$ the $SU(N)$ eigenvalues and $\Delta(u)=\prod_{i<j}(u_j-u_i)$ the $SU(N)$ Vandermonde determinant.
We will show that, with this measure, the residue sum associated with the poles of the integrand excluding the origin diverges.
Closely related to this observation, we find that the ``contribution from the origin''\footnote{See the beginning of Section \ref{sec:conv-res-sums} for a precise definition} is non-vanishing.
Somehow, this contribution should tame the divergence, but, due to its non-isolated nature, we cannot evaluate its contribution directly.
In effect, we cannot make sense of the residue sum.

We show that this issue can be resolved by using a perhaps less well--known representation of the eigenvalue integral, which takes the form
\begin{equation}
    \mathcal{I}=\prod^{N-1}_{i=1}\oint_{|u_i|=1}\frac{du_i}{2\pi iu_i}\prod_{1\leq i<j\leq N}(1-u_{ij}) \, \text{tr}_{\mathcal{H}^{\text{ext}}_{g=0}} (\cdots)\,,
\end{equation}
with $u_{ij}=u_iu_j^{-1}$.
This version of the integral can be viewed as resulting from a further ``gauge fixing'' of the residual $S_N$ Weyl symmetry of \eqref{eq:gen-int-weyl-sym}.
We show that when evaluating the contour integrals with this \emph{reduced measure}, the residue sum for the poles excluding the origin converges and that the ``contribution from the origin'' can be argued to vanish.

This simple method allows us to write down residue sums for general superconformal indices of $\mathcal{N}=1,2,4$ supersymmetric gauge theories.
We will show that, in the context of the Macdonald index of the $\mathcal{N}=4$ $SU(2)$ theory, we can use mathematical identities for basic hypergeometric series to simplify the residue sum. 
The resulting expression manifests aspects of the (conjectured) strongly coupled 1/8 BPS spectrum of the theory.
We discuss various implications of this result.
Firstly, we will argue that it provides strong evidence that there are no non-graviton operators in the Macdonald sector of the $\mathcal{N}=4$ $SU(2)$ theory, consistent with previous observations in the literature.
Secondly, our expression manifests the analytic dependence on the two superconformal fugacities $(q,t)$, and also the flavor fugacity $v$.
This allows us to analytically continue the index beyond its original domain of convergence and deduce new specializations for which the index simplifies.
For example, we find that the Macdonald index turns into a product formula for $t=q^{\frac{1}{2}}$.
Thirdly, our expression relates to another known closed form expression for the Macdonald index, the so-called TQFT expression of \cite{Gadde:2011uv}, in a non-trivial and interesting manner.
In particular, the TQFT expression takes the form of a (Laurent) series in the flavor fugacity $v$.
Our expression can be viewed as a resummation of this series, manifesting the analytic dependence on $v$.
This generalizes similar observations made for the Schur index \cite{Gadde:2011uv,Pan:2021mrw}.

Motivated by this, we evaluate the residue sums for both the Macdonald and full superconformal indices of supersymmetric $SU(2)$ gauge theories which either are or flow to an SCFT.
The closed form formulas we obtain have various interesting properties, including the aforementioned analytic dependence on the fugacities.
We also find various interesting connections with the mathematics of basic and elliptic hypergeometric integrals and series.
For example, the residue evaluations of the Macdonald index of $\mathcal{N}=2$ $SU(2)$ superconformal SQCD leads to new transformation formulas for basic hypergeometric series.

For the full superconformal index, the reduced measure allows us, for the first time, to evaluate elliptic hypergeometric integrals in terms of residue sums.
The result takes the form of a novel, convergent double infinite sum over a bilinear combination of two elliptic hypergeometric \emph{summands} with a non-factorizing prefactor.
In particular, we find that the expression does not take the form of a product of two elliptic hypergeometric series, unlike the suggested formulas in \cite{Peelaers:2014ima,spiri2024}.
Our formulas, combined with Seiberg duality (in the $\mathcal{N}=1$ case) and generalized $S$-duality (in the $\mathcal{N}=2$ case), lead us to non-terminating summation and transformation formulas involving the bilinear combinations of elliptic hypergeometric summands.

Given the relation between our formula for the Macdonald index of the $\mathcal{N}=4$ $SU(2)$ theory and the TQFT formula, we speculate that a simplification of the residue sum for the full index may shed light on its TQFT formulation, which remains an open problem \cite{Gadde:2011uv,Razamat:2013qfa}.
Finally, we demonstrate that our method extends to higher rank gauge theories by evaluating the Macdonald index of $\mathcal{N}=4$ $SU(3)$ SYM theory through residues.

\paragraph{Previous approaches}

Let us briefly mention some of the previous approaches to evaluate the superconformal index as a residue sum.
As explained above, if one wants to make sense of the residue sum for a gauge theory index, one needs to argue that the contribution at the origin vanishes.
It turns out that this \emph{can} be done for a specific class of 4d $\mathcal{N}=1$ gauge theories whose gauge group can be fully Higgsed.
In this class of theories, the residue evaluation of the index has been related to a localization scheme known as Higgs branch localization, in which the path integral can be localized to a discrete set of Higgs branch vacua \cite{Peelaers:2014ima} (see also \cite{Yoshida:2014qwa,Nieri:2015yia}).
Such theories require sufficient fundamental matter and a gauge group with a $U(1)$ factor for which a non-zero Fayet-Iliopoulos (FI) parameter is turned on.
It is somewhat implicit in the above references, but it turns out that the non-zero FI parameter is crucial to remove a potential contribution from the origin and, relatedly, to ensure convergence of the residue sum.\footnote{A similar issue arise for the 3d index and was addressed in \cite{Hwang:2012jh,Benini:2013yva}.}

Another approach, which is not limited to the above class of theories, is known as the Bethe Ansatz approach \cite{Closset:2017bse,Benini:2018mlo}.
In this approach, one avoids the non-isolated singularity at the origin altogether through a careful contour deformation, which turns the original product of unit circles into contours that bound annuli just outside the unit circles.
A particularly interesting feature of this method is that a specific residue dominates the index at large $N$ and can be interpreted holographically in terms of the exact on-shell action of the supersymmetric $AdS_5$ black hole  \cite{Benini:2018ywd}.
Further progress in this direction, specifically in interpreting other parts of the expression holographically, can be found in \cite{Benini:2020gjh,GonzalezLezcano:2020yeb,Aharony:2021zkr,David:2021qaa,Colombo:2021kbb,Mamroud:2022msu,Aharony:2024ntg} (see also the elliptic extension method of \cite{Cabo-Bizet:2019eaf,Cabo-Bizet:2020ewf}).
A known difficulty with the Bethe Ansatz (BA) formula is the classification of poles, which requires solving a set of transcendental equations, and, for $SU(N)$ gauge theories with $N>2$, the presence of continuous families of poles \cite{ArabiArdehali:2019orz}  (see also \cite{Lezcano:2021qbj,Benini:2021ano}).\footnote{These difficulties were recently addressed in \cite{Cabo-Bizet:2024kfe}.}
We will find that our formulas for the full superconformal index are in a precise sense complementary to the BA formulas. 
Whereas the BA formula requires a specific relation between the superconformal fugacities $p,q$, our formulas are valid precisely when these conditions are \emph{not} met!

Finally, let us also contrast our formulas with the so-called giant graviton expansion of the superconformal index \cite{Arai:2020qaj,Imamura:2021ytr,Gaiotto:2021xce,Murthy:2022ien,Lee:2022vig,Imamura:2022aua,Choi:2022ovw,Liu:2022olj,Eniceicu:2023uvd,Beccaria:2023zjw,Lee:2023iil,Eleftheriou:2023jxr,Deddo:2024liu,Lee:2024hef}.
These formulas are, in a sense, in the spirit of \eqref{eq:strong-coupling-index}, at least for holographic theories.
Indeed, they attempt to make manifest the gravitational interpretation of the index.
An interesting feature of (versions of) the giant graviton expansion, closely related to its holographic interpretation, is that it depends analytically on $N$. 
In particular, it manifests the large $N$ limit.
This feature also implies that for any fixed $N$ large cancellations have to occur, making the expressions highly redundant.
This seems to be a feature, rather than a bug, in that it has a natural gravitational interpretation \cite{Gaiotto:2021xce,Lee:2022vig,Choi:2022ovw,Lee:2023iil,Lee:2024hef}. 
The formulas we are after, for better or for worse, are more minimal and should reflect only the BPS spectrum at non--zero coupling.

\paragraph{Organization}

The rest of this paper is organized as follows.
In Section \ref{sec:n=2-sci}, we briefly review the construction of the superconformal index and standard expressions for the indices of superconformal gauge theories.
In Section \ref{sec:conv-res-sums}, we study in turn the convergence properties of the residue sums for the Schur, Macdonald, and full superconformal index, and demonstrate convergence when the reduced measure is used for the gauge singlet projection. 
Mathematically, these cases correspond respectively to integrals of elliptic functions, basic hypergeometric integrals and elliptic hypergeometric integrals. 
In Section \ref{sec:mcd-n=4-su2}, using the residue sum as an intermediate step, we derive a closed form expression for the Macdonald index of the $\mathcal{N}=4$ $SU(2)$ theory and study its properties.
In Section \ref{sec:examples-rank-1}, we evaluate the Macdonald and full superconformal index of $SU(2)$ gauge theories with $\mathcal{N}=4,2,1$ supersymmetry.
In Section \ref{sec:higher-rank}, we show how our method extends to a higher rank example, the Macdonald index of the $\mathcal{N}=4$ $SU(3)$ SYM theory.
We briefly summarize our findings and suggest directions for future research in Section \ref{sec:future}.

In Appendix \ref{app:spec-functions}, we collect the definitions of various special functions used in the main text and summarize some of their key properties and identities.
In Appendix \ref{app:bas-hypgeom-int}, we review the residue evaluation of basic hypergeometric integrals.
In Appendix \ref{app:ell-hypgeom-int}, we evaluate elliptic hypergeometric integrals as residue sums and discuss their convergence.
In Appendix \ref{app:Tracerelations}, we collect trace relations for the Hall-Littlewood chiral ring of the $\mathcal{N}=4$ $SU(3)$ theory.

\section{\texorpdfstring{$\mathcal{N}=2$}{N=2} superconformal index}\label{sec:n=2-sci}

In this section, we review the construction and classification of $\mathcal{N}=2$ superconformal indices, and their evaluation for superconformal gauge theories \cite{Gadde:2011uv}.\footnote{See also the reviews \cite{Rastelli:2014jja,Gadde:2020yah}.}
A reader familiar with this topic may safely skip to Section \ref{sec:conv-res-sums}.

\subsection{Definition of the superconformal index}\label{ssec:defns}

The $\mathcal{N}=2$ superconformal algebra contains eight (complex Weyl) supercharges $Q_{I\alpha}$ and $\widetilde{Q}^{I}_{\dot{\alpha}}$, with $\alpha=\pm$, $\dot{\alpha}=\dot{\pm}$ and $I=1,2$.
The algebra also contains eight superconformal charges, which in radial quantization are related to the supercharges through hermitian conjugation: $S^{I\alpha}=\left(Q_{I\alpha}\right)^{\dagger}$ and $\widetilde{S}^{\dot{\alpha}}_I=(\widetilde{Q}_{I\dot{\alpha}})^{\dagger}$.
The anti-commutators between the super- and superconformal charges are given by
\begin{align}\label{eq:n=2-sca}
    \begin{split}
        \lbrace Q_{I\alpha},S^{J\beta} \rbrace &=\tfrac{1}{2}\delta^{J}_I\delta^\beta_\alpha D +\delta^J_I M_\alpha^\beta-\delta^\beta_\alpha R_I^J+\tfrac{1}{2}\delta^{J}_I\delta^\beta_\alpha r\\
        \lbrace \widetilde{Q}^{I}_{\dot{\alpha}},\widetilde{S}^{\dot{\beta}}_J \rbrace &=\tfrac{1}{2}\delta^I_J\delta^{\dot{\beta}}_{\dot{\alpha}} D +\delta^J_I \widetilde{M}^{\dot{\beta}}_{\dot{\alpha}}-\delta^{\dot{\beta}}_{\dot{\alpha}} R_I^J-\tfrac{1}{2}\delta^{J}_I\delta^\beta_\alpha r
    \end{split}
\end{align}
Here, $D$ is the generator of dilatations on $\mathbb{R}^4$, $M_\alpha^\beta$ and $\widetilde{M}^{\dot{\beta}}_{\dot{\alpha}}$ are the generators of the rotation group $Spin(4)\cong SU(2)_1\times SU(2)_2$. $R_I^J$ and $r$ are generators for the R-symmetry group $SU(2)_R\times U(1)_r$ .
The supercharges transform as doublets under $SU(2)_R$. We take $Q_{I\alpha}$ to have charge $-\frac{1}{2}$ and $\widetilde{Q}^{I}_{\dot{\alpha}}$ to have charge $+\frac{1}{2}$ under $U(1)_r$.
Finally, $Q_{I\alpha}$ and $\widetilde{Q}^{I}_{\dot{\alpha}}$ have scaling dimensions $\frac{1}{2}$, while $S^{I\alpha}$ and $\widetilde{S}^{\dot{\alpha}}_I$ have scaling dimensions $-\frac{1}{2}$.\footnote{Note that hermitian conjugation changes the sign of all charges of $Q_{I\alpha}$ and $\widetilde{Q}^{I}_{\dot{\alpha}}$.}

We now consider the following positive semi-definite anti-commutators
\begin{equation}
	\delta_{I\pm}\equiv 2\lbrace Q_{I\pm},S^{I\pm}\rbrace\,,\qquad \tilde{\delta}_{I\dot{\pm}}\equiv 2\lbrace \widetilde{Q}^I_{\dot{\pm}},\widetilde{S}^{\dot{\pm}}_I\rbrace\,,
\end{equation}
which can be thought of as a set of ``supersymmetric Hamiltonians'' for an $\mathcal{N}=2$ SCFT.
If we denote the Cartan generators by $j_{1,2}$ for $ SU(2)_1\times SU(2)_2$, and $R$ for $SU(2)_R$, the anti-commutators can be written as
\begin{align}
	\begin{split}
		\delta_{I\pm}=D\pm 2j_1+(-1)^I\,2R+r\,,\qquad \tilde{\delta}_{I\dot{\pm}}=D\pm 2j_2+(-1)^I2R-r\,,
	\end{split}
\end{align}
The generators $j_{1,2}$ can be expressed in terms of the generators of $SO(2)^2\subset SO(4)$ as
\begin{equation}
    j_1=\tfrac{1}{2}(J_{12}+J_{34})\,,\quad j_2=\tfrac{1}{2}(J_{12}-J_{34})\,,
\end{equation}
where $J_{12}$ and $J_{34}$ rotate the two independent planes in $\mathbb{R}^4$.

To define the index, we first choose an $\mathcal{N}=1$ subalgebra with supercharges $Q_{1\alpha}$ and $\widetilde{Q}^1_{\dot{\alpha}}$ (and their hermitian conjugates).
The superconformal R-symmetry of this subalgebra can be read off from the relevant anti-commutator in \eqref{eq:n=2-sca}:
\begin{equation}\label{eq:n=1-r-sym-gen}
    r_{1}=\tfrac{2}{3}(-2R+r)\,.
\end{equation}
The general $\mathcal{N}=2$ superconformal index is now defined similarly to the $\mathcal{N}=1$ superconformal index of \cite{Romelsberger:2005eg,Kinney:2005ej}.
In particular, we choose to define it with respect to $\mathcal{Q}\equiv Q_{1-}$ and $\mathcal{Q}^\dagger\equiv S^{1-}$.
From the above definitions, one easily checks that the commutant of $\mathcal{Q}$ in the superconformal algebra has a Cartan that is generated by
\begin{equation}
	\delta_{1-}\,,\, \delta_{2+}\,,\, \tilde{\delta}_{1\dot{\pm}}\,.
\end{equation}
These charges are then used to define the index as the graded trace
\begin{align}\label{eq:n2-sci}
	\begin{split}
		\mathcal{I}(p,q,t)&=\mathrm{tr}\, (-1)^F \sigma^{\frac{1}{2}\tilde{\delta}_{1\dot{+}}}\,\rho^{\frac{1}{2}\tilde{\delta}_{1\dot{-}}}\,\tau^{\frac{1}{2}\delta_{2+}}\,e^{-\beta\delta_{1-}}\\
		&=\mathrm{tr}\, (-1)^F p^{j_1+j_2-r}\,q^{j_1-j_2-r}t^{R+r}\,e^{-\beta\delta_{1-}}\,,
	\end{split}
\end{align}
where the trace runs over the states of the theory on quantized on $S^3$ and the relation between the two sets of fugacities is
\begin{equation}
	p=\sigma\tau,\;\;q=\rho\tau,\;\; t=\tau^2\,,
\end{equation}
with absolute values constrained as
\begin{equation}\label{eq:convergence-n2-index}
	|p|<1\,,\quad |q|<1\,,\quad |pq|<|t|<1\,.
\end{equation}
By construction, the index only receives contributions from short multiplets comprised of states annihilated by both $\mathcal{Q}$ and $\mathcal{Q}^{\dagger}$.
It follows that the index is independent of $\beta$, and any other continuous deformations of the theory that preserve $\mathcal{Q}$.
The charges of these states obey the shortening condition
\begin{equation}
    \delta_{1-}=D-2j_1-2R+r=0\,.
\end{equation}
There are various types of short multiplets satisfying this condition, see for example \cite{Dolan:2002zh,Kinney:2005ej} for a classification.\footnote{And Appendix B of \cite{Gadde:2011uv} for a brief overview.} 

Finally, we can reparametrize the $\mathcal{N}=2$ index to make the connection with the $\mathcal{N}=1$ index of \cite{Romelsberger:2005eg,Kinney:2005ej} clear, via
\begin{equation}\label{eq:n2-sci-n1-language}
	\mathcal{I}(p,q,s)=\mathrm{tr}\, (-1)^F p^{j_1+j_2-\frac{r_{1}}{2}}\,q^{j_1-j_2-\frac{r_{1}}{2}}s^{R+r}\,e^{-\beta\delta_{1-}}\,,
\end{equation}
where $s=t(pq)^{-\frac{2}{3}}$ is a fugacity for $R+r$, which commutes with the full $\mathcal{N}=1$ subalgebra, and $r_1$ is the generator of the $\mathcal{N}=1$ R-symmetry in \eqref{eq:n=1-r-sym-gen}.

\paragraph{Macdonald index}

Of particular interest in this work will be the so-called Macdonald (limit of the) index.
It is defined by taking $\sigma\to 0$, while keeping $\rho$ and $\tau$ fixed ($p\to 0$, with $q$ and $t$ fixed).
From the expression \eqref{eq:n2-sci}, we see that only states with both $\delta_{1-}=\tilde{\delta}_{1\dot{+}}=0$ will contribute to this index, \textit{i.e.}, states that are annihilated by two supercharges $Q_{1-}$ and $\widetilde{Q}_{1\dot{+}}$, and their hermitian conjugates.
We write the corresponding index as:
\begin{align}
	\begin{split}
		\mathcal{I}(q,t)\equiv \mathcal{I}(0,q,t)&=\mathrm{tr}\, (-1)^F \rho^{\frac{1}{2}\tilde{\delta}_{1\dot{-}}}\,\tau^{\frac{1}{2}\delta_{2+}}\,e^{-\beta\delta_{1-}}=\mathrm{tr}\, (-1)^F q^{j_1-j_2-r}t^{R+r}\,e^{-\beta\delta_{1-}}\,.
	\end{split}
\end{align}
The states contributing to this index obey two charge constraints, which can be written as
\begin{equation}
    D=j_1-j_2+2R\,,\quad j_1+j_2=r\,.
\end{equation}

\paragraph{Schur index}

A further specialization is known as the Schur index, and can be obtained by taking $t=q$ (either in the full or Macdonald index).
We will write this index as
\begin{equation}
	\mathcal{I}(q)\equiv \mathcal{I}(p,q,q)=\mathcal{I}(q,q)=\mathrm{tr}\, (-1)^F q^{D-R}\,e^{-\beta\delta_{1-}}\,.
\end{equation}
The states contributing to this index correspond to the same as those that contribute to the Macdonald index.

\paragraph{Hall-Littlewood index}

Finally, the simplest index can be obtained from the Macdonald index, by also setting $\rho\to 0$, keeping $\tau$ fixed ($q\to 0$, $t$ fixed), and is known as the Hall-Littlewood index.
In this case only states with $\delta_{1-}=\tilde{\delta}_{1\dot{\pm}}=0$ contribute.
The charges of these states are constrained as
\begin{equation}
    D=2R+r\,,\quad j_1=r\,,\quad j_2=0\,.
\end{equation}
The corresponding index is expressed as
\begin{align}
	\begin{split}
		\mathcal{I}(t)\equiv \mathcal{I}(0,0,t)&=\mathrm{tr}\, (-1)^F \tau^{\frac{1}{2}\delta_{2+}}\,e^{-\beta\delta_{1-}}=\mathrm{tr}\, (-1)^F t^{R+r}\,e^{-\beta\delta_{1-}}\,.
	\end{split}
\end{align}
We now turn to the evaluation of the index for $\mathcal{N}=2$ superconformal gauge theories.

\subsection{\texorpdfstring{$\mathcal{N}=2$}{N=2} gauge theory indices}\label{ssec:sc-gauge-theory-inds}

There are two supersymmetric multiplets from which any $\mathcal{N}=2$ Lagrangian is built: the vector-- and hypermultiplet.\footnote{See, \textit{e.g.}, \cite{Tachikawa:2013kta} for a review.}
The vector multiplet can be thought of as the combination of a $\mathcal{N}=1$ vector multiplet and chiral multiplet.
As such it is comprised of a complex scalar, two complex Weyl fermions, and the gauge field strength, which we will denote collectively by $(\phi,\bar{\phi},\lambda_{i\alpha},\bar{\lambda}_{i\dot{\alpha}},F_{\alpha\beta},\bar{F}_{\dot{\alpha}\dot{\beta}})$, with $i$ being a fundamental $SU(2)_R$ index.
Note that the fermions transforms as a doublet under $SU(2)_R$, and all fields transform in the adjoint representation of the gauge group.
The hypermultiplet consists of two $\mathcal{N}=1$ chiral multiplets transforming in conjugate representations of the gauge group.
We denote these by $(q_a,\bar{q}_a,\psi_{a\alpha},\bar{\psi}_{a\dot{\alpha}})$, for $a=1,2$.
The $SU(2)_R$ doublets consist of $(q_1,\bar{q}_2)$ and $(q_2,\bar{q}_1)$.

A choice of (semi-simple) gauge group $G$, and matter representations $R$, then fully specifies the Lagrangian of a general $\mathcal{N}=2$ gauge theory.
The gauge theory will enjoy $\mathcal{N}=2$ superconformal invariance if the one-loop beta function for each factor of the gauge group vanishes.
For example, the beta function for an $SU(N)$ gauge coupling, with $M$ fundamental hypermultiplets will vanish when $M=2N$.
This theory is known as $\mathcal{N}=2$ superconformal QCD.
In addition, a vector multiplet coupled to an adjoint hypermultiplet will have vanishing beta function, as this corresponds to the $\mathcal{N}=4$ field content.
If the gauge group consists of a product of (special) unitary groups, the quiver representation of the superconformal theory admits an $ADE$ classification \cite{Gaiotto:2008ak,Gaiotto:2009we}.
These theories are part of the more general class $\mathcal{S}$ SCFTs \cite{Gaiotto:2009we}.
In this work, we consider the superconformal indices of a few basic examples of such theories.

The index of the superconformal gauge theory is easily computed in the free limit \cite{Dolan:2008qi,Gadde:2011uv}.
One first lists the letters of the vector- and hypermultiplet satisfying the $\delta_{1-}=0$ charge constraint, see Table \ref{tab:n=2-multiplets}.
\begin{table}[tbp]
\begin{centering}
\bgroup
\def\arraystretch{1.3}
\begin{tabular}{|c|c|c|c|c|c|c|c|}
\hline
Letters & $  \Delta$ & $j_1$ & $  j_2$ & $R$ & $r$ & $r_1$  & $\mathcal{I}(p, q, t)$ \tabularnewline
  \hline
$  \bar{\phi}$ & $1$ & $0$ & $0$ & $0$ & $-1$ & $-\frac{2}{3}$  & $pq/t$   \tabularnewline
  \hline
$  \lambda _{1+}$ & $  \frac{3}{2}$ & $ \frac{1}{2}$ & $0$ & $  \frac{1}{2}$ & $\frac{1}{2}$  & $-\frac{1}{3}$ & $-t$  \tabularnewline
  \hline
$  \bar{\lambda}_{2\dot{\pm}}$  & $  \frac{3}{2}$ & $0$ & $ \pm \frac{1}{2}$ & $  \frac{1}{2}$ & $  -\frac{1}{2}$  & $-1$ &  $-p$, $-q$ \tabularnewline
  \hline
$  F_{++}$ & $2$ & $1$ & $0$ & $0$ & $0$ & $0$  &  $pq$ \tabularnewline
  \hline
  $  \partial_{+\dot{-}}  \bar{\lambda}_{2\dot +} +  \partial_{+\dot{+}}  \bar{\lambda}_{2\dot -}=0$ & $  \frac{5}{2}$ & $\frac{1}{2}$ & $ 0$ & $  \frac{1}{2}$ &
 $  -\frac{1}{2}$ & $-1$ & $pq$  \tabularnewline
  \hline
$\bar{q}_{i}$ & $1$ & $0$ & $0$ & $  \frac{1}{2}$ & $0$  &  $-\frac{2}{3}$ &  $t^{\frac{1}{2}}$ \tabularnewline
  \hline
$  \psi_{i+}$ & $  \frac{3}{2}$ & $  \frac{1}{2}$ & $0$ & $0$ & $- \frac{1}{2}$ & $-\frac{1}{3}$  & $-pq/t^{\frac{1}{2}}$ \tabularnewline
  \hline
$  \partial_{+\dot{\pm}}$ & $1$ & $   \frac{1}{2}$ & $ \pm  \frac{1}{2}$ & $0$ & $0$ & $0$   & $p$, $q$ \tabularnewline
\hline
\end{tabular}
\egroup
\par\end{centering}
\caption{\label{tab:n=2-multiplets} Letters in the $\mathcal{N}=2$ vector multiplet and hypermultiplet ($i=1,2$) with $\delta_{1-}=0$. Note that only the equation of motion of the gaugino contributes as a constraint. The hypermultiplet fields $(\bar{q}_2,\psi_{2+})$ transform in the conjugate representation of the gauge/global symmetry groups as compared to $(\bar{q}_1,\psi_{1+})$.}
\end{table}
From here, one calculates the single-letter indices.
We have for the vector multiplet
\begin{align}
    \begin{split}
        i_{\text{vm}}(U;p,q,t)&=\left(\frac{pq/t-t}{(1-p)(1-q)}+\frac{2pq-p-q}{(1-p)(1-q)}\right)\chi_{\text{adj}}(U)\\
        &=\left(\frac{pq/t-t}{(1-p)(1-q)}+1-\frac{1-pq}{(1-p)(1-q)}\right)\chi_{\text{adj}}(U)\,,
    \end{split}
\end{align}
where $\chi_{\text{adj}}(U)$ is the character of the adjoint representation of the gauge group.
In this work, we only consider $SU(N)$ gauge groups, for which the adjoint character reads
\begin{equation}
    \chi_{\text{adj}}(U)=\text{tr}\, U\, \text{tr}\, U^{\dagger} -1=N-1+\sum_{i\neq j}u_{ij}\,,
\end{equation}
where $U\in SU(N)$, the traces are taken in the fundamental representation and we defined $u_{ij}=u_iu_j^{-1}$, with $u_i$ the eigenvalues of $U$ satisfying $|u_i|=1$ and the $SU(N)$ condition $\prod^N_{i=1} u_i=1$.
For the hypermultiplet one may similarly determine
\begin{align}\label{eq:single-letter-hyper}
    \begin{split}
        i_{\text{hm}}(U,V;p,q,t)&=\frac{t^{\frac{1}{2}}\chi_{\bar{R}}(U,V)-pq/t^{\frac{1}{2}}\chi_{R}(U,V)}{(1-p)(1-q)}+\frac{t^{\frac{1}{2}}\chi_{R}(U,V)-pq/t^{\frac{1}{2}}\chi_{\bar{R}}(U,V)}{(1-p)(1-q)}\\
        &=\frac{t^{\frac{1}{2}}-pq/t^{\frac{1}{2}}}{(1-p)(1-q)}(\chi_{R}(U,V)+\chi_{\bar{R}}(U,V))\,,
    \end{split}
\end{align}
with $\chi_{R}(U,V)$ the character of the representation $R=R_G\otimes R_F$ under the gauge and flavor symmetry groups:
\begin{equation}
    \chi_{R}(U,V)=\sum_{\rho\in R_G}u^\rho\sum_{\rho'\in R_F}v^{\rho'}\,,\quad u^{\rho}\equiv u_1^{\rho_1}\cdots u_N^{\rho_N}\,,\quad v^{\rho'}\equiv v_1^{\rho'_1}\cdots v_M^{\rho'_M}\,.
\end{equation}
Here, $u_i$ and $v_j$ are the eigenvalues of $U$ and $V$, $\rho=(\rho_1,\ldots,\rho_N)$ and $\rho'=(\rho'_1,\ldots,\rho'_M)$ are the weights of the representations, with $N$ and $M$ corresponding to the dimensions of their fundamental representations.

From the single-letter indices, one obtains the full index through plethystic exponentiation \cite{Aharony:2003sx,Romelsberger:2005eg,Kinney:2005ej}.
It was observed in \cite{Dolan:2008qi} (see also \cite{Gadde:2009kb,Gadde:2010te}), that the full index can be written in terms of the elliptic Gamma function $\Gamma(x;p,q)$.\footnote{See Appendix \ref{sapp:ell-Gamma} for a definition.} 
For gauge group $SU(N)$, the full vector multiplet index is then expressed as
\begin{equation}\label{eq:n=2-vm-index}
    \mathcal{I}_{\text{vm}}(U;p,q,t)=\frac{\left[(p;p)_\infty(q;q)_{\infty}\Gamma\left(\frac{pq}{t};p,q\right)\right]^{N-1}}{ \left|\Delta(u)\right|^2}\prod_{i\neq j}\frac{\Gamma\left(\frac{pq}{t}u_{ij};p,q\right)}{\Gamma\left(u_{ij};p,q\right)}\,,
\end{equation}
where $(x;q)_\infty$ is the $q$-Pochhammer symbol, $u=(u_1,\ldots,u_N)$, and
\begin{equation}
    \left|\Delta(u)\right|^2=\prod_{1\leq i\neq j\leq N}(1-u_{ij})
\end{equation}
with the $SU(N)$ Vandermonde determinant given by
\begin{equation}
    \Delta(u)\equiv \prod_{i< j}(u_j-u_i)=\sum_{\sigma\in S_N}\varepsilon(\sigma)u^0_{\sigma(1)}u^1_{\sigma(2)}\cdots u^{N-1}_{\sigma(N)}\,,
\end{equation}
with $\varepsilon(\sigma)$ the sign of the permutation.

The full index of the hypermultiplet is given by
\begin{equation}\label{eq:n=2-hm-index}                    
    \mathcal{I}_{\text{hm}}(U,V;p,q,t)=\prod_{\rho\in R_G,\,\rho'\in R_F} \Gamma\left(t^{\frac{1}{2}}u^{\bar{\rho}}v^{\bar{\rho}'};p,q\right)\Gamma\left(t^{\frac{1}{2}}u^{\rho}v^{\rho'};p,q\right)
\end{equation}
where $\bar{\rho}$ and $\bar{\rho}'$ are the weights of the conjugate representation $\bar{R}$.

The superconformal index of the gauge theory can now be obtained by combining the above ingredients for the specific gauge group and matter representations, and projecting onto gauge singlets.
The singlet projection is achieved by integrating over the gauge group with the invariant Haar measure \cite{Aharony:2003sx}.
Since the integrand is expressed in terms of $SU(N)$ characters, the integral can be reduced to an integral over eigenvalues
\begin{equation}\label{eq:full-singlet-projection}
    \int_{SU(N)}d\mu_{SU(N)}(U) f(U)=\frac{1}{N!}\prod^{N-1}_{i=1}\oint_{|u_i|=1}\frac{du_i}{2\pi iu_i}\left|\Delta(u)\right|^2 f(\mathbf{u})\,,
\end{equation}
where the measure is written in terms of the Vandermonde determinant and $f(u)$ captures the appropriate multi-letter indices listed above.
The integral over the eigenvalues can be viewed as a (multi-dimensional) contour integral in the space of complexified eigenvalues, with the contour being a product of unit circles.

An important observation for us is that the measure of the eigenvalue integral, due to the invariance of both measure and $f(u)$ under permutations of $u_i$, can be replaced by\footnote{For $SU(2)$ matrix integrals, this is an elementary fact and was noted for example in \cite{Dolan:2007rq,Dolan:2008qi}. The general case was used in \cite{Hanany:2008sb} in the context of the $\mathcal{N}=1$ adjoint SQCD Hilbert series and, more recently, in the context of the superconformal index of the $\mathcal{N}=4$ theory in \cite{Agarwal:2020zwm} (see also \cite{Choi:2021rxi,Choi:2023znd,Choi:2023tiq}) and in the context of (bosonic) matrix models in \cite{deMelloKoch:2025ngs}.}
\begin{equation}\label{eq:vandermonde-replacement}
    \left|\Delta(u)\right|^2\to N!\prod_{i<j}(1-u_{ij})\quad \text{or}\quad \left|\Delta(u)\right|^2\to N!\prod_{i<j}(1-u_{ij}^{-1})
\end{equation}
It follows that we can consider an alternative singlet projection, which takes the form
\begin{equation}\label{eq:red-singlet-projection}
    \int_{SU(N)}d\mu_{SU(N)}(U) f(U)=\prod^{N-1}_{i=1}\oint_{|u_i|=1}\frac{du_i}{2\pi iu_i}\prod_{i<j}(1-u_{ij}) f(u)\,,
\end{equation}
In the rest of this paper, we will refer to the singlet projection in \eqref{eq:full-singlet-projection} as projecting with the full (Haar) measure and to \eqref{eq:red-singlet-projection} as projecting with the reduced measure.
Similar to how the full measure arises after a ``gauge fixing'' of the matrix $U$ to a diagonal form, we can think of the reduced measure as reflecting a further gauge fixing of the residual $S_N$ Weyl symmetry.

When we consider the index of the $\mathcal{N}=4$ theory, it will turn out to convenient to consider a reparametrization of the $SU(N)$ eigenvalues \cite{deMelloKoch:2025ngs}
\begin{equation}\label{eq:reparam-ev}
    u_i= s_i\cdots s_N\,.
\end{equation}
A useful feature of this reparametrization is that $s_N$ is not present in any of the $u_{ij}$.
In a purely adjoint theory, like the $\mathcal{N}=4$ theory, it follows that the integrand only depends on the unconstrained variables $s_{1,\ldots,N-1}$.
In terms of these variables, the adjoint character is written as
\begin{equation}
    \chi_{\text{adj}}(U)=N-1+\sum_{1\leq i\leq j\leq N-1}(s_{i,j}+s^{-1}_{i,j}) \, ,
\end{equation}
where we define
\begin{equation}
    s_{i,j}=s_i\cdots s_j \, .
\end{equation}
In particular, it follows that
\begin{equation}
    \prod_{1\leq i<j\leq N}(1-u_{ij})=\prod_{1\leq i\leq j\leq N-1}(1-s_{i,j}) \, .
\end{equation}
Finally, the Jacobian for the transformation is given by
\begin{equation}
    \det \left[\frac{\partial u_i}{\partial s_j}\right]=\prod^{N-1}_{k=1}\frac{u_k}{s_k}\quad \Rightarrow \quad \prod^{N-1}_{i=1}\frac{du_i}{u_i}=\prod^{N-1}_{i=1}\frac{ds_i}{s_i} \, .
\end{equation}
We now turn to listing the indices of interest in this work.

\paragraph{\texorpdfstring{$\mathcal{N}=4$}{N=4} \texorpdfstring{$SU(N)$}{SU(N)} index}

The $\mathcal{N}=4$ theory contains an $\mathcal{N}=2$ vector multiplet and an adjoint hypermultiplet.
Its index can be expressed using the full and reduced measure, respectively, as
\begin{align}\label{eq:full-index-suN-n=4-defn}
    \begin{split}
        \mathcal{I}(v;p,q,t)&=\frac{\left[\kappa\Gamma(t^{\frac{1}{2}}v^{\pm})\Gamma\left(\tfrac{pq}{t}\right)\right]^{N-1}}{N!}\prod^{N-1}_{i=1}\oint\frac{ds_i}{2\pi is_i} \prod_{1\leq i\leq j\leq N-1} \frac{\G(t^{\frac{1}{2}}v^\pm s_{i,j}^\pm)\G(\frac{pq}t s_{i,j}^\pm)}{\G(s_{i,j}^\pm)}\\
        &=\left[\kappa\Gamma(t^{\frac{1}{2}}v^{\pm})\Gamma\left(\tfrac{pq}{t}\right)\right]^{N-1}\prod^{N-1}_{i=1}\oint\frac{ds_i}{2\pi is_i} \prod_{1\leq i\leq j\leq N-1} \frac{\G(t^{\frac{1}{2}}v^\pm s_{i,j}^\pm)\G(\frac{pq}t s_{i,j}^\pm)}{(1-s_{i,j}^{-1})\G(s_{i,j}^\pm)}
    \end{split}
\end{align}
where we integrate over the reparametrized eigenvalues defined in \eqref{eq:reparam-ev}, and use the shorthand notations $\Gamma(x)\equiv\Gamma(x;p,q)$ and $\Gamma(x^{\pm})=\Gamma(x)\Gamma(x^{-1})$.
We also define
\begin{equation}
    \kappa=(p;p)_\infty(q;q)_{\infty} \,.
\end{equation}
Note that, in the first line, the full measure of the singlet projection is contained in the denominator.
In the second line, we used the first reduction in \eqref{eq:vandermonde-replacement}.

We will also be interested in the Macdonald index, which can be obtained by taking $p\to 0$.
Using the fact that (see Appendix \ref{sapp:ell-Gamma})
\begin{equation}
    \lim_{p\to 0} \Gamma(x;p,q)=\frac{1}{(x;q)_\infty}\,,\quad \Gamma(\tfrac{pq}{x};p,q)=\frac{1}{\Gamma(x;p,q)}\,,
\end{equation}
we find 
\begin{align}\label{eq:mcd-index-suN-n=4-defn}
    \begin{split}
        \mathcal{I}(v;q,t)&=\frac{(q,t;q)^{N-1}_\infty}{N!(t^{\frac{1}{2}}v^{\pm};q)_\infty^{N-1}}\prod^{N-1}_{i=1}\oint\frac{ds_i}{2\pi is_i}\prod_{1\leq i\leq j\leq N-1}\frac{(s_{i,j}^{\pm},ts_{i,j}^{\pm};q)_\infty}{(t^{\frac{1}{2}}v^{\pm}s_{i,j}^{\pm};q)_\infty}\,,\\        
        &=\frac{(q,t;q)^{N-1}_\infty}{(t^{\frac{1}{2}}v^{\pm};q)_\infty^{N-1}}\prod^{N-1}_{i=1}\oint\frac{ds_i}{2\pi is_i}\prod_{1\leq i\leq j\leq N-1}\frac{(s_{i,j},qs_{i,j}^{-1},ts_{i,j}^{\pm};q)_\infty}{(t^{\frac{1}{2}}v^{\pm}s_{i,j}^{\pm};q)_\infty}\,,
    \end{split}
\end{align}
where we denote $(x_1\ldots,x_n;q)_\infty=(x_1;q)_\infty\cdots(x_n;q)_\infty$ and $(x^{\pm};q)_{\infty}=(x;q)_{\infty}(x^{-1};q)_{\infty}$.
Similar to the full index, in the first line the full measure of the singlet projection is contained in the numerator.
For the second line, we note that the reduced measure effectively replaces $(s_{i,j}^{-1};q)_\infty\to(qs_{i,j}^{-1};q)_\infty$.

\paragraph{\texorpdfstring{$\mathcal{N}=2$}{N=2} \texorpdfstring{$SU(N)$}{SU(N)} SQCD index}

The $\mathcal{N}=2$ $SU(N)$ SQCD theory contains a vector multiplet and $2N$ hypermultiplets in the fundamental representation of the gauge group.
The index can be expressed using the full measure as\footnote{Viewed as a class $\mathcal{S}$ theory, it is most natural to focus on an $(SU(N)\times U(1)_B)^2\subset U(2N)$ subgroup of the flavor symmetry and accordingly separate $(v_1,\ldots,v_{2N})=(ay_1,\ldots,ay_N,b^{-1}z^{-1}_1,\ldots,b^{-1}z^{-1}_N)$ with $y_i,z_i$ $SU(N)$ eigenvalues and $a,b$ as $U(1)_B$ elements \cite{Gadde:2010te}.}
\begin{align}           \label{eq:sqcd-index-defn-gen-nc}
    \begin{split}
        \mathcal{I}(v_j;p,q,t)&=\frac{\left[\kappa\Gamma\left(\tfrac{pq}{t}\right)\right]^{N-1}}{N!}\prod^{N-1}_{i=1}\oint\frac{du_i}{2\pi iu_i}\prod_{1\leq i<j\leq N}\frac{\Gamma(\frac{pq}{t}u_{ij}^{\pm})}{\Gamma(u_{ij}^{\pm})}\prod^{2N}_{j=1}\prod^{N}_{i=1}\Gamma(t^{\frac{1}{2}}(v_ju_{i})^{\pm})\,.
    \end{split}
\end{align}
where we recall that $\prod_i u_i=1$.
With the reduced measure, the index is expressed as
\begin{align}\label{eq:sqcd-index-defn-gen-nc-red}
    \begin{split}
        \mathcal{I}(v_j;p,q,t)&=\left[\kappa\Gamma\left(\tfrac{pq}{t}\right)\right]^{N-1}\prod^{N-1}_{i=1}\oint\frac{du_i}{2\pi iu_i}\prod_{1\leq i<j\leq N}\frac{\Gamma(\frac{pq}{t}u_{ij}^{\pm})}{(1-u_{ij}^{-1})\Gamma(u_{ij}^{\pm})}\prod^{2N}_{j=1}\prod^{N}_{i=1}\Gamma(t^{\frac{1}{2}}(v_ju_{i})^{\pm})\,.
    \end{split}
\end{align}
Taking the $p\to 0$ limit of either expression, we find the following two expressions for the Macdonald index
\begin{align}
    \begin{split}
        \mathcal{I}(v_j;p,q,t)&=\frac{(q,t;q)^{N-1}_\infty}{N!}\prod^{N-1}_{i=1}\oint\frac{du_i}{2\pi iu_i}\frac{\prod_{1\leq i<j\leq N}(u_{ij}^{\pm},tu_{ij}^{\pm};q)_\infty}{\prod^{2N}_{j=1}\prod^{N}_{i=1}(t^{\frac{1}{2}}(v_ju_{i})^{\pm};q)_\infty}\\
        &=(q,t;q)^{N-1}_\infty\prod^{N-1}_{i=1}\oint\frac{du_i}{2\pi iu_i}\frac{\prod_{1\leq i<j\leq N}(u_{ij},qu_{ij}^{-1},tu_{ij}^{\pm};q)_\infty}{\prod^{2N}_{j=1}\prod^{N}_{i=1}(t^{\frac{1}{2}}(v_ju_{i})^{\pm};q)_\infty}\,,
    \end{split}
\end{align}
where the reduced measure effectively replaces $(u_{ij}^{-1};q)_\infty\to(qu_{ij}^{-1};q)_\infty$ in the numerator.

\subsection{\texorpdfstring{$\mathcal{N}=1$}{N=1} gauge theory indices}\label{ssec:n=1-gauge-theory-inds}

Finally, we also collect the superconformal indices for $\mathcal{N}=1$ gauge theories with gauge group $SU(N)$ and $N_f$ fundamental chiral multiplets.\footnote{See \cite{Rastelli:2016tbz} for a review on the $\mathcal{N}=1$ superconformal index.}
Such theories are expected to flow to interacting superconformal field theories in the ``conformal window'': $3N/2<N_f<3N$.
These theories famously admit a Seiberg dual.
This is the statement that an ``electric--magnetic dual'' theory, with gauge group $SU(N_f-N)$ with $N_f$ flavors and additional decoupled mesons, flows to the same SCFT in the IR \cite{Seiberg:1994pq}.
Outside the conformal window, for $N_f\geq 3N$ the ``electric'' theory flows to a free gauge theory, while for $N+2\leq N_f\leq 3N/2$ the ``magnetic'' theory flows to a free gauge theory.
Finally, for $N_f=N+1$, the IR theory is comprised only of free mesons and quarks, without a dynamical vector multiplet.
Early works studying the superconformal indices of these theories, motivated by checks of Seiberg duality, include \cite{Romelsberger:2005eg,Romelsberger:2007ec,Dolan:2008qi,Spiridonov:2008zr,Spiridonov:2009za,Spiridonov:2010hh,Spiridonov:2011hf}.

The expressions for the $\mathcal{N}=1$ vector multiplet and chiral multiplet are easily obtained from the $\mathcal{N}=2$ expressions by stripping off the adjoint chiral multiplet and conjugate chiral multiplet contributions respectively and setting $t=(pq)^{\frac{2}{3}}$ (so that $s=1$ in \eqref{eq:n2-sci-n1-language}).
We thus find that the $\mathcal{N}=1$ vector multiplet index takes the form
\begin{equation}\label{eq:n=1-vm-index}
    \mathcal{I}_{\text{vm}}(U;p,q)=\frac{(p;p)^{N-1}_\infty(q;q)_{\infty}^{N-1}}{ \left|\Delta(u)\right|^2}\prod_{i\neq j}\frac{1}{\Gamma\left(u_{ij};p,q\right)}\,,
\end{equation}
and the $\mathcal{N}=1$ chiral multiplet index is given by
\begin{equation}\label{eq:n=1-hm-index}                    
    \mathcal{I}_{\text{cm}}(U,V;p,q)=\prod_{\rho\in R_G,\,\rho'\in R_F} \Gamma\left((pq)^{\frac{r}{2}}u^{\bar{\rho}}v^{\bar{\rho}'};p,q\right)\,.
\end{equation}
where $r$ is the $U(1)_R$ charge of the chiral multiplet (more specifically of $\bar{q}$).

One can now write down the superconformal index of the $\mathcal{N}=1$ $SU(N)$ theory with $N_f$ fundamentals as follows:
\begin{equation}\label{eq:N=1-SUN-index-defn}
    \cI_E(y_j,\tilde{y}_j;p,q) = \frac{(p;p)^{N-1}_\infty(q;q)_{\infty}^{N-1}}{N!}\oint\prod_{i=1}^{N-1}\frac{du_i}{2\pi iu_i} \frac{\prod_{j=1}^{N_f}\prod_{i=1}^{N}\Gamma(y_j^{-1}u_i^{-1})\Gamma(\tilde{y}_j u_i)}{\prod_{i<j}\Gamma(u_{ij}^{\pm})}\,  .
\end{equation}
Here $y_{j}$ and $\tilde{y}_{j}$ are defined as
\begin{equation}\label{eq:n=1-sqcd-yi}
    y_j=(pq)^{-\frac{r}{2}}bv_j\,,\quad \tilde{y}_j=(pq)^{\frac{r}{2}}b\tilde{v}_j\,,\quad r=1-\frac{N}{N_f}
\end{equation}
where $v_j$ and $\tilde{v}_j$ are eigenvalues for the $SU(N_f)\times SU(N_f)$ flavor symmetries, and $b$ is an element of the $U(1)_B$ baryonic symmetry.
The specific value of the $U(1)_R$-charge is fixed by anomaly cancellation. 
Note that
\begin{equation}\label{eq:yj-constraints}
    Y\equiv \prod_j y_j^{-1}=b^{-N_f}(pq)^{N_f-N}\,,\quad \tilde{Y} \equiv \prod_j \tilde{y}_j=b^{N_f}(pq)^{N_f-N}\,.
\end{equation}
The general statement of Seiberg duality, even outside the conformal window, corresponds to a highly non-trivial mathematical identity \cite{Rains:2003teh} between the integral in \eqref{eq:N=1-SUN-index-defn} and the index of the magnetic dual theory. 
The latter can be written as \cite{Dolan:2008qi,Spiridonov:2009za}
\begin{align}\label{eq:N=1-SUN-index-magn}
    \begin{split}
        &\cI_M(y_j,\tilde{y}_j;p,q)=\prod_{1\leq i,j\leq N_f}\Gamma(y_i/\tilde{y}_j) \\
        &\times \frac{(p;p)^{\tilde{N}-1}_\infty(q;q)_{\infty}^{\tilde{N}-1}}{\tilde{N}!}\oint\prod_{i=1}^{\tilde{N}-1}\frac{d\tilde{u}_i}{2\pi i\tilde{u}_i} \frac{\prod_{j=1}^{N_f}\prod_{i=1}^{\tilde{N}}\Gamma(Y^{\frac{1}{\tilde{N}}}y_j^{-1}\tilde{u}_i)\Gamma(\tilde{Y}^{-\frac{1}{\tilde{N}}}\tilde{y}_j \tilde{u}_i^{-1})}{\prod_{i<j}\Gamma(\tilde{u}_{ij}^{\pm})}\,  ,
    \end{split}
\end{align}
where $\tilde{N}=N_f-N$.
For $N=2$ and $N_f=3$, the equality between the electric and magnetic index is known as Spiridonov's elliptic beta integral, mentioned in the introduction.

Using the reduced measure for the singlet projection \eqref{eq:vandermonde-replacement}, we can also express the electric index as
\begin{equation}\label{eq:N=1-SUN-index-defn-red}
    \cI_E(y_j,\tilde{y}_j;p,q) = (p;p)^{N-1}_\infty(q;q)_{\infty}^{N-1}\oint\prod_{i=1}^{N-1}\frac{du_i}{2\pi iu_i} \frac{\prod_{j=1}^{N_f}\prod_{i=1}^{N}\Gamma(y_j^{-1}u_i^{-1})\Gamma(\tilde{y}_j u_i)}{\prod_{i<j}(1-u_{ij}^{-1})\Gamma(u_{ij}^{\pm})}\,  ,
\end{equation}
and similarly for the magnetic index.

\section{Convergent residue sums}\label{sec:conv-res-sums}

As reviewed in the previous section, the superconformal indices of gauge theories can be expressed as multiple contour integrals in the space of complexified gauge fugacities, with the contour being a product of unit circles.
The integrands, consisting of elliptic Gamma functions (for the full index) or infinite $q$-Pochhammer symbols (for the Macdonald index), are meromorphic functions of the gauge fugacities.
It is thus natural to attempt an evaluation of the index as a residue sum.
Using the infinite product formulas for these functions (see Appendix \ref{app:spec-functions}), it is straightforward to list all the poles of either integrand, inside or outside the unit circles.
These poles come in infinite towers and accumulate at the origin or infinity, respectively.

However, as already alluded to in the introduction, there are obstructions to a naive evaluation.
In particular, we will demonstrate that if one defines an $SU(N)$ gauge theory index with the \emph{full} measure
\begin{equation}\label{eq:full-vandermonde}
    \mathcal{I}=\frac{1}{N!}\prod^{N-1}_{i=1}\oint_{|u_i|=1}\frac{du_i}{2\pi iu_i} \left|\Delta(u)\right|^2 f(u)\,,
\end{equation}
the residue sum cannot in general be made sense of.\footnote{An exception to this turns out to be the Macdonald index of $\mathcal{N}=2$ superconformal SQCD when a specific condition on the fugacities $(q,t)$ is satisfied, as we explain at the end of Section \ref{ssec:basic-hypergeom-integrals}.}
This is closely related to the accumulation of poles and the presence of an essential singularity at the origin $u_i=0$ for each $i$.
Technically, the singularity at the origin is thus a \emph{non--isolated} essential singularity.
Let us now state more clearly why the residue sum associated with these contour integrals cannot be made sense of.
Consider a decomposition of a unit circle contour into two pieces:
\begin{equation}\label{eq:contour-decomp}
    \oint_{|u_i|=1}(\cdots)= \oint_{|u_i|<\epsilon}(\cdots) +\oint_{\epsilon\leq|u_i|\leq 1}(\cdots) \,,
\end{equation}
with $\epsilon>0$.
In the limit that $\epsilon\to 0$, in which case the first contour becomes increasingly tightly wrapped around the origin, the number of poles inside the second contour will go to infinity for all the integrals of our interest.
We will show that the corresponding residue sum does not converge.
This divergence has to be tamed by the contribution of the first contour. 
However, this contribution cannot be evaluated directly and, effectively, we cannot proceed.
In the following, we will somewhat loosely refer to the contribution of the first contour in \eqref{eq:contour-decomp} for $\epsilon\to0$ as the ``contribution from the origin'', or $I_0$, by which we really mean
\begin{equation}\label{eq:I0-defn}
    I_0\equiv \lim_{\epsilon\to 0}\oint_{|u|<\epsilon}(\cdots)\,.
\end{equation}
On the other hand, if we evaluate the gauge integral using the \emph{reduced} measure 
\begin{equation}\label{eq:reduced-vandermonde}
    \mathcal{I}=\prod^{N-1}_{i=1}\oint_{|u_i|=1}\frac{du_i}{2\pi iu_i}\prod_{1\leq i<j\leq N}(1-u_{ij}) f(u)\,,
\end{equation}
we show that the residue sum associated to the second contour in \eqref{eq:contour-decomp} converges and that the ``contribution from the origin'' vanishes!\footnote{More precisely, there are two choices of reduced measure in \eqref{eq:vandermonde-replacement}.
Using the above yields a convergent residue sum when closing the contour inside the unit circles, with vanishing residue at the origin.
Using the other will yield a convergent residue sum when closing the contour outside the unit circles, with vanishing residue at infinity.}
This gives rise to a closed form formula for the superconformal index. 
As we will see in due course, it complements various other known closed form formulas.

\subsection{Elliptic integrals}\label{ssec:ell-integrals}

The key features described above can already be demonstrated in the context of the Schur index (defined in Section \ref{ssec:defns}).
Since this case is technically the simplest, it will serve as a useful warm-up for the Macdonald and full superconformal index.

We can write the Schur index of a general $\mathcal{N}=2$ gauge theory schematically as
\begin{equation}
    \mathcal{I}(q)=\prod^{r}_{i=1}\oint_{|u_i|=1}\frac{du_i}{2\pi iu_i}I(u;q)\,,
\end{equation}
with $r$, the rank of the gauge group, and we suppress the dependence on potential flavor fugacities.
For now, we take $I(u;q)$ to contain the full measure \eqref{eq:full-vandermonde}.

A key property of the Schur index of an $\mathcal{N}=2$ super\emph{conformal} gauge theory is that $I(u;q)$ is an elliptic function of the gauge chemical potentials \cite{Razamat:2012uv}.
In terms of the gauge fugacities $u_i$, this means that $I(u;q)$ depends only on integer powers of $u_i$, and is invariant under $u_i\to q u_i$, for each $u_i$, separately.
When evaluating the contour integrals,\footnote{The most straightforward way to evaluate multiple contour integrals is to pick up residues inside a unit circle $|u_i|=1$, keeping the other $u_j$ fixed on their respective unit circles, and continue the process until no integrals are left. Some care is required when poles are located on a unit circle. This turns out to happen generically for all but the first integral when evaluating the index of rank $>1$ gauge theories. We propose and test a prescription for such poles in Section \ref{ssec:su3mcd}.} this property of $I(u;q)$ implies the following \cite{Razamat:2012uv}.
Firstly, the poles of the integrand are completely determined by its poles in a single annulus, $|q|^{k_i+1}\leq |u_i|<|q|^{k_i}$, for each $i$ and some $k_i\in \mathbb{Z}$, where we recall that $|q|<1$. 
Secondly, the sum of residues in a given annulus vanishes for a meromorphic, elliptic function.
Finally, ellipticity implies that poles accumulate at the origin $u_i=0$. 
These three points together suggest that, in terms of a residue evaluation, the entire Schur index is encoded by a contribution from the origin. But, due to the nature of the singularity there, it is not clear how to evaluate it.
To explicate these three points, let us work for concreteness with the index of the $\mathcal{N}=4$ $SU(N)$ theory. However, the arguments only require ellipticity, and therefore apply to general $\mathcal{N}=2$ superconformal gauge theories as well.

The Schur index of the $\mathcal{N}=4$ $SU(N)$ theory may be obtained from either \eqref{eq:full-index-suN-n=4-defn}, or \eqref{eq:mcd-index-suN-n=4-defn} by setting $t=q$. 
Since the $\mathcal{N}=4$ theory only contains adjoint-valued fields, it turns out to be convenient to change integration variables from the $SU(N)$ eigenvalues $u_{i}$ to the variables $s_{i}$ introduced at the end of Section \ref{ssec:sc-gauge-theory-inds}.
In terms of these variables, we have
\begin{align}\label{eq:schur-index-suN-n=4-full}
    \begin{split}
        \mathcal{I}_N(v;q)=&\frac{1}{N!}\frac{(q;q)^{2(N-1)}_{\infty}}{\theta_q(q^{\frac{1}{2}}v)^{N-1}}\prod^{N-1}_{i=1}\oint_{|s_i|=1}\frac{ds_i}{2\pi is_i}\prod_{1\leq i\leq j\le N-1}\frac{\theta_q(s_{i,j}^{\pm})}{\theta_q(q^{\frac{1}{2}}vs_{i,j}^{\pm})}
    \end{split}
\end{align}
where $\theta_q(x)=(x;q)_\infty(qx^{-1};q)_\infty$ is the reduced Jacobi or $q$-theta function (see Appendix \ref{sapp:q-poch}), and we use the shorthand notation $\theta_q(x^{\pm})=\theta_q(x)\theta_q(x^{-1})$.
We also note that the full measure, written in terms of the $s_i$ variables, is contained in the numerator of the integrand.
Finally, we recall the definition of $s_{i,j}=s_i\cdots s_j$ for $1\leq i\leq j\leq N-1$.

Two distinguishing features of the $s_i$ variables, as compared to the $u_i$ variables, are that the integrand does not depend on the $N^{\text{th}}$ variable ($s_N$) and that the Weyl symmetry is not manifest.
The latter suggests a preferred order of integration.
As it turns out, a descending order from $s_{N-1}$ has the useful feature that after each integral, the remaining integrand still depends on the $s_i$ only through the combinations $s_{i,j}$.\footnote{Alternatively but equivalently, one could take an ascending order from $s_1$.}
In the following, we stick to this order of integration.

Using the quasi-periodic property of $\theta_q(x)$ under $x\to qx$,\footnote{Collected in Appendix \ref{sapp:q-poch}.} it is easy to check that the integrand is invariant under $s_i\to qs_i$, for any $i$.
Let us now demonstrate the issues mentioned above explicitly for the first ($s_{N-1}$) integral.
To this end, we separate out the $s_{N-1}$ dependence in \eqref{eq:schur-index-suN-n=4-full}.
Writing the full integrand as $I_N(s_i)$, we have
\begin{align}\label{eq:schur-index-suN-n=4-recurs-struc}
    \begin{split}
        \mathcal{I}_N&(v;q,t)=\prod^{N-1}_{i=1}\oint\frac{ds_i}{2\pi is_i}I_N(s_{1},\ldots,s_{N-1})\\
        &=\prod^{N-2}_{i=1}\oint\frac{ds_i}{2\pi is_i} I_{N-1}(s_{1},\ldots,s_{N-2})
        \frac{(q;q)^{2}_{\infty}}{N\theta_q(q^{\frac{1}{2}}v)}\oint\frac{ds_{N-1}}{2\pi is_{N-1}}\prod^{N-1}_{i=1}\frac{\theta_q(s_{i,N-1}^{\pm})}{\theta_q(q^{\frac{1}{2}}vs_{i,N-1}^{\pm})}\,.
    \end{split}
\end{align}
Using the fact that $\theta_q(q^n)=0$, for any $n\in\mathbb{Z}$, we see that the integrand has infinite towers of poles.
Inside the $s_{N-1}$ unit circle, keeping the remaining $|s_{i}|=1$ fixed, the poles lie at
\begin{equation}\label{eq:pole-tower-suN-n=4-schur}
    s^{(j,k)}_{N-1}=v^{\pm}q^{k+\frac{1}{2}}s_{j,N-2}^{-1}\,,\quad j=1,\ldots,N-1\,,
\end{equation}
where $k\in \mathbb{Z}_{\geq 0}$, and we define $s_{N-1,N-2}\equiv 1$.
Clearly, these poles accumulate at $s_{N-1}=0$.
To make matters worse, the integrand has an essential singularity at $s_{N-1}=0$, as can be seen for example from the definition of $\theta_q(x)$ in \eqref{eq:q-theta-defn}.

We can now directly observe the previously mentioned difficulties.
First, ellipticity of the integrand in $s_{N-1}$ implies that the residues associated with the poles in \eqref{eq:pole-tower-suN-n=4-schur} do not depend on $k$.
If we were to sum the residues over $k$ first, we would thus obtain a divergent residue sum.
This divergence can be avoided by summing the residues over $j$ first, \textit{i.e.}, summing over the residues in an annulus $|q|^{k+1}\leq |s_{N-1}|< |q|^{k}$, for fixed $k$, which gives a vanishing answer.\footnote{We mention the former option since it will relate more directly to the Macdonald and full superconformal index cases, where the residues in a given annulus do not sum to zero.}
It follows that the Schur index is determined by a contribution from the origin.
But, this brings in our second issue, namely, that the poles in \eqref{eq:pole-tower-suN-n=4-schur} accumulate at the origin, such that the integrand has a non-isolated essential singularity at $s_{N-1}=0$.
Any subsequent integral has the same issue, since the integrand is elliptic in all $s_i$.

Let us rephrase the above argument, for why the ``contribution from the origin'', as defined in general by $I_0$ in \eqref{eq:I0-defn}, is non-vanishing, in a manner that will generalize to the Macdonald and full superconformal index, whose integrands are not elliptic.\footnote{The following argument is adapted from Section 4.10 of \cite{Gasper_Rahman_2004}.}
We focus on the $s_{N-1}$ integral, but the argument extends to each subsequent integral as well.
First, note that an integral on a contour $K$ is upper bounded by a Cauchy-Schwarz inequality
\begin{equation}
    \left|\oint_{K}dx\,I(x)\right|\leq \oint_{K}dx\left|I(x)\right|\,.
\end{equation}
Let us write the integrand of the $s_{N-1}$ integration in the second line of \eqref{eq:schur-index-suN-n=4-recurs-struc} as $\tilde{I}(s_{N-1})$, where we suppress the dependence on all other variables.
Now, let $C_{k}$ be the contour $|s_{N-1}|=\d|q|^{k}$, with $k\geq 0$, and $\delta$ an $\mathcal{O}(1)$ number chosen, such that $C_{k}$ does not pass through any of the poles of $\tilde{I}(s_{N-1})$.
It follows that the leading term in the integrand for small $|q|$ is of order
\begin{align}\label{eq:estimate-integrand-schur-index-full-measure}
    \begin{split}
        \left|\tilde{I}(\delta q^{k})\right|&=\left|\frac{1}{\delta q^{k}}\right|\left|\prod^{N-1}_{i=1}\frac{\theta_q((\delta s_{i,N-2})^{\pm})}{\theta_q(q^{\frac{1}{2}}v(\delta s_{i,N-2})^{\pm})}\right|=\mathcal{O}(|q|^{-k})\,,
    \end{split}
\end{align}
where we used the elliptic properties of the theta functions and the fact that $|\theta_q(x)|=\mathcal{O}(1)$ for small $|q|$, as long as $|q|<|x|<1$.
Since $C_k$ has a circumference of order $\mathcal{O}(|q|^k)$, we find that the upper bound on the contour integral is non-vanishing and of order
\begin{equation}\label{eq:estimate-integral-schur-index-full-measure}
    \left|\oint_{C_{k}}ds_{N-1}\tilde{I}(s_{N-1})\right|\leq \oint_{C_{k}}ds_{N-1}\left|\tilde{I}(s_{N-1})\right|=\mathcal{O}(1)\,,
\end{equation}
to leading order in the $q$--expansion.
In the limit $k\to \infty$, the contour becomes increasingly tightly wound around the origin, like the first contour in \eqref{eq:contour-decomp} when $\epsilon\to 0$.
Since the bound is independent of $k$, we cannot conclude that the contribution of the origin, \textit{i.e.}, $I_0$, is vanishing, as expected.

To the best of our knowledge, these issues with the residue sum have been addressed head--on only in \cite{Pan:2021mrw}, by expanding the integrand in terms of the so-called Weierstrass $\zeta$--function.\footnote{There is a variety of more indirect ways to evaluate the Schur index of $\mathcal{N}=2$ SCFTs, including the TQFT method of \cite{Gadde:2011ik,Gadde:2011uv}, the Fermi gas method \cite{Bourdier:2015wda,Bourdier:2015sga} (see also \cite{Hatsuda:2022xdv}) and the related giant graviton expansion of the Schur index \cite{Arai:2020qaj,Gaiotto:2021xce,Hatsuda:2024uwt}, an IR wall-crossing formula \cite{Cordova:2015nma}, and methods based on the connection with VOA characters and modular linear differential equations (MLDEs) \cite{Beem:2017ooy,Beem:2021zvt,Huang:2022bry,Zheng:2022zkm}.}
We now show that the reduced measure provides perhaps the most direct way of addressing the issues, with the added benefit that the method generalizes to the Macdonald and full superconformal index as well.

With the reduced measure, the index \eqref{eq:schur-index-suN-n=4-full} is expressed as
\begin{align}\label{eq:schur-index-suN-n=4-red}
    \begin{split}
        \mathcal{I}_N(v;q)= \frac{(q;q)^{2(N-1)}_{\infty}}{\theta_q(q^{\frac{1}{2}}v)^{N-1}}\prod^{N-1}_{i=1}\oint_{|s_i|=1}\frac{ds_i}{2\pi is_i}\prod_{1\leq i\leq j\le N-1}\frac{\theta_q(s_{i,j}^{\pm})}{(1-s_{i,j}^{-1})\theta_q(q^{\frac{1}{2}}vs_{i,j}^{\pm})}\,.
    \end{split}
\end{align}
Note that the modification of the integrand does not introduce new poles, since the additional factors in the denominator cancel against factors in the numerator.
We find it convenient, however, to keep the theta functions intact and thus keep the explicit factors in the denominator.
It follows that the $s_{N-1}$ integrand still has the infinite towers of poles $s^{(j,k)}_{N-1}$ as listed in \eqref{eq:pole-tower-suN-n=4-schur}.
However, the modification of the integrand spoils ellipticity, which implies two things: the residues in a given tower now depend on $k$, and the sum of residues in an annulus no longer vanishes.

To study the $k$-dependence of a residue, let us consider the $j^{\text{th}}$ tower in \eqref{eq:pole-tower-suN-n=4-schur}.
All $k$-dependence comes from the additional, non-elliptic factors in the denominator.
Evaluating just the $s_{N-1}$ dependent non-elliptic factors on the pole $s^{(j,k)}_{N-1}$, we find  
\begin{equation}\label{eq:R-n4-suN-non-ell}
    \frac{v^{N-1}q^{(N-1)(k+\frac{1}{2})}}{(vq^{k+\frac{1}{2}}-1)}\prod^{j-1}_{i=1}\frac{1}{(vq^{k+\frac{1}{2}}-s_{i,j-1}^{-1})}\prod^{N-1}_{i=j+1}\frac{1}{(vq^{k+\frac{1}{2}}-s_{j,i-1})}\,.
\end{equation}
Assuming for now that $|s_{1,\ldots,N-2}|=1$, we see that the $k^{\text{th}}$ residue is suppressed by a factor of order $\mathcal{O}(|q^{(N-1)(k+\frac{1}{2})}|)$.
This suppression ensures a convergent residue sum, at least for the $s_{N-1}$ integral.
We can also estimate the ``contribution from the origin'', as we did above for the full measure.
Separating out the $s_{N-1}$ dependent part in \eqref{eq:schur-index-suN-n=4-red},  which we call $\tilde{I}^{\text{red}}(s_{N-1})$, we can follow the same steps leading to \eqref{eq:estimate-integral-schur-index-full-measure} to find
\begin{equation}\label{eq:estimate-integral-schur-index-red-measure}
    \left|\oint_{C_{k}}ds_{N-1}\tilde{I}^{\text{red}}(s_{N-1})\right|\leq\oint_{C_{k}}ds_{N-1}\left|\tilde{I}^{\text{red}}(s_{N-1})\right|=\mathcal{O}(|q|^{(N-1)(k+\frac{1}{2})})\,,
\end{equation}
where the right hand side reflects the leading order of the integral in the $q$--expansion.
Taking $k\to \infty$, it follows that the contribution of the origin is vanishing.\footnote{Similar arguments have appeared in the context of the 3d superconformal index in \cite{Hwang:2012jh,Benini:2013yva}. However, in these works the key ingredient to the argument appears to be the presence of fundamental matter and a non-zero Fayet-Iliopoulos (FI) parameter. Such a parameter can only be turned on when the gauge group contains a $U(1)$ factor. In our case, we see that the reduced measure plays a role similar to the FI parameter in their case, but also works for $SU(N)$ gauge groups without fundamental matter.}
We thus observe the close connection between the convergence of a residue sum associated with an infinite tower of poles accumulating at the origin, and the vanishing of a potential contribution from the origin.

Let us also analyze convergence properties of the residue sum for the subsequent integration, over $s_{N-2}$, again keeping fixed all other $s_i$, at $|s_i|=1$.
To this end, we write the index after the $s_{N-1}$ integration in the schematic form
\begin{equation}\label{eq:schurindexsingleintegral}
    \mathcal{I}_N(v;q,t)=\prod^{N-2}_{i=1}\oint\frac{ds_i}{2\pi is_i} I^{\text{red}}_{N-1}(s_{1},\ldots,s_{N-2})
       \sum^{N-1}_{j=1}\left(R_{N-1}^{(j)}(v;s)+R_{N-1}^{(j)}(v^{-1};s)\right)\,,
\end{equation}
where $I^{\text{red}}_{N-1}(s_{1},\ldots,s_{N-2})$ is the part of the integrand in \eqref{eq:schur-index-suN-n=4-red}, which does not depend on $s_{N-1}$.
Furthermore, $R_{N-1}^{(j)}(v;s)$ is the residue sum of the $j^{\text{th}}$ tower in \eqref{eq:pole-tower-suN-n=4-schur}, with positive power of $v$, we abbreviated its dependence on the remaining $s_{1,\ldots,N-2}$ by $s$, and suppress its dependence on $q$.

Similar to the case of the $s_{N-1}$ integral, the poles originate from $s_{N-2}$--dependent theta functions.
They also come in infinite towers: $s^{(j',l)}_{N-2}\sim q^l$, with $l\in \mathbb{Z}_{\geq 0}$, and now labeled by $j'=1,\ldots,N-2$.
Since we only wish to demonstrate convergence for the second residue sum, we do not keep track of the detailed pole locations, leaving a full classification of poles to future work \cite{Leuven26}.
To this end, because the combined theta functions are an elliptic function of $s_{N-2}$, the only $l$--dependence of the residues results from the non-elliptic factors in the integrand.
For the present integrand, such factors appear in both $I^{\text{red}}_{N-1}(s_{1},\ldots,s_{N-2})$ and $R_{N-1}^{(j)}(v^{\pm};s)$.

Following a similar analysis as the one leading to \eqref{eq:R-n4-suN-non-ell}, we find that the $(k,l)^{\text{th}}$ residue of the combined $s_{N-1}$ and $s_{N-2}$ integrals has a leading term in the $q$--expansion of order
\begin{align}\label{eq:suN-n4-suppression-N-2}
    \begin{split}
        \begin{cases}
            \mathcal{O}\left(|q^{(N-1)k}q^{(N-2)l}q^{-\min(k,l)}|\right) \quad &\text{for}\quad j=1,\ldots,N-2\\
            \mathcal{O}\left(|q^{(N-1)k}q^{2(N-2)l}|\right) \quad &\text{for}\quad j=N-1
        \end{cases} \,,
    \end{split}
\end{align}
where we ignore unimportant factors of $q^{\frac{1}{2}}$.
Here, the dependence on $\min(k,l)$ in the first line is due to a factor in the non-elliptic part of $R_{N-1}^{(j)}(v^{\pm};s)$.
The fact that a negative power of $q$ arises, is a consequence of the assumption that all other $|s_i|=1$, made just after \eqref{eq:R-n4-suN-non-ell}, which is clearly violated for $s_{N-2}$ when picking up residues.
However, when $N>3$ we see that there are sufficient positive powers of $q$ to ensure convergence for the sum over $k$ and $l$.
To demonstrate convergence for $N=3$, we split the sum over residues as $\sum_{k,l}=\sum_{k\leq l}+\sum_{k>l}$.
For the $k\leq l$, the sum is again clearly convergent. 
For $k>l$, the suppression of the $(k,l)^{\text{th}}$ residue is $q^{2k}$.
Since $|q^{2k}|<|q^{k+l}|$ for $k>l$, convergence follows for this part of the sum as well.

Concluding, we see that the residue sum for the combined $s_{N-1}$ and $s_{N-2}$ integrations is convergent and, consequently, that the contributions at the origin of both the $s_{N-1}$ and $s_{N-2}$ planes vanishes.
We expect this pattern to continue for subsequent integrations, allowing one to evaluate the Schur index of the $\mathcal{N}=4$ $SU(N)$ SYM theory in terms of a residue sum.
In fact, let us stress that our arguments only make use of ellipticity and the reduced measure, and therefore applies to the evaluation of the Schur index of arbitrary $\mathcal{N}=2$ superconformal gauge theories.

\subsubsection{Example: \texorpdfstring{$\mathcal{N}=4$}{N=4} \texorpdfstring{$SU(2)$}{SU(2)} Schur index}

Let us demonstrate the above in the simple setting of an $SU(2)$ gauge group.
The integral we wish to evaluate is
\begin{align}\label{eq:n=4-schur-index-su2-red}
	\begin{split}
		&\mathcal{I}(v;q)=\frac{(q;q)^2_{\infty}}{\theta_q(q^{\frac{1}{2}}v)}\oint_{|s|=1} \frac{ds}{2\pi i s} \frac{\theta_q(s^{\pm})}{(1-s^{-1})\theta_q(q^{\frac{1}{2}}vs^{\pm})}\,,
	\end{split}
\end{align}
The integrand has two towers of poles inside the unit circle. In each annulus there are exactly two poles located at
\begin{equation}
    s^{(k)}_{\pm}=v^{\pm}q^{k+\frac{1}{2}}\,,\quad k\geq 0\,.
\end{equation}
The elliptic part of the residue evaluates for each $k$, to
\begin{equation}
	\text{Res}_{s=s^{(k)}_{\pm}}\left[\frac{1}{s}\frac{\theta_q(s)\theta_q(s^{-1})}{\theta_q(q^{\frac{1}{2}}vs)\theta_q(q^{\frac{1}{2}}vs^{-1})}\right]=\pm q^{-\frac{1}{2}}v\frac{\theta_q(q^{\frac{1}{2}}v)^2}{(q;q)_{\infty}^2\theta_q(v^{2})}\,.
\end{equation}
where we made use of properties of $\theta_q(x)$ collected in Appendix \ref{sapp:q-poch}.
Including the non-elliptic factor 
\begin{equation}
    \frac{1}{1-s^{-1}}=\frac{-s}{1-s}\,,
\end{equation}
we find a simple expression for the Schur index
\begin{align}\label{eq:n=4-schur-index-su2-res-sum}
	\begin{split}
		\mathcal{I}(v;q)=v\frac{\theta_q(q^{\frac{1}{2}}v)}{\theta_q(v^{2})}\sum^{\infty}_{k=0}\left(\frac{v^{-1}q^{k}}{(1-v^{-1}q^{k+\frac{1}{2}})}-\frac{vq^{k}}{(1-vq^{k+\frac{1}{2}})}\right)
	\end{split}
\end{align}
The $k^{\text{th}}$ residue is suppressed by a factor of order $\mathcal{O}(|q^{(N-1)k}|)$ with $N=2$, consistent with our general arguments.
As explained in the main part of this section, this also implies that a contribution from the origin to the residue sum, as defined by $I_0$ in \eqref{eq:I0-defn}, is vanishing.

One may recognize that our result matches the expression obtained in \cite{Pan:2021mrw}, using the Weierstrass $\zeta$--function.
In particular, the twisted Eisenstein series in terms of which \cite{Pan:2021mrw} write their result originates in our calculation from the non-elliptic part of the integrand associated with the reduced measure.
An advantage of our prescription is that it readily generalizes to the Macdonald and full superconformal index, as we will demonstrate in the subsequent sections.

\subsection{Basic hypergeometric integrals}\label{ssec:basic-hypergeom-integrals}

In this section, we extend the arguments of the previous section to the Macdonald index.
Here too, we primarily work in the context of the $\mathcal{N}=4$ $SU(N)$ SYM theory. 
But, at the end of the section we look at $\mathcal{N}=2$ $SU(N)$ superconformal QCD as well. 
Together, these are the two fundamental examples of Lagrangian class $\mathcal{S}$ theories \cite{Gaiotto:2009we}.

As in the case of the Schur index, we first show that the index defined with the full measure gives rise to a divergent residue sum, preventing an argument for the vanishing of the contribution from the origin.
Unlike in the case of the Schur index, the integrand of the Macdonald index is not an elliptic function, and therefore the details of the argument change, but only slightly so.
In particular, we will relate each consecutive integral to basic hypergeometric integrals, as defined in Appendix \ref{app:bas-hypgeom-int}, and study the convergence properties of the residue sums using standard results reviewed in the same appendix.

As in the previous section, we write the Macdonald index of the $\mathcal{N}=4$ theory in terms of the $s_i$ variables.
The expression with the full measure is given in the first line of \eqref{eq:mcd-index-suN-n=4-defn}, which we repeat here for convenience
\begin{align}\label{eq:mcd-index-suN-n=4-full}
    \begin{split}
        \mathcal{I}_N(v;q,t)&=\frac{(q,t;q)^{N-1}_\infty}{N!(t^{\frac{1}{2}}v^{\pm};q)_\infty^{N-1}}\prod^{N-1}_{i=1}\oint\frac{ds_i}{2\pi is_i}\prod_{1\leq i\leq j\leq N-1}\frac{(s_{i,j}^{\pm},ts_{i,j}^{\pm};q)_\infty}{(t^{\frac{1}{2}}v^{\pm}s_{i,j}^{\pm};q)_\infty}\,,
    \end{split}
\end{align}
To understand the convergence properties of the residue sum, and potential ``contributions from the origin'', as defined at the beginning of Section \ref{sec:conv-res-sums}, we start with the $s_{N-1}$ integration, keeping the other $|s_i|=1$.
We separate out the $s_{N-1}$ dependence and, writing the integrand in \eqref{eq:mcd-index-suN-n=4-full} as $I_N(s_i)$, we have
\begin{align}\label{eq:mcd-index-suN-n=4-recurs-struc}
    \begin{split}
        \mathcal{I}_N&(v;q,t)=\prod^{N-1}_{i=1}\oint\frac{ds_i}{2\pi is_i}I_N(s_i)\\
        &=\prod^{N-2}_{i=1}\oint\frac{ds_i}{2\pi is_i} I_{N-1}(s_i)
        \frac{(q,t;q)_\infty}{N(t^{\frac{1}{2}}v^{\pm};q)_\infty}\oint\frac{ds_{N-1}}{2\pi is_{N-1}}\prod^{N-1}_{i=1}\frac{(s_{i,N-1}^{\pm},ts_{i,N-1}^{\pm};q)_\infty}{(t^{\frac{1}{2}}v^{\pm}s_{i,N-1}^{\pm};q)_\infty}\,,
    \end{split}
\end{align}
Recalling that $s_{i,j}=s_i\cdots s_j$, it is easy to see that the $s_{N-1}$ integral is of the form of a basic hypergeometric integral \eqref{eq:gen-basic-hypergeom-integral}.
As in the case of the Schur index, the integrand has infinite towers of poles inside (and outside) the $s_{N-1}$ unit circle, which accumulate at the origin.
Explicitly, they lie at
\begin{equation}\label{eq:pole-tower-suN-n=4-mcd}
    s^{(j,k)}_{N-1}=t^{\frac{1}{2}}v^{\pm}q^{k}s_{j,N-2}^{-1}\,,\quad j=1,\ldots,N-1\,,
\end{equation}
where $k\in \mathbb{Z}_{\geq 0}$, and we recall that $s_{N-1,N-2}\equiv 1$.
As reviewed in Appendix \ref{app:bas-hypgeom-int}, the convergence of the residue sum and, relatedly, the vanishing of a contribution from the origin depend on the parameters in the integrand multiplying $s^{-1}_{N-1}$.
In particular, the $k^\text{th}$ residue in a tower of poles accumulating at the origin will be suppressed by $z^k$, with the suppression factor $z$ given in terms of the parameters of the integrand (see \eqref{eq:suppres-factor-basic-hypergeom-int}) as
\begin{equation}\label{eq:supp-mcd-n4-suN-full}
    z=\prod^{N-1}_{i=1}\lb\frac{ts_{i,N-2}^{-2}}{ts_{i,N-2}^{-2}}\rb=1\,,
\end{equation}
where we recall $s_{N-1,N-2}\equiv 1$.
As explained in the appendix, this implies a divergence of the residue sum associated to the pole tower, and moreover that the ``contribution from the origin'' cannot be argued to vanish.
The fact that $z=1$ with the full measure is a direct generalization of ellipticity of the Schur index integrand defined with the full measure.

We are thus led to study the Macdonald index with the reduced measure.
The index is then expressed as
\begin{align}\label{eq:mcd-index-suN-n=4-red}
    \begin{split}
        \mathcal{I}_N(v;q,t)&=\frac{(q,t;q)^{N-1}_\infty}{(t^{\frac{1}{2}}v^{\pm};q)_\infty^{N-1}}\prod^{N-1}_{i=1}\oint\frac{ds_i}{2\pi is_i}\prod_{1\leq i\leq j\leq N-1}\frac{(s_{i,j},qs_{i,j}^{-1},ts_{i,j}^{\pm};q)_\infty}{(t^{\frac{1}{2}}v^{\pm}s_{i,j}^{\pm};q)_\infty}\,,
    \end{split}
\end{align}
where the numerator of the integrand now only contains the reduced measure.
In this case, there is an additional factors of $q$ multiplying $s_{N-1}^{-1}$, so that the suppression factor $z$ becomes
\begin{equation}
    z=\prod^{N-1}_{i=1}\frac{qts_{i,N-2}^{-2}}{ts_{i,N-2}^{-2}}= q^{N-1}\,.
\end{equation}
Note that this suppression factor is the same as the one for the Schur index when computed with the reduced measure.
It follows that the $s_{N-1}$ residue sum converges, and that the ``contribution from the origin'' vanishes.
For subsequent integrals, one can perform a similar analysis as for the Schur index.
We will not repeat it here, and instead consider the explicit example of an $SU(3)$ gauge group in Section \ref{sec:higher-rank}, leaving general $N$ to future work.

To end this section, we mention the generalization to $\mathcal{N}=2$ $SU(N)$ superconformal QCD.
We repeat the indices with full and reduced measure from Section \ref{ssec:sc-gauge-theory-inds}:
\begin{align}
    \begin{split}
         \mathcal{I}(v_j;p,q,t)&=\frac{(q,t;q)^{N-1}_\infty}{N!}\prod^{N-1}_{i=1}\oint\frac{du_i}{2\pi iu_i}\frac{\prod_{1\leq i<j\leq N}(u_{ij}^{\pm},tu_{ij}^{\pm};q)_\infty}{\prod^{2N}_{j=1}\prod^{N}_{i=1}(v_ju_{i})^{\pm};q)_\infty}\\
        &=(q,t;q)^{N-1}_\infty\prod^{N-1}_{i=1}\oint\frac{du_i}{2\pi iu_i}\frac{\prod_{1\leq i<j\leq N}(u_{ij},qu_{ij}^{-1},tu_{ij}^{\pm};q)_\infty}{\prod^{2N}_{j=1}\prod^{N}_{i=1}(t^{\frac{1}{2}}(v_ju_{i})^{\pm};q)_\infty}
    \end{split}
\end{align}
Since the only poles come from operators in the fundamental representation, it is not convenient to change to $s_i$ variables.
Keeping in mind the $SU(N)$ condition $u_N=(u_1\cdots u_{N-1})^{-1}$, we separate out the $u_{N-1}$--dependent part of the integrand.
Using that 
\begin{equation}
    (x^2;q)_\I=(x,-x,q^{\frac{1}{2}}x,-q^{\frac{1}{2}}x;q)_\I\,,
\end{equation}  
one sees that the $u_{N-1}$ integral is of the basic hypergeometric form.
Therefore, we can readily calculate the suppression factors for both cases, using the general result \eqref{eq:suppres-factor-basic-hypergeom-int}.
With the full measure, we have $m=n=4N$, and the suppression factor becomes
\begin{equation}
    z=\frac{q^2t^{2(N-1)}(u_Nu_{N-1}^{-1})^{2N}}{t^{2N}(u_Nu_{N-1}^{-1})^{2N}}=\lb\frac{q}{t}\rb^{2}\,,
\end{equation}
If one would take $|t|=|q|$ in the Macdonald index, which is natural for the Schur index specialization $t=q$, we see that the residue sum is not guaranteed to converge and we cannot argue that the contribution from the origin vanishes.
On the other hand, assuming that $|q|<|t|$, the residue sum will converge and the contribution from the origin will vanish.

With the reduced measure we find the suppression factor
\begin{equation}
    z=\frac{q^{N+2}t^{2(N-1)}(u_Nu_{N-1}^{-1})^{2N}}{t^{2N}(u_Nu_{N-1}^{-1})^{2N}}=q^{N}\lb\frac{q}{t}\rb^{2}\,,
\end{equation}
Therefore, in this case the residue sum will be convergent even for $|t|=|q|$ and the contribution from the origin will vanish.
This feature is expected to hold for any subsequent integration as well.
In Section \ref{ssec:nf4-indices}, we consider both the calculation with the reduced measure and with the full measure when $|q|<|t|$.
This will allow us to derive an interesting identity between two distinct residue sums whenever $|q|<|t|$.

\subsection{Elliptic hypergeometric integrals}\label{ssec:ell-hypergeom-integrals}

In this section, we argue that the reduced measure can also be used to evaluate the full superconformal index.
Technically, this requires us to evaluate (multivariate) elliptic hypergeometric integrals.
Such integrals have only been evaluated in closed form in very specific cases, and not through the method of residues.
For example, as reviewed in Section \ref{ssec:n=1-gauge-theory-inds}, the $\mathcal{N}=1$ SQCD index can be evaluated in closed form when it has $SU(N)$ gauge group, with $N_f=N+1$ fundamentals, such that the Seiberg dual theory is free.
However, there appears to be no complete residue evaluation of these and more general cases in the literature, despite some (relatively recent) attempts \cite{Peelaers:2014ima,Goldstein:2020yvj,spiri2024}.
We describe the issues with these attempts and a convergent residue evaluation of general classes of univariate elliptic hypergeometric integrals in Appendix \ref{app:ell-hypgeom-int}.
In Section \ref{sec:examples-rank-1}, we use these general results to describe in detail the evaluation and convergence of residue sums for the full superconformal indices of the following rank $1$ gauge theories: $\mathcal{N}=4$ $SU(2)$ SYM, and $\mathcal{N}=1,2$ $SU(2)$ SQCD for values of $N_f$ in the conformal window and at $N_f=4$, respectively.

Here, we present a (very) schematic overview of how one would go about the higher rank case. 
As in the preceding sections, we work for concreteness with the $\cN=4$ $SU(N)$ SYM theory.
From \eqref{eq:full-index-suN-n=4-defn}, we collect the $\mathcal{N}=4$ $SU(N)$ index with the full measure and the reduced measure:
\begin{align}\label{eq:full-index-suN-n=4-def-2}
    \begin{split}
        \mathcal{I}(v;p,q,t)&=\frac{\left[\kappa\Gamma(t^{\frac{1}{2}}v^{\pm})\Gamma\left(\tfrac{pq}{t}\right)\right]^{N-1}}{N!}\prod^{N-1}_{i=1}\oint\frac{ds_i}{2\pi is_i} \prod_{1\leq i\leq j\leq N-1} \frac{\G(t^{\frac{1}{2}}v^\pm s_{i,j}^\pm)\G(\frac{pq}t s_{i,j}^\pm)}{\G(s_{i,j}^\pm)}\\
        &=\left[\kappa\Gamma(t^{\frac{1}{2}}v^{\pm})\Gamma\left(\tfrac{pq}{t}\right)\right]^{N-1}\prod^{N-1}_{i=1}\oint\frac{ds_i}{2\pi is_i} \prod_{1\leq i\leq j\leq N-1} \frac{\G(t^{\frac{1}{2}}v^\pm s_{i,j}^\pm)\G(\frac{pq}t s_{i,j}^\pm)}{(1-s_{i,j}^{-1})\G(s_{i,j}^\pm)}
    \end{split}
\end{align}
As in the previous sections, we focus on the $s_{N-1}$ integration.
The relevant part of \eqref{eq:full-index-suN-n=4-def-2} is then given by
\begin{equation}\label{eq:full-index-suN-n=4-s_N-1-integral}
     \frac{\kappa\Gamma(t^{\frac{1}{2}}v^{\pm})\Gamma\left(\tfrac{pq}{t}\right)}{N^{1-\e}} \oint \frac{ds_{N-1}}{2\pi is_{N-1}} \prod_{i=1}^{N-1} \frac{\G(t^{\frac{1}{2}}v^\pm s_{i,N-1}^\pm) \G(\frac{pq}t s_{i,N-1}^\pm)}{(1-s_{i,N-1}^{-1})^\e\:\G(s_{i,N-1}^\pm)}  
\end{equation}
where $\epsilon=0$ corresponds to the full measure, and $\epsilon=1$ to the reduced measure.
The $s_{N-1}$ integral is a generalized elliptic hypergeometric integral, as defined and analyzed in  Appendix \ref{ssec:generalized-EHI}.
In contrast with the previous sections, poles now come in doubly infinite towers, as can be seen from the product definition of the elliptic Gamma function.
For the integrand at hand, the only poles arise from the elliptic Gamma functions in the numerator, and are located at:
\begin{equation}
    s^{(j,k,l)}_{N-1}=y_ap^kq^{l}s_{j,N-2}^{-1}\,,\quad k,l\in \mathbb{Z}_{\geq 0}\,,\quad j=1,\ldots,N-1\,, 
\end{equation}
where $y_1=t^{\frac{1}{2}}v$, $y_2=t^{\frac{1}{2}}v^{-1}$ and $y_3=\frac{pq}{t}$.
The poles will be simple for generic values of the parameters.
Due to the accumulation of poles at the origin $s_{N-1}=0$, the integrand has a non-isolated essential singularity at the origin, as should be familiar by now.

As described in detail in Appendix \ref{ssec:generalized-EHI}, the integral \eqref{eq:full-index-suN-n=4-s_N-1-integral} can be explicitly evaluated as a residue sum with the schematic form 
\begin{equation}
    \sum_{\a=1}^{3(N-1)}\sum_{k,l=0}^\I\text{residue}_{\a,k,l}\,,
\end{equation}
where the $\a$ index combines the $(a,j)$ indices labeling the poles.
To establish the convergence of this residue sum, and the vanishing of a contribution from the origin, we perform a ratio test.
This requires significantly more care than for the residue sums of basic hypergeometric integrals. 
In particular, in order to concretely test for convergence, one has to specify the relative magnitudes of the fugacities as an \emph{input} to the analysis.\footnote{For basic hypergeometric integrals, there can also be conditions on the relative sizes of the fugacities, see, \textit{e.g.}, the example of the SQCD Macdonald index in Section \ref{ssec:basic-hypergeom-integrals}. However, these conditions arise as an \emph{output} of the calculation. This distinction can be traced to the fact that the residue sum of a basic hypergeometric integral depends on $q$-shifted factorials $(x;q)_k$, while the residue sum of elliptic hypergeometric integrals depends on elliptic shifted factorials $\theta_q(x;p)_k$. We refer to Appendix \ref{app:ell-hypgeom-int} for a more detailed discussion.}
As an example, we take $|p|=|q|= T^3$ and $|t|= T^4$ for $T<1$, but otherwise generic, and $|s_{1,\ldots,N-2}|=|v|=1$.\footnote{This choice ensures a symmetric scaling of the $y_a\sim T^2$. We expect that there are other ways to scale the fugacities for which the residue sum still converges, including asymmetric scalings, but we leave a systematic analysis to future work.}
In particular, in a small $T$--expansion a ratio test leads to
\begin{equation}
    \lv \frac{\text{residue}_{\a,k+1,l}}{\text{residue}_{\a,k,l}} \rv = \mathcal{O}(T^{(N-1)(3\e-1)})\,,
\end{equation}
where we only have to consider the ratio between the $k^{\text{th}}$ and $(k+1)^{\text{th}}$ residues at fixed $l$ due to a symmetry under the simultaneous exchange $k\leftrightarrow l$ and $p\leftrightarrow q$.
This result implies that the residue sum evaluated with the full measure ($\epsilon=0$) is divergent, at least for our choice of scaling of the fugacities, while evaluating with the reduced measure ($\epsilon=1$) makes the sum convergent. 
As for the Schur and Macdonald index, this also implies a vanishing of the ``contribution at the origin''.

To evaluate the full index and establish its convergence, one should now consider the subsequent integrals over $s_{1,\ldots,N-2}$.
We leave a detailed analysis to future work, and here simply conjecture that, with the reduced measure, the subsequent integrations also yield convergent residue sums with vanishing contributions from the origin, as we argued for the Schur and Macdonald indices.

To end this section, let us also briefly consider the index of the $\cN=2$ $SU(N)$ SQCD theory with $N_f=2N$ fundamental flavors with both the full and the reduced measure, given by (\ref{eq:sqcd-index-defn-gen-nc}) and (\ref{eq:sqcd-index-defn-gen-nc-red}) respectively. 
As above, we focus on a single integral, say, the $u_{N-1}$ integral. 
Solving the $SU(N)$ constraint as $u_N=(\fu u_{N-1})^{-1}$, with $\fu=u_1u_2\cdots u_{N-2}$, the $u_{N-1}$ dependent part of the integrand reads
\begin{align}\label{eq:N=2-SQCD-u-N-1}
    &\	\oint\frac{du_{N-1}}{2\pi iu_{N-1}} \frac{\G(\frac{pq}t(\fu u_{N-1}^2)^\pm)}{\G((\fu u_{N-1}^2)\pm)} \lB \prod_{i=1}^{N-2}\frac{\G(\frac{pq}t u_{i}^\pm u_{N-1}^\mp) \G(\frac{pq}t (u_i\fu u_{N-1})^\pm)}{(1-u_i^{-1}u_{N-1})^\e\G(u_i^\pm u_{N-1}^\mp)\G((u_i\fu u_{N-1})^\pm)} \rB   \\
    &\qquad\times \lB \prod_{j=1}^{2N} \G(t^{1/2}v_ju_{N-1})\G(t^{1/2}v_j^{-1}u_{N-1}^{-1})	\rB \lB \prod_{j=1}^{2N} \G(t^{1/2}v_j\fu^{-1}u_{N-1}^{-1})\G(t^{1/2}v_j^{-1}\fu u_{N-1}) \rB   \nn
\end{align}
As in the $\mathcal{N}=4$ case, only the elliptic Gamma functions in the numerator contribute poles.
Assuming all poles to be simple, we again use \eqref{eq:generalized-EHI-evaluation} to obtain a residue sum with the schematic form 
\begin{equation}
    \sum_{\a=1}^{6N+4}\sum_{k,l=0}^\I\text{residue}_{\a,k,l}\,.
\end{equation}
Scaling the fugacities as $|p|=| q|= |t|= T^2$ for some $T<1$, the ratio test yields
\begin{equation}
    \lv \frac{\text{residue}_{\a,k+1,l}}{\text{residue}_{\a,k,l}} \rv =\mathcal{O}( T^{2\e N})\,.
\end{equation}
It follows that convergence of the residue sum is only guaranteed when using the reduced measure, similar to the conclusion for the Schur index.

We conclude that even elliptic hypergeometric integrals are amenable to a residue evaluation, as long as one uses the reduced measure for the gauge singlet projection.
As mentioned at the beginning of this section, we study examples with rank $1$ gauge groups in full detail in Section \ref{sec:examples-rank-1}.
There, we will also comment on the convergence of the residue sums for different, asymmetric scalings of the parameters.

\section{Case study: Macdonald index of \texorpdfstring{$\mathcal{N}=4$}{N=4} \texorpdfstring{$SU(2)$}{SU(2)} SYM}\label{sec:mcd-n=4-su2}

In this section, we derive various closed form formulas for the Macdonald index of the $\mathcal{N}=4$ $SU(2)$ SYM theory.
We then show how the formulas encode various expected features of the operator spectrum at non--zero coupling, how specializations can lead to product formulas for the index and how to interpret such formulas.
Finally, we discuss and interpret the analytic structure and asymptotics of the formulas, and compare to the TQFT formula.

\subsection{Closed form expressions}\label{ssec:derivation}

As explained in Section \ref{ssec:basic-hypergeom-integrals}, when evaluating the $\mathcal{N}=4$ $SU(N)$ Macdonald index through residues, one has to use the reduced measure to guarantee convergence of the residue sum and to argue for the vanishing of a contribution from the origin.
The $SU(2)$ version of the integral in \eqref{eq:mcd-index-suN-n=4-defn} reads
\begin{align}\label{eq:mcd-index-su2-n=4-red}
    \begin{split}
        \mathcal{I}_2(v;q,t)&=\frac{(q,t;q)_\infty}{(t^{\frac{1}{2}}v,t^{\frac{1}{2}}v^{-1};q)_\infty}\oint\frac{ds}{2\pi is}\frac{(s,qs^{-1},ts^{\pm};q)_\infty}{(t^{\frac{1}{2}}vs^{\pm},t^{\frac{1}{2}}v^{-1}s^{\pm};q)_\infty}\,.
    \end{split}
\end{align}
This integral is of the form of the basic hypergeometric contour integral discussed in Appendix \ref{app:bas-hypgeom-int}.
Adapting the general residue sum in \eqref{eq:ressum-gen-basic-hypergeom-integral} to the case at hand, we find
\begin{align}\label{eq:ressum-mcd-index-su2-n=4}
    \begin{split}
        \mathcal{I}_2(v;q,t)    &=\frac{(t^{\frac{3}{2}}v,qt^{-\frac{1}{2}}v^{-1};q)_\I}{(tv^2,v^{-2};q)_\I} {}_{3}\phi_{2}\left[\begin{array}{ccc}
		tv^2 & t & qt^{-\frac{1}{2}}v   \\
		  qv^{2} & t^{\frac{3}{2}}v  & 
	\end{array};q,q\right]+\left(v\leftrightarrow v^{-1}\right)
    \end{split}
\end{align}
where ${}_{r+1}\phi_{r}$ is a basic hypergeometric series defined in Appendix \ref{sapp:basic-hypergeom}.
The two terms in the expression correspond to the residue sums associated with the two towers of poles at $s=t^\frac{1}{2}v^{\pm}q^k$, $k\in \mathbb{Z}_{\geq 0}$.
This provides a new closed form formula for the $\mathcal{N}=4$ $SU(2)$ Macdonald index.\footnote{Recently, a closed form formula for the Macdonald index of the $\mathcal{N}=4$ $U(N)$ SYM theory was obtained in \cite{Hatsuda:2025mvj}, closely related to the ``TQFT'' formulas developed for class $\mathcal{S}$ theories in \cite{Gadde:2011uv}. Factoring out the $U(1)$ part in their expression for $N=2$, one may observe some key distinctions with our formula. For example, unlike in their expression, the $v\leftrightarrow v^{-1}$ symmetry is manifest in ours. In Section \ref{ssec:comparison-tqft} we show in detail how our formula is related to the TQFT formula of \cite{Gadde:2011uv}.}

It would be very interesting to have a direct, physical interpretation of the two terms in our formula, perhaps along the lines of \cite{Peelaers:2014ima}.
For now, however, we note that the expression has some undesirable features:
\begin{itemize}
    \item Each term separately diverges in the unflavored limit $v=1$.\footnote{This subtlety also arises for the expression of the Schur index obtained in \cite{Pan:2021mrw}.}
    This can be traced to the fact that the integrand develops double poles when $v=1$.
    It can be checked that these divergences cancel between the two terms, as they should.
    \item Each term separately contains negative powers of $t$.
    However, the single letter operators contributing to the Macdonald index all have $R+r\geq 0$, and thus only contribute positive powers of $t$.  
    Upon expanding the index as a power series, the terms with negative powers of $t$ cancel, again as they should.
    This redundancy in individual terms is reminiscent of the Bethe Ansatz formula for the full index \cite{Lezcano:2021qbj,Benini:2021ano}.    
\end{itemize}
We now show that these issues can be resolved by making use of transformation formulas of the ${}_{3}\phi_{2}$ series.
We will do so in a few steps.

First, we use the four-term transformation formula \eqref{eq:3phi2-trans-4-term} with $a=t$, $b=qt^{-\frac{1}{2}}v^{-1}$ and $\tilde{b}=qt^{-\frac{1}{2}}v$ to rewrite the residue sum as
\begin{align}\label{eq:ressum-mcd-index-su2-n=4-1}
    \begin{split}
        \mathcal{I}_2(v;q,t)=\frac{(t^{\frac{1}{2}}v,t^{\frac{1}{2}}v^{-1};q)_\I}{(tv^2,v^{-2};q)_\I} \sum_{k=0}^{\infty}\frac{(tv^2,t;q)_{k}}{(qv^2,q;q)_{k}}\frac{q^k}{(1-q^kt^{\frac{1}{2}}v)}+\left(v\leftrightarrow v^{-1}\right)
    \end{split}
\end{align}
where $(x;q)_k$ is the finite $q$-Pochhammer symbol defined in Appendix \ref{sapp:q-poch}.
This expression makes manifest that the index does not depend on negative powers of $t$, and also its reduction to the Schur index, calculated in \eqref{eq:n=4-schur-index-su2-res-sum}, upon taking $t=q$.
However, the apparent poles at $v=1$ remain.

We can further improve the expression by making use of three-term transformation formulas collected in \eqref{eq:3phi2-trans-3-term-2}.
Setting $a$, $b$ and $\tilde{b}$, as above, we obtain three equivalent forms of \eqref{eq:ressum-mcd-index-su2-n=4}:
\begin{align}\label{eq:ressum-mcd-index-su2-n=4-2}
    \begin{split}
        \mathcal{I}_2(v;q,t)&=\frac{(qt^{\frac{1}{2}}v,t^{\frac{3}{2}}v^{-1};q)_{\infty}}{(tv^2,tv^{-2};q)_{\infty}} {}_{3}\phi_{2}\left[\begin{array}{ccc}
		t & qt^{-\frac{1}{2}}v^{-1}  & t^{\frac{1}{2}}v   \\
		t^{\frac{3}{2}}v^{-1}  & qt^{\frac{1}{2}}v  &  \\
	\end{array};q,t\right]\\
    &=\frac{(qt^{\frac{1}{2}}v,qt^{\frac{1}{2}}v^{-1},t^2/q;q)_{\infty}}{(tv^2,tv^{-2},t;q)_{\infty}} {}_{3}\phi_{2}\left[\begin{array}{ccc}
		q & qt^{-\frac{1}{2}}v^{-1}  & qt^{-\frac{1}{2}}v   \\
		qt^{\frac{1}{2}}v  & qt^{\frac{1}{2}}v^{-1}  &  \\
	\end{array};q,\frac{t^2}{q}\right]\\
    &=\frac{(t^{\frac{3}{2}}v,t^{\frac{3}{2}}v^{-1},q;q)_{\infty}}{(tv^2,tv^{-2},t;q)_{\infty}}{}_{3}\phi_{2}\left[\begin{array}{ccc}
		t^2/q & t^{\frac{1}{2}}v^{-1}  & t^{\frac{1}{2}}v   \\
		t^{\frac{3}{2}}v  & t^{\frac{3}{2}}v^{-1}  &  \\
	\end{array};q,q\right]
    \end{split}
\end{align}
The apparent pole at $v=1$ has now completely disappeared so that the unflavored limit is easily taken.
Furthermore, the expression on the third line does not contain negative powers of $t$.

Let us end this section with an arguably simpler version of \eqref{eq:ressum-mcd-index-su2-n=4-2}, upon plugging in the explicit basic hypergeometric series.
In each case, the denominator of the summand in the ${}_{3}\phi_{2}$ series cancels against part of the numerator of the prefactor.
We have for the first line
\begin{align}\label{eq:ressum-mcd-index-su2-n=4-3}
    \begin{split}
        \mathcal{I}_2(v;q,t)&=\frac{1}{(tv^2,tv^{-2};q)_{\infty}} \sum_{k=0}^{\infty}\frac{(q^{k+1}t^{\frac{1}{2}}v,q^kt^{\frac{3}{2}}v^{-1};q)_{\infty}\,(t^{\frac{1}{2}}v,qt^{-\frac{1}{2}}v^{-1},t;q)_{k}}{(q;q)_{k}}\,t^k\,.
    \end{split}
\end{align}
The second and third line of \eqref{eq:ressum-mcd-index-su2-n=4-2} can be simplified similarly.
We now turn to the physical interpretation of these expressions.

\subsection{Operator spectrum from index}\label{ssec:op-spec}

In this section, we demonstrate how our formulas for the Macdonald index can be used to deduce information about operators which remain BPS at non--zero Yang--Mills coupling in the 1/8 BPS Macdonald sector of the $\mathcal{N}=4$ $SU(2)$ theory.

To this end, we first recall the interpretation of the Hall-Littlewood (HL) limit $q\to 0$ of the Macdonald index.
This limit receives non-vanishing contributions from 1/4 BPS operators, as discussed in Section \ref{ssec:defns}, which for Lagrangian theories can be expressed in terms of the letters $\bar{q}_i$ and $\lambda_{1+}$ in Table \ref{tab:n=2-multiplets}.
The limit of the index is conveniently taken in \eqref{eq:ressum-mcd-index-su2-n=4-3} and we find\footnote{This formula can be viewed as a resummation of the TQFT formula of \cite{Gadde:2011uv}, and appeared more recently in \cite{Felder_2018} (see also Appendix B of \cite{Bonetti:2018fqz}), which also contains the generalization to $SU(N)$ gauge group.}
\begin{equation}\label{eq:HL-index-su2-n4}
    \mathcal{I}_2(v;t)=\frac{1+t-t^{\frac{3}{2}}(v+v^{-1})}{(1-tv^2)(1-tv^{-2})}\,.
\end{equation}
This expression has a simple interpretation in terms of the HL chiral ring \cite{Gadde:2011uv}.
The HL chiral ring is generated by three bosonic and two fermionic operators, which can be expressed in terms of the above letters as \cite{Beem:2013sza,Beem:2017ooy}
\begin{equation}\label{eq:su2operators}
    \mu_{A}=\text{Tr}(\bar{q}_{(i_1}\bar{q}_{i_2)})\,,\quad \omega_i=\text{Tr}(\lambda_{1+}\bar{q}_{i})\,.
\end{equation}
Here, $\mu_{A=\pm,0}$ transforms as a triplet and $\omega_i$ as a doublet of the $SU(2)$ ``flavor'' symmetry, \textit{i.e.}, the commutant of the $\mathcal{N}=2$ in the $\mathcal{N}=4$ R-symmetry.
The operator $\mu_A$ corresponds to the chiral primary of a 1/16 BPS multiplet of the $\mathcal{N}=4$ theory, also known as the $S_2$ multiplet \cite{Kinney:2005ej}.
The HL chiral ring operators can be viewed as sitting in a 1/4 BPS truncation of this multiplet.
At large $N$ (and large 't Hooft coupling), the $S_2$ multiplet can be matched with a Kaluza-Klein multiplet of supergravity on $AdS_5\times S^5$, and is therefore sometimes referred to as a (super--)graviton multiplet \cite{Witten:1998qj,Kinney:2005ej}.
We will refer to operators in this multiplet as graviton operators or, as has become common recently, as monotone operators \cite{Chang:2024zqi}.
It is expected that this multiplet is absolutely protected, \textit{i.e.}, remains BPS, for arbitrary Yang--Mills coupling even at finite $N$, although as far as we are aware this has only been proven at tree--level \cite{Grant:2008sk,Chang:2013fba}.

At the level of the (classical) chiral ring, the symmetrization of the scalars in $\mu_A$ follows from the fact their commutator is set to zero by the $F$--term equations \cite{Witten:1998qj,Kinney:2005ej}.
Or, in a cohomological language, BPS operators are in one--to--one correspondence with $\mathcal{Q}$--cohomology classes of operators, and the commutators of the scalars in the $\mathcal{N}=4$ theory turn out to be $\mathcal{Q}$--exact \cite{Grant:2008sk}.
The letters $\bar{q}_{1,2}$ can therefore be treated as mutually diagonalizable matrices.
In fact, it appears that all letters within the Macdonald sector, and therefore the Hall-Littlewood sector, can be treated as mutually diagonalizable, since \emph{any} commutator between the letters is $\mathcal{Q}$--exact \cite{Chang:2013fba,Chang:2023ywj}.\footnote{Another way of stating this is that there appear to be no non--Coulomb graviton (or monotone) operators in the Macdonald sector \cite{Chang:2023ywj}. Recently, however, an example of a non--Coulomb operator was found in the full 1/16 BPS sector of the $\mathcal{N}=4$ theory with $SO(7)$ gauge group \cite{Chang:2025mqp,Choi:2025bhi}. At present, it is not clear whether such operators exist in $SU(N)$ theories as well.}
Using the simultaneous diagonalizability, one can easily verify that the generators in \eqref{eq:su2operators} satisfy four relations among themselves.
We list both generators and relations, and their corresponding indices, in the table below.

\begin{table}[ht]
    \centering
    \bgroup
    \def\arraystretch{1.2}
    \begin{tabular}{|c|c|c|c|c|c|c|c|}
    \hline
         $\mathcal{O}$ & $\mu_{\pm}$ & $\mu_0$ & $\omega_{\pm}$ & $\mathcal{R}$ & $\mu_0^2=\mu_+\mu_-$ & $\mu_0\omega_{\pm}=\mu_{\pm}\omega_{\mp}$ &  $\omega_+\omega_-=0$  \tabularnewline
  \hline
index & $tv^{\pm 2}$  & $t$ & $-t^{\frac{3}{2}}v^{\pm}$ &  & $-t^2$  & $t^{\frac{5}{2}}v^{\pm}$ & $-t^3$    \tabularnewline
\hline 
    \end{tabular}
    \egroup
    \caption{Generators $\mathcal{O}$ and relations $\mathcal{R}$ in the HL chiral ring, with corresponding indices. Note that the relations contribute to the index as constraints and therefore have opposite sign from their statistics.}
    \label{tab:HL-chiral-ring}
\end{table}

The relations (and Fermi statistics) imply that the full ring of operators in the Hall--Littlewood sector $R_{HL}$ is spanned by four sectors, with at most a single power of either $\mu_0$ or $\omega_{\pm}$ and arbitrary powers of $\mu_{\pm}$, \textit{i.e.},
\begin{equation}\label{eq:HL-chiral-ring-solved}
    R_{HL}=\left\lbrace \mu_+^{m}\mu_-^n\left\lbrace1,\, \mu_0,\,\omega_+,\, \omega_- \right\rbrace \; | \;  m,n\geq 0\right\rbrace\,.
\end{equation}
Evaluating the index over these sectors gives precisely the expression in \eqref{eq:HL-index-su2-n4}.
The fact that the index is reproduced solely by operators in the $S_2$ multiplet strongly suggests that this multiplet is indeed absolutely protected and that there are no ``non-graviton operators''\footnote{The terminology non-graviton in the context of $SU(2)$ gauge group refers to any BPS operator which survives at $g_{YM}\neq 0$ and is not contained in the $S_2$ multiplet \cite{Chang:2022mjp,Choi:2022caq,Choi:2023znd,Choi:2023vdm,Chang:2024zqi}.} in this sector, as recently noted in \cite{Gadde:2025yoa} as well.

It is also interesting to take the $q\to 0$ limit in the third line of \eqref{eq:ressum-mcd-index-su2-n=4-3} which leads to an alternative expression for the HL index:\footnote{An almost identical expression appeared before in eqn 6.52 of \cite{Bonetti:2018fqz} in the context of the SCFT/VOA correspondence \cite{Beem:2013sza}.}
\begin{align}\label{eq:HL-index-su2-n4-2}
    \begin{split}
         \mathcal{I}_2(v;t)&=\frac{(1-t^{\frac{3}{2}}v)(1-t^{\frac{3}{2}}v^{-1})-t^2+t^{\frac{5}{2}}(v+v^{-1})-t^3}{(1-tv^2)(1-tv^{-2})(1-t)}\,.
    \end{split}
\end{align}
We recognize the first part of the numerator, together with the denominator, as the HL index if the ring of operators in \eqref{eq:su2operators} were freely generated.
The additional terms in the numerator capture precisely the contributions of the basic relations in Table \ref{tab:HL-chiral-ring}.
That the index does not take on a product form in terms of the generators and basic relations indicates the presence of relations among relations \cite{Benvenuti:2006qr}.
This is unlike the quarter-BPS sector of the theory involving the bosonic operators only, also known as the Higgs branch chiral ring, in which case the numerator would take on the simpler form $1-t^2$ \cite{Feng:2007ur,Grant:2008sk}.

We now attempt a similar interpretation for our expression of the Macdonald index.
To this end, we first note from Table \ref{tab:n=2-multiplets} that the derivatives $\partial\equiv \partial_{+\dot{-}}$ and the letter $\bar{\lambda}_{2\dot{-}}$ should also be included in the construction of operators. 
This sector includes therefore at least the additional generators in the appropriate 1/8 BPS truncation of the $S_2$ multiplet (and their $\partial$ derivatives):\footnote{Assuming that all letters within this truncation of the $S_2$ multiplet can be taken as diagonal matrices, the covariant derivative turns into an ordinary derivative.}
\begin{equation}\label{eq:addgensu2}
    \tilde{\mu}=\text{Tr} (\lambda_{1+}\bar{\lambda}_{2\dot{-}}+\epsilon^{ij}\bar{q}_i\partial \bar{q}_j)\,,\quad \tilde{\omega}_i=\text{Tr}(\bar{\lambda}_{2\dot{-}}\bar{q}_{i})\,,
\end{equation}
where $\tilde{\mu}$ is an $SU(2)$ flavor singlet, and $\tilde{\omega}_i$ transforms as a doublet.
We collect the indices of these operators and the simplest relations in the table below.
The relations can again be verified by taking each letter as a diagonal matrix.

\begin{table}[ht]
    \centering
    \bgroup
    \def\arraystretch{1.2}
    \begin{tabular}{|c|c|c|c|c|c|c|}
     \hline
         $\mathcal{O}$ & $\tilde{\mu}$ & $\tilde{\omega}_\pm$ & $\mathcal{R}$ & $\mu_0\tilde{\omega}_{\pm}=\mu_{\pm}\tilde{\omega}_{\mp}$ &  $\tilde{\omega}_+\tilde{\omega}_-=0$ 
         & $\omega_+\tilde{\omega}_-=\omega_-\tilde{\omega}_+$ \tabularnewline
   \hline
index & $qt$ & $-qt^{\frac{1}{2}}v^{\pm}$ & &  $qt^{\frac{3}{2}}v^{\pm}$ & $-q^2t$  & $-qt^2$ \tabularnewline
 \hline
    \end{tabular}
    \egroup
    \caption{Additional operators an relations contributing to the Macdonald index.}
    \label{tab:mcd-chiral-ring}
\end{table}
Further relations between the operators are given by:
\begin{equation}\label{eq:ad-mcd-relns}
    \begin{aligned}
        \tilde{\mu}^2 &= (\partial \mu_{0})^2 - \partial \mu_{+} \partial \mu_{-} + 2 ( (\partial \omega_{-}) \tilde{\omega}_{+} -  (\partial \omega_{+}) \tilde{\omega}_{-})\,,\\\tilde{\mu}\omega_{\pm}&=\partial(\mu_{\pm}\omega_{\mp})-\mu_0\partial \omega_{\pm}-\omega_{\mp}\partial\mu_\pm\\
        \omega_{\pm}\tilde{\omega}_{\pm}&=\mu_\pm\tilde{\mu}+\mu_0\partial\mu_\pm-\mu_\pm\partial \mu_0\,, \\
        \omega_{\pm} \tilde{\omega}_{\mp}& = \mu_{0} \tilde{\mu} - \tfrac{1}{2} (\mu_{\pm} \partial \mu_{\mp} - \mu_{\mp} \partial \mu_{\pm})\,,
    \end{aligned}
\end{equation}
where the second relation also applies to $\tilde{\omega}_i$.
The generators in Tables \ref{tab:HL-chiral-ring} and \ref{tab:mcd-chiral-ring} represent the lowest order, or zero-mode, generators of the Macdonald sector.
We note, however, that within this enlarged set of operators, the relations start mixing conformal primaries with descendants.
Clearly, this is caused by the presence of the level two (superconformal) descendant $\tilde{\mu}$.

From the generators and relations listed above, we deduce that we can always consider the action of
\begin{equation}\label{eq:bos-gens-mcd}
    \left\lbrace \partial^{k_1}\mu_+,\,\partial^{k_2}\mu_- \right\rbrace
    \,,\quad k_{1,2}\in \mathbb{Z}_{\geq 0} \,,
\end{equation}
to be free in the ring of operators in the Macdonald sector, similar to how the zero-modes $\mu_{\pm}$ act freely in the HL chiral ring \eqref{eq:HL-chiral-ring-solved}.
Furthermore, the only independent product among the zero-mode generators excluding $\mu_{\pm}$ can be taken to be $\omega_-\tilde{\omega}_+$.
Any other product can be expressed in terms of $\omega_-\tilde{\omega}_+$, products involving $\mu_{\pm}$, and/or conformal descendants.

Let us now see to what extent this structure is reflected in our expressions for the Macdonald index.
We first note the overall $(tv^{\pm2};q)_{\infty}$ factors in the denominator of all three expressions in \eqref{eq:ressum-mcd-index-su2-n=4-2}.
Such factors are precisely explained by a free action of the (multi-particled) bosonic generators in \eqref{eq:bos-gens-mcd}.
Continuing with just the first line of \eqref{eq:ressum-mcd-index-su2-n=4-2}, the full prefactor can be interpreted as the index of a freely generated ring by the above bosonic generators and additional fermionic generators\footnote{Due to the symmetry of the Macdonald index under $v\leftrightarrow v^{-1}$, an alternative set of fermionic generators would be given by $\left\lbrace\partial^{l_1}\omega_+ ,\,\partial^{l_2}\tilde{\omega}_- \right\rbrace$.}
\begin{equation}\label{eq:ferm-gens-mcd}
    \left\lbrace \partial^{l_1}\omega_- ,\, \partial^{l_2}\tilde{\omega}_+\right\rbrace
    \,,\quad l_{1,2}\in \mathbb{Z}_{\geq 0} \,.
\end{equation}
For $l_{1,2}=0$, these are the two additional zero-mode generators whose product can be treated as independent in the ring of operators, as identified above.
It is thus natural to interpret the full prefactor in the first line of \eqref{eq:ressum-mcd-index-su2-n=4-2} in terms of a maximal set of independent generators, corresponding to \eqref{eq:bos-gens-mcd} and \eqref{eq:ferm-gens-mcd}, which generate a free subring in the full Macdonald sector.
It also follows that the ${}_3\phi_2$ series encodes the remaining operators and the various relations they satisfy in the full ring.\footnote{It would be interesting to make contact with the free-field realization of the associated VOA \cite{Bonetti:2018fqz,Beem:2019tfp,Bonetti:2025kan}, which expresses the ring of operators as a quotient of a free $bc\beta\gamma$ module. Identifying the free module with the maximal set of independent generators suggests that the basic hypergeometric series could encode the quotient.}
The fact that the additional generators and relations are captured so efficiently by the ${}_3\phi_2$ series is remarkable, and suggests an underlying simplicity in the Macdonald sector.

To understand whether there are additional operators, beyond the $S_2$ multiplet, contributing to the index, let us expand the ${}_3\phi_2$ series. 
The first two terms read
\begin{align}\label{eq:ressum-mcd-index-su2-n=4-cf-1}
    \begin{split}
       &\mathcal{I}_2(v;q,t)=\frac{(qt^{\frac{1}{2}}v,t^{\frac{3}{2}}v^{-1};q)_{\infty}}{(tv^2,tv^{-2};q)_{\infty}}\\
       &+\frac{(q^{2}t^{\frac{1}{2}}v,qt^{\frac{3}{2}}v^{-1};q)_{\infty}}{(tv^2,tv^{-2};q)_{\infty}}\frac{(t+qt-qt^{\frac{1}{2}}v^{-1}-t^{\frac{3}{2}}v-t^2-qt^2+qt^{\frac{3}{2}}v^{-1}+t^{\frac{5}{2}}v)}{(1-q)}+\ldots\,.
    \end{split}
\end{align} 
Here, the first term simply reflects the free ring mentioned above, without any contributions from additional operators.
For the second term, we note that the numerator of the prefactor can now be associated to the operators $\lbrace\partial^{l_1+1}\omega_-,\partial^{l_2+1},\tilde{\omega}_+\rbrace$, $l_{1,2}\geq 0$, \textit{i.e.}, it excludes their zero-modes.
This is consistent with an identification of the first four terms in the numerator of the second factor with the contributions of the operators $(\mu_0,\tilde{\mu},\tilde{\omega}_-,\omega_+)$.
Such operators should not be paired with $\omega_-$ and $\tilde{\omega}_+$, because they are not independent due to the relations listed above.
It seems that the remaining terms in the numerator can be interpreted in terms of relations, which effectively remove the operators
\begin{equation}
    \mu_0^2\,, \;\mu_0\tilde{\mu}\,, \;\mu_0\tilde{\omega}_-\,, \;\mu_0\omega_+\,.
\end{equation}
These terms, therefore, anticipate contributions of $\mu_0^2$, $\mu_0\tilde{\mu}$, $\mu_0\tilde{\omega}_-$ and $\mu_0\omega_+$, which have not arisen at this stage. 
One can check that they do arise in the subsequent term of the basic hypergeometric series.
This provides a hint that the ${}_3\phi_2$ series can also be interpreted just in terms of contributions from operators in the $S_2$ multiplet.
To substantiate this further, one would ideally have an interpretation for each summand of the ${}_3\phi_2$ series separately, in terms of operators in the $S_2$ multiplet.

One may also interpret the other expressions in \eqref{eq:ressum-mcd-index-su2-n=4-2} along similar lines.
For example, the expression on the third line has prefactor given by
\begin{align}\label{eq:prefac-mcd-3}
    \begin{split}
        \frac{(t^{\frac{3}{2}}v,t^{\frac{3}{2}}v^{-1},q;q)_{\infty}}{(tv^2,tv^{-2},t;q)_{\infty}} \, .
    \end{split}
\end{align}
Apart from the $(q;q)_\I$ factor, this reflects a freely generated ring with generators\footnote{A similar interpretation holds for the prefactor of the second line with $\omega_i$ replaced by $\tilde{\omega}_i$.}
\begin{equation}
    \left\lbrace \partial^{k_1}\mu_+,\,\partial^{k_2}\mu_- ,\,\partial^{k_3}\mu_0 ,\, \partial^{l_1}\omega_+ ,\, \partial^{l_1}\omega_- \right\rbrace
    \,,\quad k_{1,2}\in \mathbb{Z}_{\geq 0} \,.
\end{equation}
However, as we have seen above, $\mu_0$ is not a free generator, nor are $\omega_{\pm}$ together.
It follows that the ${}_3\phi_2$ series multiplying the prefactor in the third expression encodes additional relations to compensate the overcounting.
Indeed, one of the arguments of the ${}_3\phi_2$ series contains a factor of $t^2$, which is the index of the relation that relates $\mu_0^2=\mu_+\mu_-$.
This may be viewed as the Macdonald generalization of the interpretation of the alternative expression for the HL index in \eqref{eq:HL-index-su2-n4-2}, although more work is needed to make this precise.
Remarkably, the additional generators and relations are still efficiently encoded in terms of a single basic hypergeometric series.

Concluding, our analysis above suggests that the full Macdonald index of the $\mathcal{N}=4$ $SU(2)$ theory can be accounted for by the (1/8 BPS truncation) of the $S_2$ multiplet.
Given that these operators are expected to remain BPS at strong coupling, this shows how our expressions for the index represent the more minimal versions advertised in Section \ref{sec:intro}, manifesting only the strongly coupled BPS spectrum.
We will provide additional evidence in the next section, where we prove that a specialization of the index is fully accounted for by the $S_2$ multiplet.

Our conclusion is consistent with various observations in the literature.
For example, in the context of the VOA/SCFT correspondence, conjecture 3 of \cite{Beem:2013sza} states that the Macdonald (or Schur) truncation of the $S_2$ multiplet fully constitutes the BPS spectrum in this sector and generalizes the statement to $SU(N)$ gauge group as well.\footnote{See \cite{Beem:2017ooy,Bonetti:2018fqz,Beem:2019tfp,Bonetti:2025kan} for further progress and \cite{Bonetti:2016nma,Costello:2018zrm} for a proof at large $N$.}
Additional evidence was recently provided in \cite{Chang:2023ywj}, motivated by the search for non-graviton operators in the Macdonald sector.
Indeed, if the Macdonald index can be fully accounted for by the $S_2$ multiplet, this would all but prove the absence of non-graviton operators in the Macdonald sector.\footnote{It falls short of a full proof since there is the possibility that non-graviton operators cancel pairwise in the index. Perfect cancellation, however, seems unlikely.}

\subsection{Product formulas and (almost) freely generated rings}\label{ssec:prod-forms}

In this section, we will show how certain specializations of $t$, and a modification of the integrand in \eqref{eq:mcd-index-su2-n=4-red}, lead to product formulas for the Macdonald index.
In both cases, the resulting index only receives non--trivial contributions from operators in the $S_2$ multiplet, strengthening the observations of the previous section.

The basic mechanism by which the Macdonald index turns into a product formula follows from an elementary property of basic hypergeometric series:
\begin{equation}\label{eq:basichypergeocollapse}
	{}_{r+1}\phi_{r}\left[\begin{array}{cccc}
		a_1 & \cdots & a_r & a_{r+1}\\
		b_1 & \cdots & b_r & \\
	\end{array};q,z\right]=1\,,
\end{equation}
when $a_i=1$, for some $i$, as is easily seen from its definition \eqref{eq:defn-basic-hyper}.
It follows immediately that the expression on the third line of \eqref{eq:ressum-mcd-index-su2-n=4-2} reduces for $t=q^{\frac{1}{2}}$ to\footnote{This can also be derived from the first line when making use of the $q$-Gauss sum \eqref{eq:q-gauss}.}
\begin{equation}\label{eq:mcd-product-k=0}
     \mathcal{I}_2(v;q,q^{\frac{1}{2}})=\frac{(q^{\frac{3}{4}}v,q^{\frac{3}{4}}v^{-1},q;q)_{\infty}}{(q^{\frac{1}{2}}v^2,q^{\frac{1}{2}}v^{-2},q^{\frac{1}{2}};q)_{\infty}} \, .
\end{equation}
We emphasize that this specialization is distinct from the Schur limit $t=q$, and to the best of our knowledge has not been studied before.
Using the results from the previous section, we interpret the formula in terms of a quotient of a freely generated ring with generators
\begin{equation}\label{eq:gens-mcd-spec}
    \left\lbrace \partial^{k_A}\mu_A,\partial^{l_i}\omega_i \right\rbrace \Big/ \left\lbrace \partial^{m}(\mu_0^2-\mu_+\mu_-)\right\rbrace
    \,,\quad k_A,l_i,m\in \mathbb{Z}_{\geq 0} \,.
\end{equation}
In particular, we note that the $(q;q)_\I$ factors in the numerator, which we could not interpret for general $t$ in the previous section, precisely incorporate the contributions of the relations by which the freely generated ring is quotiented when $t=q^{\frac{1}{2}}$.
We also note the complete absence, \emph{at the level of the index}, of the generators $\tilde{\mu}$ and $\tilde{\omega}_i$, and their conformal descendants.

This somewhat drastic simplification is likely explained by the following observation.
As we set $t=q^{\frac{1}{2}}$, the indices of the operators $\tilde{\mu}$ and $\tilde{\omega}_i$ agree, up to a sign, with the indices of the basic relations $\omega_+\omega_-=0$ and $\mu_0\omega_{\pm}=\mu_{\pm}\omega_{\mp}$ (see Tables \ref{tab:HL-chiral-ring} and \ref{tab:mcd-chiral-ring}).
If we treat $\mu_0$ and $\omega_i$ (and their derivatives) as free generators, we should compensate for the overcounting through explicit inclusion of their relations. 
However, the contributions of these relations effectively cancel, at the level of the index, against the contributions of the generators $\tilde{\mu}$ and $\tilde{\omega}_i$ (and their derivatives).
It follows that only the generators in \eqref{eq:gens-mcd-spec} remain and can be treated as free generators up to the relation for $\mu_0^2$ and its derivatives.
Accepting this interpretation, it follows that for $t=q^{\frac{1}{2}}$: the only operators contributing to the index are operators in the $S_2$ multiplet!

There are various other interesting specializations one may explore using our expressions and properties of the basic hypergeometric series.
Let us mention one further specialization: $t=1$.
Naively, this specialization may seem incompatible with convergence properties of the index, which require $|t|<1$.
However, our closed form formulas are meromorphic functions of $t$ and can be extended to $|t|\geq 1$.
In particular, one may deduce, for example from \eqref{eq:ressum-mcd-index-su2-n=4-1}, that for $t=1$ the index takes on another simple product form\footnote{\label{fn:t-spec} More generally, \eqref{eq:ressum-mcd-index-su2-n=4-1} may be used to deduce simple formulas for any $t=q^{-m}$ for $m\in \mathbb{Z}_{\geq 0}$.}
\begin{equation}\label{eq:su2t=1product}
    \mathcal{I}_2(v;q,1)=\frac{(v^{\pm};q)_\I}{(v^{\pm2};q)_\I}\,.
\end{equation}
Note that the $(1-v^{\pm2})$ factors in the denominator resum the naively divergent contributions to the index from arbitrary powers of the bosonic operators $\mu_{\pm}$ at $t=1$.

We end this section with the curious observation that a product formula also arises for a slight modification of the Macdonald index:
\begin{align}\label{eq:mcd-index-su2-n=4-mod}
    \begin{split}
        \mathcal{I}^{\text{mod}}_2(v;q,t)&=\frac{(q,t;q)_\infty}{(t^{\frac{1}{2}}v,t^{\frac{1}{2}}v^{-1};q)_\infty}\oint_{|s|=1}\frac{ds}{2\pi is}\frac{(s,qs^{-1},qts,ts^{-1};q)_\infty}{(t^{\frac{1}{2}}vs^{\pm},t^{\frac{1}{2}}v^{-1}s^{\pm};q)_\infty}\,,
    \end{split}
\end{align}
where we left out a factor $(1-ts)$ from the numerator, which corresponds to the contribution of the positive root component of the zero mode $\lambda_{1+}$.
Using our general expression for the basic hypergeometric integral \eqref{eq:ressum-gen-basic-hypergeom-integral} and simplifying the result using the non-terminating version of the $q$-Saalsch\"utz summation \eqref{eq:q-saalschutz-non-term}, we find that\footnote{To the best of our knowledge, this result has appeared before only recently, implied by Corollary 10.6 of \cite{COHL2023102517}. Although the result seems similar to the so-called Askey-Roy integral \cite{Askey-Roy-1986} (see also Section 4.11 of \cite{Gasper_Rahman_2004} and more recently \cite{Cohl_2022}), we have not been able to find a direct relation.}
\begin{align}\label{eq:ressum-mcd-index-su2-n=4-mod}
    \begin{split}
        \mathcal{I}^{\text{mod}}_2(v;q,t)&=\frac{(qt^{\frac{1}{2}}v,qt^{\frac{1}{2}}v^{-1},t^2;q)_{\infty}}{(tv^2,tv^{-2},t;q)_{\infty}}\,.
    \end{split}
\end{align}
This product formula can again be interpreted in terms of a quotient of a freely generated ring with generators
\begin{equation}\label{eq:gens-mcd-mod}
    \left\lbrace \partial^{k_A}\mu_A,\partial^{l_i}\tilde{\omega}_i, \right\rbrace \Big/ \left\lbrace \partial^{m}(\mu_0^2-\mu_+\mu_-)\right\rbrace
    \,,\quad k,l,m\in \mathbb{Z}_{\geq 0} \,,
\end{equation}
which is identical to \eqref{eq:gens-mcd-spec}, except for the exchange $\omega_i\leftrightarrow \tilde{\omega}_i$.
We can again suggest a likely explanation for the emergence of a product formula.
First, we note that removal of the positive root component of $\lambda_{1+}$ effectively removes the zero modes of $\omega_i$ and $\tilde{\mu}$.
Indeed, since $\lambda_{1+}$ can be considered diagonal inside these operators, the removal of a positive root component should remove the degree of freedom entirely.
One then observes from the Tables \ref{tab:HL-chiral-ring} and \ref{tab:mcd-chiral-ring} that the indices of the operators $\partial^{k+1}\omega$ and $\partial^{k+1}\tilde{\mu}$ cancel against the relations $\mu_0\tilde{\omega}_{\pm}=\mu_{\pm}\tilde{\omega}_{\mp}$ and $\tilde{\omega}_{+}\tilde{\omega}_{-}=0$ (and their derivatives).

Concluding, we have shown that either a specialization or a modification of the index allows us to interpret our expression for the Macdonald index in terms of a simple quotient ring, generated solely by operators in the $S_2$ multiplet.

\subsection{Analytic structure, residues and asymptotics}\label{ssec:analytic-struct}

In this section, we study the analytic properties of the Macdonald index.
For convenience, we repeat the expression on the third line of \eqref{eq:ressum-mcd-index-su2-n=4-2} here
\begin{align}\label{eq:ressum-mcd-index-su2-n=4-3-rep}
    \begin{split}
        \mathcal{I}_2(v;q,t)&=\frac{(t^{\frac{3}{2}}v,t^{\frac{3}{2}}v^{-1},q;q)_{\infty}}{(tv^2,tv^{-2},t;q)_{\infty}}{}_{3}\phi_{2}\left[\begin{array}{ccc}
		t^2/q & t^{\frac{1}{2}}v^{-1}  & t^{\frac{1}{2}}v   \\
		t^{\frac{3}{2}}v  & t^{\frac{3}{2}}v^{-1}  &  \\
	\end{array};q,q\right]\\
    &=\frac{1}{(tv^2,tv^{-2},t;q)_{\infty}}\sum_{k=0}^{\infty}(q^{k}t^{\frac{3}{2}}v^{\pm},q^{k+1};q)_{\infty}\,(t^2/q,t^{\frac{1}{2}}v^{\pm};q)_{k}\,q^k \, .
    \end{split}
\end{align}
As already mentioned in the previous subsection, this expression shows that the index is a meromorphic function of $t,v\in\mathbb{C}$, even though the index was originally defined for $|t|<1$ and $|v|=1$.
It has (simple) poles at $v^2=tq^k$, for $k\geq 0$, which reflect the contributions of arbitrarily high power of the operator $\partial^k\mu_-$ to the index (and similarly for $v\to v^{-1}$ and  $\partial^k\mu_+$).
The poles at $t=q^{-k}$, for $k\geq 0$, are only apparent.
Indeed, the alternative expression \eqref{eq:ressum-mcd-index-su2-n=4-1} shows that the index is perfectly finite at these points.

As discussed in \cite{Gaiotto:2012xa}, the residues of the index have a physical interpretation.
Namely, the residue at the pole $v^2=tq^k$, for $k=0$, corresponds to the index of the IR theory obtained by the RG flow of of the parent theory after turning on a vev $\langle\mu_-\rangle \neq 0$.
The poles for $k>0$ correspond to position-dependent vevs of $\langle\mu_-\rangle \sim z^k$ (with $z$ the coordinate associated with $\partial$). 
In this case, the residue can be interpreted as the index of theory obtained through RG flow away from $z=0$, with a surface defect located at $z=0$.
The fact that there is no pole at $t=q^{-k}$ is consistent with the fact that one cannot turn on an expectation value for just $\mu_0$, due to the relation $\mu_0^2=\mu_+\mu_-$.

Evaluating the corresponding residues turns out to be particularly simple when $k$ is even.
To demonstrate this, we recall the prescription of \cite{Gaiotto:2012xa} applied to the case at hand:
\begin{equation}
    \mathcal{I}^{(k)}_{IR}(q,t)=2\mathcal{I}_1(q,t)\text{Res}_{v=t^{\frac{1}{2}}q^{\frac{k}{2}}}\frac{\mathcal{I}_2(v;q,t)}{v}\,,
\end{equation}
where $\mathcal{I}_1(q,t)$ corresponds to the Macdonald index of the $\mathcal{N}=4$ $U(1)$ theory.
Now we let $k=2\tilde{k}$.
We then use the (terminating) $q$-Saalsch\"utz sum \eqref{eq:q-saalschutz} to find
\begin{equation}
     \mathcal{I}^{(k)}_{IR}(q,t)=\frac{(t,q;q)_\infty}{(t^{\frac{1}{2}};q)^2_\infty}\frac{(q/t,t^2;q)_{\tilde{k}}}{(t^2,q;q)_{2\tilde{k}}}(-t)^{\tilde{k}}q^{\frac{1}{2}\tilde{k}(3\tilde{k}+1)}\,.
\end{equation}
For $\tilde{k}=0$, the index equals that of a $\mathcal{N}=4$ $U(1)$ theory.
This is as expected, since a vev of any of the scalars in the $\mathcal{N}=4$ theory generates masses for all off-diagonal components of the $\mathcal{N}=4$ multiplet.
For $\tilde{k}>0$, we see an additional contribution which should be attributed to the surface defect at $z=0$. 
It would be nice to interpret these factors in terms of a 2d UV gauge theory description of a vortex worldsheet theory, which flows to the surface defect in the IR.
On the other hand, for odd $k$ the residue takes on a more complicated form due to the fact that the ${}_3\phi_2$ series does not terminate.

Finally, let us comment on the asymptotic behavior of the index for $|q|\to 1$.
The (root-of-unity) asymptotics of the $q$-Pochhammer symbol $(x;q)_{\infty}$ has been studied in the mathematical literature, see, \textit{e.g.}, \cite{Banerjee:2017,Garoufalidis:2018qds,Wheeler:phd}.\footnote{The evaluation of the asymptotics is closely related to its modular properties. More precisely, it can be viewed as a holomorphic quantum Jacobi form \cite{Garoufalidis:2022wij,Fantini:2025wap}.}
We are not aware of similar, general studies for basic hypergeometric series, although some examples have appeared in \cite{Garoufalidis:2021lcp,Garoufalidis:2023wez}.

It seems natural to assume that as $q\to 1$, the basic hypergeometric in \eqref{eq:ressum-mcd-index-su2-n=4-3-rep} will be dominated by its summand for large $k$.
In this case, the ${}_3\phi_2$ series should be well approximated by
\begin{equation}
    {}_{3}\phi_{2}\left[\begin{array}{ccc}
		t^2/q & t^{\frac{1}{2}}v^{-1}  & t^{\frac{1}{2}}v   \\
		t^{\frac{3}{2}}v  & t^{\frac{3}{2}}v^{-1}  &  \\
	\end{array};q,q\right]\stackrel{q\to 1}{\sim}\lim_{q\to 1} \frac{(t^{\frac{1}{2}}v,t^{\frac{1}{2}}v^{-1},t^2/q;q)_{\infty}}{(t^{\frac{3}{2}}v,t^{\frac{3}{2}}v^{-1},q;q)_{\infty}}\,.
\end{equation}
Substituting this in the full expression for the index, we find
\begin{equation}
     \mathcal{I}_2(v;q,t)\stackrel{q\to 1}{\sim} \lim_{q\to 1}\frac{(t^{\frac{1}{2}}v,t^{\frac{1}{2}}v^{-1},t^2/q;q)_{\infty}}{(tv^2,tv^{-2},t;q)_{\infty}} \, .
\end{equation}
The asymptotics of this expression can be analyzed using the results of \cite{Banerjee:2017,Garoufalidis:2018qds,Wheeler:phd}, including more general asymptotics, where $q$ tends to a root-of-unity.
We will not pursue the asymptotics in detail here, but just note that the asymptotics match those of the modified index \eqref{eq:ressum-mcd-index-su2-n=4-mod}.
This suggests that, with regards to the asymptotics of the index for generic $t$, the dominant contributions come from the operators $\partial^k\mu_A$ and $\partial^k\tilde{\omega}_i$.

\subsection{Comparison to TQFT formula}\label{ssec:comparison-tqft}

In the preceding sections, we have seen that our expression for the Macdonald index reduces to known expressions in the literature for the Hall-Littlewood and Schur index.
In this section, we compare our expressions for the Macdonald index with the so-called TQFT formula for the Macdonald index \cite{Gadde:2011uv}.\footnote{Incidentally, the Macdonald index inherits its name from this very formula.}

The TQFT expression for the $A_1$ class $\mathcal{S}$ theory labeled by a Riemann surface $C_{n,g}$ of genus $g$ and $n$ punctures is given by \cite{Gadde:2011uv}
\begin{equation}\label{eq:mcd-tqft-class-s}
    \mathcal{I}^{\text{TQFT}}_{2}\left[C_{n,g}\right](v;q,t)=\frac{\left[(t^2;q)_\I\right]^{2g-2+n}\left[(t;q)_\I\right]^{1-g}}{\left[(q;q)_\I\right]^{1-g}\prod^n_{i=1}(tv_i^2,tv_i^{-2},t;q)_\I}\sum^{\infty}_{\lambda=0}\frac{\prod^n_{i=1}P_{\lambda}(v_i,v_i^{-1};q,t)}{\left[P_{\lambda}(t^{\frac{1}{2}},t^{-\frac{1}{2}};q,t)\right]^{2g-2+n}}\,.
\end{equation}
Here, the $SU(2)$ Macdonald polynomial is given by
\begin{equation}\label{eq:su2-mcd-pol}
    P_{\lambda}(v,v^{-1};q,t)=N^{\frac{1}{2}}_\lambda(q,t)\sum^{\lambda}_{m=0}\frac{(t;q)_m(t;q)_{\lambda-m}}{(q;q)_m(q;q)_{\lambda-m}}v^{2m-\lambda}
\end{equation}
where the normalization factor can be written as
\begin{equation}\label{eq:norm-mcd-pols}
    N_\lambda(q,t)=(1-tq^\lambda)\frac{(q;q)_\lambda}{(t^2;q)_\lambda}\frac{(q, t^2;q)_\I}{(t;q)^2_\I}\,.
\end{equation}
When $(n,g)=(1,1)$, which corresponds to the $\mathcal{N}=4$ $SU(2)$ theory with a decoupled hypermultiplet, we have
\begin{equation}\label{eq:su2-n4-mcd-tqft}
    \mathcal{I}^{\text{TQFT}}_2\left[C_{1,1}\right](v;q,t)=\frac{(t^2;q)_\I}{(tv^2,tv^{-2},t;q)_\I}\sum^{\infty}_{\lambda=0}\frac{P_{\lambda}(v,v^{-1};q,t)}{P_{\lambda}(t^{\frac{1}{2}},t^{-\frac{1}{2}};q,t)}\,.
\end{equation}
Note that the normalization factor of the Macdonald polynomial drops out in \eqref{eq:su2-n4-mcd-tqft}.

After stripping off the decoupled hypermultiplet contribution, we equate the TQFT expression with both the second and third line of \eqref{eq:ressum-mcd-index-su2-n=4-2}.
The equality of the two expressions reduces to the following identities:
\begin{align}\label{eq:mcd-pol-3phi2-id}
    \begin{split}
        \sum^{\infty}_{\lambda=0}\frac{P_{\lambda}(v,v^{-1};q,t)}{P_{\lambda}(t^{\frac{1}{2}},t^{-\frac{1}{2}};q,t)}&=\frac{(1-\frac{t^2}{q})}{(1-t^{\frac{1}{2}}v)(1-t^{\frac{1}{2}}v^{-1})}{}_{3}\phi_{2}\left[\begin{array}{ccc}
		q & qt^{-\frac{1}{2}}v^{-1}  & qt^{-\frac{1}{2}}v   \\
		qt^{\frac{1}{2}}v  & qt^{\frac{1}{2}}v^{-1}  &  \\
	\end{array};q,\frac{t^2}{q}\right]\\
    &=\frac{(t^{\frac{3}{2}}v,t^{\frac{3}{2}}v^{-1},q;q)_{\infty}}{(t^{\frac{1}{2}}v,t^{\frac{1}{2}}v^{-1},t^2;q)_{\infty}}{}_{3}\phi_{2}\left[\begin{array}{ccc}
		t^2/q & t^{\frac{1}{2}}v^{-1}  & t^{\frac{1}{2}}v   \\
		t^{\frac{3}{2}}v  & t^{\frac{3}{2}}v^{-1}  &  \\
	\end{array};q,q\right] \, .
    \end{split}
\end{align}
Note that the second identity, together with the results from Section \ref{ssec:prod-forms}, implies a product formula for the summed ratio of Macdonald polynomials when $t=q^{\frac{1}{2}}$.

We now show that this is indeed an identity.
In fact, the identity is closely related to a known generating function of $q$-ultraspherical polynomials.
The $q$-ultraspherical polynomial is defined as
\begin{equation}\label{eq:q-ultra-sph-pol}
    C_{\lambda} (v,v^{-1};t|q)=\sum^{\lambda}_{m=0}\frac{(t;q)_m(t;q)_{\lambda-m}}{(q;q)_m(q;q)_{\lambda-m}}v^{(2m-\lambda)}
\end{equation}
and is identical to the unnormalized $SU(2)$ Macdonald polynomial.
It is known that\footnote{See, \textit{e.g.}, exercise 9.8 in \cite{Gasper_Rahman_2004}.} 
\begin{equation}
    \sum^{\infty}_{\lambda=0}\frac{(q;q)_\lambda}{(t^2;q)_\lambda} C_{\lambda} \left(v,v^{-1};t|q\right)t^{\frac{\lambda}{2}}=\frac{(t^{\frac{3}{2}}v,t^{\frac{3}{2}}v^{-1},q;q)_{\infty}}{(t^{\frac{1}{2}}v,t^{\frac{1}{2}}v^{-1},t^2;q)_{\infty}}{}_{3}\phi_{2}\left[\begin{array}{ccc}
		t^2/q & t^{\frac{1}{2}}v^{-1}  & t^{\frac{1}{2}}v   \\
		t^{\frac{3}{2}}v  & t^{\frac{3}{2}}v^{-1}  &  \\
	\end{array};q,q\right]\,.
\end{equation}
Comparing with \eqref{eq:mcd-pol-3phi2-id}, we see that the identity reduces to:
\begin{equation}\label{eq:tqft-mcd-id}
    \sum^{\lambda}_{m=0}\frac{(t;q)_m(t;q)_{\lambda-m}}{(q;q)_m(q;q)_{\lambda-m}}t^{m-\frac{\lambda}{2}}=\frac{(t^2;q)_\lambda}{(q;q)_\lambda} t^{-\frac{\lambda}{2}}\,,
\end{equation}
where the left hand side corresponds to $C_{\lambda} (t^{\frac{1}{2}},t^{-\frac{1}{2}};t|q)$.
This represents another known identity and follows, for example, from the so--called principal specialization of the $A_1$ Macdonald polynomial, see, \textit{e.g.}, p. 337 of \cite{Macdonald1995}.
Thus, we have shown the relation between our (simplified) formula for the Macdonald index and the TQFT formula.
In particular, our expression can be viewed as a resummation of the (Laurent) series in the flavor fugacity $v$, as expressed by the TQFT formula.
The resulting expression makes the analytic properties in $v$ manifest.
A similar resummation was observed in the context of the Schur index in Appendix E of \cite{Gadde:2011uv} and more recently, and for general class $\mathcal{S}$ theories, in \cite{Pan:2021mrw}.
We expect a similar relation with the TQFT--like formula recently derived in \cite{Hatsuda:2025mvj}.

\section{Further indices at rank one}\label{sec:examples-rank-1}

In this section, we evaluate the residue sums for superconformal indices of $SU(2)$ gauge theories with $\mathcal{N}=4,2,1$ supersymmetry, using the reduced measure for the gauge integral.

\subsection{\texorpdfstring{$\mathcal{N}=4$}{N=4} \texorpdfstring{$SU(2)$}{SU(2)} SYM}

In Section \ref{sec:mcd-n=4-su2}, we derived an expression for the Macdonald index of the $\mathcal{N}=4$ $SU(2)$ SYM theory, and showed that its form suggested the absence of non-graviton operators in the Macdonald sector.
On the other hand, it has recently become clear that the full superconformal index captures an $\mathcal{O}(N^2)$ entropy \cite{Cabo-Bizet:2018ehj,Choi:2018hmj,Benini:2018ywd}.
This entropy can only be accounted for if there exist additional BPS operators, beyond the $S_n$ ``super-graviton'' multiplets of \cite{Kinney:2005ej}, in the 1/16 BPS sector at strong coupling.
To find examples of such operators, one has so far relied on two crutches \cite{Kinney:2005ej,Grant:2008sk}.

First, a $1/16^{\text{th}}$ BPS operator, which is annihilated by a single complex supercharge $\mathcal{Q}$ and its hermitian conjugate $\mathcal{Q}^\dagger =\mathcal{S}$, is in one-to-one correspondence with a $\mathcal{Q}$-cohomology class of operators, \textit{i.e.}, a set of operators annihilated only by $\mathcal{Q}$ modulo the addition of $\mathcal{Q}$-exact operators $\mathcal{O}'=\left[\mathcal{Q},\mathcal{O}\right\rbrace$. 
Secondly, it was conjectured that the states which remain BPS at one-loop do not lift from the BPS spectrum at higher perturbative or non-perturbative orders.\footnote{\label{fn:1-loop-conj}In the final stages of this work, the interesting preprint \cite{Chang:2025mqp} appeared which provides a counterexample to this conjecture, based on observations of \cite{Gadde:2025yoa}. The precise implications for the cohomology at strong coupling are, as of yet, unclear.} 
This allows one to use the classical supercharge (for the interacting Langrangian) to define the $\mathcal{Q}$-cohomology, which has a reasonably simple action on the elementary fields.
In particular, based an (unsuccessful) early attempt \cite{Chang:2013fba}, this has recently led to the identification of non-graviton cohomologies \cite{Chang:2022mjp} and the explicit construction of representatives at low values of $N$ \cite{Choi:2022caq,Choi:2023znd,Choi:2023vdm,deMelloKoch:2024pcs}.
A general feature of non-graviton operators is that they are $\mathcal{Q}$-closed only up to $SU(N)$ trace relations, implying that they lift from the spectrum at sufficiently large $N$.
This feature has a natural interpretation in terms of black holes in the gravitational dual \cite{Chang:2024zqi}.
It led also led the authors to refer to these operators as ``fortuitous'', as opposed to the ``monotonous'' graviton operators which remain BPS for arbitrary $N$.

In this section, we will derive closed form expressions for two indices of the $\mathcal{N}=4$ $SU(2)$ theory which are known to receive contributions from non-graviton operators.
Such expressions represent a first step in the attempt to develop closed form expressions for the index over the non--graviton or fortuitous part of the spectrum. 
Indeed, given a closed form expression for \emph{both} the superconformal index $\mathcal{I}$ and the graviton or monotone index $\mathcal{I}_{m}$, their difference would represent the index $\mathcal{I}_f$ over the fortuitous part of the spectrum:
\begin{equation}\label{eq:If-from-I-Im}
    \mathcal{I}_f=\mathcal{I}-\mathcal{I}_m\,.
\end{equation}
So far, this procedure has only been carried out fully for the so--called BMN index (see below) of the $SU(2)$ theory \cite{Choi:2023znd} (see also \cite{Gadde:2025yoa}) and partially for the $SU(3)$ theory \cite{Choi:2023vdm}.\footnote{Restricting to the $SU(3)_R$ singlet sector of the BMN subsector, further progress can be made, see \cite{Gaikwad:2025ugk}.}

As mentioned above, we focus on finding closed form expressions for indices $\mathcal{I}$, evaluated over the full spectrum of the theory, and will not consider the problem of finding a closed form for $\mathcal{I}_m$ in the present work.
It is likely that the latter problem requires an alternative approach, given that the graviton index does not appear to have a matrix integral formulation.
Indeed, methods pursued in the literature (at finite $N$) involve explicit construction of the graviton operators and evaluate the indices through explicit ``eigenvalue counting'' \cite{Choi:2023znd,Choi:2023vdm,Gaikwad:2025ugk}. 
At large $N$, simple expressions for the graviton index are well--known and readily available \cite{Kinney:2005ej,Chang:2013fba}. 

Before turning to the full superconformal index, we first define a ``BMN Macdonald'' index, which is easier to compute than the full index.

\subsubsection{BMN Macdonald index}\label{ssec:bmn-mcd}

As observed in \cite{Chang:2013fba}, the ``Hamiltonian'' $\delta=2\lbrace \mathcal{Q},\mathcal{Q}^{\dagger}\rbrace$ at one-loop has an additional $U(1)_Y$ symmetry.
The charge of this symmetry counts the number of Lagrangian letters in an operator and can be used to refine the index.\footnote{Since this symmetry will be broken at higher loops, the refined index is in principle only protected as one moves from the free to the one-loop spectrum. Since the status of the one--loop exactness conjecture is unclear (see footnote \ref{fn:1-loop-conj}), one should be careful with the interpretation of this index at strong coupling.}
As detailed in \cite{Gadde:2025yoa}, this refinement effectively allows one to define indices in the $\mathcal{N}=4$ theory for more BPS sectors than just those described in Section \ref{ssec:defns}.
In particular, an index can be defined for the BMN sector of the $\mathcal{N}=4$ theory \cite{Choi:2023znd,Choi:2023vdm,Gadde:2025yoa}, so-called due to its connection with the BMN matrix model \cite{Berenstein:2002jq,Kim:2003rza}.
This sector gets contributions from all letters in Table \ref{tab:n=2-multiplets}, except the derivatives and the gaugino $\bar{\lambda}_{2\dot{\pm}}$, and it plays a crucial role in finding explicit representatives of non-graviton operators in the papers cited above.

In this section, we show that if we include a single derivative in the BMN sector and a component of the gaugino, the corresponding index takes on the form of a basic hypergeometric integral which we can evaluate.
This sector, which we will refer to as the BMN Macdonald sector, corresponds to one of the two types of maximal consistent truncations of the full $1/16^{\text{th}}$ BPS sector identified in \cite{Gadde:2025yoa}.
We will here simply state the single letter indices for the contributing letters, referring for full details to \cite{Chang:2013fba,Gadde:2025yoa}.

\begin{table}[ht]
    \centering
    \bgroup
    \def\arraystretch{1.2}
    \begin{tabular}{|c|c|c|c|c|c|c|c|c|c|}
    \hline
         letter & $\bar{q}_1$ & $\bar{q}_2$  & $\psi_{1+}$  & $\psi_{2+}$ & $\bar{\phi}$ & $\lambda_{1+}$ & $\bar{\lambda}_{2\dot{-}}$ & $F_{++}$ & $\partial_{+\dot{-}}$ \tabularnewline
  \hline
index & $y_1$ & $y_2$  & $-y_1y_3$ & $-y_2y_3$ & $y_3$ & $-y_1y_2$ & $-q$ & $y_1y_2y_3$ & $q$  \tabularnewline
 \hline
    \end{tabular}
    \egroup
    \caption{The letters contained in a maximal 1-loop truncation of the $1/16$ BPS sector of the $\mathcal{N}=4$ SYM theory with corresponding indices.}
    \label{tab:bmn-mcd}
\end{table}

Notice that this truncation gets contributions from almost all letters of the full $\mathcal{N}=4$ theory (see Table \ref{tab:n=2-multiplets}), except for the gaugino component $\bar{\lambda}_{2\dot{+}}$ (and its equation of motion) and the derivative $\partial_{+\dot{+}}$.
Comparing with the letters contributing to the Macdonald index, we note the addition of $\lbrace \bar{\phi},\, \psi_{1+},\,\psi_{2+},\, F_{++}\rbrace$.
To recover the Macdonald index, one sets
\begin{equation}
    y_1=t^{\frac{1}{2}}v\,,\quad y_2=t^{\frac{1}{2}}v^{-1}\,,\quad y_3=0\,.
\end{equation}
From the Table \ref{tab:bmn-mcd}, it follows that the index is given by the integral
\begin{align}\label{eq:bmn-mcd-index-su2-n=4-red}
    \begin{split}
        \mathcal{I}_2(y_1,y_2,y_3;q)&=\frac{(q,y_1y_2,y_1y_3,y_2y_3;q)_\infty}{(y_1,y_2,y_3,y_1y_2y_3;q)_\infty}\oint\frac{ds}{2\pi is}\frac{(s,qs^{-1},y_1y_2s^{\pm},y_1y_3s^{\pm},y_2y_3s^{\pm};q)_\infty}{(y_1s^{\pm},y_2s^{\pm},y_3s^{\pm},y_1y_2y_3s^{\pm};q)_\infty}\,,
    \end{split}
\end{align}
where we perform the singlet projection using the reduced measure.
We notice an interesting symmetry between the contributions of the $\mathcal{N}=2$ vector- and hypermultiplet contributions, not present for the ordinary Macdonald index.
In particular, both multiplets contribute two towers of poles and two towers of zeroes.

Using the results of Appendix \ref{app:bas-hypgeom-int}, we find that the corresponding residue sum is convergent and therefore the contribution of the origin vanishes.
We then use \eqref{eq:ressum-gen-basic-hypergeom-integral} to evaluate the residue sum as
\begin{equation}\label{eq:ressum-bmn-mcd-index-su2-n=4}
    \mathcal{I}_2(y_1,y_2,y_3;q)=\tilde{\mathcal{R}}(y_1,y_2,y_3;q)+\sum^3_{a=1}\mathcal{R}_a(y_1,y_2,y_3;q)
\end{equation}
where $\tilde{\mathcal{R}}$ corresponds to the residue sum associated with the tower of poles at $s=y_1y_2y_3q^k$, and the $\mathcal{R}_a$ to $s=y_aq^k$.
Explicitly, the functions are given by
\begin{align}
    \begin{split}
        \tilde{\mathcal{R}}(y_1,y_2,y_3;q)&=\frac{(1/y_1,1/y_2,1/y_3,y_1y_2,y_1y_3,y_2y_3,q/y_1y_2y_3,y_1^2y_2^2y_3,y_1^2y_2y_3^2,y_1y_2^2y_3^2;q)_\infty}{(y_1,y_2,y_3,1/y_1y_2,1/y_1y_3,1/y_2y_3,y_1^2y_2y_3,y_1y_2^2y_3,y_1y_2y_3^2,y_1^2y_2^2y_3^2;q)_\infty}\\
        &\times {}_{7}\phi_{6}\left[\begin{array}{cccccccc}
		y_1^2y_2^2y_3^2 & y_1^2y_2y_3 & y_1y_2^2y_3 & y_1y_2y_3^2  & qy_1 & qy_2& qy_3\\
		 qy_2y_3 &  qy_1y_3 & qy_1y_2 & y_1y_2^2y_3^2 &  y_1^2y_2y_3^2  & y_1^2y_2^2y_3 & 
	\end{array};q;q\right]\,,
    \end{split}
\end{align}
and
\begin{align}
    \begin{split}
        \mathcal{R}_1(y_1,y_2,y_3;q)&=
    \frac{(y_1^2 y_2,y_1^2 y_3,qy_1^{-1},y_2 y_3 y_1^{-1};q)_\I}{(y_1^2,y_1^2y_2y_3,y_2y_1^{-1},y_3y_1^{-1};q)_\I}\\
        &\times {}_{7}\phi_{6}\left[\begin{array}{cccccccc}
		y_1^2 & y_1y_2 & y_1y_3 & y_1^2y_2y_3  & q/y_2 & q/y_3& qy_1/ y_2 y_3\\
		qy_1/y_2 & qy_1/y_3 & q/y_2y_3   & y_1^2y_2 & y_1^2y_3 & y_1y_2y_3 &   
	\end{array};q;q\right]\,.
    \end{split}
\end{align}
The functions $\mathcal{R}_2$ and $\mathcal{R}_3$ are related to $\mathcal{R}_1$ through the exchanges of $y_1\leftrightarrow y_2$ and $y_1\leftrightarrow y_3$, respectively.
We note that the ${}_7\phi_6$ series are well-poised and $0$-balanced.
Furthermore, we note that $\tilde{\mathcal{R}}$ is by itself symmetric in the $y_a$, while the $\mathcal{R}_a$ are not separately symmetric (but, their sum is).
Finally, the unflavored limits, where $y_a\to y$, for some $a$, have to be taken with care due to apparent poles in the $\mathcal{R}_a$. 
We observed similar features for the ordinary Macdonald index in Section \ref{ssec:derivation}, which can be recovered through a (careful) $y_3\to 0$ limit.

In the case of the Macdonald index, a simplification of the residue sum helped in interpreting the expression in terms of the operator spectrum.
In particular, the use of a $3$--term transformation formula resulted in a simplified expression of the schematic form 
\begin{equation}
    \mathcal{I}_2(v;q,t)=\prod_i\frac{ (a_i;q)_\I}{ (b_i;q)_\I}\times {}_3\phi_2\,.
\end{equation}
The prefactor was then interpreted in terms of a freely generated subring.
One may ask whether a similar simplification could occur for the BMN Macdonald index.
For example, a five-term transformation formula for a ${}_7\phi_6$ series could in principle allow for a simplification of the form\footnote{It seems that related questions are explored in the recent mathematical literature, see, \textit{e.g.}, \cite{bradley-thrush:2025}.} 
\begin{equation}\label{eq:simpl-bmn-mcd-index}
    \mathcal{I}_2(y_1,y_2,y_3;q)=\prod_i\frac{ (a_i;q)_\I}{ (b_i;q)_\I}\times {}_7\phi_6\,.
\end{equation}
As it turns out, five-term transformation formulas do exist for well-poised ${}_7\phi_6$ series and represent a special case of Sears' transformation formulas (see Section 4.12 of \cite{Gasper_Rahman_2004} and references therein).
However, it appears that the transformation formula does not quite apply to our context.
Lacking appropriate transformation formulas, we list some features one would expect for a simplified formula of the form \eqref{eq:simpl-bmn-mcd-index} for the BMN Macdonald index (if it exists), based on our experience with the Macdonald index:
\begin{itemize}
    \item It should not depend on negative powers of the $y_a$. If an expression without negative powers exists, the apparent poles at $y_a=y_b$ will be removed. As such, its unflavored limit should be manifest.
    \item Related to the previous point, we expect that the denominator of the prefactor manifests the true poles of the index. 
    From the explicit forms of $\tilde{\mathcal{R}}$ and $\mathcal{R}_a$, and the previous point, we expect this denominator to contain (at least) the factors
    \begin{equation}
        (y_1^2,y_2^2,y_3^2,y_1^2y_2y_3,y_1y_2^2y_3,y_1y_2y_3^2,y_1^2y_2^2y_3^2;q)_\infty\,.
    \end{equation}
    \item A specialization of the parameters $y_a$ may lead to a simplification of the index, perhaps even a product formula.
\end{itemize}
Assuming such a simplified expression exists, it could shed light on structural properties of the non-graviton spectrum.
Another logical possibility is that the presence of non-graviton operators precludes a simplification.

\subsubsection{Full superconformal index}\label{ssec:full-indexn4-su2}

We now turn to the evaluation of the full index of the $\cN=4$ $SU(2)$ SYM theory, making use of the reduced measure.
Note that the integration for the general $SU(N)$ case was already discussed at a schematic level in Section \ref{ssec:ell-hypergeom-integrals}.
Here, we provide full details for the $N=2$ case of the discussion there.

From \eqref{eq:full-index-suN-n=4-defn}, we have the following integral
\begin{equation}\label{eq:full-index-su2-red}
	\cI_2(y_1,y_2,y_3;p,q) = (p;p)_{\infty} (q;q)_{\infty} \prod_{i=1}^3\G(y_i) \oint_{|s|=1} \frac{ds}{2\pi is} \frac{\prod_{i=1}^3\G(y_is^\pm)}{(1-s^{-1})\G(s^\pm)},
\end{equation}
where we recall the shorthand notation $\Gamma(x)\equiv \Gamma(x;p,q)$ and defined
\begin{equation}
	y_1 = t^{\frac{1}{2}}v, \qquad y_2=t^{\frac{1}{2}}v^{-1}, \qquad y_3 = pq/t, \qquad y_1y_2y_3 = pq.
\end{equation}
The evaluation of this integral corresponds to a special case of the well-poised elliptic hypergeometric integral studied in Appendix \ref{sapp:wp-ell-int} (up to the prefactor).
Including the prefactor and simplifying the explicit evaluation \eqref{eq:wp-EHI-n=4}, we find the following expression for the residue sum
\begin{align}\label{eq:ressum-su2-n4-full}
    \begin{split}
        \cI_2(y_1,y_2,y_3;p,q) =&\ \sum_{i=1}^3\q_p(y_i)\q_q(y_i) \G(y_i;p,q) 
        \G(y_i^2;p,q) \prod_{j\neq i}\G(y_jy_i^{-1};p,q)\\
        & \phantom{\sum_{i=1}^3abc} \times\sum^{\infty}_{k,l=0} \frac{{}_4W^{(k)}_3(\vec{\mathpzc{a}}_i;p;q;p)\: {}_4W^{(l)}_3( \vec{\mathpzc{b}}_i;q;p;q)}{1-y_ip^kq^l} 
    \end{split}
\end{align}
where ${}_4W_3^{(k)}$ is the $k^{\text{th}}$ summand of a well-poised elliptic hypergeometric series ${}_4W_3$, defined in Appendix \ref{sapp:ell-hypergeom}.
For convenience, we write it explicitly here:
\begin{equation}\label{eq:4w3-example}
    {}_4W^{(k)}_3(\vec{\mathpzc{a}}_1;p;q;z)=\frac{\q_{q}(y_1^2,py_1,y_1y_2,q^{-1}y_1y_3;p)_k}{\q_{q}(p,y_1,py_1y_{2}^{-1},pqy_1y_{3}^{-1};p)_k} z^k\,,
\end{equation}
with $\q_{q}(x;p)_k$ the elliptic shifted factorial defined in Appendix \ref{sapp:ell-shift-fact}.
The $\vec{\mathpzc{a}}_1$ argument is shorthand notation for the arguments of the theta functions on the top row, while the arguments on the lower row are fixed by the well-poisedness condition.
We also note that, in \eqref{eq:ressum-su2-n4-full}, the two factors have $z=p$ and $z=q$ respectively.
Furthermore, $\vec{\mathpzc{b}}_1$ is identical to $\vec{\mathpzc{a}}_1$ up to the exchange $p\leftrightarrow q$, and the parameters $\vec{\mathpzc{a}}_{2,3}$ and $\vec{\mathpzc{b}}_{2,3}$ are related to $\vec{\mathpzc{a}}_{1}$ and $\vec{\mathpzc{b}}_{1}$ through cyclic permutation of the $y_i$.
By expanding the denominator of the summand in the second line of \eqref{eq:ressum-su2-n4-full}, one can also write this line in terms of (an infinite sum over) a product of two well-poised elliptic hypergeometric series ${}_4W_3$:
\begin{align}\label{eq:inf-sum-ell-hyp-series}
    \begin{split}
        \sum^{\infty}_{k,l=0} \frac{{}_4W^{(k)}_3(\vec{\mathpzc{a}}_i;p;q;p)\: {}_4W^{(l)}_3( \vec{\mathpzc{b}}_i;q;p;q)}{1-y_ip^kq^l}=\sum^{\infty}_{m=0} y_i^{m} \:{}_4W_3(\vec{\mathpzc{a}}_i;p;q;p^{m+1})\, {}_4W_3( \vec{\mathpzc{b}}_i;q;p;q^{m+1}) \ .
    \end{split}
\end{align}
As mentioned in Section \ref{ssec:ell-hypergeom-integrals} and discussed in more detail in Appendix \ref{app:ell-hypgeom-int}, arguing for the convergence of elliptic hypergeometric series requires some care.
By performing a ratio test, we show in Appendix \ref{sapp:wp-ell-int} that the expression obtained above converges when $|y_a|=y<1$ and $|p|=|q|<1$, with the balancing condition $y_1y_2y_3=pq$ satisfied.\footnote{We assume that the phases of the fugacities are generic such that the poles of the integrand in \eqref{eq:full-index-su2-red} are simple.}
This also implies that a potential contribution from the origin ($s=0$) is in fact vanishing.
A symmetric treatment of the absolute values of $p$, $q$ and $y_a$ certainly seems natural from the perspective of the full superconformal index.
However, recovering for example the Macdonald index from the full index, one would also like to consider asymmetric regimes, for example where $|p|,|y_3|\ll |q|,|y_{1,2}|$.
We have checked in a few asymmetric cases,  for example $|q|=|p|^{\frac{1}{2}}$, that a ratio test again implies convergence.
We leave a more systematic analysis to future work.

The analytic structure of a term in \eqref{eq:ressum-su2-n4-full} for fixed $i$ is highly non-trivial. 
Apart from simple poles at unflavored limits $y_a=y$, as we are now accustomed to, it also has poles of arbitrary degree whenever $p^m=q^n$, for any $m,n\in\mathbb{N}$.
This is due to the factors of $\theta_q(p;p)_k$ and $\theta_p(q;q)_k$ in the denominators of the respective elliptic hypergeometric summands and reflects the fact that the original integrand develops higher order poles for such specializations.
In terms of the chemical potentials $\sigma\equiv \sigma_1+i\sigma_2$ and $\tau=\tau_1+i\tau_2$, with $p=e^{2\pi i\sigma}$ and $q=e^{2\pi i\tau}$, this implies that the summands hit singularities when simultaneously
\begin{equation}
    \sigma_1=\frac{n}{m}\tau_1\,,\quad \sigma_2=\frac{n}{m}\tau_2\,,
\end{equation}
for any $m,n\in\mathbb{N}$.
It follows that as long as the real and imaginary parts of $\sigma$ and $\tau$ are not related by the same rational, our expression for the index is well-defined.\footnote{Note that this restriction is complementary to the Bethe Ansatz formula for the superconformal index, which is precisely defined for $p^m=q^n$ \cite{Benini:2018mlo,Lezcano:2021qbj,Benini:2021ano}.}
A particularly simple example that guarantees this is
\begin{equation}
    \sigma=-\bar{\tau}\,.
\end{equation}
This special case is also natural from the point of view of a localization computation of the index in that the background geometry, a so--called Hopf surface, preserves four supercharges (as opposed to two) \cite{Closset:2013vra}.
We stress that the complicated analytic behavior of the summands should cancel out in the full expression, since the index should admit these specializations. 

It would be extremely interesting if our expression for the full index admits a simplification meeting (some of) the criteria specified at the end of Section \ref{ssec:bmn-mcd}.
Such an expression could perhaps simplify the analytic structure and suggest a physical interpretation, as in the case of the Macdonald index.
We also expect that a simplified expression, if it exists, will shed light on the TQFT formulation of the general superconformal index \cite{Gadde:2011uv} (see also \cite{Razamat:2013qfa}).
This expectation is based on Section \ref{ssec:comparison-tqft}, where we observed that the simplified expression for the Macdonald index can be viewed as a resummation of the TQFT expression.

We end this section with some comments on the modular properties of the expression.
First, apart from the fact that the elliptic hypergeometric summands ${}_4W^{(k)}_3(\vec{\mathpzc{a}}_i;p;q;p)$ are well-poised, they are also elliptically balanced (see Appendix \ref{sapp:ell-hypergeom}).
This implies, by Theorem 3 of \cite{Spiridonov32002}, that the functions are modular invariant in the following sense.
If we set $y_i=e^{2\pi i \Delta_i}$, $p=e^{2\pi i\sigma}$, $q=e^{2\pi i\tau}$, and define 
\begin{equation}
    \tilde{y}_i=e^{\frac{2\pi i \Delta_i}{\tau}}\,,\quad \tilde{p}=e^{\frac{2\pi i\sigma}{\tau}}\,,\quad \tilde{q}=e^{-\frac{2\pi i}{\tau}} \, .
\end{equation}
Then using the modular property of the $\theta_q(x)$ function in \eqref{eq:theta-mod-prop} we find for the elliptic hypergeometric summand in \eqref{eq:4w3-example} that
\begin{equation}\label{eq:4W3-mod-prop}
{}_4W^{(k)}_3(\tilde{y}_1^2;\tilde{p}\tilde{y}_1,\tilde{y}_1\tilde{y}_2,\tilde{p}\tilde{y}_2^{-1};\tilde{p};\tilde{q};\tilde{p})=e^{2\pi i k\frac{\sigma(1-\tau)}{\tau}} {}_4W^{(k)}_3(y_1^2;py_1,y_1y_2,py_2^{-1};p;q;p)\,,
\end{equation}
where we solved for $y_3=pq/y_1y_2$.
The ($k$--dependent) factor of automorphy is completely due to the non--trivial transformation of the final argument $z=p$ of ${}_4W^{(k)}_3$ (see \eqref{eq:4w3-example}).
Indeed, one can check that the ratio of theta functions in \eqref{eq:4w3-example} is fully invariant under the modular transformation.
Moreover, the function is manifestly invariant under $\tau\to \tau +1$, so it transforms simply under the full modular group $SL(2,\mathbb{Z})$.

The second elliptic hypergeometric summand in \eqref{eq:ressum-su2-n4-full} satisfies a similar modular property, but now with the roles of $p$ and $q$ exchanged.
The fact that the factors satisfy modular properties with different modular parameters ($\tau$ and $\sigma$, respectively), is closely related to modular factorization of 4d superconformal indices \cite{Gadde:2020bov,Jejjala:2022lrm}.
This proposal states that the index for general gauge theories can be factorized in a variety of ways labeled by the modular group and generalizes holomorphic block factorization of the 4d index \cite{Nieri:2015yia} (see also \cite{Yoshida:2014qwa,Peelaers:2014ima,Gadde:2020bov}).
Schematically, it states that
\begin{align}\label{eq:gen-factorization-sci}
    \begin{split}
        \mathcal{I}(a_i;\sigma,\tau)&\cong\sum_i\mathcal{B}_{i}\left(\tfrac{a_i}{c\tau+d};\tfrac{\sigma+\tau}{c\tau+d},\tfrac{a\tau+b}{c\tau+d}\right)\tilde{\mathcal{B}}_i\left(\tfrac{a_i}{-c\sigma+d};\tfrac{\tau+\sigma}{-c\sigma+d},\tfrac{a\sigma-b}{-c\sigma+d}\right)\,,
    \end{split}
\end{align}
for any integers $a,b,c,d$, such that $ad-bc=1$.
Using the modular properties of the elliptic hypergeometric summands and of the elliptic Gamma function \eqref{eq:Gamma-gen-mod-prop-app}, the fixed $k,l$ summand of the index in \eqref{eq:ressum-su2-n4-full} can be seen to almost reflect this structure, apart from the denominator of the summand, which does not factorize.\footnote{
Interestingly, in the Schur limit it is precisely this denominator that, when summed over, becomes a twisted Eisenstein series (see the discussion below \eqref{eq:n=4-schur-index-su2-res-sum}), which does have modular properties \cite{Pan:2021mrw}.}
The lack of exact factorization may be related to the fact that one cannot fully Higgs the gauge group in the $\mathcal{N}=4$ theory.
Indeed, as shown in \cite{Peelaers:2014ima}, full factorization of the index requires the gauge group to be fully Higgsable.
More specifically, sufficient fundamental matter is required and, in addition, a $U(1)$ factor in the gauge group with a non-zero Fayet-Iliopoulos coupling turned on.

\subsection{\texorpdfstring{$\mathcal{N}=2$}{N=2} \texorpdfstring{$SU(2)$}{SU(2)} SQCD with \texorpdfstring{$N_{f} = 4$}{Nf=4}}\label{ssec:nf4-indices}

Apart from the $\mathcal{N}=4$ $SU(2)$ SYM theory, there is another well-known superconformal field theory with $SU(2)$ gauge group: $\mathcal{N}=2$ $SU(2)$ SQCD with $N_{f} = 4$.
This theory fits into the $A_1$ class $\mathcal{S}$ theories and is labeled by a four-punctured sphere \cite{Gaiotto:2009we}.
Various indices of this theory were first analyzed in \cite{Gadde:2009kb,Gadde:2011ik,Gadde:2011uv} and, more recently, its Schur index in \cite{Pan:2021mrw}.
In this section, we will evaluate its Macdonald and full superconformal index through residues.

\subsubsection{Macdonald Index}

From Section \ref{ssec:sc-gauge-theory-inds}, we collect the contour integral expression for the Macdonald index.
With the reduced measure, we have the following expression:
\begin{align}\label{eq:mcd-su2-nf4-red}
    \begin{aligned}
    \mathcal{I}_{\text{SQCD}}(y_a;q,t) &= (q,t;q)_{\infty} \oint \frac{du}{2 \pi i u} \frac{(u^{2},qu^{-2},t u^{\pm 2};q)_{\infty}}{\prod_{a= 1}^{8} (y_{a} u^{\pm 1};q)_{\infty}} 
    \end{aligned}
\end{align}
where we define 
\begin{equation}\label{eq:ya-in-ups}
    y_{a} = \{ t^{\frac{1}{2}} \upsilon_1^{\pm } \upsilon_2^{\pm }, t^{\frac{1}{2}} \upsilon_3^{\pm } \upsilon_4^{\pm } \}
\end{equation}
with $|\upsilon_i|=1$ flavor fugacities for the $SU(2)^4\subset SO(8)$ flavor symmetry subgroup.
We now use the fact that
\begin{equation}
    (x^2;q)_\I=(\pm x,\pm q^{\frac{1}{2}}x;q)_\I
\end{equation}
to note that \eqref{eq:mcd-su2-nf4-red} is again of the basic hypergeometric form defined in \eqref{eq:gen-basic-hypergeom-integral}.
In particular, as argued at the end of Section \ref{ssec:basic-hypergeom-integrals}, the suppression factor equals $q^4/t^2$, making the residue sum well-defined for $|q|<|t^{\frac{1}{2}}|<1$. 
In addition, this implies that the contribution from the origin vanishes.
We then use the general evaluation in \eqref{eq:ressum-gen-basic-hypergeom-integral} to find the following expression for the residue sum:
\begin{align}\label{eq:mcd-nf4-red}
    \begin{aligned}
    \mathcal{I}_{\text{SQCD}}(y_a;q,t) & =  \sum_{a = 1}^{8} \frac{(t,q y_{a}^{-2},t y_{a}^{\pm 2};q)_{\infty}}{ \prod_{b\neq a}^{8}( y_{b} y_{a}^{\pm};q)_{\infty}} \\
    &\times {}_{12} \phi_{11} \left[ \begin{array}{cccc}
        y_a^2 & y_{a}y_{b}  & \pm q t^{-\frac{1}{2}} y_{a} & \pm q^{\frac{1}{2}} t^{-\frac{1}{2}} y_{a}   \\
        q y_{a} y_{b}^{-1} &  \pm  t^{\frac{1}{2}} y_{a} & \pm q^{\frac{1}{2}} t^{\frac{1}{2}} y_{a}  & 
    \end{array}; q, \frac{q^4}{t^2} \right] 
    \end{aligned} 
\end{align}
where the $y_ay_b$ and $q y_{a} y_{b}^{-1}$ arguments of the ${}_{12} \phi_{11}$ series should be understood as seven separate arguments labeled by $b\neq a$.
Similarly, the arguments with $\pm$ should be read as two separate arguments with opposite sign.
We note that the entries are such that the series is well-poised.
This provides a closed form formula for the Macdonald index of $\mathcal{N}=2$ $SU(2)$ SQCD.

We note that the expression shares similar features with the other residue sums evaluated so far: the (apparent) presence of negative powers of $y_a$, and the apparent poles in the unflavored limits $y_a=y_b$.
One may again wonder whether a simplification exists, for which some or all of these features disappear.
As in Section \ref{ssec:bmn-mcd}, a special case of Sears' transformation formulas seems closely related but not quite applicable.
We leave further investigation to future work.

Generalized $S$-duality for the present theory implies that the index is invariant under the exchange $\upsilon_1\leftrightarrow \upsilon_3$ \cite{Gaiotto:2009we,Gadde:2009kb}.
This is not obvious from our expression and, as such, should lead to an interesting transformation formula for the ${}_{12} \phi_{11}$ series.
We expect that a simplification of the formula may manifest this symmetry, like the TQFT formula does, as we will discuss at the end of this section.

As observed in Section \ref{ssec:basic-hypergeom-integrals}, one can also make sense of the residue sum for this theory using the full measure, as long as $|q|<|t|$ (which precludes the Schur limit).
We may then again use the general results in Appendix \ref{app:bas-hypgeom-int} to find the alternative expression
\begin{align}\label{eq:mcd-nf4-full}
    \begin{aligned}
    \mathcal{I}_{\text{SQCD}}(y_a;q,t) & =\frac{1}{2}\sum_{a = 1}^{8} \frac{(t,y_{a}^{-2},t y_{a}^{\pm 2};q)_{\infty}}{ \prod_{b\neq a}^{8}( y_{b} y_{a}^{\pm};q)_{\infty}} \\
    &\times {}_{14} \phi_{13} \left[ \begin{array}{ccccc}
        y_a^2 & \pm q y_a & y_{a}y_{b}  & \pm q t^{-\frac{1}{2}} y_{a} & \pm q^{\frac{1}{2}} t^{-\frac{1}{2}} y_{a}   \\
        \pm y_a & q y_{a} y_{b}^{-1} &  \pm  t^{\frac{1}{2}} y_{a} & \pm q^{\frac{1}{2}} t^{\frac{1}{2}} y_{a}  & 
    \end{array}; q, \frac{q^2}{t^2} \right] \, .
    \end{aligned} 
\end{align}
The arguments of the ${}_{14}\phi_{13}$ series are such that it is very-well-poised balanced (see Appendix \ref{sapp:basic-hypergeom}).
The present formula appears most closely related to a special case of Sears' transformation formulas, specifically the $M=7$ case in eqn.~4.12.1 of \cite{Gasper_Rahman_2004}.
However, we have not been able to use the identity to simplify the index.
Similar to the expression obtained with the reduced measure, this formula is not manifestly invariant under generalized $S$-duality.
Therefore, the exchange $\upsilon_1\leftrightarrow \upsilon_3$ should imply yet again interesting transformation formulas.
In addition, the equality of \eqref{eq:mcd-nf4-red} and \eqref{eq:mcd-nf4-full}, for $|q|<|t|$, represents a non-trivial transformation formula for basic hypergeometric series.
This demonstrates how the indices computed with the reduced and full measure, when possible, can lead to non-trivial identities for basic, and perhaps also elliptic, hypergeometric series.

\paragraph{Hall--Littlewood limit}

Let us now consider the Hall-Littlewood limit of our expression.
Taking $q\to 0$ in \eqref{eq:mcd-nf4-red}, we obtain
\begin{align}\label{eq:nf4-hl-ressum}
    \mathcal{I}(y_a;t) =  \sum_{a = 1}^{8} \frac{(1 - t)(1 - ty_{a}^{\pm 2})}{\prod_{b \neq a}^{8} (1 - y_{b}y_{a}^{\pm} )} \,.
\end{align}
We find that this expression \emph{can} be brought into a simplified form
\begin{equation}
    \mathcal{I}(y_a;t) = \frac{(1-t)(1-t^2)P_{20}(\upsilon_i;t)}{\prod_{a<b}(1-y_ay_b)}
\end{equation}
where $P_{20}(\upsilon_i;t)$ is a palindromic polynomial of order 20 in $t$, and we observe the symmetric dependence of the expression on solely positive powers of $y_a$.
As a result, the unflavored limit is easily taken, which would have been subtle in \eqref{eq:nf4-hl-ressum} due to the apparent poles.
In other words, the simplified expression satisfies all criteria listed at the end of Section \ref{ssec:bmn-mcd}.
The fully refined expression for $P_{20}(\upsilon_i;t)$, however, is very long.
Only when all $\upsilon_i=\upsilon$, the expression simplifies and can be written as
\begin{align}
    \begin{split}
        \mathcal{I}(\upsilon;t) =&\, \frac{(1+t)P_{6}(\upsilon;t)}{(1-t)^2(1-t\upsilon^{\pm 2})^4(1-t\upsilon^{\pm 4})}\\
        P_6(\upsilon;t)=& \,1+t(4\upsilon^2+7+4\upsilon^{-2})+t^2(\upsilon^4+4\upsilon^2+5+4\upsilon^{-2}+\upsilon^{-4})\\
        &-t^3(5\upsilon^4+16\upsilon^2+20+16\upsilon^{-2}+5\upsilon^{-4})+\ldots 
    \end{split}
\end{align}
where $P_6(\upsilon;t)$ is a polynomial of order $6$ in $t$, and the dots are completed using the palindromic property.

The simplified expression for the HL index can be interpreted in terms of the HL chiral ring of the theory \cite{Gadde:2011uv,Beem:2013sza}.
In particular, the bosonic generators of the ring are the mesons $M^{[ab]}=\bar{q}^{a}_i\bar{q}^{b}_j\epsilon^{ij}$, with $a,b=1,\ldots, 8$, $SO(8)$ vector indices, and $i,j=1,2$ fundamental $SU(2)$ indices.
Their contributions are contained in the denominator of the expression.
Furthermore, all fermionic generators are lifted from the ring at one-loop; their contribution to the index can be interpreted in terms of bosonic relations between the mesonic generators in the Higgs branch chiral ring \cite{Argyres:1996eh,Gadde:2011uv}, whose Hilbert series therefore matches with the HL index \cite{Benvenuti:2010pq,Hanany:2010qu}.

\paragraph{Comparison with TQFT formula}
Let us finally compare the expressions with the TQFT expression for the Macdonald index, which corresponds to the $(n,g)=(4,0)$ case of the general expression in \eqref{eq:mcd-tqft-class-s}:
\begin{equation}\label{eq:su2-nf4-mcd-tqft}
    \mathcal{I}^{\text{TQFT}}_{\text{SQCD}}(\upsilon_i;q,t)=\frac{(t^2;q)^2_\I}{(q;q)_\I(t;q)^3_\I\prod^4_{i=1}(t\upsilon_i^2,t\upsilon_i^{-2};q)_\I}\sum^{\infty}_{\lambda=0}\frac{\prod^4_{i=1}P_{\lambda}(\upsilon_i,\upsilon_i^{-1};q,t)}{\left(P_{\lambda}(t^{\frac{1}{2}},t^{-\frac{1}{2}};q,t)\right)^2}\,,
\end{equation}
where the $SU(2)$ Macdonald polynomial $P_{\lambda}(v,v^{-1};q,t)$ was defined in \eqref{eq:su2-mcd-pol}.
Unlike our residue sums, this expression is manifestly invariant under permutations of the $\upsilon_i$, reflecting the generalized $S$-duality of the $N_f=4$ theory.
Making use of \eqref{eq:norm-mcd-pols} and \eqref{eq:tqft-mcd-id}, we can rewrite the expression in terms of the $q$-ultraspherical polynomials defined in \eqref{eq:q-ultra-sph-pol}:
\begin{equation}\label{eq:su2-nf4-mcd-tqft-2}
    \mathcal{I}^{\text{TQFT}}_{\text{SQCD}}(\upsilon_i;q,t)=\frac{(t^2;q)^3_\I}{(t;q)^5_\I\prod^4_{i=1}(t\upsilon_i^2,t\upsilon_i^{-2};q)_\I}\sum^{\infty}_{\lambda=0}t^\lambda(1-tq^\lambda)\frac{(q;q)^3_\lambda}{(t^2;q)^3_\lambda}\prod^4_{i=1}C_{\lambda}(\upsilon_i,\upsilon_i^{-1};t|q)\,.
\end{equation}
The equality between the residue sums and the TQFT expression represents a highly non-trivial identity.
As in Section \ref{ssec:comparison-tqft}, the residue sums can be viewed as resumming the Laurent series in flavor fugacities expressed by the TQFT formula.
A potential simplification of the residue sums for the Macdonald index, either \eqref{eq:mcd-nf4-red} or \eqref{eq:mcd-nf4-full}, could realize a relatively simple generating function for this four-fold product of $q$-ultraspherical polynomials.

\subsubsection{Full superconformal index}\label{sec:N=2-Nf=4-SQCD-full-index}

In this section, we use the reduced measure to evaluate the residue sum for the full superconformal index.
The relevant integral is the $N=2$ case of \eqref{eq:sqcd-index-defn-gen-nc-red}, which we copy here for convenience
\begin{equation}
	\cI_{\text{SQCD}}(v_i;p,q,t) = (p;p)_\I(q;q)_\I\G(pq/t)\oint\frac{du}{2\pi iu} \frac{\G(\frac{pq}{t} u^{\pm2})}{(1-u^{-2})\G(u^{\pm 2})} \prod_{a=1}^8 \G(y_a u^\pm) \,,
\end{equation}
with the $y_a$ as in \eqref{eq:ya-in-ups} and the shorthand notation $\Gamma(x)\equiv \Gamma(x;p,q)$.
Making use of \eqref{eq:gamma-split}, this integral can be seen to correspond to the reduced measure, or the $\epsilon=1$ version of the very-well-poised elliptic hypergeometric integral for $A=16$, as defined in \eqref{eq:general-vwp-EHI}.
To evaluate its residue sum, we use the general evaluation in \eqref{eq:vwp-EHI-general-evaluation}.
In particular, we define the variables $a_\a$ as
\begin{equation}\label{eq:ai-nf4}
    a_{1,\ldots,8} = y_a\,, \quad a_{9,\cdots,16} = \pm (pq/t)^{\frac{1}{2}} p^{\frac{\d_1}{2}} q^{\frac{\d_2}{2}}\,,\quad \prod_\alpha a^2_\alpha=(pq)^{12}
\end{equation}
with $\d_{1,2}=0,1$.
We then obtain the following expression:
\begin{align}\label{eq:N=2-Nf=4-SQCD-full-index}
    \begin{split}
        \cI_{\text{SQCD}}(\upsilon_a;p,q,t)	=\ &  \G(pq/t;p,q)\sum^{16}_{i=1} \frac{\prod_{j\neq 1}^{16} \G(a_j a_i^\pm;p,q)}{\G(a_i^{-2};p,q)}\\
        &\times \sum_{k,l=0}^\I \frac{(-a_i^2) \:{}_{20}V_{19}^{(k)}(\vec{\mathpzc{a}}_i;p;q;p^2)		\, {}_{20}V_{19}^{(l)}(\vec{\mathpzc{b}}_i;q;p;q^2) }{1-a_i^2 p^{2k}q^{2l}} \,,
    \end{split}           
\end{align}
where ${}_{20}V_{19}^{(k)}$ is the $k^{\text{th}}$ summand of the very-well-poised elliptic hypergeometric series ${}_{20}V_{19}$ defined in Appendix \ref{sapp:ell-hypergeom}.
Explicitly, we have
\begin{equation}\label{eq:20V19in-thetas}
    {}_{20}V_{19}^{(k)}(\vec{\mathpzc{a}}_1;p;q;z) = \frac{\q_q(a_1^2p^{2k})}{\q_q(a_1^2)} \frac{\q_{q}(a_1^2,a_1a_2,\ldots,a_1a_{10},a_1a_{11}q^{-1},\ldots,a_1a_{16}q^{-1};p)_k}{\q_{q}(p,pa_1a_2^{-1},\ldots,pa_1a_{10}^{-1},pqa_1a_{11}^{-1},\ldots,pqa_1a_{16}^{-1};p)_k} (pz)^{k}\,,
\end{equation}
where $\vec{\mathpzc{a}}_1$ is shorthand notation for the arguments of the top-row of theta functions (see also \eqref{eq:vwp-pzc-ab-defn}) and we note that the (very-)well-poised condition fixes the arguments of the lower row.
We also note that, in \eqref{eq:N=2-Nf=4-SQCD-full-index}, the two factors have $z=p^2$ and $z=q^2$ respectively.
Furthermore, $\vec{\mathpzc{b}}_1$ is obtained from $\vec{\mathpzc{a}}_1$ through an exchange of $p\leftrightarrow q$, and, \textit{e.g.}, $\vec{\mathpzc{a}}_2$ is obtained from $\vec{\mathpzc{a}}_1$ through the cyclic permutation $(a_1,a_2,\ldots,a_{15},a_{16})\to (a_2,a_3,\ldots,a_{16},a_{1})$.

Up until this point, our expression is formal in the sense that we have not yet demonstrated its convergence.
Leaving the details of the ratio test to Appendix \ref{sapp:vwp-ell-int}, we simply state here that the expression turns out to be convergent for $|p|=|q|=|t|= T^2$ (so, that $|y_a|=T$), with $T<1$.
As usual, this implies that the contribution from the origin vanishes.
We expect that less symmetric regimes will lead to convergent expressions as well, but we leave a systematic analysis to future work.  

It can be checked that the  ${}_{20}V_{19}^{(k)}$ satisfies the same modular property as we encountered for ${}_{4}W_{3}^{(k)}$ in the $\mathcal{N}=4$ $SU(2)$ index.
To see this, one should transform each of the theta functions in \eqref{eq:20V19in-thetas} as:
\begin{equation}\label{eq:theta-mod-prop-main-text}
    \theta_{\tilde{q}}(\tilde{x})=e^{i\pi B(z;\tau)}\theta_q(x)\,,\quad \tilde{x}=e^{2\pi i \frac{z}{\tau}}\,,\: \tilde{p}=e^{2\pi i \frac{\sigma}{\tau} }\,,\: \tilde{q}=e^{-2\pi i \frac{1}{\tau}}\,,
\end{equation}
even if $x$ is $q$-dependent.\footnote{In the parametrization \eqref{eq:ai-nf4}, this is the case for theta functions containing $a_9$ and $a_{10}$ in their argument.}
With this understanding, we find that
\begin{equation}\label{eq:20V19-mod-prop}
    {}_{20}V^{(k)}_{19}(\vec{\tilde{\mathpzc{a}}}_1;\tilde{p};\tilde{q};\tilde{p}^2)=e^{2\pi i k\frac{\sigma(1-\tau)}{\tau}}{}_{20}V_{19}^{(k)}(\vec{\mathpzc{a}}_1;p;q;p^2)
\end{equation}
where $\vec{\tilde{\mathpzc{a}}}_1$ transforms each of the $x$ arguments as in \eqref{eq:theta-mod-prop-main-text}.
The non--trivial transformation under the modular transformation is purely due to the final $z=p^2$ argument of ${}_{20}V^{(k)}_{19}$.

Just like our residue sums for the Macdonald indices, this expression is not obviously invariant under generalized $S$-duality.
Therefore, generalized $S$-duality, implemented in the index by $\upsilon_1\leftrightarrow \upsilon_3$ (see \eqref{eq:ya-in-ups}), implies a non-trivial transformation formula for the sums of bilinear elliptic hypergeometric summands.
Finding a formula which manifests this symmetry may well have consequences for the understanding of the TQFT description of the full superconformal index.
We refer the reader for further comments to the text below equation \eqref{eq:inf-sum-ell-hyp-series}, which apply here as well.

\subsection{\texorpdfstring{$\cN=1$}{N=1} \texorpdfstring{$SU(2)$}{SU(2)} SQCD}\label{ssec:n=1-sqcd}

In this section, we calculate the superconformal indices of $\cN=1$ $SU(2)$ theories for arbitrary $N_f$ within the conformal window.
See Section \ref{ssec:n=1-gauge-theory-inds} for a brief review and references.

We take the integral with the reduced measure from \eqref{eq:N=1-SUN-index-defn-red}, which reads 
\begin{equation}            \label{eq:N=1-full-index-integral}
    \cI_E(y_i;p,q) = (p;p)_\I(q;q)_\I\oint\frac{du}{2\pi iu}\frac{\prod_{i=1}^{2N_f}\G(y_iu^\pm;p,q)}{(1-u^{-2})\G(u^{\pm2};p,q)}\,,
\end{equation}
where the $y_{i}$ are fugacities for the enhanced $ SU(2N_f)$ flavor symmetry.
One may now recognize that the expression corresponds to the reduced measure, or the $\epsilon=1$ version of the very-well-poised elliptic hypergeometric integral for $A=2N_f$, as defined in \eqref{eq:general-vwp-EHI}.
In particular, the parameters satisfy the balancing condition 
\begin{equation}
    \prod^{2N_f}_{i=1} y_i^2=(pq)^{2N_f-4}\,.
\end{equation}
We evaluate the integral making use of \eqref{eq:vwp-EHI-general-evaluation}, which provides us with
\begin{align} \label{eq:N=1-general-Nf}
    \begin{split}
        \cI_E(y_i;p,q) =&\ \sum^{2N_f}_{i=1} \frac{\prod_{j\neq1}^{2N_f}\G(y_jy_i^\pm)}{\G(y_i^{-2})} \\
        &\times \sum_{k,l=0}^\I \frac{(-y_i^2)\:{}_{2N_f+4}V_{2N_f+3}^{(k)}(\vec{\mathpzc{a}}_i;p;q;p^2)		\, {}_{2N_f+4}V_{2N_f+3}^{(l)}(\vec{\mathpzc{b}}_i;q;p;q^2)}{1-y_i^2 p^{2k}q^{2l}}   \,,
    \end{split}
\end{align}
where ${}_{2N_f+4}V_{2N_f+3}^{(k)}$ is the $k^{\text{th}}$ summand of the very-well-poised elliptic hypergeometric series ${}_{2N_f+4}V_{2N_f+3}$ defined in \eqref{eq:defn-vwp-elliptic-hypergeom-series}.
More precisely, we have
\begin{align}\label{eq:2nf+4V2n+3in-thetas}
    \begin{split}
        {}_{2N_f+4}&V_{2N_f+3}^{(k)}(\vec{\mathpzc{a}}_1;p;q;z) = \\
        &\frac{\q_q(y_1^2p^{2k})}{\q_q(y_1^2)} \frac{\q_{q}(y_1^2,y_1y_2,\ldots,y_1y_{N_f+2},y_1y_{N_f+3}q^{-1},\ldots, y_1y_{2N_f}q^{-1};p)_k}{\q_{q}(p,py_1y_2^{-1},\ldots,py_1y_{N_f+2}^{-1},pqy_1y_{N_f+3}^{-1},\ldots,pqy_1y_{2N_f}^{-1};p)_k} (pz)^{k}\,,
    \end{split}
\end{align}
where $\vec{\mathpzc{a}}_1$ is shorthand notation for the arguments of the top-row of theta functions (see also \eqref{eq:vwp-pzc-ab-defn}) and we note that the (very-)well-poised condition fixes the arguments of the lower row.
where the two factors in \eqref{eq:N=1-general-Nf} have $z=p^2$ and $z=q^2$ respectively.
Furthermore, $\vec{\mathpzc{b}}_1$ is obtained from $\vec{\mathpzc{a}}_1$ through an exchange of $p\leftrightarrow q$, and, \textit{e.g.}, $\vec{\mathpzc{a}}_2$ is obtained from $\vec{\mathpzc{a}}_1$ through the cyclic permutation $(y_1,y_2,\ldots,y_{2N_f-1},y_{2N_f})\to (y_2,y_3,\ldots,y_{2N_f},y_{1})$.
One can check that the elliptic hypergeometric summand satisfies the same modular property we observed in \eqref{eq:20V19-mod-prop}.

As described in detail in Appendix \ref{sapp:vwp-ell-int}, if we take the magnitude of the various fugacities as $|y_i|=T^{N_f-2}$, and $|p|=|q|= T^{N_f}$, for $T<1$, a ratio test demonstrates that the residue sum converges and, consequently, the contribution from the origin vanishes.
Interestingly, with this symmetric scaling, convergence can only be shown for $3\leq N_f\leq 7$.
We note that this range contains the conformal window  $3\leq N_f\leq 6$. But, it also includes the $N_f=7$ case, for which the theory is IR free \cite{Seiberg:1994pq}.

One may be puzzled by the fact that we cannot demonstrate convergence for $N_f=8$.
Indeed, we were able to establish convergence for the residue sum associated with the $\mathcal{N}=2$ $SU(2)$ SQCD theory with $N_f=4$ in Section \ref{ssec:nf4-indices}, which corresponds to the same elliptic hypergeometric integral!
This apparent contradiction is resolved by noting that the relative magnitudes of the fugacities are scaled differently between the two cases.
This therefore provides a concrete example of the subtleties associated with the convergence of residue sums for elliptic hypergeometric integrals or, indeed, elliptic hypergeometric series.

Let us contrast our proposed residue sum with previous attempts \cite{Peelaers:2014ima,spiri2024}.
Both references define their contour integrals with the full measure.
However, as shown in Appendix \ref{sapp:vwp-ell-int}, the residue sum as defined with the full measure diverges, at least with our choice of relative magnitudes of the fugacities.
There may exist alternative choices for which the residue sum with full measure converges. But, based on some experiments we ran, we find this unlikely.
As a consequence, for these cases one \emph{cannot} conclude that the contribution from the origin is vanishing.
We note that these issues were acknowledged in \cite{Peelaers:2014ima} and discussed in more detail in \cite{spiri2024}, but to the best of our understanding they were not resolved.

Finally, let us consider the special case of $N_f=3$, when the Seiberg dual theory is free and the index is described by a product formula (see Section \ref{ssec:n=1-gauge-theory-inds}).
Equating our residue sum with the product formula, we find
\begin{align}\label{eq:ell-hypgeom-sum}
    \begin{split}
        & \sum^{6}_{i=1} \frac{\prod_{j\neq1}^{6}\G(y_jy_i^\pm)}{\G(y_i^{-2})}  \sum_{k,l=0}^\I \frac{(-y_i^2)\:{}_{10}V_{9}^{(k)}(\vec{\mathpzc{a}}_i;p;q;p^2)		\, {}_{10}V_{9}^{(l)}(\vec{\mathpzc{b}}_i;q;p;q^2)}{1-y_i^2 p^{2k}q^{2l}}=\prod_{1\leq i<j\leq 6}\Gamma(y_iy_j)  \,.
    \end{split}
\end{align}
This can be read as a non-terminating summation formula for the product of elliptic hypergeometric summands.
One could rewrite this identity alternatively in terms of an infinite sum of products of two full elliptic hypergeometric series by expanding the denominator as a geometric series, as in \eqref{eq:inf-sum-ell-hyp-series}.
A key feature this identity is that the right hand side demonstrates how the complicated analytic structure of summands on the left hand side for fixed $i$, as described for the $\mathcal{N}=4$ $SU(2)$ case in Section \ref{ssec:full-indexn4-su2}, simplify when they are summed, as expected.
Indeed, the right hand side is well-defined in the unflavored limits $y_i=y_j$, and even for $p^m=q^n$, for $m,n\in \mathbb{N}$.
We therefore see that in this case, there indeed exists a simplification of the residue sum with the expected features listed at the end of Section \ref{ssec:bmn-mcd}.
Whether there exist simplifications for the superconformal indices of other theories remains an important open question.

As reviewed in Appendix \ref{sapp:basic-hypergeom}, for basic hypergeometric series there exist both terminating and non-terminating summation formulas.
However, in the case of elliptic hypergeometric series, only summations for terminating series are known such as the Frenkel-Turaev summation \cite{Frenkel-Turaev}.
It has been an open problem to find summations for non-terminating elliptic hypergeometric series.
The fact that such summations should take on a bilinear form, as we find, was for example suggested in \cite{spiri2024}.
However, the non-factorizing denominator, which is a direct manifestation of the reduced measure, and is therefore closely related to convergence of the residue sum, was not noted before.

Let us end this section with a small comment on the $N_f=4$ case \cite{Spiridonov:2008thf}.
As reviewed in Section \ref{ssec:n=1-gauge-theory-inds}, in this case both the electric and magnetic index can be expressed in terms of a single $(SU(2))$ contour integral. 
Therefore, by evaluating both sides as a residue sum, with the reduced measure, could lead to a non-trivial transformation formula for (bilinear summands) of elliptic hypergeometric series.
We hope to report on this identity soon.

\section{Towards higher rank indices}\label{sec:higher-rank}

All examples considered so far correspond to $SU(2)$ gauge theories.
However, as argued in Section \ref{sec:conv-res-sums}, our method generalizes to higher rank gauge theories as well.
Technically, the main complication at higher rank is that one has to evaluate multiple contour integrals.
We take a first step towards evaluating higher rank indices through residues by considering the Macdonald index of the $\mathcal{N} = 4$ $SU(3)$ SYM theory.
We find that it can be expressed in terms of a double basic hypergeometric series.
After obtaining the expression for the index, we evaluate and interpret two limits: a $t = 1$ limit, in which the index is expressed in terms of a product formula, and the Hall-Littlewood limit $q\to 0$.

\subsection{\texorpdfstring{$SU(3)$}{SU(3)} Macdonald index for \texorpdfstring{$\mathcal{N} = 4$}{N=4} SYM}\label{ssec:su3mcd}

\begin{figure}[t]
    \centering
    \subfigure[Residue sum for $s_{2}$]{\includegraphics[width=0.45\linewidth]{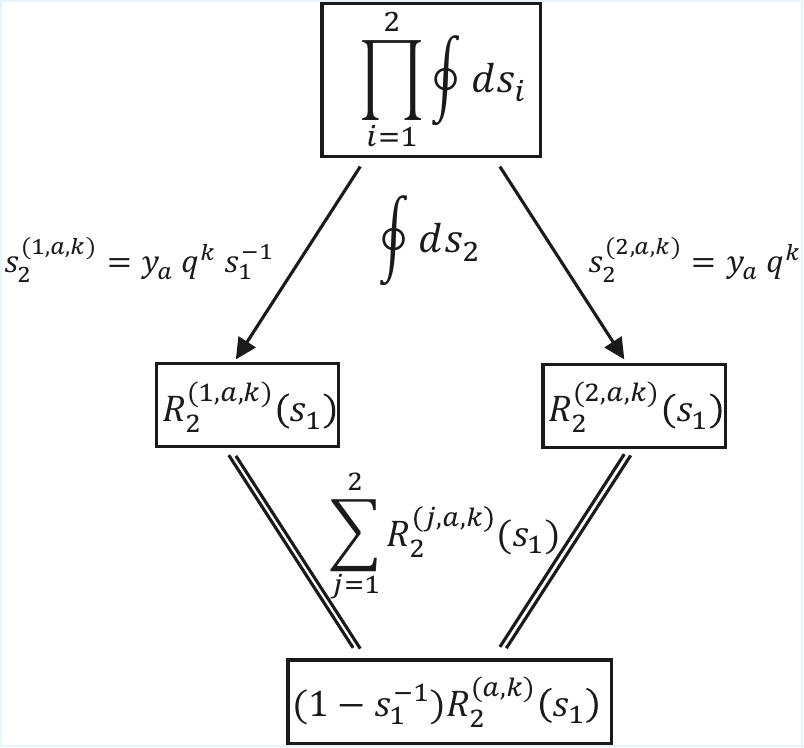}}
    \subfigure[Residue sum for $s_{1}$]{\includegraphics[width=0.53\linewidth]{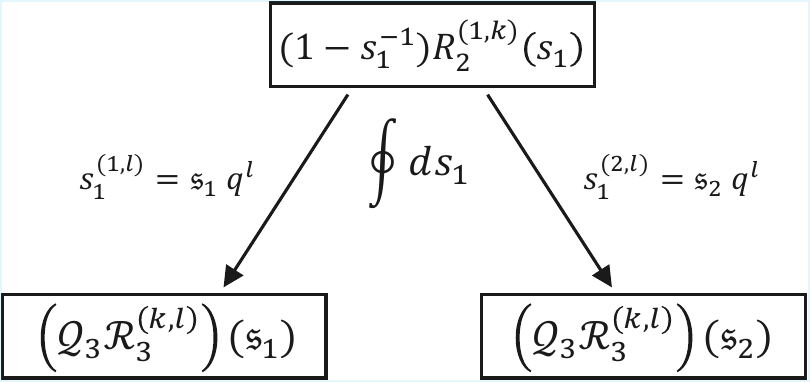}}
    \label{fig:SU3schematic}
    \caption{A schematic breakdown of the $SU(3)$ Macdonald index residue sum computation. The residue evaluations are illustrated by the single lines with arrows which are labeled with their respective towers of poles. The intermediate simplifications are depicted by double lines without arrows, instead. (a) Evaluating the poles for the $s_{2}$ contour integral, then summing them to obtain equation \eqref{eq:secondresidueintegralSU(3)trick}. In (b) we take \eqref{eq:secondresidueintegralSU(3)trick} and evaluate its $s_{1}$ contour integral for $a = 1$, with the towers of poles labeled as in (a), where $\mathfrak{s}_{b} = \{ y_{1}, y_{1}^{-1} y_{2} \}$. These provide us with two terms in the final residue sum (where the remaining two terms can be obtained with the other choice of $a = 2$ in $y_{a}$), then by subtracting the enclosed pole on the unit circle we obtain \eqref{eq:finalexpressionsI3SU(3)}.}
\end{figure}

The integral of interest is the $N = 3$ specialization of \eqref{eq:mcd-index-suN-n=4-defn} with the reduced measure for the gauge singlet projection, which we reproduce here for convenience  
\begin{align}
    \begin{split}
        \mathcal{I}_{3}(v;q,t) &= \frac{(q,t;q)_{\infty}^2}{(t^{\frac{1}{2}} v, t^{\frac{1}{2}} v^{-1};q)_{\infty}^2} \prod_{i = 1}^{2} \oint_{|s_i|=1} \frac{ds_{i}}{2 \pi i s_{i}} \prod_{1 \leq i \leq j \leq 2} \frac{(s_{i,j},q s_{i,j}^{-1}, t s_{i,j}^{\pm};q)_{\infty}}{(t^{\frac{1}{2}}vs_{i,j}^{\pm},t^{\frac{1}{2}}v^{-1} s_{i,j}^{\pm};q)_{\infty}} \, .
    \end{split}
\end{align}
To evaluate the contour integrals, we follow the strategy outlined in Section \ref{ssec:basic-hypergeom-integrals}.
In particular, we first separate out the $s_2$ dependent part of the integrand (\textit{c.f.}, \eqref{eq:mcd-index-suN-n=4-recurs-struc}):
\begin{align}\label{eq:macdonald-index-su(3)-n=4-recurs-struc}
    \begin{split}
        \mathcal{I}_{3} (y_{1,2}; q) &= \frac{(q,y_{1} y_{2};q)_{\infty}^2}{(y_{1}, y_{2};q)_{\infty}^2} \oint\frac{ds_1}{2\pi i s_1} I_{2} (s_{1})
        \oint\frac{ds_{2}}{2\pi is_{2}}\prod^{2}_{i=1} \frac{(s_{i,2},q s_{i,2}^{-1}, y_{1} y_{2} s_{i,2}^{\pm};q)_{\infty}}{(y_{1} s_{i,2}^{\pm},y_{2} s_{i,2}^{\pm};q)_{\infty}}\,,
    \end{split}
\end{align}
where we write $y_{1} = t^{\frac{1}{2}} v$ and $y_{2} = t^{\frac{1}{2}} v^{-1}$.
The $s_2$--dependent integrand has four towers of poles located at 
\begin{align}\label{eq:su3s2poles}
    s_{2}^{(1,a,k)} = y_{a} q^{k} s_{1}^{-1}, \quad s_{2}^{(2,a,k)} = y_{a} q^{k}, \quad a \in \{ 1, 2\}, \quad k \in \mathbb{Z}_{\geq 0}\,,
\end{align}
where $|s_1|=1$ and $|y_{1,2}|,|q|<1$.
After evaluating the $s_2$ residue sum, the remaining contour integral over $s_1$ takes the form
\begin{align}\label{eq:secondresidueintegralSU(3)}
    \mathcal{I}_{3} (y_{1,2}; q) &=\sum^\infty_{k=0} \frac{(y_{1} y_{2};q)_{\infty}^2}{(y_{1}, y_{2};q)_{\infty}^2} \oint \frac{ds_{1}}{2 \pi i s_{1}} \frac{(s_{1},q s_{1}^{-1}, y_{1} y_{2} s_{1}^{\pm};q)_{\infty}}{(y_{1} s_{1}^{\pm}, y_{2} s_{1}^{\pm};q)_{\infty}} \sum_{j,a = 1}^{2} R^{(j,a,k)}_{2} (y_{1,2} ;q;s_{1})\,,
\end{align}
where $R^{(j,a,k)}_{2} (y_{1,2};q;s_{1})$ corresponds to the residue sum over the poles $s_{2}^{(j,a,k)}$.
For $j=2$, it is given by
\begin{align}
    R^{(2,a,k)}_{2} (y_{1,2};q;s_{1}) = \frac{(y_{a}q^{k},qy_{a}^{-1} q^{-k},y_{1} y_{2} (y_{a} q^{k})^{\pm},s_{1} y_{a} q^{k}, q s_{1}^{-1} y_{a}^{-1} q^{-k},y_{1} y_{2} (s_{1} y_{a} q^{k})^{\pm};q)_{\infty}}{(y_{1} (y_{a} q^{k})^{\pm},y_{2} (y_{a} q^{k})^{\pm},y_{1} (s_{1} y_{a} q^{k})^{\pm},y_{2} (s_{1} y_{a} q^{k})^{\pm};q)_{\infty}} \, .
\end{align}
and the $j=1$ expression is found to be $R_{2}^{(1,a,k)}(y_{1,2};q;s_{1})=R_{2}^{(2,a,k)} (y_{1,2};q;s_{1}^{-1})$. 
This enables us to collapse the sum in equation \eqref{eq:secondresidueintegralSU(3)} and write the $s_{1}$ integral as
\begin{align}\label{eq:secondresidueintegralSU(3)trick}
    \mathcal{I}_{3} (y_{1,2}; q) &=\sum^\infty_{k=0}\sum^2_{a=1} Q_{2}^{(a,k)}(y_{1,2}; q) \oint \frac{ds_{1}}{2 \pi i s_{1}} \frac{(s_{1}^{\pm}, y_{1} y_{2} s_{1}^{\pm};q)_{\infty}}{(y_{1} s_{1}^{\pm}, y_{2} s_{1}^{\pm};q)_{\infty}} R^{(a,k)}_{2} (y_{1,2};q;s_{1})\,,
\end{align}
where we defined
\begin{align}
    \begin{aligned}
    Q_{2}^{(1,k)} (y_{1,2}; q) &= \frac{(q, y_{1} y_{2}, q y_{1}^{-1}, y_{1}^{2} y_{2};q)_{\infty}}{(y_{1},y_{2},y_{1}^2, y_{1}^{-1} y_{2};q)_{\infty}} \frac{(q y_{2}^{-1}, y_{1}^2, y_{1} y_{2};q)_{k}}{(q, y_{1}^2 y_{2}, q y_{1} y_{2}^{-1};q)_{k}} q^{k} \\
    R_{2}^{(1,k)} (y_{1,2};q;s_{1}) &= \frac{(y_{1} q^{k} s_{1},q y_{1}^{-1} q^{-k} s_{1}^{-1}, y_{1}^2 y_{2} q^{k} s_{1}, y_{2} q^{-k} s_{1}^{-1};q)_{\infty}}{(y_{1}^2 q^{k} s_{1}, q^{-k} s_{1}^{-1}, y_{1} y_{2} q^{k} s_{1}, y_{1}^{-1} y_{2} q^{-k} s_{1}^{-1};q)_{\infty}}\,. \end{aligned}
\end{align}
The functions for $a=2$ can be obtained from these by exchanging $y_1\leftrightarrow y_2$.
See Figure \ref{fig:SU3schematic} (a) for a schematic depiction.

We now continue with the $s_{1}$ integral.
For $a=1$, it can be seen that the remaining integral only has two towers of poles inside the unit circle with non-vanishing residues.
These two towers arise at
\begin{align}\label{eq:su3s2poles-2}
    s_{1}^{(1,l)} = y_{1} q^{l}\,,\quad  l\in \mathbb{Z}_{\geq 0}\,, \quad s_{2}^{(2,m-k)} = y_{1}^{-1}y_2 q^{m-k}\,,  \quad m- k\in\mathbb{Z}_{\geq 0} \,,
\end{align}
For $a=2$, the same applies but again with $y_1\leftrightarrow y_2$.
Notice that $ s_{2}^{(2,0)}$ lies on the unit circle since $y_1^{-1}y_2=v^{-2}$, with the flavor fugacity $|v|=1$.
When evaluating the residue sum, we find that the correct way of dealing with this pole involves the following contour prescription for the $s_1$ integral of the $a=1$ term\footnote{Another way to think about this prescription is as follows. While for the $a=1$ term, the pole on the unit circle lies at $s_1=v^{-2}$, for the $a=2$ term the pole lies at $s_1=y_1y_2^{-1}=v^2$. Slightly deforming $v$ such that $|v|<1$, one pole moves inside the unit circle while the other one moves out. It turns out that the residues of both poles equal, so that including just a single pole has the same effect as including half of both, as in the stated prescription. }
\begin{equation}\label{eq:contour-prescr}
    \oint_{|s_{1}| = 1} = \frac{1}{2} \left( \oint_{|s_{1}| = 1 + \epsilon} + \oint_{s_{1} = 1 - \epsilon} \right) = \left( \oint_{|s_{1}| = 1 + \epsilon} - \frac{1}{2} \oint_{s_{1} =y_{1}^{-1}y_2} \right)\,,
\end{equation}
for small $\epsilon>0$ and similarly for the $a=2$ term with $y_1\leftrightarrow y_2$.
For convenience, we relabel $l=m-k\geq 0$, so that the two towers of poles are given by
\begin{align}
    s_{1}^{(b,l)} = \mathfrak{s}_{b} q^{l}, \quad l \in \mathbb{Z}_{\geq 0}
\end{align}
with their $q$ independent coefficients $\mathfrak{s}_{b} \in \{ y_{1}, y_{1}^{-1} y_{2} \}$. 
The residues for the above poles contain a factor which is expressed as the infinite product
\begin{align}
    \mathcal{Q}_{3} (y_{1,2}; q; \mathfrak{s}_{b}) = \frac{(y_{1} y_{2}, q y_{1}^{-1}, y_{1}^{2} y_{2},\mathfrak{s}_{b}, y_{1}^2 y_{2} \mathfrak{s}_{b}, y_{1} y_{2} \mathfrak{s}_{b}^{-1},q y_{1}^{-1} \mathfrak{s}_{b}^{-1},\delta_{b,1} y_{1} \mathfrak{s}_{b}^{-1},\delta_{b,2} y_{1}^{-1} y_{2} \mathfrak{s}_{b}^{-1};q)_{\infty}}{(y_{1},y_{2},y_{1}^2, y_{1}^{-1} y_{2}, y_{2} \mathfrak{s}_{b},y_{1}^2 \mathfrak{s}_{b},y_{1} \mathfrak{s}_{b}^{-1}, y_{1}^{-1} y_{2} \mathfrak{s}_{b}^{-1};q)_{\infty}} \, .
\end{align}
The terms $\delta_{b,1} y_{1} \mathfrak{s}_{b}^{-1}$ and $\delta_{b,2} y_{1}^{-1} y_{2} \mathfrak{s}_{b}^{-1}$ are included to cancel their respective infinite product term in the denominator (when $b$ is equal to the respective value in $\mathfrak{s}_{b}$). Here, the notation $\delta_{b,1}$ represents the usual Kronecker delta symbol, which allows us to write an expression for general $b$  (note that $(0;q)_{\infty} = 1$). 

In addition, the residue contains the finite product
\begin{align}
    \mathcal{R}_{3}^{(k,l)} (y_{1,2}; q;\mathfrak{s}_{b}) &= {}_{3} \mathcal{W}_{3}^{(k)} (y_{1}^2, q y_{2}^{-1}, y_{1} y_{2};q y_{1}^2;q;q^{2}) \, {}_{4} \mathcal{W}_{4}^{(l)} (q \mathfrak{s}_{b}, q y_{1}^{-1} y_{2}^{-1} \mathfrak{s}_{b}, y_{1} \mathfrak{s}_{b}, y_{2} \mathfrak{s}_{b};q \mathfrak{s}_{b}^2;q;q)  \nn \\ &\quad \times {}_{3} \mathcal{W}_{3}^{(k+l)} (q y_{2}^{-1} \mathfrak{s}_{b}, y_{1}^2 \mathfrak{s}_{b}, y_{1} y_{2} \mathfrak{s}_{b};q y_{1}^2 \mathfrak{s}_{b}^2;q;1)\,.
\end{align}
In the above ${}_{r} \mathcal{W}_{r}^{(k)}$ is what we define as the $k$th-summand of an \textit{almost} well-poised finite basic hypergeometric series, which is expressed as (see Appendix \ref{app:double-basic})
\begin{align}
    {}_{r} \mathcal{W}_{r}^{(k)} (a_{1}, \dots, a_{r};c;q;z) = \frac{(a_{1},a_{2},\dots,a_{r};q)_{k}}{(c a_{1}^{-1},c a_{2}^{-1},\dots,c a_{r}^{-1};q)_{k}} z^{k}\,.
\end{align}
We have illustrated this second part of the residue evaluation in Figure \ref{fig:SU3schematic} (b).
Then by summing the residues we obtain the $SU(3)$ Macdonald index
\begin{align}\label{eq:finalexpressionsI3SU(3)}
    \begin{aligned}
    \mathcal{I}_{3} (y_{1,2};q) =  \sum^2_{b=1}&\sum_{k,l = 0}^{\infty} \left[(\mathcal{Q}_{3} \mathcal{R}_{3}^{(k,l)}) (y_{1,2}; q;\mathfrak{s}_{b}) + (y_{1} \leftrightarrow y_{2})\right] \\ \quad - \, &\frac{1}{2}\sum_{k = 0}^{\infty} \left[(\mathcal{Q}_{3}\mathcal{R}_{3}^{(k,0)}) (y_{1,2}; q;\mathfrak{s}_{2}) + (y_{1} \leftrightarrow y_{2})\right]\,,
    \end{aligned}
\end{align}
where the second line implements the prescription \eqref{eq:contour-prescr} for dealing with poles on the $s_1$ unit circle.
The summand for fixed $b$ in the first line of equation \eqref{eq:finalexpressionsI3SU(3)} is expressible as a double basic hypergeometric functions, defined in Appendix \ref{app:double-basic}, namely $\Phi_{3:2;3}^{3:3;4}$ and $\Phi_{3:2;4}^{3:3;5}$. 
These functions generalize basic hypergeometric series to doubly infinite series.
In contrast to the (single) basic hypergeometric series, they are far less extensively studied in mathematical literature \cite{Srivastava_Jain_1986}. 
Their study is limited to restrictive specializations, including Jackson's $q$-Appel functions \cite{Jackson1942}, with specific choices of variables \cite{Gasper_Rahman_2004}. 
So, unlike for $SU(2)$ case, we are not able to simplify our expression any further using identities of these double basic hypergeometric functions, as we did with the basic hypergeometric functions in Section \ref{sec:mcd-n=4-su2}.

However, as studied in Section \ref{ssec:prod-forms} for $SU(2)$ gauge group, the closed form expressions for the residue sum can be used to deduce interesting specializations where the index simplifies.
In particular, due to its analytic dependence on the fugacity $t$, we can consider limits outside the naive domain of convergence $|t|<1$.
In the present case, we mention a considerable simplification for $t=1$, in which case all (double) basic hypergeometric series collapse and we only keep the prefactors, \textit{i.e.}, the prefactors denoted by $\mathcal{Q}_{3}$ in \eqref{eq:finalexpressionsI3SU(3)}.
Replacing $y_{1,2}=v^{\pm}$, we find
\begin{align}
    \begin{aligned}        
    \mathcal{I}_{3} (v;q,1) &= \frac{(v,qv^{-1},v^2,q v^{-2};q)_{\infty}}{(v^{\pm 3},v^{\pm 2};q)_{\infty}}  + \frac{(qv,v^{-1},qv^2,v^{-2};q)_{\infty}}{(v^{\pm 3},v^{\pm 2};q)_{\infty}} + \frac{(qv,qv^{-1};q)_{\infty}}{(v^{\pm 3};q)_{\infty}}\,.  \end{aligned}
\end{align}
This can be simplified into a product formula, which is a clear generalization of our $SU(2)$ result given in \eqref{eq:su2t=1product}, namely
\begin{align}
    \begin{aligned}        
    \mathcal{I}_{3} (v;q,1) &= \frac{(v^{\pm};q)_{\infty}}{(v^{\pm 3};q)_{\infty}}\,. \end{aligned} 
\end{align}
It is thus tempting to conjecture that the $\mathcal{N}=4$ $SU(N)$ Macdonald index at, or rather analytically continued to, $t=1$, is given by
\begin{align}
    \mathcal{I}_{N} (v;q,1) = \frac{(v^{\pm};q)_{\infty}}{(v^{\pm N};q)_{\infty}}.
\end{align}
It would be very interesting to understand whether there are more interesting specializations of $t$, such as the specialization $t=q^{\frac{1}{2}}$ studied in Section \ref{ssec:prod-forms} for the $SU(2)$ theory.

\subsection{Hall-Littlewood chiral ring}\label{ssec:hall-littlewoodsu3}

To develop further intuition for a potential simplification of the full $SU(3)$ Macdonald index, we finally consider the Hall-Littlewood limit $q \to 0$.
We find that in this limit our expression takes a relatively simple form
\begin{align}\label{eq:HLsu3}
    \begin{aligned}
    \mathcal{I}_{3} (v;t) &= \frac{1 + t - t^2 (v^2 + v^{-2}) - t^{\frac{5}{2}} (v + v^{-1}) + t^3 (1 - v^{2} - v^{-2}) + t^{\frac{7}{2}} (v^{3} + v^{-3}) + t^4}{(1 - t v^{-2})(1 - t v^2)(1 - t^{\frac{3}{2}} v^{-3})(1 - t^{\frac{3}{2}}v^3)}\,.
    \end{aligned}
\end{align}
This represents the $SU(3)$ version of equation \eqref{eq:ressum-mcd-index-su2-n=4-3}, and can be checked to match with previous results in the literature \cite{Felder_2018,Bonetti:2018fqz}. 
As in the $SU(2)$ case, we can interpret the expression in terms of the HL chiral ring. 
The generators are given by \cite{Beem:2013sza}
\begin{align}
    \mu_{A} = \text{Tr} (\bar{q}_{(i_{1}} \bar{q}_{i_{2})}), \quad \nu_{B} = \text{Tr} (\bar{q}_{(i_{1}} \bar{q}_{i_{2}} \bar{q}_{i_{3})}), \quad \omega_{i} = \text{Tr} (\lambda_{1+} \bar{q}_{i}), \quad \sigma_{A} = (\lambda_{1+} \bar{q}_{(i_{1}} \bar{q}_{i_{2})}),
\end{align}
with $A = \{ 0, \pm \}$, $B = \{ \pm 1, \pm 3 \}$, and $i = \{ \pm \}$. 
As in the $SU(2)$ case, each letter inside the trace can be viewed as a diagonal matrix in the HL chiral ring, since commutators among the letters are set to zero by the $F$--term equations (or, in a cohomological language, the commutators are $\mathcal{Q}$--exact).
We list the operators and their indices in Table \ref{tab:HL-chiral-ring-SU(3)} below. 

\begin{table}[ht]
    \centering
    \bgroup
    \def\arraystretch{1.2}
    \begin{tabular}{|c|c|c|c|c|c|c|c|c|c|c|c|c|}
  \hline
$\mathcal{O}$ & $\mu_{0}$ & $\mu_{\pm}$ & $\nu_{\pm 3}$ & $\nu_{\pm}$ & $\omega_{\pm}$ & $\sigma_{0}$ & $\sigma_{\pm}$ \\ \hline 
index & $t$ & $t v^{\pm 2}$ & $t^{\frac{3}{2}} v^{\pm 3}$ & $t^{\frac{3}{2}} v^{\pm}$ & $- t^{\frac{3}{2}} v^{\pm}$ & $-t^2$ & $-t^2 v^{\pm 2}$ \\ 
 \hline
    \end{tabular}
    \egroup
    \caption{Single trace generators $\mathcal{O}$ in the HL chiral ring with their corresponding indices.}
    \label{tab:HL-chiral-ring-SU(3)}
\end{table}

The denominator of the HL index in equation \eqref{eq:HLsu3} reflects arbitrary powers of the bosonic generators $\mu_{\pm}$ and $ \nu_{\pm3}$, which can therefore be seen to act freely within the chiral ring. 
Instead, the numerator can be understood in terms of the remaining single--trace generators in Table \ref{tab:HL-chiral-ring-SU(3)} and the independent multi--trace operators listed in Table \ref{tab:mt-HL-chiral-ring-SU(3)} below.
The fact that these operators appear in the numerator implies that they do not act freely in the HL chiral ring. 
In particular, whenever two such operators are multiplied, they can be reduced through chiral ring relations to an operator in the original set up to multiplication by the free generators (see below).
They can be viewed as the analogues of the operators $\mu_0$ and $\omega_{\pm}$ in the $SU(2)$ HL chiral ring (see Section \ref{ssec:op-spec}).
Note that various cancellations occur in the index, for example between the contributions of $\nu_{\pm}$ and $\omega_{\pm}$.

\begin{table}[ht]
    \centering
    \begin{tabular}{|c|c|c|c|c|c|c|c|c|}
        \hline
        $\mathcal{O}' \mathcal{O}$ & $\mu_{0}^2$ & $\mu_{0} \omega_{\pm}$ & $\mu_{0} \sigma_{0}$ & $\mu_{0} \sigma_{\pm}$ & $\omega_{+} \omega_{-}$ & $\omega_{\pm} \sigma_{\pm}$ & $\omega_{\pm} \sigma_{\mp}$ & $\sigma_{+} \sigma_{-}$ \\
        \hline
        index & $t^2$ & $t^{\frac{5}{2}} v^{\pm}$ & $-t^3$ & $-t^3 v^{\pm 2}$ & $t^3$ & $t^{\frac{7}{2}} v^{\pm 3}$ & $t^{\frac{7}{2}} v^{\pm}$ & $t^{4}$ \\ \hline 
    \end{tabular} \smallbreak \begin{tabular}{|c|c|c|}
        \hline
        $\mathcal{O}'' \mathcal{O}' \mathcal{O}$ & $\mu_{0}^3$ & $\mu_{0}^2 \omega_{\pm}$  \\
        \hline
        index & $t^3$ & $- t^{\frac{7}{2}} v^{\pm}$  \\ \hline 
    \end{tabular}
    \caption{Multi--trace independent generators $\mathcal{O}$ which consist of double--trace $\mathcal{O}' \mathcal{O}$ and triple--trace $\mathcal{O}'' \mathcal{O}' \mathcal{O}$ composite operators with their corresponding indices.}
    \label{tab:mt-HL-chiral-ring-SU(3)}
\end{table} 

We have derived the above set of generators by studying the $SU(3)$ ``trace relations'', which are really relations between symmetric polynomials in the eigenvalues due to the simultaneous diagonalizability of the letters.
As it turns out, trace relations become relevant for composite operators with index $-t^{\frac{5}{2}} v^{\pm}$ onwards. 
We collect the trace relations in Appendix \ref{app:Tracerelations}. 
Let us just give two examples here.
The first bosonic, composite generator which, by a trace relation, is not independent from the operators listed above, is given by $\mu_{0} \nu_{\pm}$.
Its precise relation to the independent generators is given by
\begin{align}
    2 \mu_{0} \nu_{\pm} = \nu_{\pm 3} \mu_{\mp} + \nu_{\mp} \mu_{\pm}\,,
\end{align}
which turns out to be the only trace relation at this order.
Another example worth highlighting is the bosonic composite generator that is composed of two fermionic single trace generators, namely $\omega_{\pm} \sigma_{0}$, which becomes dependent due to the trace relation\footnote{Here we have to account for the Grassmannian nature of the fermions and the antisymmetrization of their eigenvalues. See Appendix \ref{app:Tracerelations} for more details.}
\begin{align}
    \omega_{\pm} \sigma_{0} = \omega_{\mp} \sigma_{\pm}.
\end{align}
The operator $\omega_{\mp} \sigma_{\pm}$ is linearly independent, since there is no other relation that reduces it. 
A full set of independent trace relations is listed in Table \ref{tab:HLchiralringtracerelationsSU(3)} in Appendix \ref{app:Tracerelations}. 

Concluding, we see that the structure of the HL chiral ring is significantly more complicated than in the $SU(2)$ case. 
Indeed, whereas there are 5 generators satisfying 4 relations for $SU(2)$, we obtain 12 generators and 31 relations for $SU(3)$. 
In analogy with Section \ref{ssec:prod-forms}, we hope that exhibiting this structure will help to determine further specializations of the Macdonald index, beyond $t=1$, which may or may not admit a product form.

\section{Conclusion and future directions}\label{sec:future}

In this work, we have evaluated superconformal indices of gauge theories in closed form by evaluating the contour integrals that define them through residues.
The key new insight on which our analysis relies is that in almost all examples considered, the residue sum can only be made sense of when using a ``reduced measure'' to perform the gauge singlet projection.
The reduced measure can be obtained from the usual Haar measure by gauge fixing the unitary integral to not just a diagonal $SU(N)$ matrix, but also fixing the residual $S_N$ Weyl symmetry.
We have applied this method to various indices of $\mathcal{N}=4,2,1$ $SU(2)$ gauge theories and, by doing so, derived new closed form expressions for their indices.
In the final section, we demonstrated how the method generalizes to higher rank gauge theories, by considering the example of the $\mathcal{N}=4$ $SU(3)$ Macdonald index.

The main motivation behind our work was to find  expressions for the index which, unlike the usual gauge integral expression, more faithfully represent the BPS spectrum of the corresponding SCFT at non--zero or even strong coupling.
We showed that, for the case of the $\mathcal{N}=4$ $SU(2)$ Macdonald index, the residue sum represents a useful intermediate step towards this goal.
Indeed, by expressing the index in terms of basic hypergeometric series, we were able to use known identities for such series to simplify the residue sum.
We argued that the resulting expression manifests features of the (expected) strongly coupled BPS spectrum in the Macdonald sector.
In particular, the expression provides further support for the conjecture that there are no ``non-graviton'' or fortuitous operators in the Macdonald (or Schur) sector of the $\mathcal{N}=4$ $SU(2)$ theory \cite{Beem:2013sza,Chang:2023ywj}.

Another interesting feature of the closed form expressions is that they manifest the analytic properties of the index and, in particular, can be used to analytically continue in the fugacities.
In this way, we found new specializations of the $\mathcal{N}=4$ $SU(2)$ Macdonald index, including $t=q^{\frac{1}{2}}$ and $t=1$, for which the index simplifies into a product formula.
Finally, we found an interesting relation with the TQFT expression for the index, which is written in terms of a sum over Macdonald polynomials \cite{Gadde:2011uv}.

On the mathematical side, our method has allowed us, for the first time, to evaluate elliptic hypergeometric integrals through residues.
As has been expected in the literature, the convergent residue sums are closely related to (bilinear combinations of) elliptic hypergeometric series.
Our evaluation is mostly consistent with this expectation, except for a non-factorizing prefactor, such that residues are bilinear combinations in terms of elliptic hypergeometric \emph{summands}.
Dualities of supersymmetric gauge theories, such as Seiberg duality and generalized $S$-duality, imply non-trivial summation and transformation properties of these \emph{series} (as opposed to well-known implications for elliptic hypergeometric \emph{integrals}, see, \textit{e.g.}, \cite{Dolan:2008qi,Gadde:2009kb,Spiridonov:2009za,Rains:2003teh}).
In short, our results provide new examples of the rich interplay between superconformal indices and the theory of basic and elliptic hypergeometric integrals and series. 

Let us end with some questions and directions for future research:
\begin{itemize}
    \item Our residue sums generalize the closed form expressions for the Schur index obtained in \cite{Pan:2021mrw} (see also \cite{Beem:2021zvt,Hatsuda:2022xdv}). The latter expressions manifest interesting modular properties, namely those of quasi--Jacobi forms. It would be interesting to understand if and how the Macdonald, and full, index uplift these properties. The concept of modular factorization \cite{Jejjala:2022lrm} (see also \cite{Gadde:2020bov}) for the full index may be relevant here.  
    \item For the residue sums of full superconformal indices, a systematic analysis of the domain of convergence is warranted.
    We expect such an exploration to also clarify the connection of our formula with the Bethe Ansatz formula for the index \cite{Benini:2018mlo,Lezcano:2021qbj,Benini:2021ano}.
    \item We have argued that our method applies to higher rank gauge theories as well, and demonstrated this in detail for the Macdonald index of the $\mathcal{N}=4$ $SU(3)$ theory.
    Even though the number of pole towers may appear to grow exponentially with $N$, in the $SU(3)$ case a simplification of the integrand after the first integration effectively reduced the number of towers to consider.
    It would be desirable to develop a more systematic understanding of such simplifications, in particular whether they persist to larger values of $N$.
    \item Is it possible to find simplifications\footnote{In the sense described for example at the end of Section \ref{ssec:bmn-mcd}.} of the residue sums for indices other than the $\mathcal{N}=4$ $SU(2)$ Macdonald index? 
    If there exists, for example, a simplification of the residue sums for the $\mathcal{N}=4$ $SU(2)$ BMN Macdonald or even full superconformal index, it could have tantalizing applications. 
    For one, we imagine this could lead to a more structural understanding of the non-graviton spectrum in this theory \cite{Chang:2022mjp,Choi:2022caq,Choi:2023znd,Choi:2023vdm}.\footnote{Although, one would need to independently determine the graviton index to obtain an expression for the index over purely the non--graviton part of spectrum, see the discussion around \eqref{eq:If-from-I-Im}.}
    In addition, it may yield hints about the TQFT description of the full superconformal index \cite{Gadde:2011uv,Razamat:2013qfa}.
    A simplification is also expected to manifest analytic properties of the index and shed further light on its modular properties \cite{Gadde:2020bov,Jejjala:2021hlt,Jejjala:2022lrm}.
    Finally, it may help in finding a microscopic interpretation of various ``thermodynamic'' instabilities observed for supersymmetric black holes in AdS$_5$ \cite{Choi:2025lck,Deddo:2025jrg}.
    \item We observed in Section \ref{ssec:prod-forms} that the specialization $t=q^{\frac{1}{2}}$ simplifies the Macdonald index of the $\mathcal{N}=4$ $SU(2)$ theory into a product formula.
    Could the same, or similar, specializations also simplify the Macdonald indices of other $\mathcal{N}=2$ theories? 
    Are there similar specializations that simplify the full superconformal index?
     \item Is there a physical interpretation for the residue sum itself?
    For a different class of theories, this has indeed been argued and the residue sum is known in those cases as the Higgs branch formula for the index \cite{Peelaers:2014ima}. 
    This is also closely related to the concept of holomorphic block, and more generally modular, factorization of superconformal indices \cite{Yoshida:2014qwa,Nieri:2015yia,Gadde:2020bov,Jejjala:2022lrm}.
    \item Seiberg duality (in the $\mathcal{N}=1$ case) and generalized S-duality (in the $\mathcal{N}=2$ case) give rise to non-trivial summation and transformation formulas for both basic hypergeometric series and (sums of) bilinear combinations of elliptic hypergeometric series.
    We have really only scratched the surface of this connection and it would be interesting to develop a general overview of the types of mathematical identities implied by the dualities in supersymmetric gauge theories.
\end{itemize}

\acknowledgments

We are grateful for fruitful discussions with Abhijit Gadde, Finn Larssen, Robert de Mello Koch, Shiraz Minwalla, Vineeth Talasila and Shivansh Tomar, and Yiwen Pan for correspondence.
We are especially grateful to Adwait Gaikwad and Tanay Kibe for many discussions and collaboration on a related project.
SvL would like to thank the group at TIFR, Mumbai, for hospitality and discussions that initiated this project.
He is also grateful to the group at the University of Michigan for hospitality and Antal Jevicki and George Konidaris for hosting him at Brown University during the Fall term of 2024, when part of this work was completed.
SvL acknowledges support from the DSI-NRF Centre of Excellence in Mathematical and Statistical Sciences (CoE-MaSS), South Africa, grant \#2022-59-Phy-Indices.
SvL and KM also gratefully acknowledge generous support from the Wits-IBM Quantum Computing Seed Funding Programme with reference number QCSeed003/2023. PR is supported by an NRF Freestanding Postdoctoral Grant PSTD23042899253 and the NRF grant KIC240826263500.
Opinions expressed and conclusions arrived at are those of the authors and are not necessarily to be attributed to the CoE-MaSS nor NRF.


\appendix

\section{Special functions and their properties}\label{app:spec-functions}

In this appendix, we collect the definitions of the special functions and some identities.
Most of the content in this appendix may be found in \cite{Gasper_Rahman_2004}.

\subsection{\texorpdfstring{$q$}{q}--Pochhammer symbol \& \texorpdfstring{$q$}{q}--theta funtion}\label{sapp:q-poch}

The finite $q$--Pochhammer symbol $(x;q)_n$ is defined as
\begin{equation}
    (x;q)_n=\begin{cases}
                1\,, & n=0\\
                \prod_{k=0}^{n-1} (1-x q^k)\,,&n\geq 1\\
                \frac{1}{\prod_{k=1}^{-n} (1-x q^{-k})}\,,&n\leq - 1
            \end{cases}\,,
\end{equation}
From these definitions, it follows that 
\begin{equation}
    (x;q)_{-n}=\frac{1}{(xq^{-n};q)_{n}}=\frac{(-q/x)^n}{(q/x;q)_n}q^{\frac{n(n-1)}{2}}\,,\quad (x;q^{-1})_{n}=(x^{-1};q)_{n}(-x)^n q^{-\frac{n(n-1)}{2}}\,.
\end{equation}
For $|q|<1$ and $x\in \mathbb{C}$, we can extend the definition to an infinite product:
\begin{equation}\label{eq:q-poch-defn}
	(x;q)_{\infty}=\prod_{k=0}^{\infty} (1-x q^k)=\exp\left(-\sum^{\infty}_{l=1}\frac{1}{l}\frac{x^{l}}{(1-q^{l})}\right)\,.
\end{equation}
Using the exponential expression, one can continue the infinite $q$--Pochhammer symbol to $|q|>1$ via:
\begin{equation}\label{eq:q-poch-ext}
	(x;q)_{\infty}=\frac{1}{(q^{-1}x;q^{-1})_{\infty}}\,.
\end{equation} 
One observes
\begin{equation}
    (x;q)_n=\frac{(x;q)_{\infty}}{(q^nx;q)_{\infty}}\,,
\end{equation}
which can be used to continue $(x;q)_n$ to non-integer $n$.
We will often use a shorthand notation for products of (finite or infinite) $q$-Pochhammer symbols,
\begin{equation}\label{eq:prod-q-poch-notation}
    (a_1,\dots,a_r;q)_n = \prod_{i=1}^r(a_i;q)_n\,.
\end{equation}
Two identities we will make frequent use of are
\begin{align}\label{eq:q-poch-shift}
    \begin{split}
        (q^nx;q)_{\infty}&=\frac{(x;q)_{\infty}}{(x;q)_n}\\
        (q^{-n}x;q)_{\infty}&=(x;q)_{\infty} (q/x;q)_n \left(-\frac{x}{q}\right)^nq^{-\frac{n(n-1)}{2}} \,,
    \end{split}
\end{align}
for $n\geq 0$.
Another useful identity is (using the notation \eqref{eq:prod-q-poch-notation})
\begin{equation}\label{eq:q-poch-split}
	(x^{2};q)_\I = (\pm x,\pm q^{1/2}x;q)_\I \,.
\end{equation}
The $q$--theta function $\theta(z;\tau)$, also known as the modified Jacobi theta function, is defined in terms of the $q$--Pochhammer symbols as
\begin{equation}\label{eq:q-theta-defn}
	\theta(z;\tau) = (x;q)_{\infty}(qx^{-1};q)_{\infty}=\exp\left(-\sum^{\infty}_{l=1}\frac{1}{l}\frac{(x^{l}+qx^{-l})}{(1-q^{l})}\right)\,,
\end{equation}
where $q=e^{2\pi i \tau}$ and $x=e^{2\pi i z}$.
In the main text, we often use the shorthand notation
\begin{equation}\label{eq:q-theta-notation}
	\theta_q(x) \equiv \theta(z;\tau) \,.
\end{equation}
From its definition, it is clear that $\q_q(x)$ has zeros at $x=q^k$, for $k\in\bZ$.
A few basic properties of $\theta(z;\tau)$ are
\begin{eqnarray}
	&& \label{eq:theta-ell-prop} 
	\theta(z+m\tau+n; \tau) = (-x)^{-m} q^{ - \frac{m(m-1)}{2}} \theta(z;\tau) \,, \\
	&& 
	\theta(-z;\tau) = \theta(z+\tau;\tau) \,, \\
        && \label{eq:theta-ext-prop}\theta(z;-\tau)\theta(z;\tau) = -x \,.
\end{eqnarray}
It also satisfies a modular property.
That is, for $g\in SL(2,\mathbb{Z})$, we have
\begin{align}\label{eq:theta-mod-prop}
	\theta\left(g(z;\tau)\right)  &=e^{i\pi B_g(z;\tau)} \theta(z;\tau)  \,, \quad g(z;\tau)=\left(\tfrac{z}{ m\tau+n};\tfrac{k\tau+l}{m\tau+n}\right) \,,
\end{align}
with
\begin{align}\label{eq:Bh}
	\begin{split}
		B_g(z;\tau) =& \tfrac{m}{m\tau+n} z^2 + \left(\tfrac{1}{m\tau+n}-1 \right)z +	\tfrac{1}{6} \left(\tau+ \tfrac{1}{m(m\tau+n)} \right)+ \tfrac{n}{6m} -\tfrac{1}{2} -2s(n,m)\,,
	\end{split}
\end{align}
and $s(n,m)$ the Dedekind sum.
We note that when $k=n=0$ and $m=-l=1$, \textit{i.e.}, the $S$-transformation in $SL(2,\mathbb{Z})$, we have $B(z;\tau)\equiv B_S(z;\tau)$ with
\begin{align}\label{eq:Bpol}
	\begin{split}
		B(z;\tau)= \tfrac{1}{\tau}z^2+\left(\tfrac{1}{\tau}-1\right)z+\tfrac{1}{6}\left(\tau+\tfrac{1}{\tau}\right)-\tfrac{1}{2} \,,
	\end{split}
\end{align}
where we used that $s(0,1)= 0$.
The general polynomial $B_g(z;\tau)$ can be expressed in terms of $B(z;\tau)$ as
\begin{equation}\label{eq:Bm-B-reln}
	B_g(z;\tau)= \tfrac{1}{m} B(mz,m\tau+n) + 2\sigma_1(n,1;m)\,,
\end{equation}
where $\sigma_1(n,1;m)$ is a Fourier--Dedekind sum defined for general $t=0,1,\ldots$ as
\begin{equation}\label{eq:sigmat}
	\sigma_t (n_1,n_2 \cdots,n_r;m) = \frac{1}{m} \sum_{\xi^m=1 \neq \xi} \frac{\xi^t}{(\xi^{n_1}-1) \cdots (\xi^{n_r} -1)}\,.
\end{equation}

\subsection{Basic hypergeometric series}\label{sapp:basic-hypergeom}

The basic hypergeometric series is defined as \cite{Gasper_Rahman_2004}
\begin{equation}\label{eq:defn-basic-hyper}
	{}_{r+1}\phi_{r}\left[\begin{array}{cccc}
		a_1 & \cdots & a_r & a_{r+1}\\
		b_1 & \cdots & b_r & \\
	\end{array};q,z\right]=\sum_{k=0}^{\infty}\frac{(a_1,a_2,\ldots ,a_{r+1};q)_{k} }{(q,b_1,b_2,\ldots ,b_r;q)_{k}}z^k\,,
\end{equation}
where we use the notation introduced in \eqref{eq:prod-q-poch-notation}.
Note that the series terminates when $a_i=q^{-k}$, for some $i$, and $k\in\mathbb{Z}_{\geq 0}$.
This series converges for $|q|<1$ and $|z|<1$.
It is called $k$-balanced if $z=q$, and
\begin{equation}\label{eq:k-balanced}
	b_1\cdots b_r =q^ka_1\cdots a_{r+1}\,.
\end{equation}
If the series is $1$-balanced, we will refer to it as balanced.
The series is called well-poised if 
\begin{equation}
	qa_1=a_2b_1=\ldots=a_{r+1}b_r\,,
\end{equation}
and very-well-poised if in addition
\begin{equation}
	a_2=-a_3=qa_1^{\frac{1}{2}}\,.
\end{equation}
Finally, a basic hypergeometric series is called nearly-poised of the first kind when 
\begin{equation}
	a_1q\neq a_2b_1=\ldots=a_{r+1}b_r\,,
\end{equation}
and nearly-poised of the second kind when 
\begin{equation}
	a_1q= a_2b_1=\ldots\neq a_{r+1}b_r\,.
\end{equation}
For a very-well-poised series ${}_{r+1}\phi_{r}$, we sometimes use the more compact notation
\begin{equation}\label{eq:vwp-basic-notation}
    {}_{r+1}v_{r}(a_1;a_4,\ldots,a_{r+1};q;z)\,.
\end{equation}
The very-well-poised balancing condition, for general $z$, can be written as
\begin{equation}
    (a_4a_5\cdots a_{r+1})z=(\pm (a_1q)^{\frac{1}{2}})^{r-3}\,,
\end{equation}
with either the $+$ or $-$ sign, and $(a_1q)^{\frac{1}{2}}$ indicates the principal part of the square root.

Basic hypergeometric series satisfy a variety of identities. 
These identities fall under two classes: summation and transformation formulas.
We collect, and sometimes rewrite, various useful identities from \cite{Gasper_Rahman_2004} (see, \textit{e.g.}, their appendices II and III).

\paragraph{Summation formulas}

The most elementary summation formula is known as the $q$-binomial theorem, and is given by
\begin{equation}\label{eq:q-binom}
    {}_{1}\phi_{0}\left[\begin{array}{c}
		a \\
		  \\
	\end{array};q,z\right]=\frac{(az;q)_\I}{(z;q)_\I} \,,
\end{equation}
or when $a=q^{-k}$, for a non-negative integer $k$, then
\begin{equation}\label{eq:q-binom-term}
    {}_{1}\phi_{0}\left[\begin{array}{c}
		q^{-k} \\
		  \\
	\end{array};q,z\right]=(q^{-k}z;q)_k\,.
\end{equation}
We also have 
\begin{equation}\label{eq:1phi1-non-term}
    {}_{1}\phi_{1}\left[\begin{array}{cc}
		a \\
		  b & \\
	\end{array};q,b/a\right]=\frac{(b/a;q)_\I}{(b;q)_\I}\,.
\end{equation}
The $q$-Vandermonde sum is given by
\begin{equation}\label{eq:q-vandermonde}
    {}_{2}\phi_{1}\left[\begin{array}{cc}
		a & q^{-k} \\
		  b & \\
	\end{array};q,q\right]=\frac{(b/a;q)_k}{(b;q)_k}a^k\,,
\end{equation}
and the $q$-Gauss sum by
\begin{equation}\label{eq:q-gauss}
    {}_{2}\phi_{1}\left[\begin{array}{cc}
		a_1 & a_2 \\
		  b & \\
	\end{array};q,\frac{b}{a_1a_2}\right]=\frac{(b/a_1,b/a_2;q)_\I}{(b,b/a_1a_2;q)_\I}\,.
\end{equation}
The (terminating) $q$-Saalsch\"utz sum reads for nonnegative $k$
\begin{equation}\label{eq:q-saalschutz}
    {}_{3}\phi_{2}\left[\begin{array}{ccc}
		a & b & q^{-k} \\
		  c & q^{1-k}ab/c\\
	\end{array};q,q\right]=\frac{(c/a,c/b;q)_k}{(c,c/ab;q)_k}\,.
\end{equation}
A non-terminating form of the $q$-Saalsch\"utz sum reads 
\begin{align}\label{eq:q-saalschutz-non-term}
    \begin{split}
        &\frac{(a,c/d,abcd;q)_\infty}{(ad^2,1/d^2;q)_\infty}{}_{3}\phi_{2}\left[\begin{array}{ccc}
		ad^2 & b & cd \\
		  qd^2 & abcd\\
	\end{array};q,q\right]\\&+\frac{(b,cd,abc/d;q)_\infty}{(b/d^2,d^2;q)_\infty}{}_{3}\phi_{2}\left[\begin{array}{ccc}
		b/d^2 & a &  c/d \\
		  q/d^2 & abc/d\\
	\end{array};q,q\right]
    =\frac{(acd,bc/d,ab;q)_\infty}{(ad^2,b/d^2;q)_\infty}
    \end{split}
\end{align}
We note that when $a=b$ and $ac^2=qd$, both ${}_3\phi_2$ series are well-poised.

There is a very-well-poised (balanced) ${}_6\phi_5$ summation
\begin{align}\label{eq:6phi5-summation-nonterm}
    \begin{split}
        {}_{6}\phi_{5}\left[\begin{array}{cccccc}
		a & qa^{\frac{1}{2}} & -qa^{\frac{1}{2}} & b & c & d \\
		  a^{\frac{1}{2}} & -a^{\frac{1}{2}} & aq/b & aq/c & aq/d \\
	\end{array};q,\frac{aq}{bcd}\right]=\frac{(aq,aq/bc,aq/bd,aq/cd;q)_\infty}{(aq/b,aq/c,aq/d,aq/bcd;q)_\infty}
    \end{split}
\end{align}
or, when $d=q^{-n}$, this takes the form
\begin{align}\label{eq:6phi5-summation-term}
    \begin{split}
        {}_{6}\phi_{5}\left[\begin{array}{cccccc}
		a & qa^{\frac{1}{2}} & -qa^{\frac{1}{2}} & b & c & q^{-n} \\
		  a^{\frac{1}{2}} & -a^{\frac{1}{2}} & aq/b & aq/c & aq^{n+1} \\
	\end{array};q,\frac{aq^{n+1}}{bc}\right]=\frac{(aq,aq/bc;q)_n}{(aq/b,aq/c;q)_n}
    \end{split}
\end{align}
We also have Jackson's terminating very-well-poised ${}_8\phi_7$ summation
\begin{align}\label{eq:8phi7-summation-term}
    \begin{split}
        {}_{8}\phi_{7}\left[\begin{array}{cccccccc}
		a & qa^{\frac{1}{2}} & -qa^{\frac{1}{2}} & b & c & d & e & q^{-n} \\
		  a^{\frac{1}{2}} & -a^{\frac{1}{2}} & aq/b & aq/c & aq/d & aq/e & aq^{n+1} \\
	\end{array};q,q\right]=\frac{(aq,aq/bc,aq/bd,aq/cd;q)_n}{(aq/b,aq/c,aq/d,aq/bcd;q)_n}
    \end{split}
\end{align}
when $a^2q=bcdeq^{-n}$.
Finally, we have Bailey's non-terminating extension of Jackson's very-well-poised ${}_8\phi_7$ summation
\begin{align}\label{eq:8phi7-summation-nonterm}
    \begin{split}
    &\frac{(aq/c,aq/d,aq/e,aq/f;q)_\I}{(b/a,aq;q)_\I} {}_{8}v_{7}(a;b,c,d,e,f;q,q)\\
    &+\frac{(bq/c,bq/d,bq/e,bq/f;q)_\I}{(a/b,b^2q/a;q)_\I} {}_{8}v_{7}(b^2/a;b,bc/a,bd/a,be/a,bf/a;q,q)\\
    &=\frac{(aq/cd,aq/ce,aq/cf,aq/de,aq/df,aq/ef;q)_\I}{(c,d,e,f,bc/a,bd/a,be/a,bf/a;q)_\I}
    \end{split}
\end{align}
where we use the notation \eqref{eq:vwp-basic-notation}, and $qa^2=bcdef$.

\paragraph{Transformation formulas}

We now turn to transformation formulas.
A three-term transformation formula for ${}_{3}\phi_{2}$ is given in equation III.34 of \cite{Gasper_Rahman_2004}.
We rewrite their identity in a more convenient form for our purposes as
\begin{align}\label{eq:3phi2-trans-3-term}
    \begin{split}
        &\frac{(a,b,\tilde{c};q)_{\infty}}{(a\tilde{b}/b,b/\tilde{b};q)_{\infty}}{}_{3}\phi_{2}\left[\begin{array}{ccc}
		a\tilde{b}/b & \tilde{a} & \tilde{b}\\
		q\tilde{b}/b & \tilde{c} &  \\
	\end{array};q,q\right]+\frac{(\tilde{a},\tilde{b},c;q)_{\infty}}{(\tilde{a}b/\tilde{b},\tilde{b}/b;q)_{\infty}}{}_{3}\phi_{2}\left[\begin{array}{ccc}
		\tilde{a}b/\tilde{b} & a & b\\
		qb/\tilde{b} & c &  \\
	\end{array};q,q\right]\\
    &=\frac{(\tilde{a},a\tilde{b},c;q)_{\infty}}{(a\tilde{b}/b,\tilde{a}b/\tilde{b};q)_{\infty}}{}_{3}\phi_{2}\left[\begin{array}{ccc}
		a & b & \tilde{c}/\tilde{a}\\
		c & a\tilde{b} &  \\
	\end{array};q,\tilde{a}\right]
    \end{split}
\end{align}
where $b\tilde{c}=\tilde{b}c$.
Note that our version is written in terms of 6 parameters with the constraint $b\tilde{c}=\tilde{b}c$, as opposed to the version in terms of 5 parameters in \cite{Gasper_Rahman_2004}.
In our way of writing the identity, it is clear from the left hand side that the identity is symmetric under the exchange of tilded and untilded parameters, although this is not obvious from the expression on the right hand side.
In fact, this implies another, known (two-term) transformation formula of ${}_{3}\phi_{2}$ (equation III.10 in \cite{Gasper_Rahman_2004}), which we write here as
\begin{equation}\label{eq:3phi2-trans-2-term-1}
    {}_{3}\phi_{2}\left[\begin{array}{ccc}
		a & b & \tilde{c}/\tilde{a}\\
		c & a\tilde{b} &  \\
	\end{array};q,\tilde{a}\right]=\frac{(a,\tilde{a}b,\tilde{c};q)_{\infty}}{(\tilde{a},a\tilde{b},c;q)_{\infty}}{}_{3}\phi_{2}\left[\begin{array}{ccc}
		\tilde{a} & \tilde{b} & c/a\\
		\tilde{c} & \tilde{a}b &  \\
	\end{array};q,a\right]
\end{equation}
Further transformation formulas (equation III.9 in \cite{Gasper_Rahman_2004}) we will make use of are
\begin{align}\label{eq:3phi2-trans-2-term-2}
    \begin{split}
        {}_{3}\phi_{2}\left[\begin{array}{ccc}
		a & b & \tilde{c}/\tilde{a}\\
		c & a\tilde{b} &  \\
	\end{array};q,\tilde{a}\right]&=\frac{(\tilde{a}b,c/b;q)_{\infty}}{(\tilde{a},c;q)_{\infty}}{}_{3}\phi_{2}\left[\begin{array}{ccc}
		a\tilde{a}\tilde{b}/\tilde{c} & b & \tilde{b} \\
		a\tilde{b} & \tilde{a}b &  \\
	\end{array};q,\frac{c}{b}\right]\\
    &=\frac{(a\tilde{a}\tilde{b}/\tilde{c},\tilde{c};q)_{\infty}}{(\tilde{a},a\tilde{b};q)_{\infty}}{}_{3}\phi_{2}\left[\begin{array}{ccc}
		c/b & c/a  & \tilde{c}/\tilde{a}  \\
		\tilde{c} & c &  \\
	\end{array};q,\frac{a\tilde{a}\tilde{b}}{\tilde{c}}\right]
    \end{split}
\end{align}
where in the second line we applied \eqref{eq:3phi2-trans-2-term-1}.
Note that if we replace the right hand side of \eqref{eq:3phi2-trans-3-term} with either line, one finds a manifestly symmmetric right hand side under the exchange of the tilded and untilded parameters (using that $b\tilde{c}=\tilde{b}c$).

In the main text, we only use a specialization of \eqref{eq:3phi2-trans-3-term}, when the ${}_{3}\phi_{2}$ series on the left hand side are both well-poised. 
This happens when
\begin{equation}
    \tilde{a}=a\,,\quad \tilde{b}c=aq=b\tilde{c}\,.
\end{equation}
Note that with this specialization the right hand side of \eqref{eq:3phi2-trans-2-term-2} in both lines also contain well-poised ${}_{3}\phi_{2}$ series.
In this case, the three-term transformation formula in \eqref{eq:3phi2-trans-3-term} simplifies.
Using also \eqref{eq:3phi2-trans-2-term-2} and solving for $c$ and $\tilde{c}$, in terms of $a$, $b$ and $\tilde{b}$, we write it in three ways
\begin{align}\label{eq:3phi2-trans-3-term-2}
    \begin{split}
         &\frac{(b,qa/b;q)_{\infty}}{(a\tilde{b}/b,b/\tilde{b};q)_{\infty}}{}_{3}\phi_{2}\left[\begin{array}{ccc}
		a\tilde{b}/b & a & \tilde{b}\\
		q\tilde{b}/b & qa/b &  \\
	\end{array};q,q\right]+\frac{(\tilde{b},qa/\tilde{b};q)_{\infty}}{(ab/\tilde{b},\tilde{b}/b;q)_{\infty}}{}_{3}\phi_{2}\left[\begin{array}{ccc}
		ab/\tilde{b} & a & b\\
		qb/\tilde{b} & qa/\tilde{b} &  \\
	\end{array};q,q\right]\\
    &=\frac{(a\tilde{b},qa/\tilde{b};q)_{\infty}}{(a\tilde{b}/b,ab/\tilde{b};q)_{\infty}}{}_{3}\phi_{2}\left[\begin{array}{ccc}
		a & b & q/b\\
		qa/\tilde{b} & a\tilde{b} &  \\
	\end{array};q,a\right]\\
    &=\frac{(a\tilde{b},ab,qa/b\tilde{b};q)_{\infty}}{(a\tilde{b}/b,ab/\tilde{b},a;q)_{\infty}}{}_{3}\phi_{2}\left[\begin{array}{ccc}
		ab\tilde{b}/q & b & \tilde{b} \\
		a\tilde{b} & ab &  \\
	\end{array};q,\frac{qa}{b\tilde{b}}\right]\\
    &=\frac{(qa/\tilde{b},qa/b,ab\tilde{b}/q;q)_{\infty}}{(a\tilde{b}/b,ab/\tilde{b},a;q)_{\infty}}{}_{3}\phi_{2}\left[\begin{array}{ccc}
		qa/b\tilde{b} & q/\tilde{b}  & q/b   \\
		qa/b & qa/\tilde{b} &  \\
	\end{array};q,\frac{ab\tilde{b}}{q}\right]
    \end{split}
\end{align}
Interestingly, from the right hand sides of these equations, we observe the invariance under
\begin{equation}
    b\to \frac{q}{\tilde{b}}\,,\quad \tilde{b}\to \frac{q}{b}\,.
\end{equation}
This symmetry is not manifest on the left hand side, and implies the following four-term transformation formula
\begin{align}\label{eq:3phi2-trans-4-term}
    \begin{split}
         &\frac{(b,qa/b;q)_{\infty}}{(a\tilde{b}/b,b/\tilde{b};q)_{\infty}}{}_{3}\phi_{2}\left[\begin{array}{ccc}
		a\tilde{b}/b & a & \tilde{b}\\
		q\tilde{b}/b & qa/b &  \\
	\end{array};q,q\right]+\frac{(\tilde{b},qa/\tilde{b};q)_{\infty}}{(ab/\tilde{b},\tilde{b}/b;q)_{\infty}}{}_{3}\phi_{2}\left[\begin{array}{ccc}
		ab/\tilde{b} & a & b\\
		qb/\tilde{b} & qa/\tilde{b} &  \\
	\end{array};q,q\right]\\
    &=\frac{(q/\tilde{b},a\tilde{b};q)_{\infty}}{(ab/\tilde{b},\tilde{b}/b;q)_{\infty}}{}_{3}\phi_{2}\left[\begin{array}{ccc}
		ab/\tilde{b} & a & q/b\\
		qb/\tilde{b} & a\tilde{b} &  \\
	\end{array};q,q\right]+\frac{(q/b,ab;q)_{\infty}}{(a\tilde{b}/b,b/\tilde{b};q)_{\infty}}{}_{3}\phi_{2}\left[\begin{array}{ccc}
		a\tilde{b}/b & a & q/\tilde{b}\\
		q\tilde{b}/b & ab &  \\
	\end{array};q,q\right]
    \end{split}
\end{align}
Finally, we collect two transformations of a very-well-poised ${}_8v_7$ series
\begin{align}\label{eq:8phi7-trans}
    \begin{split}
       &{}_{8}v_{7}\left(a;b,c,d,e,f;q,\frac{a^2q^2}{bcdef}\right)\\
       &=\frac{(aq,aq/ef,\lambda q/e,\lambda q/f;q)_\I}{(aq/e,aq/f,\lambda q,\lambda q/ef;q)_\I}{}_{8}v_{7}\left(\lambda;\lambda b/a,\lambda c/a,\lambda d/a,e,f;q,\frac{aq}{ef}\right)\\
    &=\frac{(aq,b,bc\mu/a,bd\mu/a,be\mu/a,bf\mu/a;q)_\I}{(aq/c,aq/d,aq/e,aq/f,\mu q,b\mu/a;q)_\I}{}_{8}v_{7}\left(\mu;\frac{aq}{bc},\frac{aq}{bd},\frac{aq}{be},\frac{aq}{bf},\frac{b\mu}{a};q,b\right)
    \end{split}
\end{align} 
with $\lambda=qa^2/bcd$ and $\mu=q^2a^3/b^2cdef$.

\subsection{Double basic hypergeometric series}\label{app:double-basic}

Double basic hypergeometric series are defined as \cite{Gasper_Rahman_2004}
\begin{align}\label{eq:doublebasic}
    \begin{aligned}
    \Phi_{D:E;F}^{A:B;C} \left[ \begin{matrix}
        a_{A} \, : \, b_{B} \, ; \, c_{C} \\ d_{D} \, : \, e_{E} \, ; f_{F}
    \end{matrix}; q; x, y \right] &= \sum_{m,n = 0}^{\infty} \frac{(a_{A};q)_{m+n} (b_{B};q)_{m} (c_{C};q)_{n}}{(d_{D};q)_{m+n} (q, e_{E};q)_{m} (q, f_{F};q)_{n}} \\ &\quad \times [(-1)^{m+n} q^{\frac{(m+n)(m+n-1)}{2}}]^{D-A} \\ &\quad \times [(-1)^{m} q^{\frac{m(m-1)}{2}}]^{1+E-B} [(-1)^{n} q^{\frac{n(n-1)}{2}}]^{1+F-C} x^{m} y^{n}.
    \end{aligned}
\end{align}
Here, in $\Phi^{A:B;C}_{D:E;F}$, we have that $A, \dots, F$, which refer to the total number of parameters that $a_{A}$, $b_{B}$, $c_{C}$, $d_{D}$, $e_{E}$ and $f_{F}$ run upto in each finite $q$-Pochhammer (they are implicitly listed inside the function, and taken as a product of finite $q$-Pochhammers on the right hand side).  
One property of these double basic hypergeometric series is that they collapse into a basic hypergeometric series if one of the parameters $b_{1}, \dots, b_{B}$, $c_{1}, \dots, c_{C}$ equals 1. 
Furthermore, they collapse entirely if one or more parameters from both $b_{1}, \dots, b_{B}$ and $c_{1}, \dots, c_{C}$ are equal to 1. 

In the two cases we consider in Section \ref{ssec:su3mcd}, (specifically, $\Phi_{3:2;3}^{3:3;4}$ and $\Phi_{3:2;4}^{3:3;5}$), both happen to observe the condition $A = D$, $B = E + 1$, and $C = F + 1$, with $|x|, |y| < 1$ (as we have that $|x| = |q^n|$ and $|y| = |q^m|$, with $n, m \in \{ 1,2 \}$), which coincides with the case where the series converges absolutely. 
In this case, we can write the series alternatively in terms of an \textit{almost} well-poised basic hypergeometric summands ${}_{r} \mathcal{W}_{r}^{(k)}$ which we define as
\begin{align}\label{eq:almost-well-poised-finite-bsc-hypergeom-series}
    {}_{r} \mathcal{W}_{r}^{(k)} (a_{1}, \dots, a_{r};c;q;z) = \frac{(a_{1},a_{2},\dots,a_{r};q)_{k}}{(c a_{1}^{-1},c a_{2}^{-1},\dots,c a_{r}^{-1};q)_{k}} z^{k}\,.
\end{align}
where $c$ is a fixed ratio that relates the terms in the numerator and denominator. They are well-poised when $c$ has the form $c = qa_{1}$ (wherein, its denominator contains $(q;q)_{k}$). 

Indeed, one may observe that in the case that it has this $\textit{almost}$ well-poised structure for the $m$, $n$ and $m+n$ finite $q$-Pochhammers, then one may express it as
\begin{align}\label{eq:DBHGFalmost-well-poised}
    \begin{aligned}
    &\Phi^{D:E+1;F+1}_{D:E;F} \left[ \begin{matrix}
        a_{D} \, : \, b_{E+1} \, ; \, c_{F+1} \\ \mathfrak{c}_{1} a_{D}^{-1} \, : \, \mathfrak{c}_{2} b_{E}^{-1} \, ; \mathfrak{c}_{3} c_{F}^{-1}
    \end{matrix}; q; x, y \right] \\ &= \sum_{m,n = 0}^{\infty} {}_{D} \mathcal{W}_{D}^{(m+n)} (a_{D};\mathfrak{c}_{1};q,1) \, {}_{E+1} \mathcal{W}_{E+1}^{(m)} (b_{E+1};\mathfrak{c}_{2};q;x) \, {}_{F+1} \mathcal{W}_{F+1}^{(n)} (c_{F+1};\mathfrak{c}_{3};q;y),
    \end{aligned}
\end{align}
where $\mathfrak{c}_{1,2,3}$ refer to the fixed ratios. We follow the same abbreviated notation as in the double basic hypegeometric function definition in \eqref{eq:doublebasic}, taking $a_{D}$, $b_{E+1}$ and $c_{F+1}$ to run over all the terms inside the finite $q$-Pochhammers in its numerator (as in $a_{1},\dots,a_{D}$, for each summand \eqref{eq:almost-well-poised-finite-bsc-hypergeom-series}). In the function ${}_{D} \mathcal{W}_{D}^{(m+n)}$ stated in the above expression, we also take $D$ to specify the total number of parameters, as we did stating $\Phi^{A:B;C}_{D:E;F}$.  

For products containing ${}_{r} \mathcal{W}_{r}^{(k+l)}$, ${}_{s} \mathcal{W}_{s}^{(k)}$ and ${}_{t} \mathcal{W}_{t}^{(l)}$, we can shift the $z$ term between the almost basic hypergeometric function summands as follows
\begin{align}
    \begin{aligned}
    &{}_{r} \mathcal{W}_{r}^{(k+l)} (\cdots \, ; \, \cdot \,;q;z_{3}) \, {}_{s} \mathcal{W}_{s}^{(k)} (\cdots \, ; \, \cdot \, ;q;z_{2}) \, {}_{t} \mathcal{W}_{t}^{(l)} (\cdots \, ; \, \cdot \, ;q;z_{1}) \\ &= {}_{r} \mathcal{W}_{r}^{(k+l)} (\cdots \, ; \, \cdot \, ;q;\mathfrak{b}_{3}) \, {}_{s} \mathcal{W}_{s}^{(k)} (\cdots \, ; \, \cdot \, ;q;\mathfrak{a}_{3} z_{2}) \,  {}_{t} \mathcal{W}_{t}^{(l)} (\cdots \, ; \, \cdot \, ;q;\mathfrak{a}_{3} z_{1}),
    \end{aligned}
\end{align}
where we can factor $z_{3}$ as $\mathfrak{a}_{3} \mathfrak{b}_{3} = z_{3}$. By the property of the finite $q$-Pochhammers
\begin{align}
    (a;q)_{m+n} = (a;q)_{m} (aq^{m};q)_{n},
\end{align}
we can also break up summed summation indices, say $k + l$, in these almost well-poised summands as
\begin{align}
    \begin{aligned}
    {}_{r} \mathcal{W}_{r}^{(k+l)} (a_{1}, \dots, a_{r};c;q;z) &= {}_{r} \mathcal{W}_{r}^{(k)} (a_{1}, \dots, a_{r};c;q;z) {}_{r} \mathcal{W}_{r}^{(l)} (a_{1} q^{k}, \dots, a_{r} q^{k};c q^{2k};q;z) \\
    &= {}_{r} \mathcal{W}_{r}^{(l)} (a_{1}, \dots, a_{r};c;q;z) {}_{r} \mathcal{W}_{r}^{(k)} (a_{1} q^{l}, \dots, a_{r} q^{l};c q^{2l};q;z).        \label{eq:almostwp-summands-identity}
    \end{aligned}
\end{align}

\subsection{Elliptic shifted factorials}\label{sapp:ell-shift-fact}

The elliptic (or theta) shifted factorials $\theta_{q}(x;p)_k$ are defined for $k\geq1$ in terms of the $q$-theta function (see \eqref{eq:q-theta-defn}) as
\begin{equation}
    \q_q(x;p)_k \defeq \prod_{m=0}^{k-1}\q_q(xp^m), \qquad \q_q(x;p)_{-k}\defeq \prod_{m=1}^k\q_q(xp^{-m}).
\end{equation}
We set $\q_q(x;p)_0=1$. For $k\geq0$, these elliptic shifted factorials are denoted by $(x;p,q)_k$ in \cite{Gasper_Rahman_2004}.\footnote{Our definition of these factorials differs from \cite{Gasper_Rahman_2004} for negative $k$.}
We note that $\q_{q}(x;p)_k$ has zeros at $x=p^{-m}q^k$, for $m\in\{0,1,\dots,k-1\},$ and $k\in\Z$.

Making use of the properties of the $q$-theta function in Appendix \ref{sapp:q-poch}, one easily verifies the following identities (for $k>0$)
\begin{align}
    \q_q(x;p)_{-k}  =&\ (-x)^{k}p^{-k(k+1)/2}\q_q(p/x;p)_{k}	= \q_q(p^{-k}x;p)_{k},	\\
	\q_q (q^lx;p)_k	=&\ (-x)^{-kl} \lb p^{(k-1)/2}q^{(l-1)/2} \rb^{-kl} \q_q (x;p)_k.
\end{align}
Another useful identity that can be derived is
\begin{equation}\label{eq:elliptic-shifted-shift}
    \frac{\q_q(a_\a a_\b;p)_k}{\q_q(pa_\a a_\b^{-1};p)_k} = a_\a^{-2k}p^{-k^2}q^k\frac{\q_q(a_\a a_\b q^{-1};p)_k}{\q_q(qpa_\a a_\b^{-1};p)_k}\,,
\end{equation}
displaying almost periodic behaviour under $a_{\b}\to q a_{\b}$.
In analogy with the $q$-Pochhammer symobols, we introduce the shorthand notation
\begin{equation}
    \q_q(a_1,\dots,a_{r};p)_k = \prod_{i=1}^r\q_q(a_i;p)_k \,.
\end{equation}

\subsection{Elliptic hypergeometric series}\label{sapp:ell-hypergeom}

The elliptic hypergeometric series is defined in analogy with the basic hypergeometric series. But, now in terms of the elliptic shifted factorials, as described in the previous subsection.
In particular, we define \cite{Spiridonov32002} (see also Chapter 11 of \cite{Gasper_Rahman_2004})
\begin{align}
    \begin{split}
        {}_{r+1}E_r\left[\begin{array}{cccc}
		a_1 & \cdots & a_r & a_{r+1}\\
		b_1 & \cdots & b_r & \\
	\end{array};p;q;z\right]&\equiv \sum_{k=0}^{\infty} {}_{r+1}E^{(k)}_r\left[\begin{array}{cccc}
		a_1 & \cdots & a_r & a_{r+1}\\
		b_1 & \cdots & b_r & \\
	\end{array};p;q;z\right]\\
    &=\sum_{k=0}^{\infty}\frac{\q_q(a_1,\dots,a_{r+1};p)_k}{\q_q(p,b_1,\dots,b_r;p)_k}z^k \,.
    \end{split}
\end{align}
We note that in the limit $q\to 0$, we recover (termwise) the basic hypergeometric series, 
\begin{equation}
    \lim_{q\to 0}{}_{r+1}E_r \left[\begin{array}{cccc}
		a_1 & \cdots & a_r & a_{r+1}\\
		b_1 & \cdots & b_r & \\
	\end{array};p;q;z\right] = {}_{r+1}\f_r \left[\begin{array}{cccc}
		a_1 & \cdots & a_r & a_{r+1}\\
		b_1 & \cdots & b_r & \\
	\end{array};p;z\right]\,.
\end{equation}
assuming $z,a_i,b_i$ are all independent of $q$.
${}_{r+1}E_r$ is said to be elliptically balanced if
\begin{equation}
    a_1a_2\cdots a_{r+1} = pb_1b_2\cdots b_r,
\end{equation}
well-poised if 
\begin{equation}
    pa_1 = a_2b_1 = a_3b_2 = \cdots = a_{r+1}b_r,
\end{equation}
and very-well-poised if in addition $r\geq 4$, and
\begin{equation}
    a_2 = qa_1^{\frac{1}{2}}, \quad a_3= - qa_1^{\frac{1}{2}}, \quad a_4 = qa_1^{\frac{1}{2}}/p^{\frac{1}{2}}, \quad a_5 = -qa_1^{\frac{1}{2}}p^{\frac{1}{2}}.
\end{equation}
We will use a separate symbol for the well-poised and very-well-poised elliptic hypergeometric series.
We define a well-poised elliptic hypergeometric series ${}_{r+1}W_r$ as
\begin{align}\label{eq:defn-wp-elliptic-hypergeom-series}
    \begin{split}
    {}_{r+1}W_r(a_1;a_2,\ldots,a_{r+1};p;q;z) & \equiv  \sum_{k=0}^\I {}_{r+1}W^{(k)}_r(a_1;a_2,\ldots,a_{r+1};p;q;z)   \\ 
    &= \sum_{k=0}^\I \frac{\q_q(a_1,a_2,\ldots,a_{r+1};p)_k}{\q_q(p,pa_1a_{2}^{-1},\ldots,pa_1a_{r+1}^{-1};p)_k} z^k.
    \end{split}
\end{align}
We define a very-well-poised elliptic hypergeometric series ${}_{r+1}V_r$ as
\begin{align}\label{eq:defn-vwp-elliptic-hypergeom-series}
    \begin{split}
    &{}_{r+1}V_r(a_1;a_6,a_7,\ldots,a_{r+1};p;q;z) \equiv  \sum_{k=0}^\I {}_{r+1}V^{(k)}_r(a_1;a_6,a_7,\ldots,a_{r+1};p;q;z)   \\ 
    &= \sum_{k=0}^\I \frac{\q_q(a_1p^{2k})}{\q_q(a_1)} \frac{\q_{q}(a_1,a_6,a_7,\ldots,a_{r+1};p)_k}{\q_{q}(p,a_1p/a_6,a_1p/a_7,\ldots,a_1p/a_{r+1};p)_k} (pz)^k.
    \end{split}
\end{align}

\subsection{Elliptic Gamma function}\label{sapp:ell-Gamma}
 
The elliptic Gamma function is defined by \cite{Felder_2000}
\begin{equation}\label{eq:Gamma-product-defn}
	\Gamma(z;\sigma,\tau)=\prod^{\infty}_{k,l=0}\frac{1-x^{-1}p^{k+1}q^{l+1}}{1-xp^{k}q^{l}}=\exp\left(\sum^{\infty}_{l=1}\frac{1}{l}\frac{x^{l}-(pqx^{-1})^l}{(1-p^{l})(1-q^{l})}\right)\,,
\end{equation}
where $p=e^{2\pi i\sigma}$, $q=e^{2\pi i\tau}$ and $x=e^{2\pi i z}$.
It is a meromorphic function of $z\in \mathbb{C}$, and the product formula is defined for $|p|,|q|<1$.
In particular, $\G(z;\s,\t)$ has zeros at $z=(j+1)\t+(k+1)\s+l$, and poles at $z=-j\t-k\s+l$, where $j,k\geq 0,l\in\bZ$. In terms of $x,p,q$, the zeros are at $x=p^{j+1}q^{k+1}$ and the poles are at $x=p^{-j}q^{-k}$, for $j,k\geq 0$. 

In the main text, we will use various shorthand notations
\begin{equation}
   \Gamma(x)= \Gamma(x;p,q)\equiv \Gamma(z;\sigma,\tau)\,.
\end{equation}
We also define the notation
\begin{equation}
    \G(a_1,\dots,a_{r};p,q) = \prod_{i=1}^r\G(a_i;p,q)\, .
\end{equation}
The elliptic Gamma function is symmetric in its $\sigma$ and $\tau$ arguments and satisfies a number of properties, among which
\begin{equation}\label{eq:Gamma-shift-prop}
	\G(px;p,q)=\theta_q(x)\G(x;p,q)\,,\quad \G(qx;p,q)=\theta_p(x)\G(x;p,q)\,.
\end{equation}
and
\begin{equation}\label{eq:Gamma-inversion-prop}
	\Gamma(pqx^{-1};p,q)=\frac{1}{\Gamma(x;p,q)}\,.
\end{equation}
From the definition of the elliptic Gamma function in terms of the exponential, it follows that its definition can be extended to $|p|,|q|>1$ through
\begin{align}\label{eq:Gamma-ext-prop}
	\begin{split}
		\Gamma(x;p^{-1},q)&=\Gamma(qx^{-1};p,q)=\frac{1}{\Gamma(px;p,q)}\,,\\
		 \Gamma(x;p,q^{-1})&=\Gamma(px^{-1};p,q)=\frac{1}{\Gamma(qx;p,q)}
	\end{split}
\end{align}
In the limit $p\to0$, we get
\begin{equation}
    \G(x;p,q) \xrightarrow{p\to0} \frac{1}{(x;q)_\I}.
\end{equation}
Similar to (\ref{eq:q-poch-split}), we have the identity
\begin{equation}\label{eq:gamma-split}
    \G(x^2;p,q) = \G(\pm x, \pm p^{1/2}x,\pm q^{1/2}x,\pm p^{1/2}q^{1/2}x; p,q).
\end{equation}
We also have               
\begin{equation}\label{eq:gamma-xpm-theta}
    \G(x;p,q)\G(x^{-1};p,q) = \frac1{\q_p(x)\q_q(x^{-1})}.
\end{equation}
We will make frequent us of the following shift identities
\begin{equation}\label{eq:Gamma-shift}
    \begin{split}
        \G(p^kq^lx) =&\ \G(x) \lb -xp^{(k-1)/2}q^{(l-1)/2} \rb^{-kl} \q_{q}(x;p)_k\,\q_{p}(x;q)_l	\ ,	\\
	    \G(xp^{-k}q^{-l})	=&\ \G(x) \frac{(-x)^{-kl-k-l} p^{k(k+1)(l+1)/2} q^{l(l+1)(k+1)/2} }{\q_{q}(p/x;p)_k\,\q_{p}(q/x;q)_l}.
    \end{split}
\end{equation}

The elliptic Gamma function also satisfies a modular property \cite{Felder_2000,Jejjala:2022lrm}. 
We can write this property as
\begin{align}\label{eq:Gamma-gen-mod-prop-app}
	\begin{split}
		\Gamma(z;\sigma,\tau)&=e^{-i\pi Q_{(c;d,\tilde{d})}(z;\sigma,\tau)}\Gamma\left(\tfrac{z}{c\tau+d};\tfrac{\sigma-\tilde{d}(a\tau+b)}{c\tau+d},\tfrac{a\tau+b}{c\tau+d}\right)\Gamma\left(\tfrac{z}{c\sigma+\tilde{d}};\tfrac{\tau-d(\tilde{a}\sigma+\tilde{b})}{c\sigma+\tilde{d}},\tfrac{\tilde{a}\sigma+\tilde{b}}{c\sigma+\tilde{d}}\right)\,,
	\end{split}
\end{align}
where $ad-bc=\tilde{a}\tilde{d}-\tilde{b}c=1$, and when $c\neq 0$
\begin{equation}
	Q_{(c;d,\tilde{d})}(z;\sigma,\tau)=\tfrac{1}{c}Q(cz;c\sigma+\tilde{d},c\tau+d)+2\sigma_1(d,\tilde{d},1;c)\,,
\end{equation}
where $Q(z;\sigma,\tau)$ is a cubic polynomial in $z$ given by
\begin{align}\label{eq:Qpol}
	\begin{split}
		Q(z;\sigma,\tau) =& \frac{z^3}{3\sigma\tau} -\frac{\sigma+\tau-1}{2\sigma\tau} z^2 + \frac{\sigma^2+\tau^2 +3 \sigma\tau -3\sigma-3\tau+1}{6\sigma\tau} z \\
		&+ \frac{(\sigma+\tau-1)(\sigma^{-1}+\tau^{-1}-1)}{12} \,.
	\end{split}
\end{align}
Furthermore, $\sigma_1(d,\tilde{d},1;c)$ was defined in \eqref{eq:sigmat}.
We also note that for $(c;d,\tilde{d})=(1;0,0)$ we have $Q_{(1;0,0)}(z;\sigma,\tau)=Q(z;\sigma,\tau)$.

\subsection{Zeros, poles, and residues}

For quick reference, we list the zeros and poles of the functions defined in this appendix. 
The infinite $q$--Pochhammer symbol $(x;q)_\I$ has zeros at $x=q^{-k}$, for $k=0,1,\dots$, and $(q/x;q)_\I$ has zeros at $x=q^k$, for $k=1,2,\dots$. $\q(x;q)$ has zeros at $x=q^k$, for $k\in\bZ.$ 
Furthermore, the elliptic shifted factorial $\q_{q}(x;p)_k$ has zeros at $x=p^{-m}q^k$, for $m=\{0,1,\dots,k-1\},$ and $k\in\bZ$. 
Finally, the elliptic Gamma function $\G(z,\t,\s)$ has zeros at $z=(j+1)\t+(k+1)\s+l$ and poles at $z=-j\t-k\s+l$, where $j,k\geq 0,l\in\bZ$. 
In terms of $x,p,q$, the zeros are at $x=p^{j+1}q^{k+1}$ and the poles are at $x=p^{-j}q^{-k}$, for $j,k\geq 0$. 

We will also frequently use the residue of both the infinite $q$--Pochhammer symbol and the elliptic Gamma function, given by
\begin{equation}\label{eq:q-Poch-residue}
	\Res_{x=tq^k}\frac1x\frac1{(tx^{-1};q)_\I}	= \frac{(-1)^k q^{k(k+1)/2}}{(q;q)_k(q;q)_\I}\,,
\end{equation}
and
\begin{equation}\label{eq:Gamma-residue}
    \Res_{x=tp^kq^l} \frac{\G(tx^{-1})}{x} = \frac1{(p;p)_\I(q;q)_\I} \frac{(-1)^{kl+k+l} p^{k(k+1)(l+1)/2} q^{l(l+1)(k+1)/2} }{\q_{q}(p;q)_k\q_{p}(q;p)_l}\, .
\end{equation}

\section{Residue sum for the basic hypergeometric integral}\label{app:bas-hypgeom-int}

In this appendix, we review the evaluation of basic hypergeometric integrals through residues, following Section 4.10 of \cite{Gasper_Rahman_2004}.

Consider the basic hypergeometric integral
\begin{align}\label{eq:gen-basic-hypergeom-integral}
    \begin{split}
        I&=\oint_{|x|=1}\frac{dx}{2\pi i x}P(x)
    \end{split}
\end{align}
with (using the notation \eqref{eq:prod-q-poch-notation})
\begin{equation}
    P(x)=\frac{(a_1x,\ldots,a_n x,b_1x^{-1},\ldots,b_n x^{-1};q)_\infty}{(c_1x,\ldots,c_m x,d_1x^{-1},\ldots,d_m x^{-1};q)_\infty}\,.
\end{equation}
We assume that $|q|<1$, $|c_i|<1$ and $|d_i|<1$, for all $i=1,\ldots,m$, and that all poles of $P(x)$ are simple.
Up to some constraints we describe below, the contour integral can be evaluated through residues.
Closing the contour inside the unit circle, one finds towers of poles at $x=d_iq^k$, for $k\geq 0$.
We also note that these poles accumulate at the origin $x=0$ and that the integrand has an essential singularity at $x=0$, preventing a direct evaluation of its contribution.
We turn to this issue momentarily.

Let us evaluate the residue of $P(x)$ at $x=d_1q^k$.
Using \eqref{eq:q-poch-shift} and \eqref{eq:q-Poch-residue}, we find
\begin{align}\label{eq:basic-hypergeom-integral-result}
        &\Res_{x=d_1q^k}\left[\frac{1}{x}\frac{(a_1x,\ldots,a_n x,b_1x^{-1},\ldots,b_n x^{-1};q)_\infty}{(c_1x,\ldots,c_m x,d_1x^{-1},\ldots,d_m x^{-1};q)_\infty}\right] \nn\\
        &=\frac{(a_1d_1,\ldots,a_n d_1,b_1d_1^{-1},\ldots,b_nd_1^{-1} ;q)_\infty}{(c_1d_1,\ldots,c_m d_1,q,d_2d_1^{-1},\ldots,d_md_1^{-1} ;q)_\infty} \frac{(c_1d_1,\ldots,c_md_1,qd_1/b_1,\ldots,qd_1/b_n;q)_k}{(a_1d_1,\ldots,a_nd_1,q,qd_1/d_2,\ldots, qd_1/d_m;q)_k}    \nn\\
        &\times \lb\frac{b_1\cdots b_m}{d_1\cdots d_m}\rb^{k} \lb d_1^kq^{k(k+1)/2}\rb^{m-n}   \,.
\end{align}
It follows that the convergence of the residue sum for the integral \eqref{eq:gen-basic-hypergeom-integral} depends on the magnitude of
\begin{equation}\label{eq:k-suppres-factor-basic-hypergeom-int}
    \lb\frac{b_1\cdots b_n}{d_1\cdots d_m}\rb^{k} \lb d_1^kq^{k(k+1)/2}\rb^{m-n}
\end{equation}
since all other factors give rise to $\mathcal{O}(|q|^0)$ contributions for any $k$.
It follows that the residue sum is convergent if either $m>n$, or $m=n$ and
\begin{equation}\label{eq:convergence-basic-hypergeom-ressum}
    \lv \frac{b_1\cdots b_m}{d_1\cdots d_m}\rv <1\,.
\end{equation}
In the context of the Macdonald index of a gauge theory, we always have $m=n$.
It follows that convergence of the index depends on the magnitude of
\begin{equation}\label{eq:suppres-factor-basic-hypergeom-int}
    z=\frac{b_1\cdots b_m}{d_1\cdots d_m}\,,
\end{equation}
which we refer to as the \emph{suppression factor}.

As mentioned above, we cannot directly evaluate the contribution at $x=0$.
However, in some cases one can nonetheless argue that it vanishes by deducing an upper bound as follows. 
First, let $C_k$ be the circle $|x|=\d|q|^k$, with $k\in\bZ_{\geq0}$, and $\d$ chosen such that $C_k$ does not pass through any pole of $P(x)$. 
Then,
\begin{align}
    \begin{split}
    &\lv \frac{P(x)}{x} \rv_{x=\d q^k} =  \lv \frac{(a_1\d,\dots,a_n\d,b_1/\d,\dots,b_n/\d;q)_\I}{(c_1\d,\dots,c_m\d,d_1/\d,\dots,d_m/\d;q)_\I} \rv \\
    &\times \lv \frac{(q\d/b_1,\dots,q\d/b_n,c_1\d,\dots,c_m\d;q)_k}{(a_1\d,\dots,a_n\d,q\d/d_1,\dots,q\d/d_m;q)_k} \rv  \lv \frac{1}{\delta q^k} \bfrac{b_1\cdots b_n}{d_1\cdots d_n}^{k} \lb\d^k q^{k(k+1)/2}\rb^{m-n} \rv   . 
    \end{split}
\end{align}
The scaling with $k$ is only determined by the last factor, and we can write
\begin{equation}
	\lv \frac{P(x)}{x} \rv_{x=\d q^k} = \cO\lb \lv \frac{1}{\delta q^k} \bfrac{b_1\cdots b_n}{d_1\cdots d_n}^{k} \lb\d^k q^{k(k+1)/2}\rb^{m-n} \rv  \rb\,,
\end{equation}
where we used that the (infinite) $q$--Pochhammer symbols are $\mathcal{O}(1)$ in the $q$--expansion.
Since the contour wraps increasingly tightly around the origin for $k\to\infty$, with circumference $\mathcal{O}(\d|q|^k)$, we find that the ``contribution from the origin'', $I_0$, as defined in \eqref{eq:I0-defn}, is upper bounded by:
\begin{equation}
    \left|I_0\right| \leq \lim_{k\to\I}\oint_{C_k}\frac{dx}{2\pi i}\lv \frac{P(x)}{x} \rv\,.
\end{equation}
It follows that if either $m>n$ or \eqref{eq:convergence-basic-hypergeom-ressum} is satisfied, $\left|I_0\right|\to 0$.
Note that these are precisely the conditions for convergence of the residue sum.

Assuming that $n=m$ and that \eqref{eq:convergence-basic-hypergeom-ressum} holds, we are thus able to evaluate the basic hypergeometric integral through residues.
One finds:
\begin{align}\label{eq:ressum-gen-basic-hypergeom-integral}
    \begin{split}
        I=\sum^m_{i=1}&\frac{\prod^m_{j=1}(a_jd_i,b_jd_i^{-1};q)_\infty}{(q;q)_{\infty}\prod^m_{j=1}(c_jd_i;q)_\I \prod_{j\neq i}(d_jd_i^{-1};q)_\infty}\\
        &\times {}_{2m}\phi_{2m-1}\left[\begin{array}{cccccc}
		c_1d_i & \cdots & c_md_i & qd_i/b_1 & \cdots & qd_i/b_m\\
		a_1d_i  & \cdots & a_md_i & qd_i/d_1 & \widehat{qd_i/d_{j=i}} & qd_i/d_m
	\end{array};q;\frac{b_1\cdots b_m}{d_1\cdots d_m}\right]
    \end{split}
\end{align}
where the hatted entry $qd_i/d_{j=i}$ is to be omitted and the basic hypergeometric series ${}_{2m}\phi_{2m-1}$ is defined in Appendix \ref{sapp:basic-hypergeom}.

\section{Residue sums for elliptic hypergeometric integrals}\label{app:ell-hypgeom-int}

In this appendix, we review the definition of a class of elliptic hypergeometric integrals and evaluate them through residues. 

The general well-poised elliptic hypergeometric integral is defined in \cite{Spiridonov:2008thf} and takes the form\footnote{See also the earlier \cite{Spi203} for the slightly more general totally elliptic hypergeometric integral and \cite{Spiridonov2020} for a recent review on special cases of this integral.}
\begin{equation}\label{eq:well-poised-elliptic-hypergeom}
	I = \oint \frac{dx}{2\pi ix} \prod_{\a=1}^A\G(a_\a x^\pm;p,q),
\end{equation}
where the elliptic Gamma function $\Gamma(x;p,q)$ was defined in Appendix \ref{sapp:ell-Gamma} and the parameters satisfy the balancing condition 
\begin{equation}
	\prod_{\a=1}^Aa_\a^2=(pq)^A\,.
\end{equation}
The connection between the elliptic hypergeometric integrals and elliptic hypergeometric series (defined in Appendix \ref{sapp:ell-hypergeom}) is expected to be through residues, see, \textit{e.g.}, the early work \cite{Spi203}.
However, it is well-known that convergence of elliptic hypergeometric series is marred by issues; see for example \cite{Spiridonov32002} and more recently \cite{Krotkov:2023inf,spiri2024}. 
As a result, the evaluation of elliptic hypergeometric integrals in terms of convergent elliptic hypergeometric series is also unclear. 
Two previous attempts at evaluating the residue sum of, \textit{e.g.}, the superconformal index of the $\cN=1$ $SU(2)$ theory with $N_f=3$ were made in \cite{Peelaers:2014ima,spiri2024},\footnote{This index can be alternatively evaluated in closed form making use of Spiridonov's elliptic beta integral \cite{Spiridonov_2001}, as first observed in \cite{Dolan:2008qi}.} and for the $\mathcal{N}=4$ theory in \cite{Goldstein:2020yvj}.
However, assuming that all parameters have modulus less than 1 with a particular symmetric scaling (specified below), we will argue that these residue sums, associated with poles inside the unit circle but \emph{excluding} the origin, do not converge.
As in the basic hypergeometric case reviewed in Appendix \ref{app:bas-hypgeom-int}, the lack of convergence also obstructs the argument for the vanishing of a ``contribution from the origin'',\footnote{See the beginning of Section \ref{sec:conv-res-sums} for a definition.} where a non-isolated essential singularity resides.
We note that this issue was acknowledged in both \cite{Peelaers:2014ima} and \cite{spiri2024}, but not resolved.

In this appendix, we evaluate classes of elliptic hypergeometric integrals as residue sums, by relating the integrals to superconformal indices of gauge theories and using the reduced measure for the gauge singlet projection.
The result takes the form of a novel double infinite sum over the product of summands of two elliptic hypergeometric series with, however, a non-factorizing prefactor.
As such, the expression cannot be directly expressed in terms of the product of two elliptic hypergeometric series, as was suggested in \cite{Peelaers:2014ima,spiri2024}, although one \emph{can} express the result in terms of an infinite sum over products of elliptic hypergeometric series.
We explicitly demonstrate the convergence of these residue sums, which also implies the vanishing of a potential contribution from the origin, in the spirit of Theorem 2.1 of \cite{Askey_Wilson_1985} and Section 4.10 of \cite{Gasper_Rahman_2004}.

\subsection{Very-well-poised integrals}\label{sapp:vwp-ell-int}

We begin with a slight specialization of the general integral in \eqref{eq:well-poised-elliptic-hypergeom}, which applies to the superconformal indices of general $\mathcal{N}=1$ gauge theories with $SU(2)$ gauge group.
As reviewed in Section \ref{ssec:n=1-gauge-theory-inds}, the contribution to the integrand of an $\mathcal{N}=1$ vector multiplet, combined with the full Haar measure, takes the form
\begin{equation}
    \Delta(x)\Delta(x^{-1})\mathcal{I}_V(x;p,q)=\frac{1}{\G(x^{2};p,q)\G(x^{-2};p,q)}=\theta_p(x^2)\theta_q(x^{-2})
\end{equation}
where we used \eqref{eq:gamma-xpm-theta} and left out the Cartan factors which do not depend on the integration variable $x$. 
Therefore, we will be specifically interested in elliptic hypergeometric integrals of the form\footnote{Note that the theta functions do not contribute any poles to integrand.}
\begin{equation}\label{eq:vwp-ell-hyper-integral}
	I_{\text{vwp}} = \oint \frac{dx}{2\pi ix} \theta_p(x^2)\theta_q(x^{-2}) \prod_{\a=1}^A\G(a_\a x^\pm;p,q),
\end{equation}
with balancing condition 
\begin{equation}\label{eq:balancing-vwp-ell-hyp-int}
	\prod_{\a=1}^Aa_\a^2=(pq)^{A-4}\,.
\end{equation}
This modified balancing condition can be obtained by using the reflection property of the elliptic Gamma function $\G(x;p,q)=1/\G(pqx^{-1};p,q)$ and (\ref{eq:gamma-split}).
The integral \eqref{eq:vwp-ell-hyper-integral} is studied in \cite{Spiridonov:2008thf}, where it is called the very-well-poised elliptic hypergeometric integral. 
As discussed there, $A=2m+6$ with $m\in \mathbb{Z}_{\geq 0}$ for very-well-poised integrals. 
We will retain the notation $A$, keeping in mind that it is always even.

Reinstating the Cartan factors, we consider the integral
\begin{align}\label{eq:general-vwp-EHI}
    \begin{split}
        \cI_{\text{vwp}}& = \frac{(p;p)_\I(q;q)_\I}2 \oint \frac{dx}{2\pi ix} \lb\frac{-2x^2}{1-x^2}\rb^{\e} \theta_p(x^2)\theta_q(x^{-2}) \prod_{\a=1}^A\G(a_\a x^\pm;p,q)\\
        &\equiv \frac\k2 \oint\frac{dx}{2\pi ix} \cP(x)
    \end{split}
\end{align}
with balancing condition as in \eqref{eq:balancing-vwp-ell-hyp-int}. 
We emphasize that the evaluation of the integral is independent of $\epsilon=0,1$.
The case $\epsilon=0$ corresponds to the usual elliptic hypergeometric integral, which contains the full measure of the gauge singlet projection in the product of theta functions, while $\epsilon=1$ corresponds to the singlet projection with the reduced measure.
Note that the included factor for $\epsilon=1$ simply cancels off part of the measure as contained in the theta functions.
We refer the reader to Section \ref{ssec:sc-gauge-theory-inds} for more details.

We assume that the parameters $a_\a$ and $p,q$ are sufficiently generic such that $\cP(x)$ has simple poles inside the unit circle at $a_\a p^kq^l$, for $k,l\in\Z_{\geq0}$.
We note that this requires $|a_\a|<1$ and distinct for all $\a$, and also that $p^m\neq q^n$, for any $m,n\in \mathbb{N}$.
As in the basic hypergeometric case, the convergence of the residue sums will depend on details of the relative sizes of $|a_\a|$ and $|p|,|q|$, and also the value of $A$.
However, to argue for convergence in the elliptic hypergeometric case, one is required to choose the relative sizes \emph{a priori}, unlike in the basic hypergeometric case.
This complicates a general analysis, and for this reason, we focus here on some special cases.
Unsurprisingly, also in the elliptic hypergeometric case, the integrand has a non-isolated essential singularity at the origin $x=0$.
The vanishing of its contribution, which we defined in \eqref{eq:I0-defn}, can again be argued when the residue sum associated with all other poles converges.
As we will see, convergence can only be demonstrated for $\epsilon=1$, at least for our choice of scaling of the fugacities, so that only in this case we can conclude that the contribution from the origin vanishes.

Using the shift identities \eqref{eq:Gamma-shift} for $\G(xp^{\pm k}q^{\pm l})$ and the expression (\ref{eq:Gamma-residue}) for residues, the residue sum associated with the poles at $x=a_\a p^kq^l$ can be evaluated to be 
\begin{align}\label{eq:vwp-EHI-general-evaluation-unsimpl}
    \begin{split}
        &\cI_{\text{vwp}}	=\ \frac{2^\e}{2}\sum_{\a=1}^A \frac{\prod_{\b\neq \a} \G(a_\b a_\a^\pm)}{\G(a_\a^{-2})} \sum_{k,l=0}^\I \lb \frac{-a_\a^2p^{2k}q^{2l}}{1-a_\a^2p^{2k}q^{2l}} \rb^\e (-a_\a)^{(A-4)(k+l)} p^k q^l 	\\
	&\times p^{\frac{(A-4)(k^2-l)}{2}} q^{\frac{(A-4)(l^2-k)}{2}}\frac{\q_q(a_\a^2p^{2k})}{\q_q(a_\a^2)} \frac{\q_p(a_\a^2q^{2l})}{\q_p(a_\a^2)} \prod_{\b=1}^A \frac{\q_{q}(a_\a a_\b;p)_k}{\q_{q}(pa_\a a_\b^{-1};p)_k} \frac{\q_{p}(a_\a a_\b;q)_l}{\q_{p}(qa_\a a_\b^{-1};q)_l} \,,
    \end{split}
\end{align}
which is formally valid for odd $A$ as well.
Using \eqref{eq:elliptic-shifted-shift}, we observe that for even $A$ we can absorb the explicit factors of $a_\alpha$, $p$ and $q$ into the elliptic shifted factorials.
Using the definition of the very-well-poised elliptic hypergeometric summand \eqref{eq:defn-vwp-elliptic-hypergeom-series}, we can then rewrite the expression in the simpler form
\begin{align}\label{eq:vwp-EHI-general-evaluation}
    \begin{split}
        \cI_{\text{vwp}}	=&\ \frac{2^\e}{2}	\frac{\prod_{\b\neq1}^A\G(a_\b a_1^\pm)}{\G(a_1^{-2})} \sum_{k,l=0}^\I \frac{(-a_1^{2})^\e\:{}_{A+4}V_{A+3}^{(k)}(\vec{\mathpzc{a}}_1;p;q;p^{2\e})		\: {}_{A+4}V_{A+3}^{(l)}(\vec{\mathpzc{b}}_1;q;p;q^{2\e}) }{(1-a_1^2 p^{2k}q^{2l})^\e}               \\
 &\ + \text{cyclic}(a_1,a_2,a_3,\cdots,a_A)	\,.	      
    \end{split}
\end{align}
where we have defined
\begin{equation}            \label{eq:vwp-pzc-ab-defn}
    \begin{split}
        \vec{\mathpzc{a}}_1 =&\ (a_1^2;a_1a_2,\cdots,a_1a_{2+A/2},a_1a_{3+A/2}q^{-1},\cdots, a_1a_{A}q^{-1})\,,\\
    \vec{\mathpzc{b}}_1 =&\ (a_1^2;a_1a_2,\cdots,a_1a_{2+A/2},a_1a_{3+A/2}p^{-1},\cdots, a_1a_{A}p^{-1})\,.
    \end{split}
\end{equation}
The arguments shifted by $q^{-1}$ and $p^{-1}$ is an arbitrary choice; we choose to shift the final $(A-4)/2$ arguments.
The symbol $\text{cyclic}(a_1,a_2,a_3,\cdots,a_A)$ after the expression indicates that the full index is a sum over the $A$ expressions obtained by cyclically permuting the variables $a_1,\cdots, a_A$ in the first term.  

We highlight that to the best of our knowledge, this appears to be the first time in the literature the series (\ref{eq:vwp-EHI-general-evaluation}) (or equivalently (\ref{eq:vwp-EHI-general-evaluation-unsimpl})) has been considered. We now demonstrate that the series is in fact convergent for some natural choices of the magnitudes of the various parameters. 

Since the summand is symmetric in $k$ and $l$, and since the denominator of the summand expands as a convergent geometric series, we can demonstrate convergence using a ratio test for fixed $\alpha$ and $l$: 
\begin{align}\label{eq:EHI-ratio-test}
    \begin{split}
        \lv \frac{{}_{A+4}V_{A+3}^{(k+1)}(\vec{\mathpzc{a}}_\a;p;q;p^{2\e})}{{}_{A+4}V_{A+3}^{(k)}( \vec{\mathpzc{a}}_\a;p;q;p^{2\e})} \rv =&\, |p|^{1+2\e+k(A-4)}|a_\a|^{A-4} |p/q|^{(A-4)/2}\\
        &\times\lv  \frac{\q_q(a_\a^2p^{2k+2})}{\q_q(a_\a^2p^{2k})} \prod_\b \frac{\q_q(a_\a a_\b p^k)}{\q_q(a_\a a_\b^{-1}p^{k+1})} \rv  \,,
    \end{split}
\end{align}
where we made use of the expression in \eqref{eq:vwp-EHI-general-evaluation-unsimpl}.
To make an estimate for this ratio, we need to specify an appropriate scaling for the absolute values of the fugacities $a_\b$, $p$ and $q$. 
The complication with a convergence analysis we mentioned above is seen here explicitly: through the theta functions $\theta_q(x)=(x,qx^{-1};q)_\I$, the residues depend on negative powers of $p$ and $q$.
This is unlike the basic hypergeometric case and necessitates a choice of the scaling of the parameters as an input to the convergence analysis.

Due to the balancing condition, any consistent scaling is necessarily $A$ dependent. 
This $A$--dependence affects the details of the scaling estimate for the above ratio. 
To provide an example of an integral that is convergent, we set $A=6$, and scale the fugacities as $|a_\a|= T$, and $|p|,|q|= T^3$ with $T<1$, compatible with the balancing condition.
This case corresponds to the index of the $\cN=1$ $SU(2)$ theory with $N_f=3$ and the original instance of Spiridonov's elliptic beta integral \cite{Spiridonov_2001}. 
Taking careful account of negative powers of $T$ appearing in the various $\q$-functions, the ratio (\ref{eq:EHI-ratio-test}) can be seen to scale as
\begin{equation}\label{eq:nf3-ratio-test}
    \lv \frac{{}_{10}V_{9}^{(k+1)}(\vec{\mathpzc{a}}_\a;p;q;p^{2\e})}{{}_{10}V_{9}^{(k)}( \vec{\mathpzc{a}}_\a;p;q;p^{2\e})} \rv \sim T^{-2+6\e}.
\end{equation}
Therefore, the residue sum is convergent if we use the reduced measure ($\e=1$), but, not if we use the full measure ($\e=0$). 

As in the basic hypergeometric case, the convergence of the residue sum implies a vanishing contribution from the origin.
More precisely, the increased suppression of residues for increasing $k,l$, as follows from the above ratio test, implies that the integral over a contour $C_{k,l}$ of radius $\d p^kq^l$, tends to 0, as $k,l\to\I$.
Since this integral provides an upper bound on the contribution of the origin, we see that it vanishes with the reduced measure.

The cases $A=8,10,12$ correspond to the superconformal indices of $\cN=1$ $SU(2)$ SCFTs with $N_f=\frac{A}{2}=4,5,6$ (see Section \ref{ssec:n=1-gauge-theory-inds}). 
With a symmetric scaling for the $|a_\a|$ and the scaling of $|p|,|q|$ fixed by the balancing condition, the ratio test \eqref{eq:EHI-ratio-test} can again be seen to lead to the same conclusion: the residue sum converges when we use the reduced measure and consequently the contribution from the origin vanishes. With this symmetric scaling, the precise scaling of the ratio test (\ref{eq:EHI-ratio-test}) is $T^{8\e}$ for $A=8$, and $T^{16-(2-\e)A}$ for $A>8$. Note in particular that this demonstrates that the residue sum (\ref{eq:vwp-EHI-general-evaluation}) diverges for $A\geq 16$ if one chooses a symmetric scaling for the $a_\a$.

The $A=16$ case is in fact an interesting boundary case, as the integral is now also of the form of the superconformal index of the $\cN=2$ $SU(2)$ theory with $N_f=4$ (see Section \ref{ssec:sc-gauge-theory-inds}). 
For this case, it can be checked for each $\alpha$ that the (asymmetric) scaling of fugacities $a_\alpha$, as in Section \ref{sec:N=2-Nf=4-SQCD-full-index}, does lead to a convergent residue sum. 
In fact, we find that the ratio of subsequent terms in $k$ in each sum over a (doubly-infinite) tower of poles $(\propto p^kq^l)$ scales as $T^{4\e}$.
This demonstrates the sensitivity of the convergence of residue sums to the precise scaling of the parameters.
A full analysis of convergence is warranted, but beyond the scope of this work.

\subsection{Well-poised integrals}\label{sapp:wp-ell-int}

In this section, we consider a slightly different set of elliptic hypergeometric integrals, which are well-poised but not very-well-poised.
More specifically, we look at the following special cases of \eqref{eq:well-poised-elliptic-hypergeom}:\footnote{The distinction with the very-well-poised integral is that $1/\G(x^{\pm2})$ is replaced by $1/\G(x^{\pm})$.}
\begin{equation}			\label{eq:EHI-N=4-general-N}
	\cI_{\text{wp}} = \frac{(p;p)_\I(q;q)_\I}2 \oint_{|x|=1} \frac{dx}{2\pi ix} \lb\frac{-2x}{1-x}\rb^\epsilon \frac{\prod_{\a=1}^A\G(a_\a x^\pm;p,q)}{\G(x^\pm;p,q)},
\end{equation}
with the balancing condition $\prod_{\a=1}^Aa_\a^2=(pq)^{A-1}$.
As in the previous section, we note that the integral is independent of $\epsilon=0,1$, where $\e=0$ corresponds to a gauge singlet projection with the full measure and $\epsilon=1$ with the reduced measure.
Evaluating this general integral is mostly motivated by the example of the $\cN=4$ $SU(2)$ SYM theory, which corresponds to the case $A=3$ (see, \textit{e.g.}, Section \ref{ssec:sc-gauge-theory-inds}). 

Similar to the previous section, we evaluate this integral through residues.
We make the same assumptions on the parameters as described there, such that $\cP(x)$ has simple poles inside the unit circle at $a_\a p^kq^l$, for $k,l\in\Z_{\geq0}$, which accumulate at the origin.
Ignoring the origin for now, we evaluate the residue sum as
\begin{align}
    \cI_{\text{wp}} &=\ \frac{2^\e}{2} \sum_{\a=1}^A  \frac{\G(a_\a^2) \prod_{\b\neq\a} \G(a_\b a_\a^\pm)}{\G(a_\a^{\pm})}\sum_{k,l=0}^\I (-a_\a)^{(A-1)(k+l)} p^{(k^2-l)(A-1)/2}q^{(l^2-k)(A-1)/2} \nn   \\
    &\ 
    \times \lb \frac{-a_\a p^kq^l}{1-a_\a p^kq^l}\rb^\e \frac{\q_{q}(a_\a p^{k})}{\q_q(a_\a )} \frac{\q_{p}(a_\a q^{l})}{\q_p(a_\a )} \prod_{\b=1}^{A} \frac{\q_{q}(a_\a a_\b;p)_k}{\q_{q}(pa_\a a_\b^{-1};p)_k} \frac{\q_{p}(a_\a a_\b;q)_l}{\q_{p}(qa_\a a_\b^{-1};q)_l}	\,.
\end{align} 
For odd $A=2B+1$, we can express this residue sum as
\begin{align}\label{eq:wp-EHI-n=4}
    \cI_{\text{wp}} =&\ \frac{2^\e}{2} \frac{\G(a_1^2) \prod_{\b\neq1} \G(a_\b a_1^\pm)}{\G(a_1^{\pm})} \sum_{k,l=0}^\I\frac{(-a_1)^\e\:  {}_{A+1}W_A^{(k)}(\vec{\mathpzc{a}}_1;p;q;p^\e) \:{}_{A+1}W_A^{(l)}(\vec{\mathpzc{b}}_1;q;p;q^\e)}{(1-a_1p^kq^l)^\e}   \nn \\
     &\ + \text{cyclic}(a_1,a_2,a_3,\cdots,a_A)	\,,  
\end{align} 
where we used \eqref{eq:elliptic-shifted-shift} and the definition of the well-poised elliptic hypergeometric summand \eqref{eq:defn-wp-elliptic-hypergeom-series}, and we defined
\begin{equation}
    \begin{split}
        \vec{\mathpzc{a}}_1 =&\ (a_1^2,pa_1,a_1a_2,\cdots,a_1a_{B+1},a_1a_{B+2}q^{-1},\cdots,a_1a_{A}q^{-1}),       \\
        \vec{\mathpzc{b}}_1 =&\ (a_1^2,qa_1,a_1a_2,\cdots,a_1a_{B+1},a_1a_{B+2}p^{-1},\cdots,a_1a_{A}p^{-1}).
    \end{split}
\end{equation}

To test the convergence of the residue sum, we consider a symmetric scaling for the absolute values of $a_\a$ and $p,q$: $|a_\a|= T^{A-1}$, so that $|p|,|q|= T^A$, with $0<T<1$, as compatible with the balancing condition. 
For the same reasons stated above \eqref{eq:EHI-ratio-test}, it suffices to perform the ratio test for only a part of the summand:
\begin{align}
\begin{split}
    \lv \frac{{}_{A+1}W_{A}^{(k+1)}(\vec{\mathpzc{a}}_\a;p;q;p)}{{}_{A+1}W_{A}^{(k)}(\vec{\mathpzc{a}}_\a;p;q;p)} \rv    =&\ |p|^{\e A}|a_\a|^{A-1}|p|^{k(A-1)} |p/q|^{(A-1)/2}\\
     &\ \times \lv  \frac{\q_q(a_\a p^{k+1})}{\q_q(a_\a p^k)} \prod_{\b=1}^A \frac{\q_q(a_\a a_\b p^k)}{\q_q(a_\a a_\b^{-1}p^{k+1})} \rv   
    \sim\ T^{2+(\e-1)A}\,.
\end{split}
\end{align}
We see that for all values of $A>1$ (for $A=1$ the integral (\ref{eq:EHI-N=4-general-N}) is trivial), the residue sum with the full measure $(\e=0)$ is divergent, while the residue sum with the reduced measure ($\e=1$) is convergent.
As in Appendix \ref{sapp:vwp-ell-int}, it can be argued that in the latter case the contribution of the origin vanishes.
We conclude that the expression \eqref{eq:wp-EHI-n=4} for $\epsilon=1$ represents a well-defined residue evaluation of the well-poised elliptic hypergeometric integral in \eqref{eq:EHI-N=4-general-N}, at least for our choice of scaling of the parameters.

\subsection{Generalized integrals: a formal evaluation}\label{ssec:generalized-EHI}

We have dealt above with elliptic hypergeometric integrals that appear for the superconformal indices of $SU(2)$ gauge theories. 
When considering an $SU(N)$ gauge group, the superconformal indices are expressed as $N-1$-dimensional contour integrals over $SU(N)$ eigenvalues (see, \textit{e.g.}, Section \ref{ssec:sc-gauge-theory-inds}). 
Focusing on one integral at a time, say the integral over the $(N-1)^{\text{th}}$ eigenvalue $x$, the $x$--dependent part of the integrands takes on the following general form,
\begin{equation}            \label{eq:generalized-EHI}
    \cP(x) = \frac{\prod_{\a=1}^A\G(a_\a x)\G(b_\a x^{-1})}{\prod_{\g=1}^C\G\lb(c_\g x)^{\pm}\rb}\,,
\end{equation}
where we assume that (the $x$-dependent part of) the full measure is contained in $\cP(x)$.
We now evaluate the integral corresponding to this more general $\cP(x)$,
\begin{equation}\label{eq:generalized-EHI0}
    I = \oint\frac{dx}{2\pi ix} \cP(x).
\end{equation}
Note that we do not assume any balancing condition for this general integral. The integral is generally not well-poised; it becomes well-poised if we set $a_\a=b_\a$, for all $\a$. 

We provide below a formal evaluation of this integral as a sum over residues, without caring about convergence. 
In some physical theories, choosing an appropriate scaling of the parameters will lead to convergent residue sum even for these more general integrals, provided one uses the reduced measure (see Section \ref{ssec:sc-gauge-theory-inds}). 

Let us denote by $\mu_{\text{red}}(x)$ the $x$--dependent part of the additional factors in the denominator of the integrand that appear in the integral when using the reduced measure. 
For example, for the $\cN=4$ SYM index (\ref{eq:full-index-suN-n=4-defn}), 
\begin{equation}
    \mu_{\text{red}}(s_{N-1}) = \prod_{i=1}^{N-1}\frac1{1-s_{i,N-1}^{-1}}.
\end{equation}
where we think of $s_{N-1}$ as $x$ in this appendix.
We assume that $\cP(x)$ has only simple poles and that the absolute values of the fugacities satisfy the conditions $|p|=|q|<1$, and $|a_\a|,|b_\a|<1$, for all $\a$.
We then consider the contour integral
\begin{equation}            \label{eq:generalized-EHI-2}
    I = (p;p)_\I(q;q)_\I\oint\frac{dx}{2\pi ix} \mu_{\text{red}}^\e(x) \cP(x),
\end{equation}
where, as above, $\e=0$ corresponds to using the full measure, and $\e=1$ to using the reduced measure. 
We can express \eqref{eq:generalized-EHI-2} as the following formal residue sum:
\begin{equation}\label{eq:generalized-EHI-evaluation}
\begin{split}
	\cI	=&\ \sum_{i=1}^A \frac{\prod_{\a,\b'}^A\G(a_\a b_i)\G(b_\b b_i^{-1})}{\prod_{\g=1}^C \G((c_\g b_i)^\pm)} \mu_{\text{red}}^\e (-b_i)^{(A-C)(k+l)} 		\\
	&\ \qquad \times (pq)^{(A-C)kl} \lb p^{k(k+1)/2} q^{l(l+1)/2}\rb^{A-C} \lB \prod_\a^A (a_\a b_\a)^{-kl} b_\a^{-k-l} \rB \lB \prod_\g^C c_\g^{-k-l} \rB       \\
    &\ \qquad \times \lB \prod_{\a}^A \frac{\q_{k,p,q}(a_\a b_i)}{\q_{k,p,q}(pb_i b_\a^{-1})}	\frac{\q_{l,q,p}(a_\a b_i)}{\q_{l,q,p}(qb_i b_\a^{-1})}	\rB \lB \prod_\g^C \frac{\q_{q}(c_\g b_i p^k)}{\q_{q}(c_\g b_i)} \frac{\q_{p}(c_\g b_i q^l)}{\q_{p}(c_\g b_i)} \rB,
\end{split}
\end{equation}
where $\prod_{\b'}\equiv\prod_{\b\neq i}^A$.
We note that all the residue sums for full superconformal indices evaluated in this work follow from the formal evaluation (\ref{eq:generalized-EHI-evaluation}) of (\ref{eq:generalized-EHI-2}). 
Depending on the values of the fugacities $a_\a,b_\a$, one can check in individual cases whether the formal sum is convergent. 
An interesting example is discussed in Section \ref{ssec:ell-hypergeom-integrals}, where we consider the $s_{N-1}$ integral (\ref{eq:full-index-suN-n=4-s_N-1-integral}) for $\cN=4$ $SU(N)$ SYM.

\section{Trace relations for \texorpdfstring{$\mathcal{N}=4$}{N=4} \texorpdfstring{$SU(3)$}{SU(3)} HL chiral ring}\label{app:Tracerelations}

In Section \ref{ssec:hall-littlewoodsu3}, we take the Hall-Littlewood (HL) $q \to 0$ limit of our expression for the $\mathcal{N}=4$ $SU(3)$ Macdonald Index \eqref{eq:finalexpressionsI3SU(3)}. 
The resulting formula \eqref{eq:HLsu3} can be understood in terms of the Hall-Littlewood chiral ring of the theory.
In this section, we derive the relations satisfied among the generators given in Tables \ref{tab:HL-chiral-ring-SU(3)}. 

Since the generators are expressed as traces of letters valued in the adjoint representation of $SU(3)$, the corresponding relations can be thought of as trace relations.
To this end, let us recall that trace relations for matrices that transform in the adjoint representation of $SU(N)$ can be obtained using the Cayley-Hamilton theorem, which states that any $N \times N$ matrix satisfies 
\begin{align}
    M^{N} + c_{N-1} (M) M^{N-1} + \cdots + c_{1} (M) M + c_{0} (M) \mathbb{I}_{N \times N} = 0,
\end{align}
where the coefficients $c_{n} (M)$ are determined in terms of $\text{Tr} \, (M^{k})$, with $k \leq N$. As shown in \cite{Choi:2023znd}, we can select $M = M_{1} + M_{2}$, $M = M_{1} + M_{2} + M_{3}$, $\dots$, $M = M_{1} + \cdots + M_{N}$, where $M_{1}, \dots, M_{N}$ are independent matrices. By applying the Cayley-Hamilton theorem to these, we obtain an identity for the symmetrized product $M_{(1} \dots M_{N)}$. When these matrices are traceless, with $N = 3$, we obtain
\begin{align}\label{eq:relationtr}
    M_{(1} M_{2} M_{3)} = \frac{1}{2} \text{Tr} \, (M_{(1} M_{2}) M_{3)} + \frac{1}{3} \text{Tr} \, (M_{(1} M_{2} M_{3)}) \mathbb{I}_{3 \times 3}.
\end{align}
By multiplying the above by $M_{4}$ and taking a trace we get
\begin{align}\label{eq:tracerelationfour1}
    \text{Tr} \, (M_{(1} M_{2} M_{3)} M_{4}) &= \frac{1}{2} \text{Tr} \, (M_{(1} M_{2}) \text{Tr} \, (M_{3)} M_{4}),
\end{align}
Similarly, we get the following non-trivial relation by multiplying \eqref{eq:relationtr} by both $M_{4}$ and $M_{5}$ 
\begin{align}
    \text{Tr} \, (M_{(1} M_{2} M_{3)} M_{4} M_{5}) = \frac{1}{2} \text{Tr} \, (M_{(1} M_{2}) \text{Tr} \, (M_{3)} M_{4} M_{5}) + \frac{1}{3} \text{Tr} \, (M_{(1} M_{2} M_{3)}) \text{Tr} \, (M_{4} M_{5})\,.
\end{align}
By swapping $M_1$ and $M_4$ and using cyclicity of the trace, we find the following non-trivial relation
\begin{align}
    \begin{aligned}
    &\frac{1}{2} \text{Tr} \, (M_{(1} M_{2}) \text{Tr} \, (M_{3)} M_{4} M_{5}) + \frac{1}{3} \text{Tr} \, (M_{(1} M_{2} M_{3)}) \text{Tr} \, (M_{4} M_{5}) \\ &= \frac{1}{2} \text{Tr} \, (M_{(4} M_{2}) \text{Tr} \, (M_{3)} M_{1} M_{5}) + \frac{1}{3} \text{Tr} \, (M_{(4} M_{2} M_{3)}) \text{Tr} \, (M_{1} M_{5}).
    \end{aligned}
\end{align}
Similarly, we can generate trace relations for an arbitrary number of matrices $M_{1}, \dots, M_{n}$ in a single trace. An equivalent way of generating these relations is using the relation $T_{3} (M_{1}, M_{2}, M_{3}, M_{4}) = 0$, whose explicit form can be found in \cite{deMelloKoch:2025ngs}. 

Using this set of relations and choosing $M_{1}, \dots, M_{n}$, with $n \in \mathbb{Z}_{\geq 1}$, to be elements within the set of single letters $\bar{q}_{1,2}$ and $\lambda_{1+}$ contributing to the HL index, we can construct the full set of trace relations. 
However, not all of the trace relations generated in this way are independent. 
We identify an independent set of relations for a given index as follows. 
Using our set of single letters we generate the full set of multi-traces with that index. 
These can be used as a basis for a linear vector space and the relations generated above are vectors in this space. 
The relations can be organized into a matrix, whose null space can be computed. 
Using the null space we can then generate the set of linearly independent trace relations.

\begin{table}[ht]
    \centering
    \begin{tabular}{|c|c|c|}
        \hline
        $\mathcal{O}$ & $\mathcal{R}$ & index
         \\ \hline
        $\mu_{0} \nu_{\pm}$ & $2 \mu_{0} \nu_{\pm} = \nu_{\pm 3} \mu_{\mp} + \nu_{\mp} \mu_{\pm}$ & $-t^{\frac{5}{2}} v^{\pm}$ \\
        \hline $\nu_{\mp} \omega_{\pm}$ & $\nu_{\pm} \omega_{\mp} + \mu_{0} \sigma_{0} = \nu_{\mp} \omega_{\pm} + \mu_{\mp} \sigma_{\pm}$ & $t^{3}$ \\
        \hline $\nu_{-} \nu_{+}$ & $\mu_{0}^3 + 3 \nu_{-} \nu_{+} = 3 \nu_{-3} \nu_{+3} + \mu_{0} \mu_{-} \mu_{+}$ & $-t^{3}$ \\
        \hline $\nu_{\pm} \omega_{\pm}$ & $\nu_{\pm 3} \omega_{\mp} + \mu_{\pm} \sigma_{0} = \nu_{\pm} \omega_{\pm} + \mu_{0} \sigma_{\pm}$ & $t^{3} v^{\pm 2}$ \\
        \hline $\nu_{\pm}^2$ & $6 \nu_{\mp} \nu_{\pm 3} + \mu_{\mp} \mu_{\pm}^2 = 6 \nu_{\pm}^2 + \mu_{0}^2 \mu_{\pm}$ & $-t^{3} v^{\pm 2}$ \\
        \hline $\nu_{\pm} \sigma_{0}$ & $3 \nu_{\pm 3} \sigma_{\mp} + \mu_{\mp} \mu_{\pm} \omega_{\pm} + 3 \nu_{\mp} \sigma_{\pm} = \mu_{0}^2 \omega_{\pm} + 6 \nu_{\pm} \sigma_{0}$ & $t^{\frac{7}{2}} v^{\pm}$ \\
        \hline $\nu_{\mp} \sigma_{\pm}$ & $\mu_{0} \mu_{\pm} \omega_{\mp} + 6 \nu_{\pm} \sigma_{0} = \mu_{\mp} \mu_{\pm} \omega_{\pm} + 6 \nu_{\mp} \sigma_{\pm}$ & $t^{\frac{7}{2}} v^{\pm}$ \\
        \hline $\omega_{\pm} \sigma_{0}$ & $\omega_{\pm} \sigma_{0} = \omega_{\mp} \sigma_{\pm}$ & $-t^{\frac{7}{2}} v^{\pm}$ \\
        \hline $\nu_{\pm} \sigma_{\pm}$ & $\mu_{\pm}^2 \omega_{\mp} + 6 \nu_{\pm 3} \sigma_{0} = \mu_{0} \mu_{\pm} \omega_{\pm} + 6 \nu_{\pm} \sigma_{\pm}$ & $t^{\frac{7}{2}} v^{\pm 3}$  \\ \hline $\mu_{0}^2 \sigma_{0}$ & $\mu_{0} \mu_{+} \sigma_{-} + \mu_{0} \mu_{-} \sigma_{+} = 2 \mu_{0}^2 \sigma_{0}$ & $t^{4}$ \\
        \hline $\mu_{0}^4$ & $2\mu_{0}^4 + 3 \nu_{+}^2 \mu_{-} + 3 \nu_{-3} \nu_{+} \mu_{+} = 6 \mu_{0} \nu_{-3} \nu_{+3} + 2 \mu_{0}^2 \mu_{-} \mu_{+}$ & $-t^{4}$ \\ \hline
        $\mu_{0} \omega_{+} \omega_{-}$ & $ \mu_{0} \omega_{+} \omega_{-} = 3 \sigma_{-} \sigma_{+}$ & $-t^{4}$ \\
        \hline $\mu_{0}^2 \sigma_{\pm}$ & $\nu_{\pm 3} \mu_{\mp} \omega_{\pm} + 2 \mu_{0}^2 \sigma_{\pm} + \nu_{\mp} \mu_{\pm} \omega_{\pm} = 2 \mu_{0} \nu_{\pm 3} \omega_{\mp} + 2 \mu_{0} \mu_{\pm} \sigma_{0}$ & $t^{4} v^{\pm 2}$ \\
        \hline
        $\sigma_{\pm} \sigma_{0}$ & $6 \sigma_{\pm} \sigma_{0} = \mu_{\pm} \omega_{\mp} \omega_{\pm}$ & $-t^{4} v^{\pm 2}$ \\
        \hline $\mu_{0}^3 \omega_{\pm}$ & $\mu_{0}^3 \omega_{\pm} + 3 \nu_{\mp} \nu_{\pm} \omega_{\pm} = \mu_{0} \mu_{\mp} \mu_{\pm} \omega_{\pm} + 3 \nu_{\mp 3} \nu_{\pm 3} \omega_{\pm}$ & $t^{\frac{9}{2}} v^{\pm}$ \\
        \hline $\mu_{0} \sigma_{+} \sigma_{-}$ & $\mu_{0} \sigma_{+} \sigma_{-} = \nu_{+} \omega_{+} \sigma_{-} + \nu_{-3} \sigma_{+} \omega_{+}$ & $-t^{5}$ \\ 
        \hline $\mu_{0} \sigma_{\pm} \omega_{\mp}$ & $\mu_{\mp} \sigma_{\pm} \omega_{\pm} + \nu_{\pm} \omega_{\pm} \omega_{\mp} = \mu_{0} \sigma_{\pm} \omega_{\mp}$ & $-t^{\frac{9}{2}} v^{\pm}$ \\
        \hline $\mu_{0} \sigma_{\pm} \omega_{\pm}$ & $\nu_{\pm 3} \omega_{\mp} \omega_{\pm} + \mu_{\pm} \sigma_{0} \omega_{\pm} =  \mu_{0} \sigma_{\pm} \omega_{\pm}$ & $-t^{\frac{9}{2}} v^{\pm 3}$ \\
        \hline  
    \end{tabular}
    \caption{Relations $\mathcal{R}$ satisfied between the HL chiral ring generators. In the left--most column, we list the (multi--trace) operators $\mathcal{O}$ which we do not consider as independent due to the corresponding relation. This leads to the set of independent operators in Tables \ref{tab:HL-chiral-ring-SU(3)} and \ref{tab:mt-HL-chiral-ring-SU(3)}.}
    \label{tab:HLchiralringtracerelationsSU(3)}
\end{table}

These relations follow through directly for bosonic multi--trace operators. When multiple fermionic matrices are involved, the symmetrization changes to an anti--symmetrization to account for their statistics. Since for $SU(3)$ with one fermionic single letter $\lambda_{1+}$ there are only two fermionic eigenvalues, any multi--trace structure containing more than two fermionic single traces is trivially zero. 

All in all, this provides us with 31 different linearly independent trace relations for $SU(3)$, which are listed in Table \ref{tab:HLchiralringtracerelationsSU(3)}.
These can be used to find the maximal set of independent generators of the $\mathcal{N}=4$ $SU(3)$ Hall--Littlewood chiral ring. 
We note that we only need to account for up to 8 matrices in a single trace, since beyond this point we find no further independent trace relations.

\bibliographystyle{JHEP}
\bibliography{bib-mindex}

\end{document}

%% file: custom.tex
\newcommand{\defeq}{\mathrel{\mathop:}=}

\newcommand{\Rmnum}[1]{\expandafter\@slowromancap\romannumeral #1@}

\newcommand{\nn}{\nonumber}

\newcommand{\be}{\begin{equation}}
\newcommand{\ee}{\end{equation}}

\newcommand{\lB}{\left [}
\newcommand{\rB}{\right ]}
\newcommand{\lb}{\left (}
\newcommand{\rb}{\right )}
\newcommand{\lv}{\left\vert}
\newcommand{\rv}{\right\vert}

\newcommand{\bfrac}[2]{\lb\frac{#1}{#2}\rb}  	

\renewcommand{\a}{\alpha}	
\renewcommand{\b}{\beta}		
\renewcommand{\d}{\delta}	
\newcommand{\e}{\epsilon}

\newcommand{\f}{\phi}

\newcommand{\g}{\gamma}

\renewcommand{\k}{\kappa}	

\newcommand{\q}{\theta}	
	
\newcommand{\s}{\sigma}

\renewcommand{\t}{\tau}		

\newcommand{\G}{\Gamma}
\newcommand{\I}{\infty}

\newcommand{\Z}{\mathbb{Z}}

\newcommand{\cI}{\mathcal{I}}

\newcommand{\cN}{\mathcal{N}}
\newcommand{\cO}{\mathcal{O}}
\newcommand{\cP}{\mathcal{P}}

\newcommand{\bZ}{\mathbb{Z}}

\newcommand{\fu}{\mathfrak{u}}


%% file: bib-mindex.bib
@book{Gasper_Rahman_2004, 
	place={Cambridge}, 
	edition={2}, 
	series={Encyclopedia of Mathematics and its Applications}, 
	title={Basic Hypergeometric Series}, 
	publisher={Cambridge University Press}, 
	author={Gasper, George and Rahman, Mizan}, 
	year={2004}, 
	collection={Encyclopedia of Mathematics and its Applications}
	}

@article{Benini:2018mlo,
    author = "Benini, Francesco and Milan, Elisa",
    title = "{A Bethe Ansatz type formula for the superconformal index}",
    eprint = "1811.04107",
    archivePrefix = "arXiv",
    primaryClass = "hep-th",
    reportNumber = "SISSA 46/2018/FISI",
    doi = "10.1007/s00220-019-03679-y",
    journal = "Commun. Math. Phys.",
    volume = "376",
    number = "2",
    pages = "1413--1440",
    year = "2020"
}

@article{Benini:2018ywd,
    author = "Benini, Francesco and Milan, Elisa",
    title = "{Black Holes in 4D $\mathcal{N}$=4 Super-Yang-Mills Field Theory}",
    eprint = "1812.09613",
    archivePrefix = "arXiv",
    primaryClass = "hep-th",
    reportNumber = "SISSA 56/2018/FISI",
    doi = "10.1103/PhysRevX.10.021037",
    journal = "Phys. Rev. X",
    volume = "10",
    number = "2",
    pages = "021037",
    year = "2020"
}

@article{Wheeler:phd,
    author = "Wheeler, Campbell",
    title = "{Modular q–difference equations and quantum invariants of hyperbolic three–manifolds}",
    journal = "PhD thesis",
    year = "2023",
    doi = {https://hdl.handle.net/20.500.11811/10811}
}

@article{Bonetti:2016nma,
    author = "Bonetti, Federico and Rastelli, Leonardo",
    title = "{Supersymmetric localization in AdS$_{5}$ and the protected chiral algebra}",
    eprint = "1612.06514",
    archivePrefix = "arXiv",
    primaryClass = "hep-th",
    doi = "10.1007/JHEP08(2018)098",
    journal = "JHEP",
    volume = "08",
    pages = "098",
    year = "2018"
}

@article{Leuven26,
    author = "Bath, Jarryd and van Leuven, Sam",
    title = "{To appear}"
}

@article{Witten:1998qj,
    author = "Witten, Edward",
    title = "{Anti de Sitter space and holography}",
    eprint = "hep-th/9802150",
    archivePrefix = "arXiv",
    reportNumber = "IASSNS-HEP-98-15",
    doi = "10.4310/ATMP.1998.v2.n2.a2",
    journal = "Adv. Theor. Math. Phys.",
    volume = "2",
    pages = "253--291",
    year = "1998"
}

@article{Costello:2018zrm,
    author = "Costello, Kevin and Gaiotto, Davide",
    title = "{Twisted holography}",
    eprint = "1812.09257",
    archivePrefix = "arXiv",
    primaryClass = "hep-th",
    doi = "10.1007/JHEP01(2025)087",
    journal = "JHEP",
    volume = "01",
    pages = "087",
    year = "2025"
}

@article{Bonetti:2025kan,
    author = "Bonetti, Federico and Meneghelli, Carlo",
    title = "{A universal W-algebra for N=4 super Yang-Mills}",
    eprint = "2506.15678",
    archivePrefix = "arXiv",
    primaryClass = "hep-th",
    month = "6",
    year = "2025"
}

@article{Romelsberger:2007ec,
    author = "Romelsberger, Christian",
    title = "{Calculating the Superconformal Index and Seiberg Duality}",
    eprint = "0707.3702",
    archivePrefix = "arXiv",
    primaryClass = "hep-th",
    reportNumber = "DIAS-STP-07-13",
    month = "7",
    year = "2007"
}

@article{Spiridonov:2008zr,
    author = "Spiridonov, V. P. and Vartanov, G. S.",
    title = "{Superconformal indices for N = 1 theories with multiple duals}",
    eprint = "0811.1909",
    archivePrefix = "arXiv",
    primaryClass = "hep-th",
    doi = "10.1016/j.nuclphysb.2009.08.022",
    journal = "Nucl. Phys. B",
    volume = "824",
    pages = "192--216",
    year = "2010"
}

@article{Seiberg:1994pq,
    author = "Seiberg, N.",
    title = "{Electric - magnetic duality in supersymmetric nonAbelian gauge theories}",
    eprint = "hep-th/9411149",
    archivePrefix = "arXiv",
    reportNumber = "RU-94-82, IASSNS-HEP-94-98",
    doi = "10.1016/0550-3213(94)00023-8",
    journal = "Nucl. Phys. B",
    volume = "435",
    pages = "129--146",
    year = "1995"
}

@article{Hwang:2012jh,
    author = "Hwang, Chiung and Kim, Hee-Cheol and Park, Jaemo",
    title = "{Factorization of the 3d superconformal index}",
    doi = "10.1007/JHEP08(2014)018",
    journal = "JHEP",
    volume = "08",
    pages = "018",
    year = "2014"
}

@article{Gadde:2025yoa,
    author = "Gadde, Abhijit and Lee, Eunwoo and Raj, Rajat and Tomar, Shivansh",
    title = "{Probing Non-Graviton Spectra in $\mathcal{N}=4$ SYM via BMN truncation and S-Duality}",
    eprint = "2506.13887",
    archivePrefix = "arXiv",
    primaryClass = "hep-th",
    month = "6",
    year = "2025"
}

@article{Bourdier:2015wda,
    author = "Bourdier, Jun and Drukker, Nadav and Felix, Jan",
    title = "{The exact Schur index of $\mathcal{N}=4$ SYM}",
    eprint = "1507.08659",
    archivePrefix = "arXiv",
    primaryClass = "hep-th",
    doi = "10.1007/JHEP11(2015)210",
    journal = "JHEP",
    volume = "11",
    pages = "210",
    year = "2015"
}

@article{Hatsuda:2024uwt,
    author = "Hatsuda, Yasuyuki and Lin, Hai and Okazaki, Tadashi",
    title = "{Giant graviton expansions and ETW brane}",
    eprint = "2405.14564",
    archivePrefix = "arXiv",
    primaryClass = "hep-th",
    reportNumber = "RUP-24-9",
    doi = "10.1007/JHEP09(2024)181",
    journal = "JHEP",
    volume = "09",
    pages = "181",
    year = "2024"
}

@article{Lee:2023iil,
    author = "Lee, Ji Hoon",
    title = "{Trace relations and open string vacua}",
    eprint = "2312.00242",
    archivePrefix = "arXiv",
    primaryClass = "hep-th",
    doi = "10.1007/JHEP02(2024)224",
    journal = "JHEP",
    volume = "02",
    pages = "224",
    year = "2024"
}

@article{Dolan:2002zh,
    author = "Dolan, F. A. and Osborn, H.",
    title = "{On short and semi-short representations for four-dimensional superconformal symmetry}",
    eprint = "hep-th/0209056",
    archivePrefix = "arXiv",
    reportNumber = "DAMTP-02-114",
    doi = "10.1016/S0003-4916(03)00074-5",
    journal = "Annals Phys.",
    volume = "307",
    pages = "41--89",
    year = "2003"
}

@article{Gaiotto:2008ak,
    author = "Gaiotto, Davide and Witten, Edward",
    title = "{S-Duality of Boundary Conditions In N=4 Super Yang-Mills Theory}",
    eprint = "0807.3720",
    archivePrefix = "arXiv",
    primaryClass = "hep-th",
    doi = "10.4310/ATMP.2009.v13.n3.a5",
    journal = "Adv. Theor. Math. Phys.",
    volume = "13",
    number = "3",
    pages = "721--896",
    year = "2009"
}

@article{deMelloKoch:2025ngs,
    author = "de Mello Koch, Robert and Jevicki, Antal",
    title = "{Structure of loop space at finite N}",
    eprint = "2503.20097",
    archivePrefix = "arXiv",
    primaryClass = "hep-th",
    doi = "10.1007/JHEP06(2025)011",
    journal = "JHEP",
    volume = "06",
    pages = "011",
    year = "2025"
}

@article{Gaiotto:2009we,
    author = "Gaiotto, Davide",
    title = "{N=2 dualities}",
    eprint = "0904.2715",
    archivePrefix = "arXiv",
    primaryClass = "hep-th",
    doi = "10.1007/JHEP08(2012)034",
    journal = "JHEP",
    volume = "08",
    pages = "034",
    year = "2012"
}

@article{Dolan:2008qi,
    author = "Dolan, F. A. and Osborn, H.",
    title = "{Applications of the Superconformal Index for Protected Operators and q-Hypergeometric Identities to N=1 Dual Theories}",
    eprint = "0801.4947",
    archivePrefix = "arXiv",
    primaryClass = "hep-th",
    reportNumber = "DAMTP-08-07, DIAS-STP-08-02, SHEP-08-06",
    doi = "10.1016/j.nuclphysb.2009.01.028",
    journal = "Nucl. Phys. B",
    volume = "818",
    pages = "137--178",
    year = "2009"
}

@article{Dolan:2007rq,
    author = "Dolan, F. A.",
    title = "{Counting BPS operators in N=4 SYM}",
    eprint = "0704.1038",
    archivePrefix = "arXiv",
    primaryClass = "hep-th",
    reportNumber = "DIAS-STP-07-05",
    doi = "10.1016/j.nuclphysb.2007.07.026",
    journal = "Nucl. Phys. B",
    volume = "790",
    pages = "432--464",
    year = "2008"
}

@article{Romelsberger:2005eg,
    author = "Romelsberger, Christian",
    title = "{Counting chiral primaries in N = 1, d=4 superconformal field theories}",
    eprint = "hep-th/0510060",
    archivePrefix = "arXiv",
    doi = "10.1016/j.nuclphysb.2006.03.037",
    journal = "Nucl. Phys. B",
    volume = "747",
    pages = "329--353",
    year = "2006"
}

@article{Gadde:2010te,
    author = "Gadde, Abhijit and Rastelli, Leonardo and Razamat, Shlomo S. and Yan, Wenbin",
    title = "{The Superconformal Index of the $E_{6}$ SCFT}",
    eprint = "1003.4244",
    archivePrefix = "arXiv",
    primaryClass = "hep-th",
    reportNumber = "YITP-SB-10-7",
    doi = "10.1007/JHEP08(2010)107",
    journal = "JHEP",
    volume = "08",
    pages = "107",
    year = "2010"
}

@article{Hanany:2008sb,
    author = "Hanany, Amihay and Mekareeya, Noppadol and Torri, Giuseppe",
    title = "{The Hilbert Series of Adjoint SQCD}",
    eprint = "0812.2315",
    archivePrefix = "arXiv",
    primaryClass = "hep-th",
    reportNumber = "IMPERIAL-TP-08-AH-11",
    doi = "10.1016/j.nuclphysb.2009.09.016",
    journal = "Nucl. Phys. B",
    volume = "825",
    pages = "52--97",
    year = "2010"
}

@article{Spiridonov:2008thf,
   title={Essays on the theory of elliptic hypergeometric functions},
   eprint = "0805.3135",
   archivePrefix = "arXiv",
   primaryClass = "math.CA",
   volume={63},
   ISSN={1468-4829},
   url={http://dx.doi.org/10.1070/RM2008v063n03ABEH004533},
   DOI={10.1070/rm2008v063n03abeh004533},
   number={3},
   journal={Russian Mathematical Surveys},
   publisher={Steklov Mathematical Institute},
   author={Spiridonov, V P},
   year={2008},
   month=jun, pages={405–472} 
}

@article{Spiridonov:2009za,
    author = "Spiridonov, V. P. and Vartanov, G. S.",
    title = "{Elliptic Hypergeometry of Supersymmetric Dualities}",
    eprint = "0910.5944",
    archivePrefix = "arXiv",
    primaryClass = "hep-th",
    reportNumber = "AEI-2009-106",
    doi = "10.1007/s00220-011-1218-9",
    journal = "Commun. Math. Phys.",
    volume = "304",
    pages = "797--874",
    year = "2011"
}

@article{Choi:2021rxi,
    author = "Choi, Sunjin and Jeong, Saebyeok and Kim, Seok and Lee, Eunwoo",
    title = "{Exact QFT duals of AdS black holes}",
    eprint = "2111.10720",
    archivePrefix = "arXiv",
    primaryClass = "hep-th",
    reportNumber = "KIAS-P21054, SNUTP21-002",
    doi = "10.1007/JHEP09(2023)138",
    journal = "JHEP",
    volume = "09",
    pages = "138",
    year = "2023"
}

@article{Choi:2023znd,
    author = "Choi, Sunjin and Kim, Seok and Lee, Eunwoo and Lee, Siyul and Park, Jaemo",
    title = "{Towards quantum black hole microstates}",
    eprint = "2304.10155",
    archivePrefix = "arXiv",
    primaryClass = "hep-th",
    doi = "10.1007/JHEP11(2023)175",
    journal = "JHEP",
    volume = "11",
    pages = "175",
    year = "2023"
}

@article{Choi:2023tiq,
    author = "Choi, Sunjin and Kim, Seunggyu and Song, Jaewon",
    title = "{Large N universality of 4d $ \mathcal{N} $ = 1 superconformal index and AdS black holes}",
    eprint = "2309.07614",
    archivePrefix = "arXiv",
    primaryClass = "hep-th",
    reportNumber = "KIAS-P23034",
    doi = "10.1007/JHEP08(2024)105",
    journal = "JHEP",
    volume = "08",
    pages = "105",
    year = "2024"
}

@article{Zheng:2022zkm,
	author = "Zheng, Haocong and Pan, Yiwen and Wang, Yufan",
	title = "{Surface defects, flavored modular differential equations, and modularity}",
	eprint = "2207.10463",
	archivePrefix = "arXiv",
	primaryClass = "hep-th",
	doi = "10.1103/PhysRevD.106.105020",
	journal = "Phys. Rev. D",
	volume = "106",
	number = "10",
	pages = "105020",
	year = "2022"
}

@article{Berenstein:2002jq,
    author = "Berenstein, David Eliecer and Maldacena, Juan Martin and Nastase, Horatiu Stefan",
    title = "{Strings in flat space and pp waves from N=4 superYang-Mills}",
    eprint = "hep-th/0202021",
    archivePrefix = "arXiv",
    doi = "10.1088/1126-6708/2002/04/013",
    journal = "JHEP",
    volume = "04",
    pages = "013",
    year = "2002"
}

@article{Kim:2003rza,
    author = "Kim, Nakwoo and Klose, Thomas and Plefka, Jan",
    title = "{Plane wave matrix theory from N=4 superYang-Mills on R x S**3}",
    eprint = "hep-th/0306054",
    archivePrefix = "arXiv",
    reportNumber = "AEI-2003-048",
    doi = "10.1016/j.nuclphysb.2003.08.019",
    journal = "Nucl. Phys. B",
    volume = "671",
    pages = "359--382",
    year = "2003"
}

@article{deMelloKoch:2024pcs,
    author = "de Mello Koch, Robert and Kim, Minkyoo and Kim, Seok and Lee, Jehyun and Lee, Siyul",
    title = "{Brane-fused black hole operators}",
    eprint = "2412.08695",
    archivePrefix = "arXiv",
    primaryClass = "hep-th",
    reportNumber = "SNUTP24-004",
    doi = "10.1007/JHEP07(2025)216",
    journal = "JHEP",
    volume = "07",
    pages = "216",
    year = "2025"
}

@article{Gaiotto:2012xa,
    author = "Gaiotto, Davide and Rastelli, Leonardo and Razamat, Shlomo S.",
    title = "{Bootstrapping the superconformal index with surface defects}",
    eprint = "1207.3577",
    archivePrefix = "arXiv",
    primaryClass = "hep-th",
    doi = "10.1007/JHEP01(2013)022",
    journal = "JHEP",
    volume = "01",
    pages = "022",
    year = "2013"
}

@article{Askey-Roy-1986,
author = {Richard Askey and Ranjan Roy},
title = {{More q-beta integrals}},
volume = {16},
journal = {Rocky Mountain Journal of Mathematics},
number = {2},
publisher = {Rocky Mountain Mathematics Consortium},
pages = {365 -- 372},
year = {1986},
doi = {10.1216/RMJ-1986-16-2-365},
URL = {https://doi.org/10.1216/RMJ-1986-16-2-365}
}

@article{Chang:2013fba,
    author = "Chang, Chi-Ming and Yin, Xi",
    title = "{1/16 BPS states in $\mathcal N=$ 4 super-Yang-Mills theory}",
    eprint = "1305.6314",
    archivePrefix = "arXiv",
    primaryClass = "hep-th",
    doi = "10.1103/PhysRevD.88.106005",
    journal = "Phys. Rev. D",
    volume = "88",
    number = "10",
    pages = "106005",
    year = "2013"
}

@article{Benvenuti:2006qr,
    author = "Benvenuti, Sergio and Feng, Bo and Hanany, Amihay and He, Yang-Hui",
    title = "{Counting BPS Operators in Gauge Theories: Quivers, Syzygies and Plethystics}",
    eprint = "hep-th/0608050",
    archivePrefix = "arXiv",
    doi = "10.1088/1126-6708/2007/11/050",
    journal = "JHEP",
    volume = "11",
    pages = "050",
    year = "2007"
}

@article{Feng:2007ur,
    author = "Feng, Bo and Hanany, Amihay and He, Yang-Hui",
    title = "{Counting gauge invariants: The Plethystic program}",
    eprint = "hep-th/0701063",
    archivePrefix = "arXiv",
    doi = "10.1088/1126-6708/2007/03/090",
    journal = "JHEP",
    volume = "03",
    pages = "090",
    year = "2007"
}

@article{Benini:2021ano,
    author = "Benini, Francesco and Rizi, Giovanni",
    title = "{Superconformal index of low-rank gauge theories via the Bethe Ansatz}",
    eprint = "2102.03638",
    archivePrefix = "arXiv",
    primaryClass = "hep-th",
    reportNumber = "SISSA 07/2021/FISI",
    doi = "10.1007/JHEP05(2021)061",
    journal = "JHEP",
    volume = "05",
    pages = "061",
    year = "2021"
}

@article{Grant:2008sk,
    author = "Grant, Lars and Grassi, Pietro A. and Kim, Seok and Minwalla, Shiraz",
    title = "{Comments on 1/16 BPS Quantum States and Classical Configurations}",
    eprint = "0803.4183",
    archivePrefix = "arXiv",
    primaryClass = "hep-th",
    reportNumber = "IMPERIAL-TP-08-SK-01",
    doi = "10.1088/1126-6708/2008/05/049",
    journal = "JHEP",
    volume = "05",
    pages = "049",
    year = "2008"
}

@article{Chang:2022mjp,
    author = "Chang, Chi-Ming and Lin, Ying-Hsuan",
    title = "{Words to describe a black hole}",
    eprint = "2209.06728",
    archivePrefix = "arXiv",
    primaryClass = "hep-th",
    doi = "10.1007/JHEP02(2023)109",
    journal = "JHEP",
    volume = "02",
    pages = "109",
    year = "2023"
}

@article{Choi:2022caq,
    author = "Choi, Sunjin and Kim, Seok and Lee, Eunwoo and Park, Jaemo",
    title = "{The shape of non-graviton operators for SU(2)}",
    eprint = "2209.12696",
    archivePrefix = "arXiv",
    primaryClass = "hep-th",
    reportNumber = "KIAS-P22052",
    doi = "10.1007/JHEP09(2024)029",
    journal = "JHEP",
    volume = "09",
    pages = "029",
    year = "2024"
}

@article{Chang:2023ywj,
    author = "Chang, Chi-Ming and Lin, Ying-Hsuan and Wu, Jingxiang",
    title = "{On $\frac{1}{8}$-BPS black holes and the chiral algebra of $\mathcal{N}=4$ SYM}",
    eprint = "2310.20086",
    archivePrefix = "arXiv",
    primaryClass = "hep-th",
    doi = "10.4310/atmp.241031230723",
    journal = "Adv. Theor. Math. Phys.",
    volume = "28",
    number = "7",
    pages = "2431--2489",
    year = "2024"
}

@article{Choi:2023vdm,
    author = "Choi, Jaehyeok and Choi, Sunjin and Kim, Seok and Lee, Jehyun and Lee, Siyul",
    title = "{Finite N black hole cohomologies}",
    eprint = "2312.16443",
    archivePrefix = "arXiv",
    primaryClass = "hep-th",
    reportNumber = "SNUTP23-002, KIAS-P23070, LCTP-23-20, SNUTP23-002; KIAS-P23070; LCTP-23-20;",
    doi = "10.1007/JHEP12(2024)029",
    journal = "JHEP",
    volume = "12",
    pages = "029",
    year = "2024"
}

@article{Chang:2024zqi,
    author = "Chang, Chi-Ming and Lin, Ying-Hsuan",
    title = "{Holographic covering and the fortuity of black holes}",
    eprint = "2402.10129",
    archivePrefix = "arXiv",
    primaryClass = "hep-th",
    month = "2",
    year = "2024"
}

@article{Choi:2025lck,
    author = "Choi, Sunjin and Jain, Diksha and Kim, Seok and Krishna, Vineeth and Kwon, Goojin and Lee, Eunwoo and Minwalla, Shiraz and Patel, Chintan",
    title = "{Supersymmetric Grey Galaxies, Dual Dressed Black Holes and the Superconformal Index}",
    eprint = "2501.17217",
    archivePrefix = "arXiv",
    primaryClass = "hep-th",
    reportNumber = "TIFR/TH/25-3, LCTP-25-02",
    journal = "SciPost Phys.",
    doi = "10.21468/SciPostPhys.19.3.072",
    month = "1",
    year = "2025"
}

@book{Macdonald1995,
  author    = {Macdonald, I. G.},
  title     = {Symmetric Functions and {H}all Polynomials},
  edition   = {2},
  publisher = {Oxford University Press},
  address   = {New York},
  year      = {1995}
}

@article{Bonetti:2018fqz,
    author = "Bonetti, Federico and Meneghelli, Carlo and Rastelli, Leonardo",
    title = "{VOAs labelled by complex reflection groups and 4d SCFTs}",
    eprint = "1810.03612",
    archivePrefix = "arXiv",
    primaryClass = "hep-th",
    doi = "10.1007/JHEP05(2019)155",
    journal = "JHEP",
    volume = "05",
    pages = "155",
    year = "2019"
}

@article{Beem:2019tfp,
    author = "Beem, Christopher and Meneghelli, Carlo and Rastelli, Leonardo",
    title = "{Free Field Realizations from the Higgs Branch}",
    eprint = "1903.07624",
    archivePrefix = "arXiv",
    primaryClass = "hep-th",
    doi = "10.1007/JHEP09(2019)058",
    journal = "JHEP",
    volume = "09",
    pages = "058",
    year = "2019"
}

@article{Felder_2018,
   title={Analyticity of Nekrasov Partition Functions},
   volume={364},
   ISSN={1432-0916},
   url={http://dx.doi.org/10.1007/s00220-018-3270-1},
   DOI={10.1007/s00220-018-3270-1},
   number={2},
   journal={Communications in Mathematical Physics},
   publisher={Springer Science and Business Media LLC},
   author={Felder, Giovanni and Müller-Lennert, Martin},
   year={2018},
   month=oct, pages={683–718} ,
    eprint = "1709.05232",
    archivePrefix = "arXiv",
    primaryClass = "hep-th",
}

@article{Eniceicu:2023uvd,
    author = "Eniceicu, Dan Stefan",
    title = "{Comments on the Giant-Graviton Expansion of the Superconformal Index}",
    eprint = "2302.04887",
    archivePrefix = "arXiv",
    primaryClass = "hep-th",
    month = "2",
    year = "2023"
}

@article{Murthy:2022ien,
    author = "Murthy, Sameer",
    title = "{Unitary matrix models, free fermions, and the giant graviton expansion}",
    eprint = "2202.06897",
    archivePrefix = "arXiv",
    primaryClass = "hep-th",
    doi = "10.4310/PAMQ.2023.v19.n1.a12",
    journal = "Pure Appl. Math. Quart.",
    volume = "19",
    number = "1",
    pages = "299--340",
    year = "2023"
}

@article{Murthy:2020scj,
    author = "Murthy, Sameer",
    title = "{Growth of the $\frac {1} {16}$-BPS index in 4d $N=4$ supersymmetric Yang-Mills theory}",
    eprint = "2005.10843",
    archivePrefix = "arXiv",
    primaryClass = "hep-th",
    doi = "10.1103/PhysRevD.105.L021903",
    journal = "Phys. Rev. D",
    volume = "105",
    number = "2",
    pages = "L021903",
    year = "2022"
}

@article{Agarwal:2020zwm,
    author = "Agarwal, Prarit and Choi, Sunjin and Kim, Joonho and Kim, Seok and Nahmgoong, June",
    title = "{AdS black holes and finite N indices}",
    eprint = "2005.11240",
    archivePrefix = "arXiv",
    primaryClass = "hep-th",
    reportNumber = "QMUL-PH-20-11, SNUTP20-001, KIAS-P20020",
    doi = "10.1103/PhysRevD.103.126006",
    journal = "Phys. Rev. D",
    volume = "103",
    number = "12",
    pages = "126006",
    year = "2021"
}

@article{Argyres:1996eh,
    author = "Argyres, Philip C. and Plesser, M. Ronen and Seiberg, Nathan",
    title = "{The Moduli space of vacua of N=2 SUSY QCD and duality in N=1 SUSY QCD}",
    eprint = "hep-th/9603042",
    archivePrefix = "arXiv",
    reportNumber = "RU-96-07, WIS-96-1-PH",
    doi = "10.1016/0550-3213(96)00210-6",
    journal = "Nucl. Phys. B",
    volume = "471",
    pages = "159--194",
    year = "1996"
}

@book{Tachikawa:2013kta,
    author = "Tachikawa, Yuji",
    title = "{N=2 supersymmetric dynamics for pedestrians}",
    eprint = "1312.2684",
    archivePrefix = "arXiv",
    primaryClass = "hep-th",
    reportNumber = "UT-13-42, IPMU-13-0234, UT-13-42, IPMU-13-0234",
    doi = "10.1007/978-3-319-08822-8",
    volume = "890",
    month = "12",
    year = "2013",
    publisher= "Springer"
}

@article{Cohl_2022,
   title={Utility of integral representations for basic hypergeometric functions and orthogonal polynomials},
   volume={61},
   ISSN={1572-9303},
   url={http://dx.doi.org/10.1007/s11139-021-00509-5},
   DOI={10.1007/s11139-021-00509-5},
   number={2},
   journal={The Ramanujan Journal},
   publisher={Springer Science and Business Media LLC},
   author={Cohl, Howard S. and Costas-Santos, Roberto S.},
   year={2022},
   month=jun, pages={649–674},
eprint = "2108.03275",
    archivePrefix = "arXiv",
    primaryClass = "hep-th"}

@article{Huang:2022bry,
	author = "Huang, Min-xin",
	title = "{Modular anomaly equation for Schur index of $ \mathcal{N} $ = 4 super-Yang-Mills}",
	eprint = "2205.00818",
	archivePrefix = "arXiv",
	primaryClass = "hep-th",
	reportNumber = "USTC-ICTS/PCFT-22-14",
	doi = "10.1007/JHEP08(2022)049",
	journal = "JHEP",
	volume = "08",
	pages = "049",
	year = "2022"
}

@article{Aharony:2021zkr,
	author = "Aharony, Ofer and Benini, Francesco and Mamroud, Ohad and Milan, Elisa",
	title = "{A gravity interpretation for the Bethe Ansatz expansion of the $\mathcal{N}=4$ SYM index}",
	eprint = "2104.13932",
	archivePrefix = "arXiv",
	primaryClass = "hep-th",
	reportNumber = "SISSA 01/2021/FISI",
	doi = "10.1103/PhysRevD.104.086026",
	journal = "Phys. Rev. D",
	volume = "104",
	pages = "086026",
	year = "2021"
}

@article{Cabo-Bizet:2019eaf,
	author = "Cabo-Bizet, Alejandro and Murthy, Sameer",
	title = "{Supersymmetric phases of 4d $ \mathcal{N} $ = 4 SYM at large $N$}",
	eprint = "1909.09597",
	archivePrefix = "arXiv",
	primaryClass = "hep-th",
	doi = "10.1007/JHEP09(2020)184",
	journal = "JHEP",
	volume = "09",
	pages = "184",
	year = "2020"
}

@article{Fantini:2025wap,
    author = "Fantini, Veronica and Rella, Claudia",
    title = "{Modular resurgence, $q$-Pochhammer symbols, and quantum operators from mirror curves}",
    eprint = "2506.08265",
    archivePrefix = "arXiv",
    primaryClass = "hep-th",
    month = "6",
    year = "2025"
}

@article{Banerjee:2017,
author = {Banerjee, Shubho and Wilkerson, Blake},
title = {Asymptotic expansions of Lambert series and related q-series},
journal = {International Journal of Number Theory},
volume = {13},
number = {08},
pages = {2097-2113},
year = {2017},
doi = {10.1142/S1793042117501135},
URL = {https://doi.org/10.1142/S1793042117501135},
eprint = {https://arxiv.org/abs/1602.010855}
}

@article{COHL2023102517,
title = {Nonterminating transformations and summations associated with some q-Mellin–Barnes integrals},
journal = {Advances in Applied Mathematics},
volume = {147},
pages = {102517},
year = {2023},
issn = {0196-8858},
doi = {https://doi.org/10.1016/j.aam.2023.102517},
url = {https://www.sciencedirect.com/science/article/pii/S0196885823000350},
author = {Howard S. Cohl and Roberto S. Costas-Santos}
}

@article{Lezcano:2021qbj,
	author = "Lezcano, Alfredo Gonz\'alez and Hong, Junho and Liu, James T. and Pando Zayas, Leopoldo A.",
	title = "{The Bethe-Ansatz approach to the $ \mathcal{N} $ = 4 superconformal index at finite rank}",
	eprint = "2101.12233",
	archivePrefix = "arXiv",
	primaryClass = "hep-th",
	reportNumber = "LCTP-21-02",
	doi = "10.1007/JHEP06(2021)126",
	journal = "JHEP",
	volume = "06",
	pages = "126",
	year = "2021"
}

@article{GonzalezLezcano:2020yeb,
	author = "Gonz\'alez Lezcano, Alfredo and Hong, Junho and Liu, James T. and Pando Zayas, Leopoldo A.",
	title = "{Sub-leading Structures in Superconformal Indices: Subdominant Saddles and Logarithmic Contributions}",
	eprint = "2007.12604",
	archivePrefix = "arXiv",
	primaryClass = "hep-th",
	reportNumber = "LCTP-20-16",
	doi = "10.1007/JHEP01(2021)001",
	journal = "JHEP",
	volume = "01",
	pages = "001",
	year = "2021"
}

@article{Cassani:2021fyv,
	author = "Cassani, Davide and Komargodski, Zohar",
	title = "{EFT and the SUSY Index on the 2nd Sheet}",
	eprint = "2104.01464",
	archivePrefix = "arXiv",
	primaryClass = "hep-th",
	doi = "10.21468/SciPostPhys.11.1.004",
	journal = "SciPost Phys.",
	volume = "11",
	pages = "004",
	year = "2021"
}

@article{Goldstein:2020yvj,
	author = "Goldstein, Kevin and Jejjala, Vishnu and Lei, Yang and van Leuven, Sam and Li, Wei",
	title = "{Residues, modularity, and the Cardy limit of the 4d $ \mathcal{N} $ = 4 superconformal index}",
	eprint = "2011.06605",
	archivePrefix = "arXiv",
	primaryClass = "hep-th",
	doi = "10.1007/JHEP04(2021)216",
	journal = "JHEP",
	volume = "04",
	pages = "216",
	year = "2021"
}

@article{Gadde:2009kb,
	author = "Gadde, Abhijit and Pomoni, Elli and Rastelli, Leonardo and Razamat, Shlomo S.",
	title = "{S-duality and 2d Topological QFT}",
	eprint = "0910.2225",
	archivePrefix = "arXiv",
	primaryClass = "hep-th",
	reportNumber = "YITP-SB-09-30",
	doi = "10.1007/JHEP03(2010)032",
	journal = "JHEP",
	volume = "03",
	pages = "032",
	year = "2010"
}

@article{Closset:2013vra,
	author = "Closset, Cyril and Dumitrescu, Thomas T. and Festuccia, Guido and Komargodski, Zohar",
	title = "{The Geometry of Supersymmetric Partition Functions}",
	eprint = "1309.5876",
	archivePrefix = "arXiv",
	primaryClass = "hep-th",
	reportNumber = "WIS-09-13-SEP-DPPA",
	doi = "10.1007/JHEP01(2014)124",
	journal = "JHEP",
	volume = "01",
	pages = "124",
	year = "2014"
}

@article{Gadde:2020bov,
	author = "Gadde, Abhijit",
	title = "{Modularity of supersymmetric partition functions}",
	eprint = "2004.13490",
	archivePrefix = "arXiv",
	primaryClass = "hep-th",
	reportNumber = "TIFR/TH/20-13",
	doi = "10.1007/JHEP12(2021)181",
	journal = "JHEP",
	volume = "12",
	pages = "181",
	year = "2021"
}

@article{Gutowski:2004ez,
    author = "Gutowski, Jan B. and Reall, Harvey S.",
    title = "{Supersymmetric AdS(5) black holes}",
    eprint = "hep-th/0401042",
    archivePrefix = "arXiv",
    reportNumber = "NSF-KITP-04-02",
    doi = "10.1088/1126-6708/2004/02/006",
    journal = "JHEP",
    volume = "02",
    pages = "006",
    year = "2004"
}

@article{Chong:2005da,
    author = "Chong, Z. W. and Cvetic, Mirjam and Lu, H. and Pope, C. N.",
    title = "{Five-dimensional gauged supergravity black holes with independent rotation parameters}",
    eprint = "hep-th/0505112",
    archivePrefix = "arXiv",
    reportNumber = "MIFP-05-11, UPR-1123-T",
    doi = "10.1103/PhysRevD.72.041901",
    journal = "Phys. Rev. D",
    volume = "72",
    pages = "041901",
    year = "2005"
}

@article{ArabiArdehali:2019orz,
    author = "Arabi Ardehali, Arash and Hong, Junho and Liu, James T.",
    title = "{Asymptotic growth of the 4d $ \mathcal{N} $ = 4 index and partially deconfined phases}",
    eprint = "1912.04169",
    archivePrefix = "arXiv",
    primaryClass = "hep-th",
    reportNumber = "LCTP 19-32",
    doi = "10.1007/JHEP07(2020)073",
    journal = "JHEP",
    volume = "07",
    pages = "073",
    year = "2020"
}

@article{Benini:2020gjh,
    author = "Benini, Francesco and Colombo, Edoardo and Soltani, Saman and Zaffaroni, Alberto and Zhang, Ziruo",
    title = "{Superconformal indices at large $N$ and the entropy of AdS$_5$ $\times$ SE$_5$ black holes}",
    eprint = "2005.12308",
    archivePrefix = "arXiv",
    primaryClass = "hep-th",
    reportNumber = "SISSA 11/2020/FISI",
    doi = "10.1088/1361-6382/abb39b",
    journal = "Class. Quant. Grav.",
    volume = "37",
    number = "21",
    pages = "215021",
    year = "2020"
}

@article{David:2021qaa,
    author = "David, Marina and Lezcano Gonz{\'a}lez, Alfredo and Nian, Jun and Pando Zayas, Leopoldo A.",
    title = "{Logarithmic corrections to the entropy of rotating black holes and black strings in AdS$_{5}$}",
    eprint = "2106.09730",
    archivePrefix = "arXiv",
    primaryClass = "hep-th",
    reportNumber = "LCTP-21-14",
    doi = "10.1007/JHEP04(2022)160",
    journal = "JHEP",
    volume = "04",
    pages = "160",
    year = "2022"
}

@article{Colombo:2021kbb,
    author = "Colombo, Edoardo",
    title = "{The large-N limit of 4d superconformal indices for general BPS charges}",
    eprint = "2110.01911",
    archivePrefix = "arXiv",
    primaryClass = "hep-th",
    doi = "10.1007/JHEP12(2022)013",
    journal = "JHEP",
    volume = "12",
    pages = "013",
    year = "2022"
}

@article{Aharony:2024ntg,
    author = "Aharony, Ofer and Mamroud, Ohad and Nowik, Shimon and Weissman, Meir",
    title = "{Bethe Ansatz for the superconformal index with unequal angular momenta}",
    eprint = "2402.03977",
    archivePrefix = "arXiv",
    primaryClass = "hep-th",
    doi = "10.1103/PhysRevD.109.085015",
    journal = "Phys. Rev. D",
    volume = "109",
    number = "8",
    pages = "085015",
    year = "2024"
}

@article{Cabo-Bizet:2024kfe,
    author = "Cabo-Bizet, Alejandro and Li, Wei",
    title = "{Generalized Bethe expansions of superconformal indices}",
    eprint = "2411.12018",
    archivePrefix = "arXiv",
    primaryClass = "hep-th",
    doi = "10.1007/JHEP07(2025)206",
    journal = "JHEP",
    volume = "07",
    pages = "206",
    year = "2025"
}

@article{Cabo-Bizet:2020ewf,
    author = "Cabo-Bizet, Alejandro",
    title = "{From multi-gravitons to Black holes: The role of complex saddles}",
    eprint = "2012.04815",
    archivePrefix = "arXiv",
    primaryClass = "hep-th",
    month = "12",
    year = "2020"
}

@article{Kunduri:2006ek,
    author = "Kunduri, Hari K. and Lucietti, James and Reall, Harvey S.",
    title = "{Supersymmetric multi-charge AdS(5) black holes}",
    eprint = "hep-th/0601156",
    archivePrefix = "arXiv",
    reportNumber = "DAMTP-2006-10",
    doi = "10.1088/1126-6708/2006/04/036",
    journal = "JHEP",
    volume = "04",
    pages = "036",
    year = "2006"
}

@article{Choi:2022ovw,
    author = "Choi, Sunjin and Kim, Seok and Lee, Eunwoo and Lee, Jehyun",
    title = "{From giant gravitons to black holes}",
    eprint = "2207.05172",
    archivePrefix = "arXiv",
    primaryClass = "hep-th",
    reportNumber = "KIAS-P22048, SNUTP22-001",
    doi = "10.1007/JHEP11(2023)086",
    journal = "JHEP",
    volume = "11",
    pages = "086",
    year = "2023"
}

@article{Gaikwad:2025ugk,
    author = "Gaikwad, Adwait and Kibe, Tanay and van Leuven, Sam and Mathieson, Kayleigh",
    title = "{To gauge or to double gauge? Matrix models, global symmetry, and black hole cohomologies}",
    eprint = "2512.02103",
    archivePrefix = "arXiv",
    primaryClass = "hep-th",
    month = "12",
    year = "2025"
}

@article{Choi:2025bhi,
    author = "Choi, Jaehyeok and Lee, Eunwoo",
    title = "{Konishi lifts a black hole}",
    eprint = "2511.09519",
    archivePrefix = "arXiv",
    primaryClass = "hep-th",
    month = "11",
    year = "2025"
}

@article{Eager:2018czl,
    author = "Eager, Richard",
    title = "{Local Operators from the Space of Vacua of Four Dimensional SUSY Gauge Theories}",
    eprint = "1810.01192",
    archivePrefix = "arXiv",
    primaryClass = "hep-th",
    month = "10",
    year = "2018"
}

@article{Hosseini:2017mds,
    author = "Hosseini, Seyed Morteza and Hristov, Kiril and Zaffaroni, Alberto",
    title = "{An extremization principle for the entropy of rotating BPS black holes in AdS$_{5}$}",
    eprint = "1705.05383",
    archivePrefix = "arXiv",
    primaryClass = "hep-th",
    doi = "10.1007/JHEP07(2017)106",
    journal = "JHEP",
    volume = "07",
    pages = "106",
    year = "2017"
}

@article{Beccaria:2023zjw,
    author = "Beccaria, Matteo and Cabo-Bizet, Alejandro",
    title = "{On the brane expansion of the Schur index}",
    eprint = "2305.17730",
    archivePrefix = "arXiv",
    primaryClass = "hep-th",
    doi = "10.1007/JHEP08(2023)073",
    journal = "JHEP",
    volume = "08",
    pages = "073",
    year = "2023"
}

@article{Eleftheriou:2023jxr,
    author = "Eleftheriou, Giorgos and Murthy, Sameer and Rossell{\'o}, Mart{\'\i}",
    title = "{The giant graviton expansion in $AdS_5 \times S^5$}",
    eprint = "2312.14921",
    archivePrefix = "arXiv",
    primaryClass = "hep-th",
    doi = "10.21468/SciPostPhys.17.4.098",
    journal = "SciPost Phys.",
    volume = "17",
    number = "4",
    pages = "098",
    year = "2024"
}

@article{Deddo:2024liu,
    author = "Deddo, Evan and Liu, James T. and Pando Zayas, Leopoldo A. and Saskowski, Robert J.",
    title = "{Giant Graviton Expansion from Bubbling Geometry: Discreteness from Quantized Geometry}",
    eprint = "2402.19452",
    archivePrefix = "arXiv",
    primaryClass = "hep-th",
    reportNumber = "LCTP-24-04",
    doi = "10.1103/PhysRevLett.132.261501",
    journal = "Phys. Rev. Lett.",
    volume = "132",
    number = "26",
    pages = "261501",
    year = "2024"
}

@article{Lee:2024hef,
    author = "Lee, Ji Hoon and Stanford, Douglas",
    title = "{Bulk thimbles dual to trace relations}",
    eprint = "2412.20769",
    archivePrefix = "arXiv",
    primaryClass = "hep-th",
    month = "12",
    year = "2024"
}

@article{Liu:2022olj,
    author = "Liu, James T. and Rajappa, Neville Joshua",
    title = "{Finite N indices and the giant graviton expansion}",
    eprint = "2212.05408",
    archivePrefix = "arXiv",
    primaryClass = "hep-th",
    reportNumber = "LCTP-22-16",
    doi = "10.1007/JHEP04(2023)078",
    journal = "JHEP",
    volume = "04",
    pages = "078",
    year = "2023"
}

@article{Imamura:2022aua,
    author = "Imamura, Yosuke",
    title = "{Analytic continuation for giant gravitons}",
    eprint = "2205.14615",
    archivePrefix = "arXiv",
    primaryClass = "hep-th",
    reportNumber = "TIT/HEP-689",
    doi = "10.1093/ptep/ptac127",
    journal = "PTEP",
    volume = "2022",
    number = "10",
    pages = "103B02",
    year = "2022"
}

@article{Lee:2022vig,
    author = "Lee, Ji Hoon",
    title = "{Exact stringy microstates from gauge theories}",
    eprint = "2204.09286",
    archivePrefix = "arXiv",
    primaryClass = "hep-th",
    doi = "10.1007/JHEP11(2022)137",
    journal = "JHEP",
    volume = "11",
    pages = "137",
    year = "2022"
}

@article{Imamura:2021ytr,
    author = "Imamura, Yosuke",
    title = "{Finite-N superconformal index via the AdS/CFT correspondence}",
    eprint = "2108.12090",
    archivePrefix = "arXiv",
    primaryClass = "hep-th",
    reportNumber = "TIT/HEP-686",
    doi = "10.1093/ptep/ptab141",
    journal = "PTEP",
    volume = "2021",
    number = "12",
    pages = "123B05",
    year = "2021"
}

@article{Mamroud:2022msu,
    author = "Mamroud, Ohad",
    title = "{The SUSY index beyond the Cardy limit}",
    eprint = "2212.11925",
    archivePrefix = "arXiv",
    primaryClass = "hep-th",
    doi = "10.1007/JHEP01(2024)111",
    journal = "JHEP",
    volume = "01",
    pages = "111",
    year = "2024"
}

@article{Gutowski:2004yv,
    author = "Gutowski, Jan B. and Reall, Harvey S.",
    title = "{General supersymmetric AdS(5) black holes}",
    eprint = "hep-th/0401129",
    archivePrefix = "arXiv",
    reportNumber = "NSF-KITP-04-10",
    doi = "10.1088/1126-6708/2004/04/048",
    journal = "JHEP",
    volume = "04",
    pages = "048",
    year = "2004"
}

@article{Cvetic:2004ny,
    author = "Cvetic, Mirjam and Lu, H. and Pope, C. N.",
    title = "{Charged rotating black holes in five dimensional U(1)**3 gauged N=2 supergravity}",
    eprint = "hep-th/0407058",
    archivePrefix = "arXiv",
    reportNumber = "MIFP-04-13, UPR-1084-T, USTC-ICTS-04-16",
    doi = "10.1103/PhysRevD.70.081502",
    journal = "Phys. Rev. D",
    volume = "70",
    pages = "081502",
    year = "2004"
}

@article{Chong:2005hr,
    author = "Chong, Z. -W. and Cvetic, Mirjam and Lu, H. and Pope, C. N.",
    title = "{General non-extremal rotating black holes in minimal five-dimensional gauged supergravity}",
    eprint = "hep-th/0506029",
    archivePrefix = "arXiv",
    reportNumber = "MIFP-05-13, UPR-1125-T",
    doi = "10.1103/PhysRevLett.95.161301",
    journal = "Phys. Rev. Lett.",
    volume = "95",
    pages = "161301",
    year = "2005"
}

@article{Garoufalidis:2022wij,
	author = "Garoufalidis, Stavros and Wheeler, Campbell",
	title = "{Modular $q$-holonomic modules}",
	eprint = "2203.17029",
	archivePrefix = "arXiv",
	primaryClass = "math.GT",
	reportNumber = "MPIM-Bonn-2022",
	month = "3",
	year = "2022"
}

@article{Jejjala:2021hlt,
	author = "Jejjala, Vishnu and Lei, Yang and van Leuven, Sam and Li, Wei",
	title = "{SL(3, \ensuremath{\mathbb{Z}}) Modularity and New Cardy limits of the $ \mathcal{N} $ = 4 superconformal index}",
	eprint = "2104.07030",
	archivePrefix = "arXiv",
	primaryClass = "hep-th",
	doi = "10.1007/JHEP11(2021)047",
	journal = "JHEP",
	volume = "11",
	pages = "047",
	year = "2021"
}

@article{Garoufalidis:2021lcp,
    author = "Garoufalidis, Stavros and Zagier, Don",
    title = "{Knots, Perturbative Series and Quantum Modularity}",
    eprint = "2111.06645",
    archivePrefix = "arXiv",
    primaryClass = "math.GT",
    doi = "10.3842/SIGMA.2024.055",
    journal = "SIGMA",
    volume = "20",
    pages = "055",
    year = "2024"
}

@article{Garoufalidis:2023wez,
    author = "Garoufalidis, Stavros and Zagier, Don",
    title = "{Knots and Their Related $q$-Series}",
    eprint = "2304.09377",
    archivePrefix = "arXiv",
    primaryClass = "math.GT",
    doi = "10.3842/SIGMA.2023.082",
    journal = "SIGMA",
    volume = "19",
    pages = "082",
    year = "2023"
}

@article{Felder_2000,
	doi = {10.1006/aima.2000.1951},	
	url = {https://doi.org/10.1006%2Faima.2000.1951},	
	year = 2000,
	month = {dec},	
	publisher = {Elsevier {BV}},	
	volume = {156},
	number = {1},
	pages = {44--76},
	author = {Giovanni Felder and Alexander Varchenko},
	title = {The Elliptic Gamma Function and {SL}(3,~Z)$\ltimes$Z3},
	journal = {Advances in Mathematics}
}

@article{Rastelli:2016tbz,
	author = "Rastelli, Leonardo and Razamat, Shlomo S.",
	title = "{The supersymmetric index in four dimensions}",
	eprint = "1608.02965",
	archivePrefix = "arXiv",
	primaryClass = "hep-th",
	doi = "10.1088/1751-8121/aa76a6",
	journal = "J. Phys. A",
	volume = "50",
	number = "44",
	pages = "443013",
	year = "2017"
}

@article{Gadde:2020yah,
	author = "Gadde, Abhijit",
	title = "{Lectures on the Superconformal Index}",
	eprint = "2006.13630",
	archivePrefix = "arXiv",
	primaryClass = "hep-th",
	reportNumber = "TIFR/TH/20-20",
	doi = "10.1088/1751-8121/ac42ac",
	journal = "J. Phys. A",
	volume = "55",
	number = "6",
	pages = "063001",
	year = "2022"
}

@article{Spi203,
  author       = {V. P. Spiridonov},
  title        = {Theta hypergeometric integrals},
  journal      = {Algebra i Analiz},
  year         = {2003},
  volume       = {15},
  number       = {6},
  pages        = {161--215},
  url          = {http://mi.mathnet.ru/aa829},
    eprint={math/0303205},
      archivePrefix={arXiv},
      primaryClass={math.CA},
doi          = {10.1090/S1061-0022-04-00839-8}
}

@Inbook{Spiridonov2020,
author="Spiridonov, Vyacheslav P.",
editor="Gritsenko, Valery A.
and Spiridonov, Vyacheslav P.",
title="Introduction to the Theory of Elliptic Hypergeometric Integrals",
bookTitle="Partition Functions and Automorphic Forms",
year="2020",
publisher="Springer International Publishing",
address="Cham",
pages="271--318",
doi="10.1007/978-3-030-42400-8_6",
url="https://doi.org/10.1007/978-3-030-42400-8_6",
eprint = "1912.12971",
    archivePrefix = "arXiv",
    primaryClass = "math.CA",
}

@article{Cabo-Bizet:2018ehj,
    author = "Cabo-Bizet, Alejandro and Cassani, Davide and Martelli, Dario and Murthy, Sameer",
    title = "{Microscopic origin of the Bekenstein-Hawking entropy of supersymmetric AdS$_{5}$ black holes}",
    eprint = "1810.11442",
    archivePrefix = "arXiv",
    primaryClass = "hep-th",
    doi = "10.1007/JHEP10(2019)062",
    journal = "JHEP",
    volume = "10",
    pages = "062",
    year = "2019"
}

@Inbook{Spiridonov32002,
author="Spiridonov, V. P.",
editor="Malyshev, Vadim
and Vershik, Anatoly",
title="Theta Hypergeometric Series",
bookTitle="Asymptotic Combinatorics with Application to Mathematical Physics",
year="2002",
publisher="Springer Netherlands",
address="Dordrecht",
pages="307--327",
isbn="978-94-010-0575-3",
doi="10.1007/978-94-010-0575-3_15",
url="https://doi.org/10.1007/978-94-010-0575-3_15",
        eprint={math/0303204},
      archivePrefix={arXiv},
      primaryClass={math.CA},
}

@article{spiri2024,
       author = {{Spiridonov}, Vyacheslav P.},
        title = "{Elliptic hypergeometric functions: integrals versus series}",
      journal = {arXiv e-prints},
         year = 2024,
        month = dec,
          eid = {arXiv:2412.12673},
        pages = {arXiv:2412.12673},
          doi = {10.48550/arXiv.2412.12673},
archivePrefix = {arXiv},
       eprint = {2412.12673},
 primaryClass = {math.CA},
       adsurl = {https://ui.adsabs.harvard.edu/abs/2024arXiv241212673S}
}

@article{Closset:2017bse,
    author = "Closset, Cyril and Kim, Heeyeon and Willett, Brian",
    title = "{$ \mathcal{N} $ = 1 supersymmetric indices and the four-dimensional A-model}",
    eprint = "1707.05774",
    archivePrefix = "arXiv",
    primaryClass = "hep-th",
    reportNumber = "CERN-TH-2017-180",
    doi = "10.1007/JHEP08(2017)090",
    journal = "JHEP",
    volume = "08",
    pages = "090",
    year = "2017"
}

@article{Gadde:2011ik,
    author = "Gadde, Abhijit and Rastelli, Leonardo and Razamat, Shlomo S. and Yan, Wenbin",
    title = "{The 4d Superconformal Index from q-deformed 2d Yang-Mills}",
    eprint = "1104.3850",
    archivePrefix = "arXiv",
    primaryClass = "hep-th",
    reportNumber = "YITP-SB-11-13",
    doi = "10.1103/PhysRevLett.106.241602",
    journal = "Phys. Rev. Lett.",
    volume = "106",
    pages = "241602",
    year = "2011"
}

@article{Gaiotto:2021xce,
    author = "Gaiotto, Davide and Lee, Ji Hoon",
    title = "{The giant graviton expansion}",
    eprint = "2109.02545",
    archivePrefix = "arXiv",
    primaryClass = "hep-th",
    doi = "10.1007/JHEP08(2024)025",
    journal = "JHEP",
    volume = "08",
    pages = "025",
    year = "2024"
}

@article{Bourdier:2015sga,
    author = "Bourdier, Jun and Drukker, Nadav and Felix, Jan",
    title = "{The $\mathcal{N}=2$ Schur index from free fermions}",
    eprint = "1510.07041",
    archivePrefix = "arXiv",
    primaryClass = "hep-th",
    doi = "10.1007/JHEP01(2016)167",
    journal = "JHEP",
    volume = "01",
    pages = "167",
    year = "2016"
}

@article{Cordova:2015nma,
    author = "Cordova, Clay and Shao, Shu-Heng",
    title = "{Schur Indices, BPS Particles, and Argyres-Douglas Theories}",
    eprint = "1506.00265",
    archivePrefix = "arXiv",
    primaryClass = "hep-th",
    doi = "10.1007/JHEP01(2016)040",
    journal = "JHEP",
    volume = "01",
    pages = "040",
    year = "2016"
}

@article{Hanany:2010qu,
    author = "Hanany, Amihay and Mekareeya, Noppadol",
    title = "{Tri-vertices and SU(2)'s}",
    eprint = "1012.2119",
    archivePrefix = "arXiv",
    primaryClass = "hep-th",
    reportNumber = "IMPERIAL-TP-10-AH-07",
    doi = "10.1007/JHEP02(2011)069",
    journal = "JHEP",
    volume = "02",
    pages = "069",
    year = "2011"
}

@article{Benvenuti:2010pq,
    author = "Benvenuti, Sergio and Hanany, Amihay and Mekareeya, Noppadol",
    title = "{The Hilbert Series of the One Instanton Moduli Space}",
    eprint = "1005.3026",
    archivePrefix = "arXiv",
    primaryClass = "hep-th",
    doi = "10.1007/JHEP06(2010)100",
    journal = "JHEP",
    volume = "06",
    pages = "100",
    year = "2010"
}

@article{Yoshida:2014qwa,
	author = "Yoshida, Yutaka",
	title = "{Factorization of 4d N=1 superconformal index}",
	eprint = "1403.0891",
	archivePrefix = "arXiv",
	primaryClass = "hep-th",
	reportNumber = "KEK-TH-1708",
	month = "3",
	year = "2014"
}

@article{Benini:2013yva,
    author = "Benini, Francesco and Peelaers, Wolfger",
    title = "{Higgs branch localization in three dimensions}",
    eprint = "1312.6078",
    archivePrefix = "arXiv",
    primaryClass = "hep-th",
    reportNumber = "YITP-SB-13-46",
    doi = "10.1007/JHEP05(2014)030",
    journal = "JHEP",
    volume = "05",
    pages = "030",
    year = "2014"
}

@article{Peelaers:2014ima,
	author = "Peelaers, Wolfger",
	title = "{Higgs branch localization of $ \mathcal{N} $ = 1 theories on S$^{3}$ x  S$^{1}$}",
	eprint = "1403.2711",
	archivePrefix = "arXiv",
	primaryClass = "hep-th",
	reportNumber = "YITP-SB-14-06",
	doi = "10.1007/JHEP08(2014)060",
	journal = "JHEP",
	volume = "08",
	pages = "060",
	year = "2014"
}

@article{Nieri:2015yia,
	author = "Nieri, Fabrizio and Pasquetti, Sara",
	title = "{Factorisation and holomorphic blocks in 4d}",
	eprint = "1507.00261",
	archivePrefix = "arXiv",
	primaryClass = "hep-th",
	reportNumber = "DMUS--MP--15-09",
	doi = "10.1007/JHEP11(2015)155",
	journal = "JHEP",
	volume = "11",
	pages = "155",
	year = "2015"
}

@article{Beem:2013sza,
    author = "Beem, Christopher and Lemos, Madalena and Liendo, Pedro and Peelaers, Wolfger and Rastelli, Leonardo and van Rees, Balt C.",
    title = "{Infinite Chiral Symmetry in Four Dimensions}",
    eprint = "1312.5344",
    archivePrefix = "arXiv",
    primaryClass = "hep-th",
    reportNumber = "YITP-SB-13-45, CERN-PH-TH-2013-311, HU-EP-13-78",
    doi = "10.1007/s00220-014-2272-x",
    journal = "Commun. Math. Phys.",
    volume = "336",
    number = "3",
    pages = "1359--1433",
    year = "2015"
}

@article{Beem:2021zvt,
	author = "Beem, Christopher and Razamat, Shlomo S. and Singh, Palash",
	title = "{Schur indices of class S and quasimodular forms}",
	eprint = "2112.10715",
	archivePrefix = "arXiv",
	primaryClass = "hep-th",
	doi = "10.1103/PhysRevD.105.085009",
	journal = "Phys. Rev. D",
	volume = "105",
	number = "8",
	pages = "085009",
	year = "2022"
}

@article{Beem:2017ooy,
	author = "Beem, Christopher and Rastelli, Leonardo",
	title = "{Vertex operator algebras, Higgs branches, and modular differential equations}",
	eprint = "1707.07679",
	archivePrefix = "arXiv",
	primaryClass = "hep-th",
	reportNumber = "YITP-SB-17-27",
	doi = "10.1007/JHEP08(2018)114",
	journal = "JHEP",
	volume = "08",
	pages = "114",
	year = "2018"
}

@article{Pan:2021mrw,
	author = "Pan, Yiwen and Peelaers, Wolfger",
	title = "{Exact Schur index in closed form}",
	eprint = "2112.09705",
	archivePrefix = "arXiv",
	primaryClass = "hep-th",
	doi = "10.1103/PhysRevD.106.045017",
	journal = "Phys. Rev. D",
	volume = "106",
	number = "4",
	pages = "045017",
	year = "2022"
}

@article{Garoufalidis:2018qds,
	author = "Garoufalidis, Stavros and Zagier, Don",
	title = "{Asymptotics of Nahm sums at roots of unity}",
	eprint = "1812.07690",
	archivePrefix = "arXiv",
	primaryClass = "math.GT",
	doi = "10.1007/s11139-020-00266-x",
	journal = "Ramanujan J.",
	volume = "55",
	number = "1",
	pages = "219--238",
	year = "2021"
}

@article{Kinney:2005ej,
	author = "Kinney, Justin and Maldacena, Juan Martin and Minwalla, Shiraz and Raju, Suvrat",
	title = "{An Index for 4 dimensional super conformal theories}",
	eprint = "hep-th/0510251",
	archivePrefix = "arXiv",
	doi = "10.1007/s00220-007-0258-7",
	journal = "Commun. Math. Phys.",
	volume = "275",
	pages = "209--254",
	year = "2007"
}

@article{Choi:2018hmj,
	author = "Choi, Sunjin and Kim, Joonho and Kim, Seok and Nahmgoong, June",
	title = "{Large AdS black holes from QFT}",
	eprint = "1810.12067",
	archivePrefix = "arXiv",
	primaryClass = "hep-th",
	reportNumber = "SNUTP18-005, KIAS-P18097",
	month = "10",
	year = "2018"
}

@article{Spiridonov:2010hh,
    author = "Spiridonov, V. P. and Vartanov, G. S.",
    title = "{Supersymmetric dualities beyond the conformal window}",
    eprint = "1003.6109",
    archivePrefix = "arXiv",
    primaryClass = "hep-th",
    doi = "10.1103/PhysRevLett.105.061603",
    journal = "Phys. Rev. Lett.",
    volume = "105",
    pages = "061603",
    year = "2010"
}

@article{Aharony:2003sx,
    author = "Aharony, Ofer and Marsano, Joseph and Minwalla, Shiraz and Papadodimas, Kyriakos and Van Raamsdonk, Mark",
    editor = "Doebner, H. D. and Dobrev, V. K.",
    title = "{The Hagedorn - deconfinement phase transition in weakly coupled large N gauge theories}",
    eprint = "hep-th/0310285",
    archivePrefix = "arXiv",
    reportNumber = "WIS-29-03-DPP",
    doi = "10.4310/ATMP.2004.v8.n4.a1",
    journal = "Adv. Theor. Math. Phys.",
    volume = "8",
    pages = "603--696",
    year = "2004"
}

@article{bradley-thrush:2025,
   title={Factorization of Basic Hypergeometric Series},
   ISSN={1815-0659},
   url={http://dx.doi.org/10.3842/SIGMA.2025.051},
   DOI={10.3842/sigma.2025.051},
   journal={Symmetry, Integrability and Geometry: Methods and Applications},
   publisher={SIGMA (Symmetry, Integrability and Geometry: Methods and Application)},
   author={Bradley-Thrush, Jonathan G.},
   year={2025},
   month=jul ,
eprint = "2507.04313",
	archivePrefix = "arXiv",
	primaryClass = "math.CO",

}

@article{Spiridonov:2011hf,
    author = "Spiridonov, V. P. and Vartanov, G. S.",
    title = "{Elliptic hypergeometry of supersymmetric dualities II. Orthogonal groups, knots, and vortices}",
    eprint = "1107.5788",
    archivePrefix = "arXiv",
    primaryClass = "hep-th",
    reportNumber = "AEI-2011-049",
    doi = "10.1007/s00220-013-1861-4",
    journal = "Commun. Math. Phys.",
    volume = "325",
    pages = "421--486",
    year = "2014"
}

@article{Arai:2020qaj,
	author = "Arai, Reona and Fujiwara, Shota and Imamura, Yosuke and Mori, Tatsuya",
	title = "{Schur index of the ${\cal N}=4$ $U(N)$ supersymmetric Yang-Mills theory via the AdS/CFT correspondence}",
	eprint = "2001.11667",
	archivePrefix = "arXiv",
	primaryClass = "hep-th",
	reportNumber = "TIT/HEP-677",
	doi = "10.1103/PhysRevD.101.086017",
	journal = "Phys. Rev. D",
	volume = "101",
	number = "8",
	pages = "086017",
	year = "2020"
}

@article{Razamat:2012uv,
	author = "Razamat, Shlomo S.",
	title = "{On a modular property of N=2 superconformal theories in four dimensions}",
	eprint = "1208.5056",
	archivePrefix = "arXiv",
	primaryClass = "hep-th",
	doi = "10.1007/JHEP10(2012)191",
	journal = "JHEP",
	volume = "10",
	pages = "191",
	year = "2012"
}

@InProceedings{Frenkel-Turaev,
author="Frenkel, Igor B.
and Turaev, Vladimir G.",
editor="Arnold, V. I.
and Gelfand, I. M.
and Retakh, V. S.
and Smirnov, M.",
title="Elliptic solutions of the Yang-Baxter equation and modular hypergeometric functions",
booktitle="The Arnold-Gelfand Mathematical Seminars",
year="1997",
publisher="Birkh{\"a}user Boston",
address="Boston, MA",
pages="171--204",
isbn="978-1-4612-4122-5"
}

@article{Jejjala:2022lrm,
    author = "Jejjala, Vishnu and Lei, Yang and van Leuven, Sam and Li, Wei",
    title = "{Modular factorization of superconformal indices}",
    eprint = "2210.17551",
    archivePrefix = "arXiv",
    primaryClass = "hep-th",
    doi = "10.1007/JHEP10(2023)105",
    journal = "JHEP",
    volume = "10",
    pages = "105",
    year = "2023"
}

@article{Gadde:2011uv,
	author = "Gadde, Abhijit and Rastelli, Leonardo and Razamat, Shlomo S. and Yan, Wenbin",
	title = "{Gauge Theories and Macdonald Polynomials}",
	eprint = "1110.3740",
	archivePrefix = "arXiv",
	primaryClass = "hep-th",
	reportNumber = "YITP-SB-11-30",
	doi = "10.1007/s00220-012-1607-8",
	journal = "Commun. Math. Phys.",
	volume = "319",
	pages = "147--193",
	year = "2013"
}

@article{Hatsuda:2022xdv,
	author = "Hatsuda, Yasuyuki and Okazaki, Tadashi",
	title = "{$ \mathcal{N} $ = 2$^{*}$ Schur indices}",
	eprint = "2208.01426",
	archivePrefix = "arXiv",
	primaryClass = "hep-th",
	reportNumber = "RUP-22-17, KIAS-P22059",
	doi = "10.1007/JHEP01(2023)029",
	journal = "JHEP",
	volume = "01",
	pages = "029",
	year = "2023"
}

@inbook{Rastelli:2014jja,
	author = "Rastelli, Leonardo and Razamat, Shlomo S.",
	publisher = "Springer",
	title = "{The Superconformal Index of Theories of Class $\mathcal {S}$}",
	booktitle = "{New Dualities of Supersymmetric Gauge Theories}",
	eprint = "1412.7131",
	archivePrefix = "arXiv",
	primaryClass = "hep-th",
	doi = "10.1007/978-3-319-18769-39",
	pages = "261--305",
	year = "2016"
}

@book{Askey_Wilson_1985, 
	title={Some basic hypergeometric orthogonal polynomials that generalize Jacobi polynomials}, 
	publisher={Memoirs of the American Mathematical Society}, 
	author={Askey, Richard and Wilson, James}, 
	year={1985}, 
	doi={10.1090/memo/0319}
	}

@article{Hatsuda:2025mvj,
    author = "Hatsuda, Yasuyuki",
    title = "{Deformed Schur indices and Macdonald polynomials}",
    eprint = "2503.03952",
    archivePrefix = "arXiv",
    primaryClass = "hep-th",
    reportNumber = "RUP-25-6",
    month = "3",
    year = "2025"
}

@misc{Rains:2003teh,
      title={Transformations of elliptic hypergometric integrals}, 
      author={Eric M. Rains},
      year={2005},
      eprint={math/0309252},
      archivePrefix={arXiv},
      primaryClass={math.QA},
      url={https://arxiv.org/abs/math/0309252}, 
}

@article{Krotkov:2023inf,
      title={Infinite elliptic hypergeometric series: convergence and difference equations}, 
      author={D. I. Krotkov and V. P. Spiridonov},
      year={2023},
      eprint={2307.08002},
      archivePrefix={arXiv},
      primaryClass={math.CA},
      url={https://arxiv.org/abs/2307.08002}, 
}

@article{Spiridonov_2001,
doi = {10.1070/RM2001v056n01ABEH000374},
url = {https://doi.org/10.1070/RM2001v056n01ABEH000374},
year = {2001},
month = {feb},
publisher = {},
volume = {56},
number = {1},
pages = {185},
author = {V P Spiridonov},
title = {On the elliptic beta function},
journal = {Russian Mathematical Surveys}
}

@article{Chang:2025mqp,
    author = "Chang, Chi-Ming and Lin, Ying-Hsuan",
    title = "{Violation of S-duality in classical $Q$-cohomology}",
    eprint = "2510.24008",
    archivePrefix = "arXiv",
    primaryClass = "hep-th",
    month = "10",
    year = "2025"
}

@article{Razamat:2013qfa,
    author = "Razamat, Shlomo S.",
    title = "{On the $\mathcal{N} =$ 2 superconformal index and eigenfunctions of the elliptic RS model}",
    eprint = "1309.0278",
    archivePrefix = "arXiv",
    primaryClass = "hep-th",
    doi = "10.1007/s11005-014-0682-5",
    journal = "Lett. Math. Phys.",
    volume = "104",
    pages = "673--690",
    year = "2014"
}

@article{Festuccia:2011ws,
    author = "Festuccia, Guido and Seiberg, Nathan",
    title = "{Rigid Supersymmetric Theories in Curved Superspace}",
    eprint = "1105.0689",
    archivePrefix = "arXiv",
    primaryClass = "hep-th",
    doi = "10.1007/JHEP06(2011)114",
    journal = "JHEP",
    volume = "06",
    pages = "114",
    year = "2011"
}

@article{Deddo:2025jrg,
    author = "Deddo, Evan and Pando Zayas, Leopoldo A. and Zhou, Wenjie",
    title = "{The superconformal index and black hole instabilities}",
    eprint = "2502.01614",
    archivePrefix = "arXiv",
    primaryClass = "hep-th",
    doi = "10.1007/JHEP05(2025)170",
    journal = "JHEP",
    volume = "05",
    pages = "170",
    year = "2025"
}

@article{Srivastava_Jain_1986, title={q-Series Identities and Reducibility of Basic Double Hypergeometric Functions}, volume={38}, DOI={10.4153/CJM-1986-010-3}, number={1}, journal={Canadian Journal of Mathematics}, author={Srivastava, H. M. and Jain, V. K.}, year={1986}, pages={215–231}}

@article{Jackson1942,
    author = {JACKSON, F. H.},
    title = {ON BASIC DOUBLE HYPERGEOMETRIC FUNCTIONS},
    journal = {The Quarterly Journal of Mathematics},
    volume = {os-13},
    number = {1},
    pages = {69-82},
    year = {1942},
    month = {01},
    issn = {0033-5606},
    doi = {10.1093/qmath/os-13.1.69},
    url = {https://doi.org/10.1093/qmath/os-13.1.69},
    eprint = {https://academic.oup.com/qjmath/article-pdf/os-13/1/69/4364506/os-13-1-69.pdf},
}
